\documentclass[linenumbers,default]{aastex631}
\nolinenumbers

\usepackage{lipsum}
\usepackage{graphicx}

\begin{document}
\nolinenumbers

\title{27 years of Spaceborne IR Astronomy: An ISO, Spitzer, WISE and NEOWISE Survey for Large-Amplitude Variability in Young Stellar Objects}

\correspondingauthor{S. T. Megeath}
\email{s.megeath@utoledo.edu}
%%%%%%%%%%%%%%

\author[0009-0002-3026-0172]{Chinmay S. Kulkarni}
\affiliation{Ritter Astrophysical Research Center, Department of Physics and Astronomy, University of Toledo, Toledo, OH 43606, USA}

\author[0009-0003-7384-6170]{Thomas Behling}
\affiliation{Ritter Astrophysical Research Center, Department of Physics and Astronomy, University of Toledo, Toledo, OH 43606, USA}

\affiliation{Department of Physics and Astronomy, Michigan State University, East Lansing, MI 48824, USA}

\author[0009-0008-1090-2215]{Elisabeth E. Banks}
\affiliation{Ritter Astrophysical Research Center, Department of Physics and Astronomy, University of Toledo, Toledo, OH 43606, USA}

\author{Jason Jones}
\affiliation{Ritter Astrophysical Research Center, Department of Physics and Astronomy, University of Toledo, Toledo, OH 43606, USA}

\author[0009-0000-2384-5397]{Tyler Robbins}
\affiliation{Ritter Astrophysical Research Center, Department of Physics and Astronomy, University of Toledo, Toledo, OH 43606, USA}
\affiliation{Department of Aerospace, Physics and Space Sciences, Florida Institute of Technology, Melbourne, FL 32901, USA}

\author[0009-0000-2051-8120]{Nathanael Burns-Watson}
\affiliation{Ritter Astrophysical Research Center, Department of Physics and Astronomy, University of Toledo, Toledo, OH 43606, USA}

%\affiliation{Department of Astronomy, Case Western Reserve University, Sears Library, 2083 MLK Jr. Dr., Cleveland, OH 44106, USA}

\affiliation{Department of Astronomy, University of Texas at Austin, Physics Math and Astronomy Building, Austin TX, 78712, USA}

\author[0000-0001-7629-3573]{S. Thomas Megeath}
\affiliation{Ritter Astrophysical Research Center, Department of Physics and Astronomy, University of Toledo, Toledo, OH 43606, USA}

\author[0000-0002-6447-899X]{Robert Gutermuth}
\affiliation{Department of Astronomy, Lederle Graduate Research Tower, University of Massachusetts, Amherst, MA 01003, USA}

\author[0000-0002-6136-5578]{Samuel Federman}
\affiliation{Ritter Astrophysical Research Center, Department of Physics and Astronomy, University of Toledo, Toledo, OH 43606, USA}

\affiliation{INAF-Osservatorio Astronomico di Capodimonte Napoli, IT}

====
\author[0000-0002-1842-8131]{Savio B. Oliveira}
\affiliation{Ritter Astrophysical Research Center, Department of Physics and Astronomy, University of Toledo, Toledo, OH 43606, USA}

\author[0000-0001-9030-1832]{Wafa Zakri}
\affiliation{Physical Sciences Department, Jazan University, Jazan, Saudi Arabia}

\author[0000-0002-3747-2496]{William J. Fischer}
\affiliation{Space Telescope Science Institute, 3700 San Martin Drive, Baltimore, MD 21218, USA}

\author[0000-0002-0557-7349]{Riwaj Pokhrel}
\affiliation{Ritter Astrophysical Research Center, Department of Physics and Astronomy, University of Toledo, Toledo, OH 43606, USA}


%%ywo can be used either before or after the list of corresponding authors. The

\newcommand\totalSources{163}

\begin{abstract}
\nolinenumbers
Infrared observations can probe photometric variability across the full evolutionary range of young stellar objects (YSOs), from deeply embedded protostars to pre-main-sequence stars with dusty disks. We present 3-8~$\mu$m light curves extending 27 years from 1997 to 2024 obtained with  three space-based IR telescopes: ISO, Spitzer and WISE. Although unevenly sampled with large gaps in coverage, these light curves show variability on time scales ranging from days to decades. We focus on the Spitzer-identified YSOs with disks and envelopes that exhibit variations of a factor of two or more in this wavelength range. We identified seven YSOs where the light curves are dominated by bursts of sustained ($> 5$~yr) high flux, including four that show a steep decay ending the burst and three that are ongoing as of the final observation. We find six YSOs that are undergoing declines, which may be the end of bursts that began before 1997. The most common form of variability, exhibited by 26 YSOs in our sample, show variations over time intervals of years to months but do not exhibit sustained bursts or fades. The Spitzer [3.6]-[4.5] and WISE [3.5]-[4.6] colors either increase or remain constant with increasing brightness, inconsistent with dust extinction as being the primary source of the large-amplitude variability.  
\end{abstract}


%% Observe the use of the LaTeX \
%% and \citet commands to identify citationsy   .  The citations are

\section{Introduction} \label{sec:intro}

Young stellar objects (YSOs) exhibit variability across the electromagnetic spectrum, from X-rays to radio waves \citep[e.g.][]{2014AJ....147...83S,2014AJ....148...92R,2017ApJ...844..109F,2020A&A...638L...4G}. Observations from the Spitzer space telescope and Wide-field Infrared Survey Explorer (WISE) show that YSOs with dusty disks and envelopes commonly show variability in the 3-8~$\mu$m regime \citep[e.g.][]{2011ApJ...733...50M,2012AJ....144..192M,2013AJ....145...66F,2021ApJ...920..132P}. This variability is manifested through a rich range of behaviors, from small-amplitude periodic fluctuations to large-amplitude bursts \citep[e.g.][]{2012ApJ...756...99F,2015AJ....150..118P,2016ApJ...833..104F,2021ApJ...920..132P}. In contrast to visible or even near-infrared data, 3-8~$\mu$m observations can detect deeply embedded protostars, and thus can be used to observe variability over the full evolutionary sequence of YSOs from Class 0 protostars to evolved pre-main sequence stars with disks \citep[e.g.][]{2014AJ....148...92R,2015ApJ...800L...5S,2022ApJ...924L..23Z}.

Large-amplitude variability, which we define to occur when the flux changes by a factor $\ge 2$, may be of particular importance for the formation of stars. These variations can be driven by large changes in the mass accretion rate, where the luminosity released by accretion is reprocessed into infrared radiation \citep{2023ASPC..534..355F}. The discovery of luminous outbursts, or ''bursts'', from young stars that lasted decades, and the observation that the typical luminosities of protostars are lower than the expected accretion luminosity, led to a picture where most of a star's mass is assembled during sustained bursts of rapid accretion \citep{1996ARA&A..34..207H,2014prpl.conf..387A,2010ApJ...710..470D,2016ARA&A..54..135H,2019ApJ...872..183F,2023ASPC..534..355F,2024ApJ...962L..16N,2025MNRAS.541.4025H}. This picture is supported by models in which disks regulate the accretion of infalling gas onto protostars, leading to episodes of rapid accretion \citep{Vorobyov2006,2012ApJ...747...52D}. Despite the support for this picture, alternative models for the lower luminosities of protostars remain viable \citep{2011ApJ...736...53O,2017ApJ...840...69F}.  

%The luminosities of protostars suggest that their accretion rates are not high enough to assemble stars over the duration of the protostellar phase \citep{1990AJ.....99..869K}. This ``luminosity'' problem can be resolved if stars accrete most of their mass during brief episodes of high accretion \citep{2010ApJ...710..470D}, or it can be resolved if the steady mass accretion rates scale with the final masses of the protostars \citep{2011ApJ...736...53O,2017ApJ...840...69F}. Since the luminosity problem has multiple solutions, 

Determining the importance of episodic accretion thus requires direct measurements of time-varying accretion, and in particular the frequency, amplitudes, and durations of accretion bursts \citep{2019ApJ...872..183F,2022ApJ...924L..23Z}. Since most mass is accreted when protostars are deeply embedded in dense envelopes, measurements in the infrared or sub-millimeter are necessary \citep[e.g.][]{2015ApJ...800L...5S,2024ApJ...966..215M}. For these sources, bursts can be detected using IR and sub-millimeter light curves, and the durations and amplitudes of bursts constrained from the curves \citep{2019ApJ...872..183F,2021ApJ...920..132P,2024ApJ...966..215M}. Current estimates using $3-5$~$\mu$m multi-epoch photometry range from one burst per 1000 years for protostars to one burst per $> 10,000$ years for more evolved pre-main sequence stars with disks \citep{2013MNRAS.430.2910S,2019ApJ...872..183F,2022ApJ...924L..23Z}; however, comparisons of these studies are complicated by the different criteria and data used to identify bursts. Due to the small number of bursts found in surveys of well characterized populations of YSOs, and the uncertainties in the rates, amplitudes, and durations of the bursts, the contribution of bursts to the total  masses of stars remains poorly constrained \citep{2022ApJ...924L..23Z}.

Directly measuring infrared variability at $\lambda >3$~$\mu$m requires the sensitivity and photometric stability of space-based IR telescopes.
%Large-amplitude variability tends to occur on timescales of years to decades \citep[e.g.][]{2022ApJ...924L..23Z,2023ASPC..534..355F}. 
Due to the limited and inhomogeneous  $\lambda >3$~$\mu$m photometry of YSOs over these timescales, previous work has often made simplifying assumptions, for instance that most large-amplitude variations are sustained outbursts \citep{2019ApJ...872..183F}. To better characterize the light curves of large-amplitude variables, we construct 3-8~$\mu$m light curves that span 27 years, the longest possible baseline to measure the mid-IR variability of these objects. These observations start in 1997 with ISOCAM on the Infrared Space Observatory \citep[ISO,][]{1996A&A...315L..32C}. Launched in 1995, ISO was the first space-based IR telescope with the angular resolution needed to separate individual YSOs in nearby star-forming regions \citep{2003ESASP1262.....B,2004A&A...421..623K}. Starting in 2004, the IRAC camera on the Spitzer Space Telescope provided high sensitivity with 2'' angular resolution at 3-8~$\mu$m \citep{2004ApJS..154...10F}. Observations at 3-5~$\mu$m with IRAC continued after the depletion of liquid helium in 2009 until the end of the Spitzer mission in 2019 \citep{10.1117/12.857814}. The WISE mission launched in 2010, scanned the sky once in all four bands and then, after the cryogen was depleted, a second time in three of its original wavebands \citep{2010AJ....140.1868W}. In 2013, the satellite was revived for observations in the 3.5 and 4.6~$\mu$m bands as part of the Near-Earth Object Wide-Field Infrared Survey Explorer (NEOWISE) mission, which scanned the entire sky twice a year \citep{2014ApJ...792...30M}. The last NEOWISE epoch was in July 2024, after which the mission ended\footnote{\url{https://neowise.ipac.caltech.edu/news/neowise20240807/}}. 
%Although JWST has been used to identify protostars between bursts \citep[e.g.][]{2024ApJ...962L..16N}, the samples of YSOs in nearby clouds observed by this telescope are currently small in number.

The combined ISOCAM, Spitzer, and (NEO)WISE light curves give the longest time interval over which YSO variability can be measured at $> 3$~$\mu$m, although they suffer from large gaps in time coverage and the poor photometric accuracy of ISOCAM \citep{2024RNAAS...8...64K}. In this contribution, we publish the light curves of YSOs with full 27~year coverage. These YSOs were identified by Spitzer surveys of clouds and clusters within 500~pc, and all show infrared emission from dusty disks and envelopes \citep{2009ApJS..181..321E,2009ApJS..184...18G,2012AJ....144..192M,2012AJ....144...31K,2015ApJS..220...11D,2016ApJS..224....5F,2023ApJS..266...32P}. Given the limitations of the data, we focus on YSOs showing large-amplitude variations in the 3-4~$\mu$m wavelength range covered by the shorter bands of Spitzer/IRAC, WISE and NEOWISE, or the 4-8~$\mu$m range covered by the longer wavelength filters of (NEO)WISE and the IRAC and ISOCAM cameras. Despite these limitations, the long time coverage of these observations can give our best constraints on the durations of bursts.

%\textbf{In Sec.~\ref{sec:assemble}, we overview the data sets and methodology used to cross-calibrate the photometry from the different observatories. In Sec.~\ref{sec:lc}, we present light curves that exhibit $\ge 2 \times$ flux variations and have data from ISO, Spitzer, and (NEO)WISE. 
%We classify the light curves into four categories: completed bursts, ongoing bursts, fades, and fluctuators. We then present a brief description of the YSOs and light curves in each of these categories.
%We compare these classification schemes to those implemented by other authors (Park, Lee). 
%In Sec.~\ref{sec:discussion}, we discuss the implications for our understanding of YSO outbursts. 
%We use the light curves to assess the duration and properties of outbursts and, by extension, the mechanisms driving these variations. Finally, we discuss the importance of the fluctuators as they relate to YSO variability. }

\section{Assembling Light Curves} \label{sec:assemble}

\begin{figure*}
    \plotone{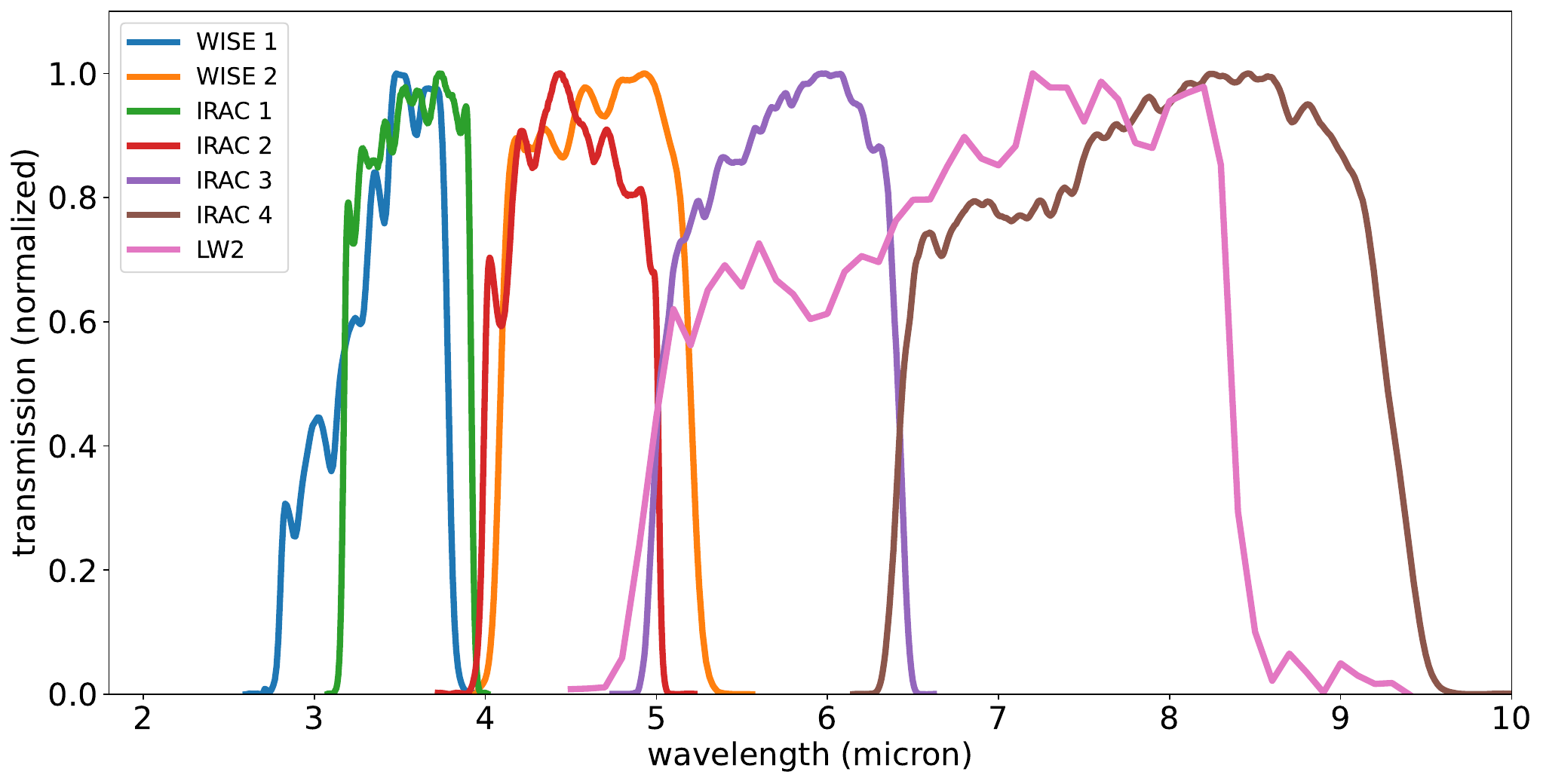} % [width=\textwidth]
    \caption{Transmission profiles of the six filters discussed in this work. While the WISE W1 and W2 filters significantly overlap with the IRAC I1 and I2 filters, the LW2 filter used for most of the ISO data overlaps with the I3 and I4 filters, with only minor overlap with the I2 filter. Thus, we use cryogenic Spitzer observations made in the I2, I3 and I4 filters to determine a typical offset of the LW2 filter to the I2 filter (\ref{sec:ISOCAM}). The bandpasses are sourced from \url{https://svo2.cab.inta-csic.es/theory/fps/} \citep{2012ivoa.rept.1015R}.}
    \label{fig:filters}
\end{figure*}

To generate light curves that extend up to 27 years, we assemble photometric data obtained from three space telescopes over five missions. The first epoch was obtained in 1997 with the Infrared Space Observatory (ISO), which was in operation from 1995 to 1998. The ISOCAM camera surveyed 15 star-forming regions in the LW2 (5-8.5~$\mu$m) and LW3 (12-18~$\mu$m) bands \citep{2000ESASP.445..201O,2004A&A...421..623K}. The second is the Spitzer Space Telescope, which was in operation between 2003 and 2020. The cryogenic mission of Spitzer lasted from 2003 to 2009. During this mission, many star-forming regions were surveyed using the 3.6 (IRAC~1, or I1), 4.5 (IRAC~2 or I2), 5.8 (IRAC~3 or I3), and 8~$\mu$m (IRAC~4 or I4) bands of the IRAC camera, and the 24~$\mu$m band of the MIPS instrument as part of the c2d survey, Orion survey, and GTO cluster survey \citep{2009ApJS..181..321E,2012AJ....144..192M,2009ApJS..184...18G}. Following the depletion of the cryogen in 2009, the warm mission of the telescope began, during which only the 3.6 and 4.5~$\mu$m bands of the IRAC camera were operated. This phase of the mission included several programs designed to re-observe star-forming regions surveyed in the cryo-mission, the most extensive of which was the YSOVAR exploration program \citep{2011ApJ...733...50M,2013AJ....145...66F,2014AJ....148...92R,2016ApJ...833..104F,2022ApJ...924L..23Z}. WISE surveyed the sky first in four bands, 3.4~$\mu$m (WISE~1 or W1), 4.6~$\mu$m (WISE~2 or W2), 12~$\mu$m (WISE~3 or W3) and 23~$\mu$m (WISE~4 or W4), and then a second time in three bands, W1, W2 and W3. The telescope began operating in 2010 and was put in hibernation in 2011 after the depletion of the cryogen. In 2013, the satellite was reactivated as part of the NEOWISE mission. In this warm phase of the mission, only the W1 and W2 bands were operable. The NEOWISE mission scanned the sky twice a year until it ended in 2024 \citep{2014ApJ...792...30M}. During each of these scans, each object is observed approximately ten times over intervals up to 4 days. In total, the three telescopes provide coverage in the 3-8~$\mu$m regime over 27 years, although not continuously. In Fig.~\ref{fig:filters}, we show the filter profiles for the bandpasses used in this paper.  Sources in this study satisfy the following criteria: i.) they are identified as a YSO with a disk or envelope, ii.)  their ISO photometry is not affected by source confusion or artifacts, iii.) they have Spitzer photometry from the cold and warm mission, and iv.) they have NEOWISE photometry extending to 2020 at minimum. 

\subsection{Spitzer YSO Catalog}
\label{sec:Spitzer_YSO}

We focus on YSOs identified by Spitzer using methods developed to identify pre-main sequence stars with disks and protostars \citep{2009ApJS..184...18G,2012AJ....144...31K,2012AJ....144..192M,2016AJ....151....5M}. For the Orion clouds, we use the YSO catalog generated from the cryogenic Spitzer Orion survey \citep{2012AJ....144..192M,2016AJ....151....5M}. For the remaining clouds, we use the SESNA reprocessing of the Spitzer data \citep{2020ApJ...896...60P}. Since Spitzer has a higher sensitivity and a finer angular resolution than ISOCAM or WISE, these data provide a complete sample of the dusty YSOs that can be detected in the ISOCAM and (NEO)WISE surveys. These catalogs also provide photometry in the IRAC I1, I2, I3, and I4 bands determined from either the mosaics of the combined images from the 2004 cryogenic epochs, or the median photometry values from all the 2004 cryogenic epochs \citep{2009ApJS..184...18G,2012AJ....144..192M}.  

\subsection{ISOCAM Photometry} \label{sec:ISOCAM}

To obtain photometry spanning 27 years, we only look at Spitzer-identified YSOs that are detected with ISOCAM. We extracted ISOCAM photometry from images downloaded from the ESO ISO Data Science Archive\footnote{https://nida.esac.esa.int/nida-cl-web/}. The images we used were all obtained in the LW2 (6.7~$\mu$m) band in either the $3''$ or $6''$ pixel scale. We used the fits files prefixed with "CSP". The images come from the ISOCAM survey of nearby star-forming regions \citep{2000ESASP.445..201O}. Here we chose the star-forming regions within 500~pc that had been observed over multiple epochs by Spitzer, specifically Orion~B, L1641N in Orion~A, Ophiuchus, the NGC~1333 cluster in the Perseus cloud, the Serpens Main cluster in the Aquilla cloud, Corona Australis, and parts of the Chameleon cloud. We did not include Taurus due to the limited multi-epoch Spitzer photometry for that region.   

To find sources in the ISOCAM images, we first used DAOStarFinder in the Python Photutils package\footnote{\url{https://photutils.readthedocs.io/en/stable/citation.html}}, which is adapted from DAOFIND \citep{1987PASP...99..191S}. We modified this routine to calculate a local background noise level in a $11\times 11$~pixel square centered on each source, based on the approach of the IDL routine PhotVis \citep{2009ApJS..184...18G}, and searched for sources that were $\ge 3-5 \sigma$ over the noise, where the exact limit was chosen for each individual image. We then inspected the images, first editing out sources that were visually identified artifacts, multiples resolved in the higher-resolution Spitzer data that are not resolved by ISOCAM, or sources too close to either the edges of the images or bright nebulosity. We then added point sources that were missed by DAOStarFinder and found by eye. 

We matched these sources to point source catalogs from the Spitzer~Orion program or the SESNA reprocessing of the Spitzer data \citep{2012AJ....144..192M,2016AJ....151....5M,2020ApJ...896...60P}. To do this, we first shifted the ISOCAM images to better match the Spitzer astrometry. After this refinement of the ISOCAM astrometry, we matched the ISOCAM sources to the Spitzer sources. There are numerous artifacts in the ISOCAM data, often induced by the presence of a bright source in a scan or noise near the edges of the images. Since the sensitivity and angular resolution of the IRAC camera are higher than those of ISOCAM, if there was no source in the Spitzer catalog within 15'' of a given ISO source, the ISO source was considered an artifact and deleted. Sources near the ISOCAM detection limit would have to decline by $\ge 4$ magnitudes to evade detection by Spitzer. Since we only consider sources identified as YSOs by Spitzer, such sources would not be in our YSO catalog. A total of 256 sources were found.

We improved the positions of the ISOCAM point sources by fitting 2D quadratic polynomials to the PSFs using the centroid\_quadratic function based in the Python PhotUtils package, with fit box and search box sizes of 5~pixels and 9~pixels, respectively. Centered on these positions, we performed aperture photometry with the Python PhotUtils package, adopting a circular aperture with a radius of 3~pixels. The background contribution was measured using the MMM statistic in an annulus extending from 4~pixels to 10~pixels. We converted the background-subtracted ISO fluxes to Vega magnitudes using the zero-point of 90.2 Jy/pixel \citep{2003ESASP1262.....B}. 
%In Orion~A and B, Ophiuchus, Serpens, NGC~1333 (Perseus), Chameleon and Corona Australis combined, we determined ISO fluxes for \totalSources\ YSOs. 

%[need comment on sizes in arcseconds]. . 
%This photometry is catalogued in <table x>.
%here

To cross-calibrate the ISOCAM data with Spitzer, we compared the ISOCAM LW2 photometry with that of the Spitzer I3 (5.8~$\mu$m) and I4 (8~$\mu$m) bands for all  cross matched sources in a given cloud. The LW2 filter covers the entire I3 filter and part of the I4 filter (Fig.~\ref{fig:filters}). For this reason, we examined the offsets between the ISOCAM LW2 photometry and the I3 photometry, between LW2 and the I4 photometry, and between LW2 and a combination of the I3 and I4 photometry. To calculate the combination of the I3 and I4 photometry, we used

\begin{equation}
I_{\rm comb} = -2.5 log_{10}\left(F_{0}(I3)  \cdot 10^{-m(I3)/2.5} \cdot \frac{c  \Delta \lambda_{I3} }{\lambda_{I3}^ 2} + F_{0}(I4)\cdot  10^{-m(I4)/2.5}  \cdot \frac{ c \Delta \lambda_{I4} }{\lambda_{I4}^2} \right) + ZP,
\end{equation}

\noindent where $F_{0} = 115$ and 64.13~Jy are the zero flux densities for the I3 and I4 bands, respectively \citep{2005PASP..117..978R}, $c$ is the speed of light, $\lambda_{I3}  = 5.731$ ~$\mu$m and $\lambda_{I3} = 7.872$~$\mu$m are the nominal wavelengths of the I3 and I4 bands, respectively \citep[as defined in][]{2005PASP..117..978R}, and $\Delta\lambda_{I3} = 1.42$~$\mu$m and $\Delta\lambda_{I4} =2.93$~$\mu$m are the filter widths for the two bands, respectively (IRAC instrument handbook\footnote{ https://irsa.ipac.caltech.edu/data/SPITZER/docs/irac/iracinstrumenthandbook/}). The zero point (ZP) is set so that $I_{\rm comb}$ equals zero when $m(I3)$ and $m(I4) = 0$, which gives $ZP = 38.451$~mag.

\begin{figure*}
\includegraphics[width=0.9\textwidth]{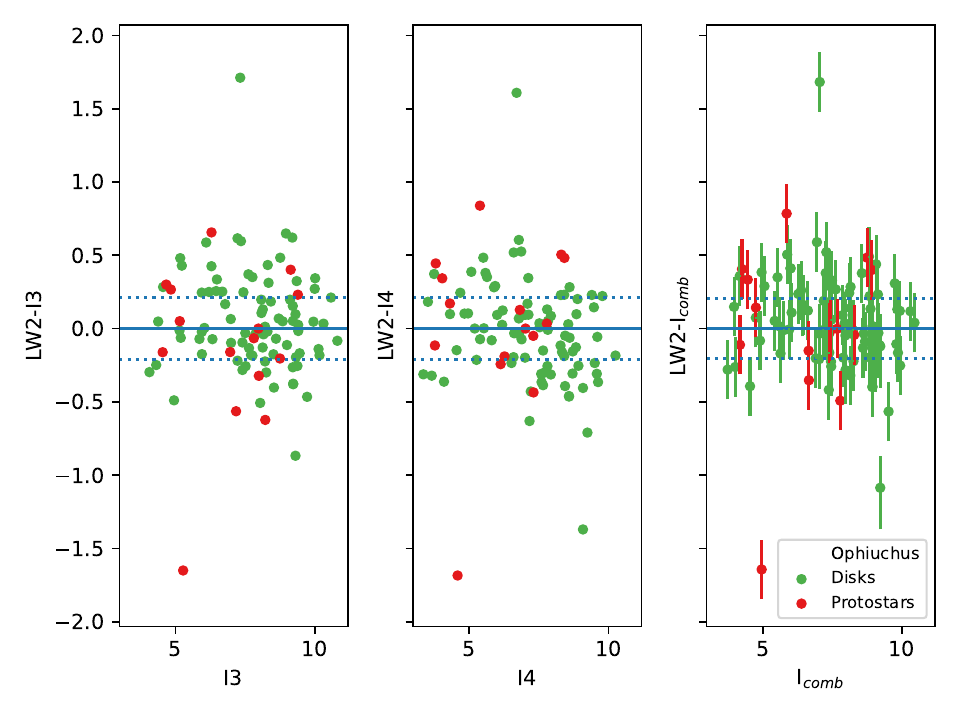}
\caption{ISOCAM - IRAC magnitude differences for each of the clouds. The red dots are YSOs classified as protostars and the green dots are those classified as pre-main sequence stars with disks. The median differences between the ISOCAM and IRAC data presented in Table~\ref{table:median} have been subtracted so that the points have a median value of 0, as shown by the solid blue line. The values of the median absolute deviations (MADs) are given by the dotted blue lines. Sources toward the top of the plots show large increases in magnitudes between the ISO and Spitzer epochs, while points near the bottom show large decreases. In the LW2-I$_{\rm comb}$ plot (3rd panel), the error bars extends between the LW2-I3 and the LW2-I4 points. If that value is less than the MAD value for the combined plot, we set the length of each of the error bars equal to the MAD. Note that some of the data points showing large differences are not in our sample of large-amplitude variables due to the lack of the necessary Spitzer and (NEO)WISE data.}
\label{fig:iso_oph}
\end{figure*}

In Fig.~\ref{fig:iso_oph}, the differences between the ISOCAM and IRAC magnitudes are plotted for the sources in each of the clouds after the subtraction of a median offset. The median offsets and the median aboslute deviation of the offets (MADs) for each field are given in Table~\ref{table:median}. Since the sources bright enough to detect in ISO are typically YSOs, they may have intrinsic variability. For this reason, we cannot separate noise and calibration issues from intrinsic source variability in the $\sim 25$\% scatter given by the MADs. We therefore consider this scatter as an upper limit to the uncertainties due to noise and calibration.

\begin{table*}[t]
\centering
    \caption{ISO vs Spitzer median offsets and MADs}
    \hskip -0.8 in
    \begin{tabular}{lrrrrrr}
 & \multicolumn{3}{c}{Medians} & \multicolumn{3}{c}{MADs} \\
 % & & Medians & & & MADs & \\
Region & 5.8 $\mu$m & 8.0 $\mu$m & Combined & 5.8 $\mu$m & 8.0 $\mu$m & Combined\\
\hline
Orion B & 0.053 & 0.854 & 0.444 & 0.314 & 0.232 & 0.256\\
Orion A & 0.396 & 1.049 & 0.707 & 0.197 & 0.226 & 0.198\\
Ophiuchus & 0.311 & 1.034 & 0.618 & 0.211 & 0.214 & 0.202\\
Perseus & -0.031 & 0.810 & 0.361 & 0.278 & 0.235 & 0.253\\
Serpens & 0.191 & 0.901 & 0.517 & 0.248 & 0.248 & 0.237\\
CrA & 0.059 & 0.699 & 0.376 & 0.241 & 0.186 & 0.199\\
Chameleon & 0.131 & 0.791 & 0.411 & 0.172 & 0.235 & 0.166\\
\end{tabular}

    \label{table:median}
\end{table*}

%In addition, we use the range between the LW2-I3 difference and LW2-I4 difference as the full uncertainty, LW2-comb valuebeing the nominal difference. The large scatter in the differences remaining between the IRAC and ISOCAM data is a primary motivation for focusing on larger-amplitude variatoions that are much greater than the level of scatter.  

\subsection{Spitzer Photometry} \label{sec:Spitzer}

%he light curve is assembled from photometry with filters having central wavelengths between 4 and 8~$\mu$m. 
%The Spitzer photometry \citep{sha2020} includes 
The Spitzer photometry was extracted from archival data downloaded from the Spitzer Heritage Archive. The extraction method is described in \citet{2022ApJ...924L..23Z}. We obtain Spitzer photometry from the observations of the cryogenic survey of Orion (PID 43: Mar and Oct 2004), the warm-mission Orion: The Final Epoch program (PID 13010: Jan and Jun 2017), and observations taken during the Spitzer Final Voyage mission (PID 14301: Jul 2019). We use the cryogenic data taken in 2004 from the c2d survey for Ophiuchus (PID~177), NGC~1333 (PID 178), Serpens (PID~174), and Chameleon (PID 176), and from the Spitzer GTO Cluster Survey (PID~4) of Ophiuchus, NGC~1333, and Serpens \citep{2009ApJS..181..321E,2009ApJS..184...18G}. In addition, we have cryogenic photometry of Corona Australis from the Spitzer Gould Belt program (PID~30574 May 2008). We use warm-mission YSOVAR data for Ophiuchus (PID~61024), NGC~1333 (PID~61026), and Serpens (PID~61029); see \citet{2014AJ....148...92R} for the time sampling of these data. We also use high cadence data from Flaherty on Chameleon (PID~90141 and PID~12090).  Finally, we use data from a warm-mission proper motion survey by Kraus (PID~90071) that observed Chameleon, Corona Australis, Ophiuchus, and Perseus in 2013/2014.

%we use IRAC sub-array data at 4.5~$\mu$m from programs PID 224, 228 and 3716 (covering four epochs in 2004 and 2005), and the 0.6 second full array data obtained in high dynamic range mode by PID 43 (Mar and Oct 2004), PID 60167 (Oct and Nov in 2009), PID 13010 (Jan and Jun 2017) and PID 14301 \citep[Jul 2019][]{sha2020}. We present the median magnitude taken within a five hour interval using the approach of \citet{2022ApJ...924L..23Z}. 

\subsection{WISE/NEOWISE Photometry} \label{sec:WISE}

The (NEO)WISE data included the analysis are collected from two archives; the NEOWISE Reactivation Releases available on the Infrared Science Archive (IRSA) and the unTimely catalog \citep{2023AJ....165...36M}.Although they originate from the same data, we use photometry from both catalogs as separate data points in our analyses. The NEOWISE Reactivation Releases, available from the InfraRed Science Archive \citep[IRSA,][] {2015nwis.rept....1C}, contains profile-fit, single-exposure photometry released on a yearly basis starting in 2013 and continuing to the deactivation of the mission in 2024. These data include passes across each source every six months as the satellite scanned the sky, with each pass containing multiple exposures. Since our focus is large-amplitude variability on longer timescales and since the individual single exposure frames suffer from a lower sensitivity and redundancy that can make them sensitive to artifacts, we combine the multiple exposures. To obtain the data from IRSA, we used the Python package PyVO to access the IVOA API available from the IRSA website. Initially, a simple-cone-search (SCS) operation was performed to retrieve any photometric detections within 5'' of the given target position. We then imposed the restriction that the detection must be $1''$ from the source position in the Spitzer SESNA catalog. These detections were filtered using the criteria shown in Table~\ref{tab:IRSA_filtering}. The detections were then  divided into passes spaced by approximately 6 months. Within a given pass, we use all detections within $1''$ from the Spitzer location given in the SESNA catalogs. After removing detections containing saturated pixels, a median magnitude and standard deviation were determined from the remaining photometry points and used for that specific epoch. The small standard deviations show no indication of substantial variability over the few day intervals of each NEOWISE scan.

The unTimely catalog \citet{2023AJ....165...36M} was generated with an alternative reduction of both the WISE and NEOWISE data, and includes original WISE mission data along with the reactivation releases up to 2020. The unTimely archive consists of a set of distinct catalogs each covering a different region of the sky \citep{2023AJ....165...36M}. Using the catalog index for unTimely, any catalog centered within 2 degrees of a Spitzer YSO detected by ISOCAM were downloaded. The detections in these catalogs were sorted by distance from the SESNA position of the source, and we used the closest detection within $1''$ of the SESNA position. Only the detections within $1''$ of the SESNA position in any given epoch are included.  For each six month pass, a single epoch of photometry is tabulated. Any detection with an UNWISE Saturation flag of True are marked as saturated.

\begin{table*}[]
    \caption{Filtering Criteria for (NEO)WISE data from IRSA.}
    \centering
    \begin{tabular}{l l l}
    \hline \hline
    Filtering Property & Filtering Variable & Value Constraint \\
    \hline
    Reduced Chi-Squared & w1rchi2, w2rchi2 & Both less than 2.5 \\
    
    Signal-to-Noise Ratio & w1snr, w2snr & Both greater than 2.0 \\

    Frame Quality & qual\_frame & Not equal to 0 \\
   
    Image Quality & qi\_fact & Not equal to 0 \\
   
    SAA Separation & saa\_sep & Greater than 5 \\
    
    Image Saturation & w1sat, w2sat & Both equal to 0 \\
    \hline
    \end{tabular}
    \label{tab:IRSA_filtering}
\end{table*}

Since many of the YSOs are bright, saturation must be considered in the data. As a source is scanned by (NEO)WISE, the sources may be saturated or unsaturated in the individual data frames depending on how the point source is divided by the under-sampled pixels. For the IRSA reactivation data, we removed any detections which were flagged as containing saturated pixels. If there are no detections without unsaturated data, we do not show that data point in the light curves. For unTimely, we included sources that were flagged as having saturated pixels in any of the passes, but marked those points as potentially being affected by saturation. For these sources, some of the data used in the PSF fits used to determine the unTimely magnitudes may be saturated. 

\begin{figure*}
    \centering
    \epsscale{1}
    \plotone{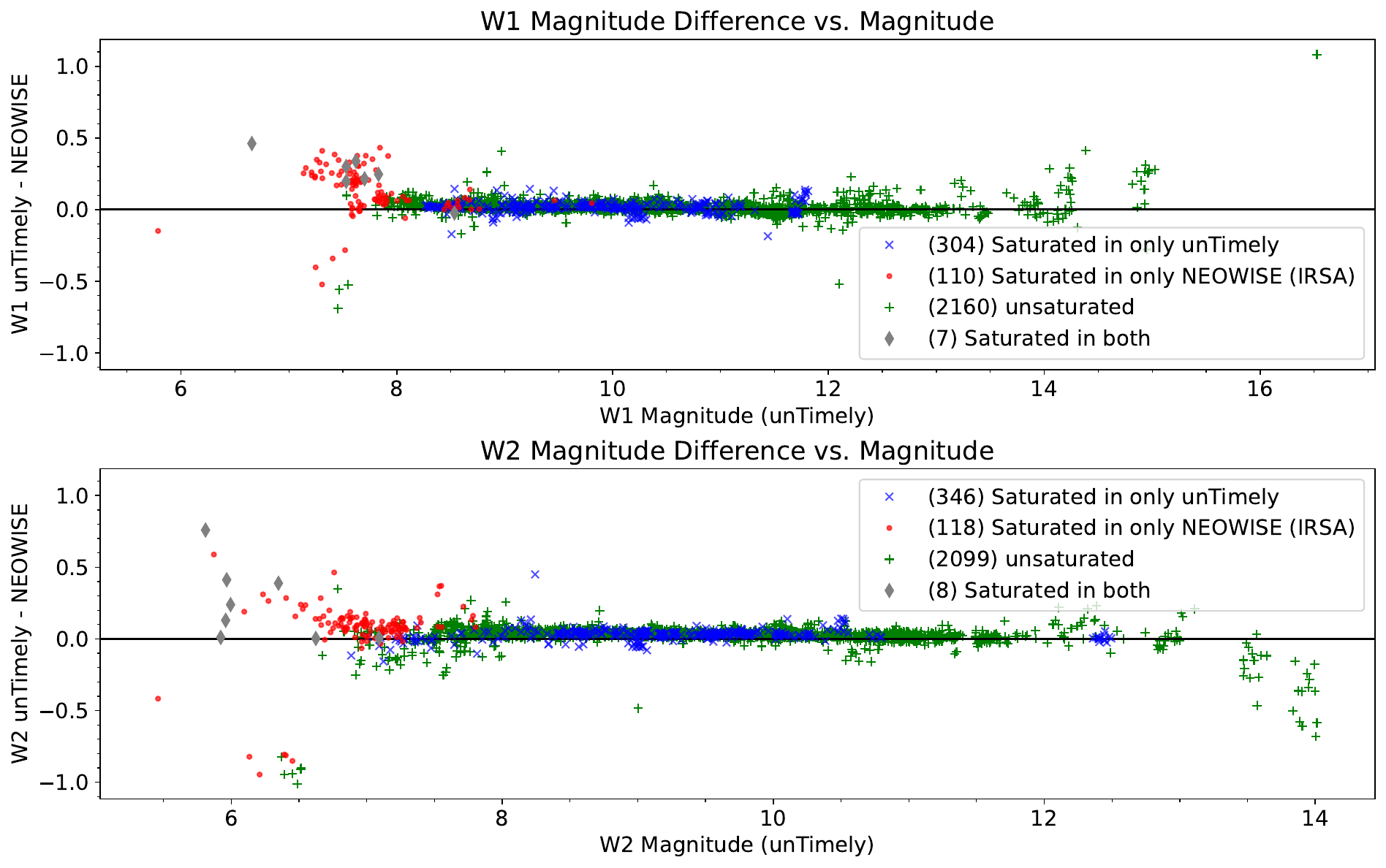}
    \caption{Difference between the NEOWISE photometry found in the IRSA and unTimely archives, with each point representing a matched set of unTimely and IRSA photometry for a specific source and epoch. These data are collected from all of the sources examined for variability in this work.}
    \label{fig:uT_NW_diff}
\end{figure*}

Figure \ref{fig:uT_NW_diff} compares the photometry returned from the two archives for all sources which also have ISOCAM and Spitzer detections. A single source is represented by a cluster of points consisting of the different epochs of that source. A small offset between the IRSA and unTimely magnitudes is apparent, with the IRSA magnitudes being fainter by a median value of 0.017~mag for W1 and 0.033~mag for W2. For the sources fainter than 14th magnitude in W1 and 13th magnitude in W2, larger deviations between the IRSA and unTimely magnitudes are apparent.

To evaluate the effect of saturation on Fig.~\ref{fig:uT_NW_diff}, we examined each pass in the IRSA NEOWISE data for saturation. Within a single pass, frames may be saturated if the source is centered on a single pixel and unsaturated if the peak is divided between four pixels, for example. For each pass, we split the frames into samples with and without saturated pixels. We then produced two different photometry points by taking the median of each sample and plotted these points separately. We also marked the unTimely sources that are flagged as having saturated pixels. We find that in most cases, photometry from saturated IRSA data shows slightly brighter magnitudes than unTimely photometry of the same sources, including those cases where the unTimely data are not flagged as saturated. We also find that there are a number of sources that are flagged as saturated in unTimely, but do not have saturated pixels in the IRSA data and have modest magnitudes.
 
In total, this analysis shows that sources brighter than 8th magnitude in the W1 band, and 7th~mag in the W2 band, must be viewed with caution. To assess the effects of saturation on the light curves in Sec.~\ref{sec:lc}, we mark unTimely (and Spitzer) data points that are potentially affected by saturation, compare the IRSA and unTimely data when possible, check that the W1 and W2 data show the same trends, and report the magnitudes of the sources in the legends of the light curves.

\begin{figure*}
    \centering
    \epsscale{1}
    \plotone{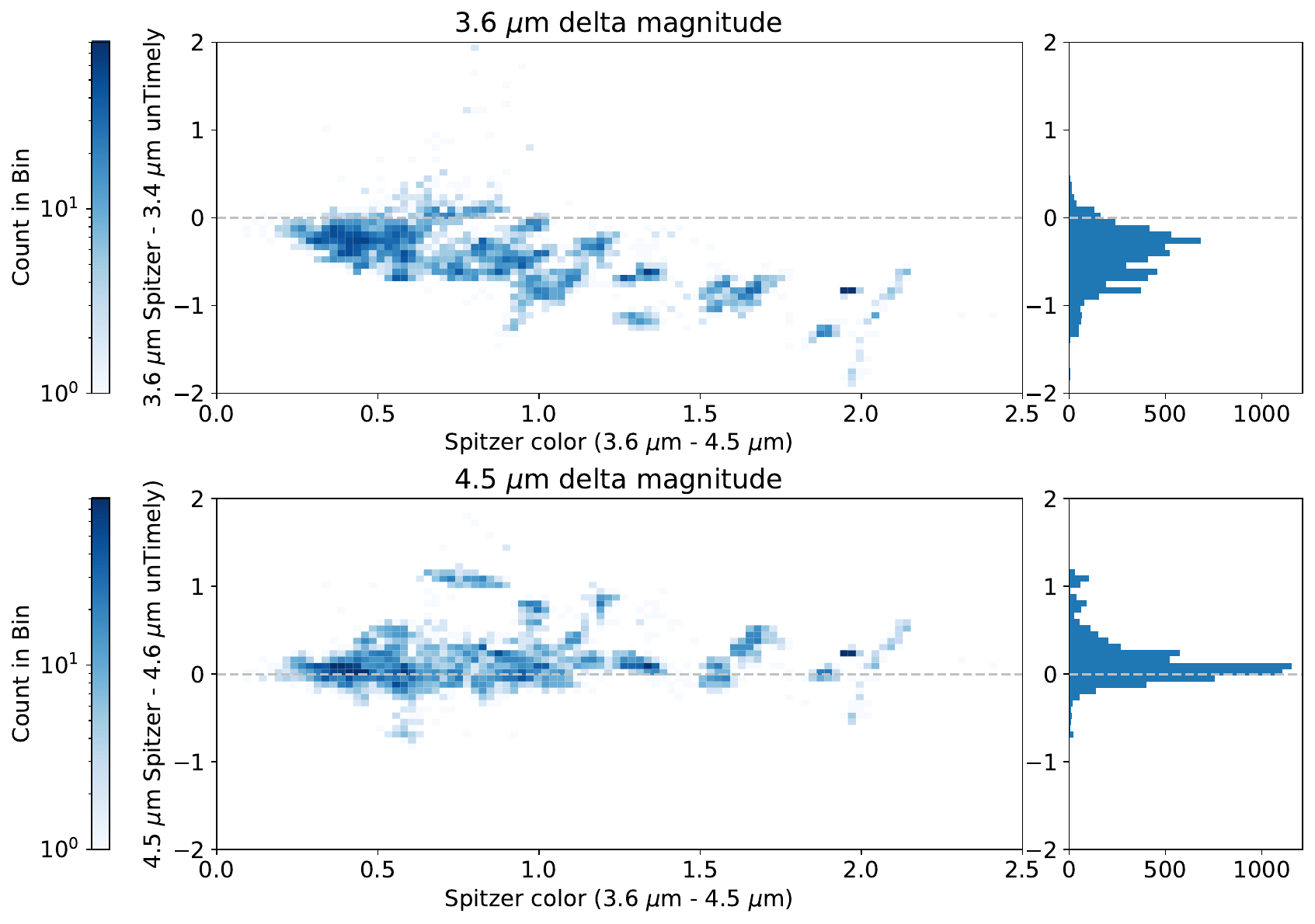}
    \caption{Offsets between concurrent Spitzer and unTimely photometry due to differences between the W1 and I1 bandpasses and between the W2 and I2 bandpasses as a function of the Spitzer 3.6~$\mu$m - 4.5~$\mu$m color. In total, 256 sources with ISOCAM photometry are used. Each epoch is considered independently, and distinct clusters of points are typically from a single source. In particular, the W1 and I1 magnitudes diverge for sources with large I1-I2 colors. In addition, there is significant scatter; this may be due to both the range of spectral shapes of the sources and variability between the WISE and Spitzer epochs. }
    \label{fig:Sp_uT_diff}
\end{figure*}

\subsection{Composite Light Curves}

We assemble, for each source, two composite light curves from the photometry described above. The first composite light curve is assembled from the Spitzer I1 and WISE W1 bands. Since we reference everything to the Spitzer I1 band, we refer to this as the 3.6~$\mu$m light curve. The second composite light curve is constructed from the I2, W2, and ISOCAM LW2 bands; this is referenced to the I2 band and is referred to as the 4.5~$\mu$m light curve. 

In this approach, the Spitzer I1 and I2 photometry are the ultimate reference points, and hence we cross-calibrate and correct the ISOCAM and WISE photometry so that it better agrees with the IRAC photometry. For the Spitzer photometry, we simply set the magnitudes of the light curve to the photometry at each observed epoch:

\begin{equation}
    m_{3.6}(t) = I1(t),~m_{4.5}(t) = I2(t)
\end{equation}

Offsets are found between the I1 and W1 photometry and, to a lesser extent, the I2 and W2 photometry \citep[Fig.~\ref{fig:Sp_uT_diff}, also see][]{2014ApJ...782...51A}. These are due to the different shapes of the filters (Fig.~\ref{fig:filters}), especially the extended short wavelength wing of W1, and the different SEDs of the YSOs in this wavelength regime, as characterized in these plots by the Spitzer color. Given the scatter in these plots, we account for the differences between the Spitzer and WISE filters for each light curve individually by offsetting the W1 and W2 photometry to better align with the Spitzer photometry. The offset is determined from epochs of Spitzer photometry that are bracketed between two (NEO)WISE epochs. Using the (NEO)WISE photometry closest in time to the Spitzer point, either the unTimely and IRSA data, as end points of a linear interpolation, we determine a W1 and W2 magnitude at the Spitzer epoch $t$, $W1_{int}(t)$ and $W2_{int}(t)$. The offset for each light curve is then determined by averaging the offsets between the Spitzer and interpolated NEOWISE photometry in that light curve,

\begin{equation}
\delta W1 = \frac{1}{n_{sp}} \sum_{i=1}^{n_{sp}} \left[I1(t_i)-W1_{int}(t_i)\right],~\delta W2 = \frac{1}{n_{sp}} \sum_{i=1}^{n_{sp}} \left[ I2(t_i)-W2_{int}(t_i)\right],
\end{equation}

\noindent
where $n_{sp}$ is the number of Spitzer epochs bracketed by NEOWISE epochs, and $I1(t_i)$ and $I2(t_i)$ are the Spitzer photometry at those points. The (NEO)WISE photometry in that light curve are then determined using the offset:  

\begin{equation}
m_{3.6}(t) = W1(t)+\delta W1,~m_{4.5}(t) = W2(t)+\delta W2.
\end{equation}

\noindent
These offsets are given for each light curve individually in the legends of the light curves  (Figs.~\ref{fig:lc_comp_bursts} - ~\ref{fig:lc_fluctuators}). We apply the same offset to the IRSA and unTimely photometry and plot both separately. 

%This is done for all the Spitzer points taken during the NEOWISE mission, and the  median offset is then subtracted from all the W1 data for that source.   For the 4.5~$\mu$m light curve, we  offset the W2 data using the identical procedure as was done for W1 in the 3.6~$\mu$m light curve. 

Each of the 4.5~$\mu$m curves includes a data point from the ISOCAM LW2 photometry. Here we use the offsets between the ISOCAM photometry and the I3, I4 and I$_{\rm comb}$ Spitzer magnitudes determined in Sec.~\ref{sec:ISOCAM}. For the light curves, we use the middle of the three offset values, typically the offset from I$_{\rm comb}$, to mark the data point and the maximum and minimum offsets to set the error bars. If the offsets are less than the MAD values in Table~\ref{table:median}, we use the MAD values for the error bars instead. Since the I3 and I4 photometry used to cross-calibrate the ISO photometry are from the combined 2004 data in the SESNA or Spitzer Orion catalogs, we calculate the ISOCAM magnitude using 

\begin{equation}
  m_{4.5}(t_{ISO}) = \left[LW2-I_{\rm spitzer}\right] + I2_{\rm 2004},
\end{equation}

\noindent
where $\left[LW2-I_{\rm spitzer}\right]$ is the offset between the ISO and Spitzer 2004 photometry obtained in Sec.~\ref{sec:ISOCAM} and $I2_{\rm 2004}$ is the 4.5~$\mu$m magnitude from the SESNA or Spitzer Orion catalogs (Sec.~\ref{sec:Spitzer_YSO}). Following the convention of \citet{2022ApJ...924L..23Z}, we then present the light curves in terms of delta magnitude,

\begin{equation}
    \Delta m_{3.6}(t) = m_{3.6}(t) - \min(m_{3.6}),~\Delta m_{4.5}(t) = m_{4.5}(t) - \min(m_{4.5}),
\end{equation} 

\noindent
where $min(I1)$ and $min(I2)$ are the minimum magnitudes of the 3.6~$\mu$m and 4.5~$\mu$m light curves, respectively. The values of these offsets are given in the legends of each light curves (Figs.~\ref{fig:lc_comp_bursts} - ~\ref{fig:lc_fluctuators}).

%We now have two catalogs with magnitudes; one from the 1997 ISOCAM data that we developed, and the second from the Spitzer catalog provided by Megeath et al 2012. We now compare the magnitudes between these two. Figure \ref{fig:Orion} shows three plots comparing the ISO 6.7 $\mu$m to the Spitzer 5.8 $\mu$m, 8.0 $\mu$m, and a combined magnitude, in Orion.

% todo -> recategorize
% Burst-Fade
% Burst-Sustained -> (rapid or slow)
% Fades
% Fluctuators (weird stuff)

% I moved the figures to separate .tdetectionsex files under the LC_figure_texts folder.
%   (mostly to clean up the main file)

%How many ISOCAM sources good sources + how additonal doubles: 
%    X348 Sources (All listed sources, no constraints) 
%These sources have data from everything
%X221 Sources with data from every source (could be saturated or unsaturated)
%**256 Good ISO Sources (ISO Quality is marked "Good") + 2 sources
%**163 Constrained Sources (Unsaturated WISE data on either side of a Spitzer Point)
%XX64 Constrained Sources w/ Large Range (Same constraint as above, but with a >0.75 magnitude range in W1 or W2) 
%**42 Final Sources (Placed in varying repository, marked good, and should be in the paper.)
%3 comp bursts + 4 burst + 3 app bursts + 6 fades + 26 fluctuators = 42 

%We have found 8 fluctuators with YSOVAR, 18 fluctuators without YSOVAR. 
%Total sample has 11 sources with YSOVAR, 31 sources without YSOVAR. About 30 percent of identified fluctuators have YSOVAR data, and about 26 percent of the total sample has YSOVAR data.

\section{Results} \label{sec:lc}

In total, 221 sources have data from ISOCAM, Spitzer and NEOWISE spanning from 1997 to at least 2020. In Table~\ref{tab:var_src}, we list the 39 YSOs of these that show variations of at least a factor of two in flux, or 0.75~mag in either the 3.6 or 4.6~$\mu$m light curves, as measured between their minimum and maximum $\Delta m$ values. This threshold is applied to focus on the multi-decade, large-amplitude variations that can be studied with this combined data set. In Table~\ref{tab:var_src}, we tabulate the source name, RA and Dec., the star-forming region in which the YSO is located, the 2004 3.6~$\mu$m and 4.5~$\mu$m magnitudes from Spitzer, and the SED class. This class was determined from the Spitzer photometry \citep{2009ApJS..184...18G, 2012AJ....144...31K, 2012AJ....144..192M} or by the combined 2MASS+Spitzer+Herschel+sub-mm photometry in the case of protostars \citep{2016ApJS..224....5F,2023ApJS..266...32P}. We also note  sources with high cadence Spitzer data from YSOVAR or other programs and include identifiers used in the \citet{2021ApJ...920..132P} paper and other commonly used names. Thirteen of these have NEOWISE photometry out to the final unTimely data point in 2020; the remainder have IRSA photometry that extends to 2024. In Table~\ref{tab:var_src} we also include an additional protostar located in the Orion~B cloud, J05473657+0020062, which has ISOCAM and Spitzer data, but no (NEO)WISE data. We exclude this source from most of our analysis.

For each source, the table also includes the maximum $\Delta m$ values of the 3.6 and 4.5~$\mu$m light curves and a classification based on the light curves. We classify the YSOs into four provisional categories: completed bursts, ongoing bursts, fades, and fluctuators, which we define in the following sections. Many of the light curves show combinations of traits that can defy simple classification, and we will discuss the relationships between these categories in Sec.~\ref{sec:discussion}.

\begin{table*}[t]
\caption{Table of Large-Amplitude Variables}
\hskip -0.7 in
\resizebox{1.05\textwidth}{!}{ 
    \begin{tabular}{llllrrrrllll}
    \tableline 
   \tableline
     & RA & Dec  &  & 3.6 $\mu$m\tablenotemark{1} & 4.5 $\mu$m\tablenotemark{1}  & 3.6  & 4.5  & Var.& SED  & YSO & Alt.\\
    Name & (deg) & (deg) & Region & Mag. & Mag. &  Range &  Range & Class & Class &  VAR & Name\tablenotemark{2}\\
    \hline
%    J03285841+3122176 & 52.2434 & 31.3716 & N1333 & 9.45 & 8.00 & 2.38 & 1.74 & Fluc. & Proto. & Y & D2415, eHOPS-per-25, GMM2008 25\\ \\ Flagged as "Multiple"
    J03283708+3113309 & 52.1545 & 31.2252 & N1333 & 9.85 & 7.96 & 1.51 & 1.88 & Fade & Proto. & N & D2375, eHOPS-per-15, EES2009 Per-emb 35\\
    J03290473+3111349 & 52.2697 & 31.1930 & N1333 & 11.91 & 10.99 & 0.50 & 0.76 & Fade & Disk & Y & D2433, GMM2008 85\\
    J03285955+3121467 & 52.2481 & 31.3630 & N1333 & 8.88 & 8.43 & 1.30 & 1.13 & Fluc. & Disk & Y & D2418, EM* LkHA 353\\
    J03283452+3107055 & 52.1438 & 31.1182 & N1333 & 12.04 & 10.51 & 1.38 & 1.09 & Fluc. & Proto. & N & D2372, eHOPS-per-13, GMM2008 14\\
    J03285102+3118184 & 52.2126 & 31.3051 & N1333 & 8.39 & 7.89 & 0.99 & 1.19 & Fluc. & Disk & Y & D2390, GMM2008 50\\
    J03292042+3118343 & 52.3351 & 31.3095 & N1333 &  8.54 & 7.99 & 0.89 & 0.89 & Fluc. & Disk & Y & D2473, HH 17\\
    J03284529+3105419 & 52.1887 & 31.0950 & N1333 & 13.19 & 11.95 & 0.77 & 0.73 & Fluc. & Proto. & N & D2384, eHOPS-per-20, IRAS 03256+3055\\
    J03285120+3119549 & 52.2133 & 31.3319 & N1333 & 9.06 & 8.71 & 0.64 & 0.82 & Fluc. & Disk & Y & D2392, GMM2008 52\\
    J03285216+3122453 & 52.2174 & 31.3793 & N1333 & 10.16 & 9.87 & 0.64 & 0.92 & Fluc. & Disk & Y & D2396, GMM2008 54\\
    \hline
    J05362461-0622413 & 84.1025 & -6.3781 & Ori. A & 9.37 & 6.95 & 1.69 & 2.52 & Burst & Proto. & N & M933, H178\\
    \hline
    J05461313-0006045 & 86.5547 & -0.1013 & Ori. B & 5.96 & 4.84 & 3.44 & 3.21 & Burst & Proto. & N & M3180, H388, V1647~Ori, McNeil's Nebula\\ %M3180, 
    J05462569+0023416 & 86.6071 & 0.3949 & Ori. B & 8.61 & 8.01 & 1.55 & 1.66 & Burst & Disk & N & M3394\\
    J05460477-0014163 & 86.5199 & -0.2379 & Ori. B & 7.22 & 6.30 & 1.02 & 0.92 & Fluc. & Proto. & N & H385, IRAS 05435-0015\\
    J05473657+0020062 & 86.9024 & 0.3351 & Ori. B & 7.84 & 7.04 & 0.57 & 1.38 & Burst & Proto. & N & M3341, H364\\
    J05465358+0000061 & 86.7232 & 0.0017 & Ori. B & 8.77 & 8.21 & 1.64 & 1.62 & Fade & Disk & N & M3196\\
    J05464312+0000525 & 86.6797 & 0.0146 & Ori. B & 6.39 & 5.19 & 0.51 & 1.03 & Fade & Proto. & N & M3208, H363\\
    J05460363-0014492 & 86.5151 & -0.2470 & Ori. B & 9.14 & 7.63 & 1.83 & 1.16 & Fluc. & Proto. & N & M3159, H315\\
    J05405172-0226486 & 85.2155 & -2.4468 & Ori. B & 8.22 & 7.58 & 1.16 & 0.98 & Fluc. & Proto. & N & M2827, H294\\
    J05413972-0202241 & 85.4155 & -2.0400 & Ori. B & 9.46 & 9.03 & 1.18 & 1.15 & Fluc. & Disk & N & M2873\\
    J05471062+0021141 & 86.7942 & 0.3539 & Ori. B & 7.77 & 6.61 & 0.92 & 0.74 & Fluc. & Proto. & N & M3359, H365\\
    J05455630+0007085 & 86.4846 & 0.1190 & Ori. B & 10.12 & 9.91 & 0.95 & 1.23 & Fluc. & Disk & N & M3264\\
    J05415555-0223405 & 85.4815 & -2.3946 & Ori. B & 10.00 & 9.52 & 0.83 & 1.15 & Fluc. & Disk & N & M2829\\
    J05464741+0012594 & 86.6975 & 0.2165 & Ori. B & 10.19 & 9.75 & 0.84 & 0.93 & Fluc. & Disk & N & M3295\\
    \hline
    J16271758-2405139 & 246.8233 & -24.0872 & Oph. & 8.32 & 7.78 & 2.14 & 2.28 & Burst & Disk & N & D2206, ISO-Oph 123\\
    J16271383-2443318 & 246.8076 & -24.7255 & Oph. & 9.20 & 8.71 & 1.45 & 1.34 & Burst & Disk & N & D2197, ISO-Oph 117\\
    J16264216-2431031 & 246.6757 & -24.5175 & Oph. & 9.43 & 8.75 & 1.05 & 1.2 & Burst & Disk & N & D2154, ISO-Oph 59\\
    J16265196-2430396 & 246.7165 & -24.5110 & Oph. & 10.77 & 9.78 & 0.66 & 1.03 & Fade & Disk & N & D2162, ISO-Oph 75\\
    J16272180-2429536 & 246.8408 & -24.4982 & Oph. & 6.94 & 5.72 & 2.01 & 2.72 & Fluc. & Proto. & Y & eHOPs-oph-24, ISO-Oph 134\\
    J16264421-2434487 & 246.6842 & -24.5802 & Oph. & 7.67 & 6.01 & 1.99 & 1.79 & Fluc. & Proto. & N & D2156, eHOPS-oph-9, ISO-Oph 65\\
    J16274162-2446450 & 246.9234 & -24.7792 & Oph. & 11.33 & 10.3 & 0.99 & 0.93 & Fluc. & Proto. & N & D2246, eHOPS-oph-38, ISO-Oph 170\\
    J16264048-2427146 & 246.6687 & -24.4541 & Oph. & 9.91 & 9.04 & 0.85 & 1.03 & Fluc. & Proto. & N & D2151, eHOPS-oph-7, ISO-Oph 54\\
    J16262755-2441538 & 246.6148 & -24.6983 & Oph. & 8.95 & 8.67 & 0.80 & 1.06 & Fluc. & Disk & N & D2146, ISO-Oph 43\\
    J16274709-2445353 & 246.9462 & -24.7598 & Oph. & 9.54 & 8.93 & 0.65 & 0.98 & Fluc. & Disk & N & D2254, ISO-Oph 177\\
    \hline
%    J18295225+0115476 & 277.4677 & 1.2632 & Serp. & 11.57 & 9.63 & 1.82 & 1.49 & Fluc. & Proto. & Y & D2880, eHOPS-aql-71, GMM2009 Serpens 10\\ 
    J18300772+0112044 & 277.5321 & 1.2012 & Serp. & 8.96 & 8.48 & 0.46 & 0.83 & Burst & Disk & Y & D2951, GMM2008 Serpens 84\\
    J18295117+0116404 & 277.4632 & 1.2779 & Serp. & 10.00 & 8.69 & 1.52 & 1.72 & Fluc. & Proto. & Y & D2876, eHOPS-aql-68, EES2009 Ser-emb 21\\
    J18295954+0111583 & 277.4981 & 1.1995 & Serp. & 9.04 & 7.30 & 1.00 & 1.11 & Fluc. & Proto. & Y & D2926, eHOPS-aql-88, EES2009 Ser-emb 24\\
    \hline
    J19025867-3707361 & 285.7444 & -37.1267 & CrA & 11.44 & 10.04 & 0.56 & 1.02 & Fade & Proto. & N & D1715, eHOPS-cra-10, GMM2009 CrA 2\\
    J19001555-3657578 & 285.0648 & -36.9660 & CrA & 13.65 & 12.00 & 0.99 & 1.07 & Fluc. & Disk & N & D1676, ISO-CrA 76\\
    \hline
    J11094191-7634585 & 167.4246 & -76.5829 & Cham. & 8.58 & 7.98 & 1.30 & 1.09 & Fluc. & Disk & Y & D1608, ISO-ChaI 199\\
    J11095438-7631114 & 167.4766 & -76.5198 & Cham. & 11.23 & 10.38 & 1.09 & 0.80 & Fluc. & Proto. & Y & D1617, eHOPS-cha-9, ISO-ChaI 225\\
    \label{tab:var_src}
     \end{tabular}}
    \tablenotetext{1}{This photometry comes from \citet{2012AJ....144..192M} or \citet{2016AJ....151....5M} for the Orion sources and from the SESNA catalog for the remainder of the regions \citep{2020ApJ...896...60P}.}
    \tablenotetext{2}{Alternative names starting with M and D are from \citet{2021ApJ...920..132P} paper and refer to sources from \citet{2012AJ....144..192M} and \citet{2015ApJS..220...11D}, respectively. Names starting with H are HOPS sources from \citet{2016ApJS..224....5F}. GMM2008 sources are from \citet{2008ApJ...674..336G} and GMM2009 sources are from \citet{2009ApJS..184...18G}. eHOPS sources are from the eHOPS catalog on IRSA \url{https://irsa.ipac.caltech.edu/data/Herschel/eHOPS/overview.html}, also see \citep{2023ApJS..266...32P}. EES2009 sources are from \citet{2009ApJ...692..973E}. ISO-Oph, ISO-CrA, and ISO-ChaI source are from \citet{2001A&A...372..173B}, \citet{1999A&A...350..883O}, and \citet{2000A&A...357..219P}, respectively.}
%    \caption{Table of large amplitude variables}

%Alternative names from \citet{2021ApJ...920..132P} paper nomenclature: M, D, and EL are used for the YSOs listed in Megeath et al. (2012), Dunham et al. (2015), and Esplin \& Luhman (2019), respectively. For M and D, source numbers are the same as those in their original catalogs. For EL, the source number is the same as the source order listed in Table 1 of Esplin & Luhman (2019).}

% Park paper nomenclature: M, D, and EL are used for the YSOs listed in Megeath et al. (2012), Dunham et al. (2015), and Esplin & Luhman (2019), respectively. For M and D, source numbers are the same as those in their original catalogs. For EL, the source number is the same as the source order listed in Table 1 of Esplin & Luhman (2019).

\end{table*}

\newpage 
\subsection{Completed Bursts} \label{subsec:comp_bursts}

\begin{figure*}
    \centering
    \includegraphics [height=10.5cm]{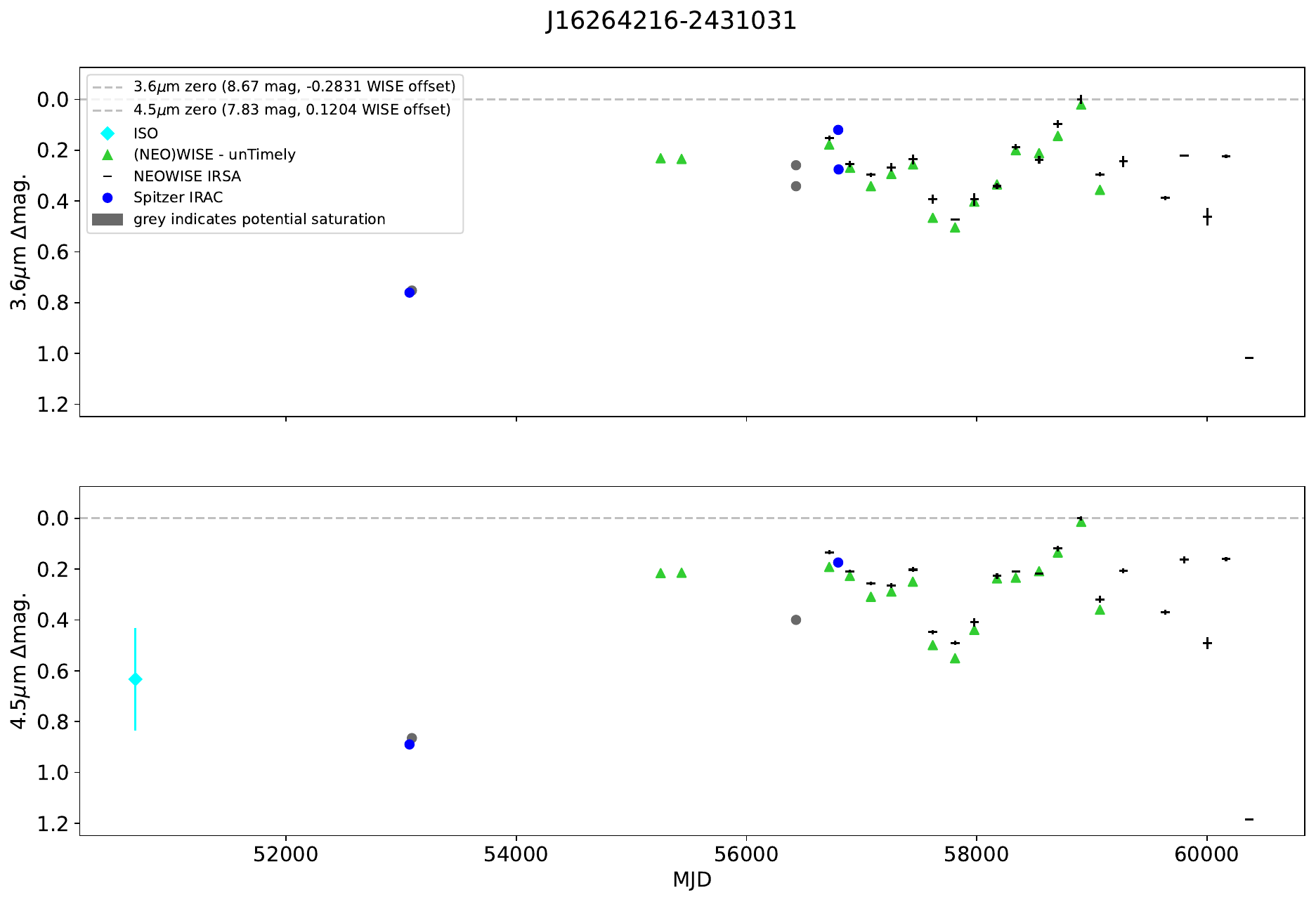}
    \caption{The 3.6 and 4.5~$\mu$m light curves of the completed bursts. We label the sources of the photometry in legend. Data that may be affected by saturation are shown with a grey color. The legends also contain the Spitzer magnitudes corresponding to $\Delta m = 0$, and the offsets added to the (NEO)WISE magnitudes to align those better to the Spitzer photometry.} 
    \label{fig:lc_comp_bursts}
\end{figure*}

Completed bursts are defined by i.)~a decrease in $\Delta m$ of $\ge 0.75$~mag after the initial ISOCAM measurement (an increase of $\ge 2\times$ in flux), ii.)~a sustained interval of low $\Delta m$ (high flux) that lasts $> 5$~yrs, and iii.)~a return in the $\Delta m$ to a value close to or below the initial ISOCAM value by the final epoch of the light curve. We have identified four such completed bursts in the 27-year time span of our data (Fig.~\ref{fig:lc_comp_bursts}): the flat spectrum protostar J05461313-0006045 in the Orion~B cloud, the pre-main sequence star with disk J05462569+0023416 also in Orion~B, and the pre-main sequence stars with disks J16264216-2431031 and J16271758-2405139 in the Ophiuchus cloud.

The outburst of J05461313-0006045 began in the fall of 2003 \citep{2004IAUC.8284....1M,2004ApJ...606L.123B}. Also known as V1647~Ori \citep{2004IAUC.8354....1S} and HOPS 388 \citep{2016ApJS..224....5F}, the protostar shows a flat spectrum SED consistent with a more evolved protostar \citep{2005ApJ...620L.107M, 2016ApJ...833..104F}. An ISOCAM, Spitzer, and (NEO)WISE light curve for J05461313-0006045 was presented by \citet{2024RNAAS...8...64K} which included 3.6~$\mu$m and 4.5~$\mu$m IRAC subarray photometry not included in our light curve. \citet{2024RNAAS...8...64K} found an amplitude for the burst of 3.1~mag at 4.5~$\mu$m and a duration of 16 years, using the I-band light curve to establish the time of the outburst \citep{2004ApJ...606L.123B}. During its burst, J05461313-0006045 (V1647~Ori) shows $\sim 1$~mag variations, as also found by \citet{2024RNAAS...8...64K}. At visible wavelengths, it underwent a large decline in 2006 before returning to the previous level in 2008. This dip was not sampled in our 3-8~$\mu$m light curves, and we do not know its depth in our wavelength range.  

%Due to the lack of data between 1997 and 2004, we do not know if these YSOs underwent the rapid rises characteristic of FU Ori outbursts \citep{2023ASPC..534..355F}. In all three of the YSOs, the rise occurred between the 1997 ISOCAM observation and the 2004 cryogenic Spitzer/IRAC observations.  In addition, the sustained low magnitude phase of these bursts are poorly sampled, and we do not have good data on variations during the burst.  The curves of two of the YSOs, J05461313-0006045 and J05462569+0023416, sampled the decline in the NEOWISE data which was taken every 6 months, and the post burst phase for all three YSOs were also observed by NEOWISE.  \textbf{}

The remaining three completed bursts do not have independent measurements of their bursts in other wavelength bands. This limits our ability to determine when and how quickly they rose. J05462569+0023416 and J16271758-2405139 show a rise between the ISOCAM 1997 epoch and the first Spitzer/IRAC 2004 epoch while J16264216-2431031 shows a rise between  Spitzer/IRAC 2004 epoch and the first WISE epoch. For J05462569+0023416, the initial $\Delta m$ is lower than the post-burst values; this suggests either different pre- and post-outburst accretion luminosities, or that the ISOCAM measurement caught this burst during its rise. During the bursts, the sparsely sampled light curves of J05462569+0023416 and J16271758-2405139 show modest variations of $< 0.5$~mag, while  J16264216-2431031 shows 0.5-0.75~mag multi-year fluctuations. All four YSOs show rapid declines in flux from their burst phases to the post-burst phases that last less than 2 years, and less than 6 months in the case of J16264216-2431031. In the post-burst phase, the NEOWISE photometry shows continued variability as high as 1~mag for J05461313-0006045.

\subsection{Ongoing Bursts} \label{subsec:ongo_bursts}

\begin{figure*}
    \centering
    \includegraphics[ height=10 cm]{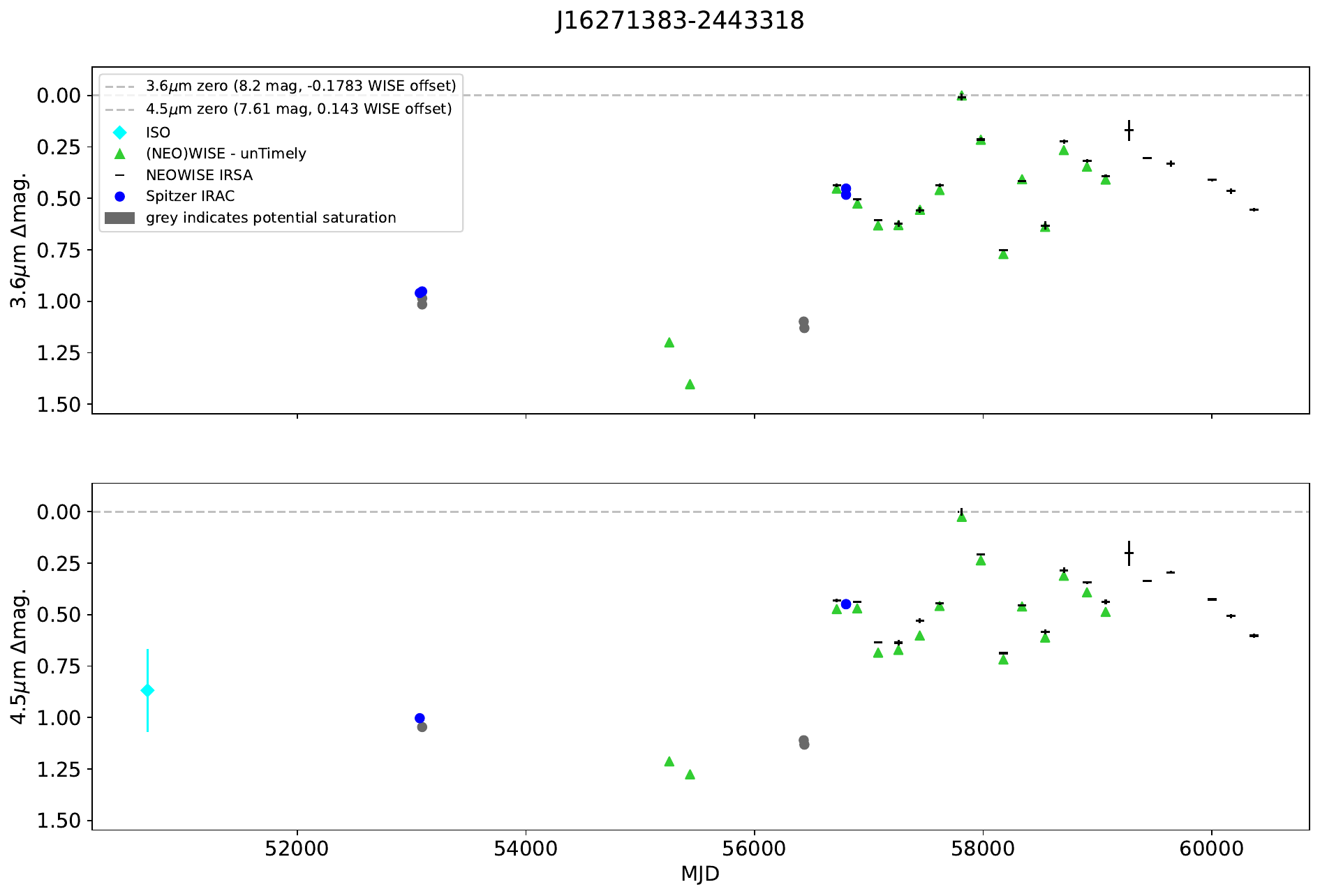}
    \caption{The 3.6 and 4.5~$\mu$m light curves of the ongoing bursts.  We label the sources of the photometry in legend. Data that may be affected by saturation are shown with a grey color. The legends also contain the Spitzer magnitudes corresponding to $\Delta m = 0$, and the offsets added to the (NEO)WISE magnitudes to align those better to the Spitzer photometry.}
    \label{fig:lc_ongoing_bursts}
\end{figure*}

%The total number bursts detected constrains the interval between bursts and the prevalence of bursts among all large amplitude variables \citep{2019ApJ...872..183F,2019MNRAS.486.4590C,2022ApJ...924L..23Z}.  
%This includes both the three completed bursts described in Sec.~\ref{subsec:comp_bursts} and ongoing bursts where the light curve did no return to the pre-burst level by the final epoch. 
Ongoing bursts are defined by i.) a decrease in $\Delta m$ of $\ge 0.75$~mag after the ISOCAM measurement followed by ii.)~a sustained low magnitude (high flux) through the final epoch of our light curves. We have identified three ongoing bursts in our sample (Fig.~\ref{fig:lc_ongoing_bursts}). These are the protostar J05362461-0622413, a Class~I protostar in the Orion A cloud, also known as HOPS~178 \citep{2016ApJS..224....5F}, and the pre-main sequence stars with disks J16271384-2443318 in the Ophiuchus cloud and J18300772+0112044 in the Serpens Main cluster.

The pre-ms star J18300772+0112044 rises between the ISOCAM 1997 epoch and the first Spitzer/IRAC 2004 epoch. There is evidence of slower, multi-year rises for the remaining two YSOs. For the protostar J05362461-0622413, the 2004 $\Delta m$ is between the pre-burst and burst levels, suggesting a slow rise over 1-2 decades. For the pre-main sequence star J16271382-2443316, the burst occurs between the WISE (2010) and NEOWISE (2013) points and takes 6 years to reach the burst maximum. The NEOWISE photometry captures part of this rise, and shows that it is not monotonic. 

In all three of these cases, the sustained burst phase shows significant variability. The protostar J05362461-0622413 undergoes a 0.75~mag dimming between 2010 and 2013, after which there is lower amplitude variability. The pre-ms star with disk  J16271384-2443318 shows 0.5-0.75~mag multi-year fluctuations after the onset of its burst. The pre-ms star with disk J18300772+0112044 underwent a slow dip in flux and then a rise from 2004-2015, although the data is sparsely sampled. In total, the data show that the burst phase is characterized by multi-year variability that can be a significant fraction of the amplitude of the burst.

Finally, we found one additional ongoing burst, J05473657+0020062 or HOPS~364, which rises slowly between 1997 and 2017. For this source, all the (NEO)WISE data has been rejected by the filters, and hence we do not include this source in our analysis. The light curve is shown in Fig.~\ref{fig:HOPS_364} of Appendix~A.

\subsection{Fading YSOs} \label{subsec:fades}

\begin{figure*}
    \centering
    \includegraphics[width=15cm, height=10.5cm]{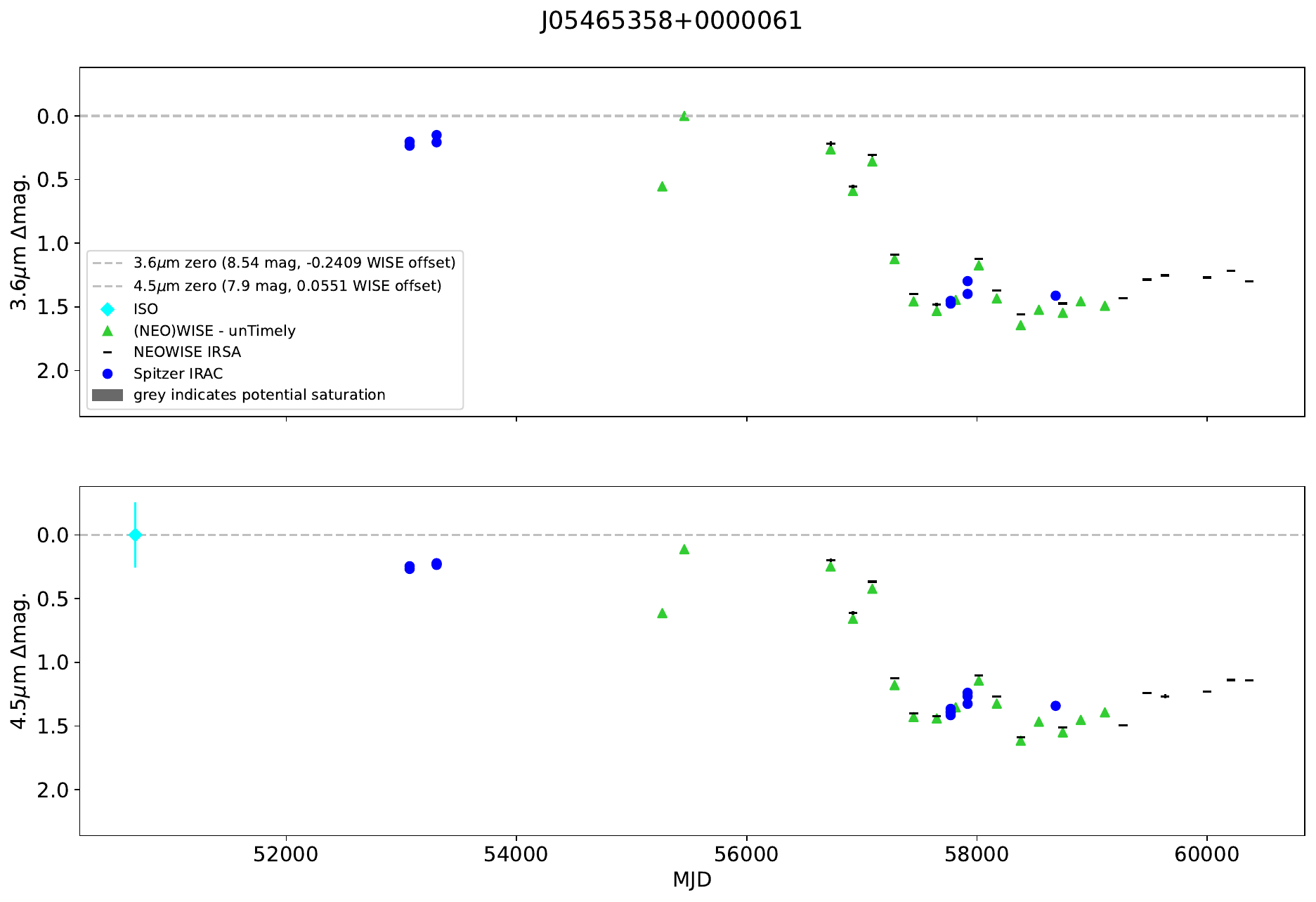}
    \caption{The 3.6 and 4.5~$\mu$m light curves of the fading YSOs.  We label the sources of the photometry in legend. Data that may be affected by saturation are shown with a grey color. The legends also contain the Spitzer magnitudes corresponding to $\Delta m = 0$, and the offsets added to the (NEO)WISE magnitudes to align those better to the Spitzer photometry.}
    \label{fig:lc_fades}
\end{figure*}

%Te detection of ongoing bursts motivates a search for light curves showing the tail end of bursts, characterized by a factor of $\ge 2$ decrease in flux and an $\ge 0.75$~mag increase in $\Delta m$. 
Fading sources, or fades, are defined by i.)~an initial low magnitude (high flux) level, ii.)~a subsequent increase in $\Delta m$ of $\ge 0.75$~mag across the light curve, and iii.)~no return to the initial high flux level by the final epoch. We find three fading protostars and three fading pre-main sequence stars with disks (Fig.~\ref{fig:lc_fades}). In the NGC~1333 cluster, the J03283708+3113309 protostar and pre-ms star J03290473+3111349 are observed to fade. In Orion~B, the protostar J05464312+0000525 (HOPS~363) and the pre-ms star J05465357+0000061 are observed to fade. In Ophiuchus and Corona Australis, the pre-ms star J16265196-2430396 and protostar J19025867-3707361, respectively, show fades.  

The fades of J03283708+3113309, J03290473+3111349 and J05464312+0000525 show a gradual decline over 15 years, although J03290473+3111349 and J05464312+0000525 show significant ($\sim 0.5$~mag) variability in the NEOWISE data, with a rise in flux in the last few epochs. The pre-ms star J05465357+0000061 shows a rapid decline in flux over two years, similar to the declines of the completed bursts in Sec.~\ref{subsec:comp_bursts}. The pre-ms star J16265196-2430396  shows a slow decline in flux over 15 years, and then a rapid decline in flux. Finally, J19025867-3707361 shows a large decline in flux between 1997 and 2004 where there is no data to sample the shape of the decline. In total, these show that the fades can be either slow or rapid, and they can exhibit significant variability superimposed on top of the overall fading.  

\subsection{Fluctuating YSOs} 
\label{subsec:fluctuators}

\begin{figure*}
    \centering
    \includegraphics[width=15cm, height=10.5cm]{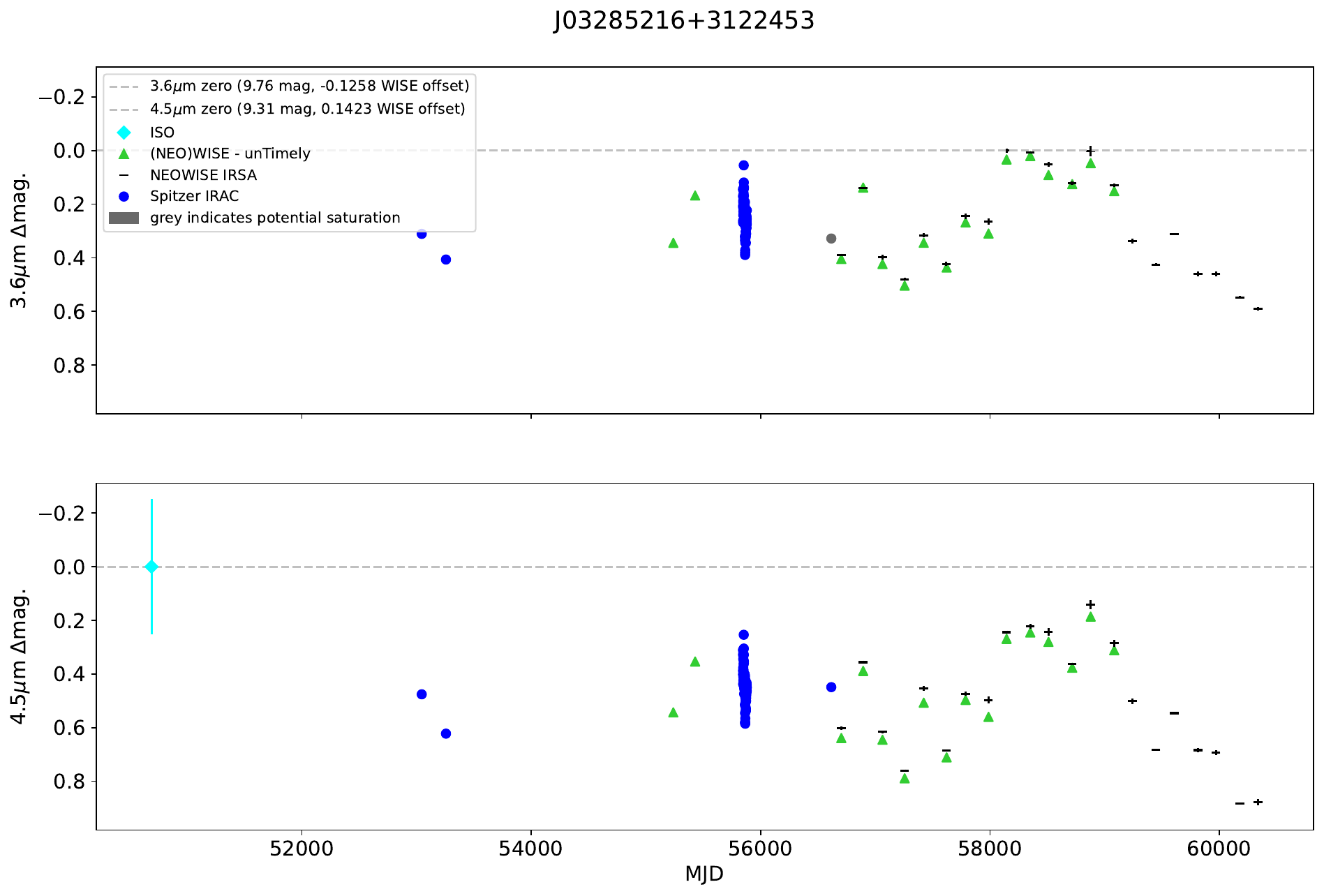}
    \caption{The 3.6 and 4.5~$\mu$m light curves of the fluctuators. We label the sources of the photometry in legend. Data that may be affected by saturation are shown with a grey color. The legends also contain the Spitzer magnitudes corresponding to $\Delta m = 0$, and the offsets added to the (NEO)WISE magnitudes to align those better to the Spitzer photometry.}
%  XJ0328346+310704 in the NGC 1333 region.}
   \label{fig:lc_fluctuators}
\end{figure*}

The category of fluctuators is defined by i.) the presence of large fluctuations in magnitude with values of $\Delta m$ $\ge 0.75$~mag over time intervals of $<5$ years, and ii.) no evident sustained burst or fade. We find 26 such YSOs that we classify as fluctuators; 13 are protostars and 13 are pre-main sequence stars with disks (Fig.~\ref{fig:lc_fluctuators}.)

%The protostar J18295225+0115476 shows a rise between the ISO point and the fluctuations in the Spitzer and (NEO)WISE epochs; however, due to the lack of reliable ISOCAM photometry, we drop the ISO point and classify this object as a fluctuator.  Similarly, due to the lack of a reliable ISOCAM point for the protostar J03285841+3122176, and due to multiple dips in the photometry after 2010, we classify this source as a fluctuator. 

This category encompasses a wide range of behaviors. Many fluctuating YSOs show large, poorly sampled fluctuations on $< 1$~year time intervals as shown in the NEOWISE or Spitzer YSOVAR data. Rapid variations are often apparent in the YSOVAR data; the most extreme example is J18295225+0115476 which shows a 1.5~mag variation over the YSOVAR data that spans 35~days. A few fluctuators show coherent, fluctuating behavior on multi-year to decade intervals. The pre-ms star J03292042+3118343 shows two sharp peaks in flux separated by about 13 years. Three peaks are evident for J05405172-0226486, J03284529+3105419 and J16274709-2445353; periods of 4.4 and 5.0~yrs, respectively, were found by \citet{2021ApJ...920..132P} for the last two of these sources in the  NEOWISE light curves. Rises, potentially suggestive of non-monotonically increasing bursts, are possible for J05460363-0014492, J16274709-2445353, and J16262755-2441538. J16262755-2441538 in particular exhibits a clear rise and may be a slow outburst, but it decays in flux in its last epochs. In all three cases, the variations are more than half the putative burst amplitude. Finally, J05471062+0021141 does not show rapid variations, and it could be categorized as a fading YSO if not for the steep rise during the last three unTimely epochs. 

The fluctuators are more likely to contain data from the high-cadence Spitzer observations, either from the YSOVAR program or the study of Chameleon by \citet{2016ApJ...833..104F}. While 1/7 bursts and and 1/6 fades have YSOVAR data, 10/26 of the fluctuators have high cadence data. All ten of the fluctuators with high-cadence data would have been identified as fluctuators without that data on the basis of the NEOWISE data. Future work will explore how the inclusion of high-cadence observations biases the classification of the light curves.

\begin{figure*}
    \centering
    \includegraphics[width=1\textwidth]{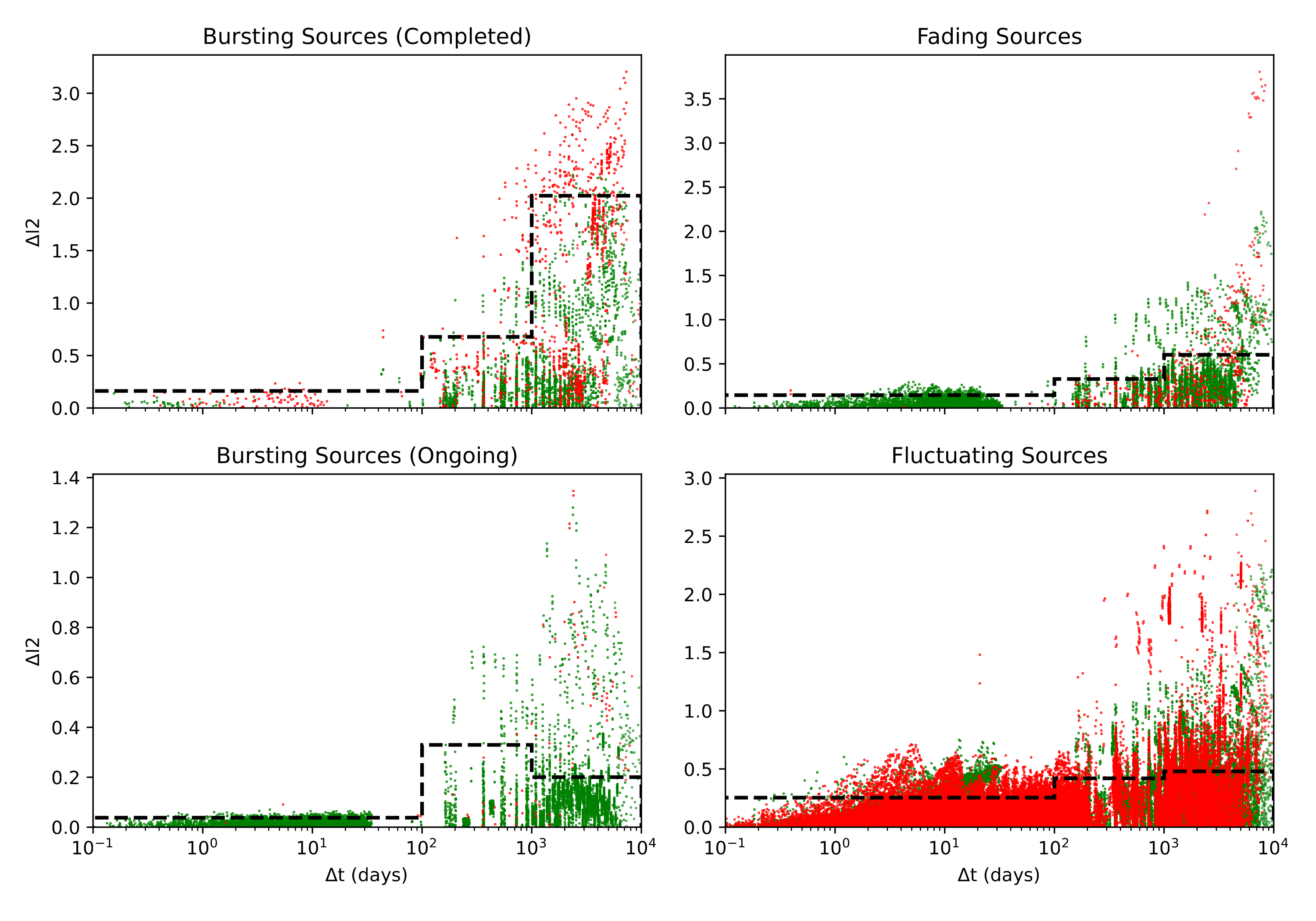}
    \caption{The $\Delta$m$_{4.5~\mu m}$ vs. $\Delta$t diagrams for the ISO, Spitzer and WISE data of all of the large-amplitude variables. These are separated into the four different categories, with protostars (red) and pre-ms stars with disks (green) distinguished by color. The dashed lines shows the 90th percentile of $\Delta$m for bins of $10^3$-$10^4$~days, $10^2$-$10^3$~days, and $< 10^2$~days.}
    \label{fig:dtvdmag}
\end{figure*}

\subsection{$\Delta t$ vs $\Delta m$ Diagrams } \label{sec:lagplots}

To compare the timescales of the variability present in our different categories, we use $\Delta t$ vs $\Delta m$ plots \citep{2011BASI...39..387M, 2015ApJ...798...89F,2021MNRAS.504..830G,2025MNRAS.543.1133R}; which are related to the structure function \citep{2015ApJ...798...89F,2016IAUS..314..209T}.
We plot $|\Delta t|$ vs $|\Delta m|$ for each pair of points in the 4.5~$\mu$m light curve (Fig.~\ref{fig:dtvdmag}). We combine the values of each pair of points from all the light curves within a given category. Over the irregular and sparse sampling of the light curve, these diagrams show the observed variation in magnitude over five orders of magnitude in time lag, from 0.1 days to 27 years. This large dynamic range in time comes from the different data sets used in this work; the YSOVAR data dominate for $|\Delta t| < 100$ days, the (NEO)WISE and Spitzer photometry dominate the $|\Delta t| = 100-10000$ day range, and data paired with the ISOCAM photometry contributes to the $|\Delta t| = 1000-10000$ day range.  

%Instead, we show the value for each pair of photometry points in a scatter plot, which gives the distribution of values. In the case of bursts, this approach has the advantage of showing the large $\Delta m$ values for pairs where one point is taken before the onset of the outburst and the other during the outburst, without these values being averaged down by pairs of points that only sample time intervals before, during, or after the bursts.; here 

Dividing the diagrams into three bins, $<100~{\rm days}$, $100-1000~{\rm days}$, and $> 1000~{\rm days}$, we find that there is an increase in $|\Delta m|$ with $|\Delta t|$. For all four categories, the largest values of $|\Delta m|$ occur for $|\Delta t| \ge 1000~{\rm days}$. For the completed bursts, ongoing bursts, and fades, only $|\Delta t| > 100~{\rm days}$ values show 90th percentile of $|\Delta m|> 0.75~{\rm mag}$ (i.e. a factor of 2 change in flux). For the fluctuators, the $|\Delta t| < 100~{\rm days}$ bin has one pair of points with $|\Delta m|> 0.75~{\rm mag}$, and there are many pairs close to this value. Higher amplitudes in the 3-5~$\mu$m variability at longer timescales were also found in a Spitzer and WISE study by \citet{2024ApJ...962...38L}.

We also show the 90th percentile values of $|\Delta m|$ for the three bins. Both the completed bursts and fades show amplitudes at lags of $> 1000~{\rm days}$ that are a factor of two times higher in magnitude (i.e. a factor of 6 higher in flux) than those at shorter lags. For the ongoing bursts, the 90\% limit of the $>1000~{\rm days}$ bin is lower than that of the $100-1000~{\rm days}$ bin. This is due to the large number of points in the YSOVAR data of J18300772+0112044 that were located in the middle of burst; they are evident as a dense cluster of points in that plot (Fig.~\ref{fig:dtvdmag}). For both the bursts and fades, the 90\% limit for the $<100~{\rm days}$ bin is very low ($\le 0.2$~mag). Finally, the 90\% limit for the fluctuators has the highest value out of all the categories in the $<100~{\rm days}$ bin, with only a factor of two change in magnitude between them and those in the $>1000~{\rm day}$ bin. Furthermore, the 90\% limits of the $100-1000~{\rm days}$ bin and $> 1000~{\rm days}$ bin are similar. The fluctuators show the lowest degree of change across the three bins, indicating that significant variability is observed across all sampled time intervals.  

\subsection{Color Plots and the Role of Extinction} \label{sec:colorplots}

\begin{figure*}[t]
    \centering
    \includegraphics[width=0.8\textwidth] {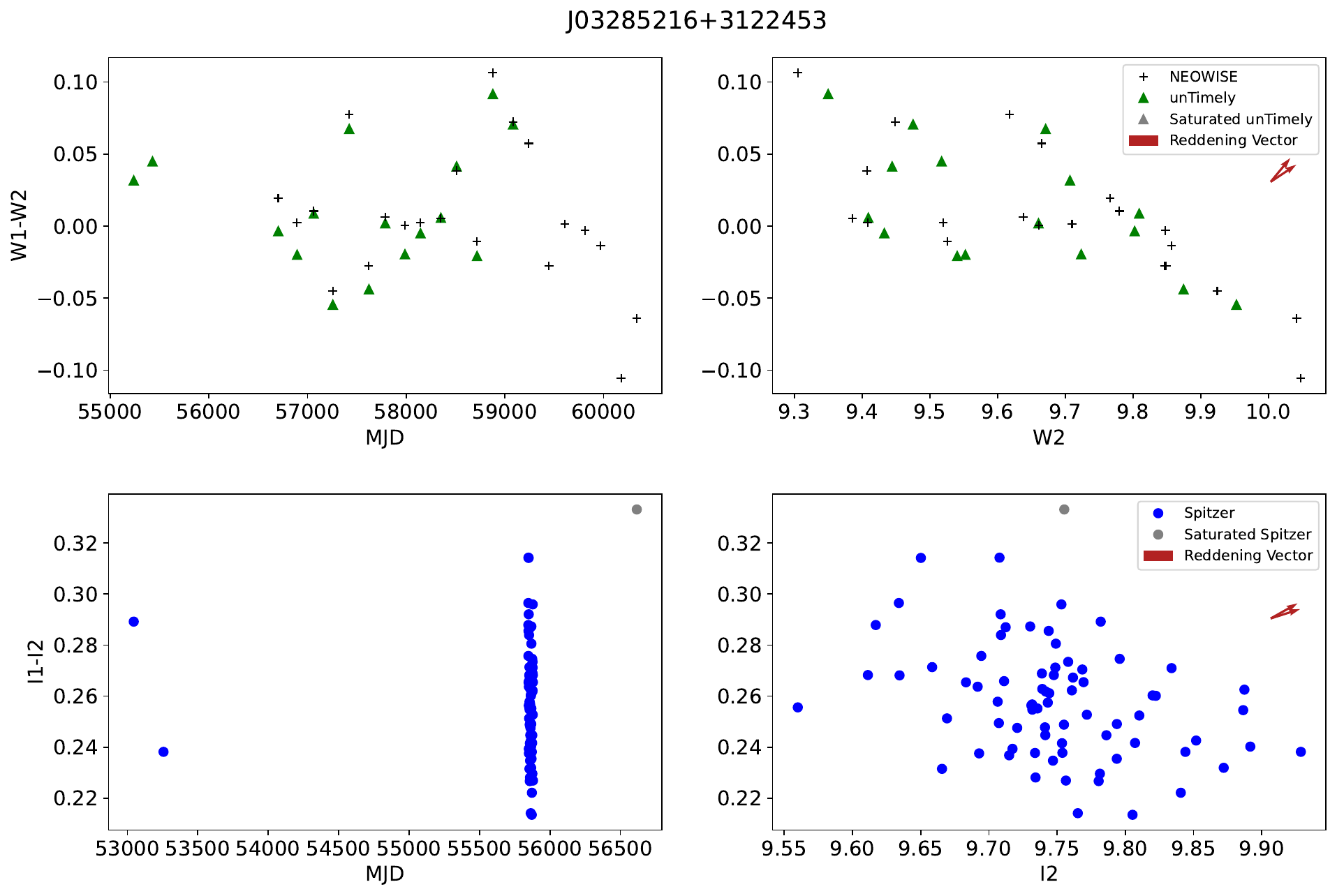}
     \caption{Left panels: color vs MJD plots; the WISE photometry is in the top panel and the Spitzer photometry is in the bottom panel. The time range covered in the plots is different. Right panels: color vs 4.5~$\mu$m mag plots. On the top row, we display the (NEO)WISE W1-W2 vs W2 diagram, while on the bottom row we show the Spitzer I1-I2 vs I2 diagram. For comparison, extinction vectors are shown for both of these plots to show the direction that extinction would shift the photometry. In the WISE plots, we show the IRSA and unTimely data separately. We also mark points that may be affected by saturation.}
     \label{fig:color_fluctuate}
\end{figure*}

Colors are a diagnostic of the processes responsible for variability and, in particular, whether changes in extinction are responsible for changes in the brightness \citep[e.g.][]{2019AJ....158..240H,2021ApJ...920..132P}. In Fig.~\ref{fig:color_fluctuate}, we show the Spitzer I1-I2 and WISE W1-W2 colors for the light curve of the protostar J16274709-2445353. In addition, we plot color-magnitude diagrams (hereafter: CMDs) showing the I1-I2 color vs the I2 magnitude for each point in the light curve of this protostar from the Spitzer data, and the W1-W2 color vs the W2 magnitude for each point in the light curve from the (NEO)WISE photometery. We split the Spitzer and (NEO)WISE data since differences in the Spitzer/IRAC and WISE bandpasses can result in large differences in color (see Fig.~\ref{fig:filters} for the bandpass profiles). We plot extinction vectors in the CMDs; this gives the direction of changes in the color and magnitude due to extinction. For the I1-I2 vs I2 plots, we use the tabulation of extinction laws from from \citet{2009ApJ...690..496C}. We adopt their tabulated $1 < A_K \le 2$ and \citet{2007ApJ...663.1069F} laws; the \citet{2009ApJ...690..496C} $A_K \ge 2$ law is between the two, and the laws for lower extinctions have high uncertainties and higher slopes. For the W1-W2 vs W2 CMDs, we adopt the $0.5 < A_K \le 1$ and $A_K > 1$ extinction laws from \citet{2014ApJ...791..131K}. In each CMD, we plot two vectors showing the different extinction laws for the moderate and high extinction regimes. The overall distribution of points in the CMDs are not aligned with these vectors, which suggests that the variations in magnitude are not due to changes in extinction. 

\begin{figure*}
    \centering
    \includegraphics[width=1\textwidth]{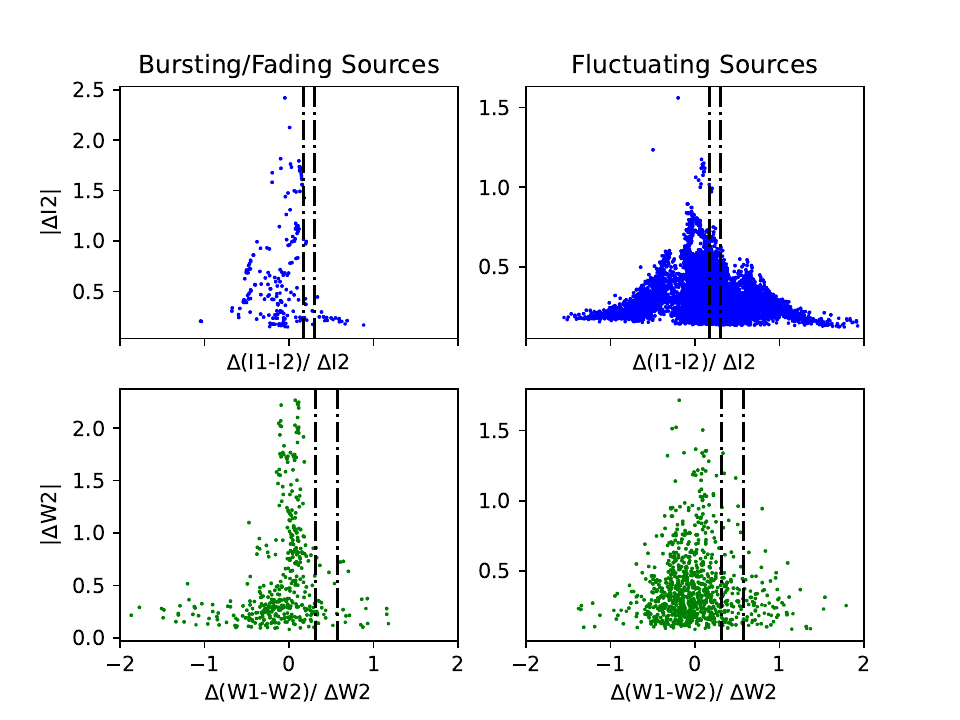}
    \caption{The change in magnitude vs the slopes between pairs of points in the CMDs. {\bf Top:} $|\Delta I2|$ vs $\Delta (I1-I2)/\Delta I2$ plots for the combined bursts and fades (left panel) and fluctuators (right panel). The vertical lines mark the slopes of the molecular cloud extinction laws for the Spitzer filters \citep{2007ApJ...663.1069F,2009ApJ...690..496C}; points found on or between these lines have the slopes bracketed by the extinction vectors in the CMD described in the text. {\bf Bottom:} $\Delta (W1-W2)/\Delta W2$ vs $|\Delta W2|$ plots for the combined bursts and fades (left panel) and fluctuators (right panel). Here the lines show the slopes for the extinction law found the WISE filters for regions of moderate (right) and high (left) extinction \citep{2014ApJ...791..131K}; points on or between these lines are consistent with changes due to extinction.} %difference must be greater 0.1}
    %0.071, 0.173, 0.130 and 0.364 - what are thse
    \label{fig:slope_colormag}
\end{figure*}

%A fundamental question is whether the variability is due to extinction. In particular, orbiting material near the protostar may result in episodes of high extinction, as are observed for dippers and UX Ori objects.  

To further assess whether changes in magnitude are consistent with variations in extinction, we calculate the $\Delta (I1-I2)/\Delta I2$ and $\Delta (W1-W2)/\Delta W2$ slopes for each pair of photometry points in the CMDs. We plot $|\Delta I2|$ or $|\Delta W2|$ vs these slopes in Fig.~\ref{fig:slope_colormag}, comparing them to the slopes of the extinction laws used in the CMDs. For small values of $|\Delta I2|$ or $|\Delta W2|$, the slopes become dominated by the uncertainties in the colors, and the distributions spread out in slope. For variations due to extinction, we expected that the slope will be positive, i.e. increases in the I2 or W2 magnitudes will be accompanied by increasingly red colors. We find that for the bursts and fades, the points are concentrated at neutral to negative slopes for $|\Delta I2|$ and $|\Delta W2|$ $> 0.75$~mag, ruling out reddening for most of the points. For the Spitzer data, we do find a chain of points that have the same slope as the reddening vector. On inspection of the light curve, these slopes are for non-sequential pairs of points obtained before and after the decline of the completed burst of the J05461313-0006045 protostar, and they do not seem to be following the extinction vector. We also find that for the fluctuators, slopes are concentrated at neutral to positive values, similar to the bursts and fades. Most are not consistent with extinction, although there is more scatter in these diagrams than for the bursts and fades, and there could be a prominent, but not dominant, component of the variability due to extinction.

%For all dusty YSOs, the emission in these bands is thought to be dominated by emission from dust in the inner regions of their disks. This is a combination of passive heating of the disk by the central, accreting (proto)star and internal heating of the disk by the rapid flow of gas through the midplane.  Only during episodes of rapid accretion can the internal heating dominate over passive heating.

%\subsection{Bursts} \label{subsec:cbursts}
%\input{C_figure_texts/C_burst_figures}

\section{Discussion} 
\label{sec:discussion}

%\textbf{We have identified 40 YSOs which vary by a factor of $\ge 2$ in flux in our 3-5~$\mu$m light curves, 39 of which have ISOCAM, Spitzer, and (NEO)WISE data spanning 27 years (the remaining one is discussed in Appendix~A).} 

In this section, we discuss the classification of protostars into the four different types of variability and assess whether they represent fundamentally different types of variability or whether they are simply the result of the sampling of our data. We then explore the implications of our work for the mechanisms responsible for YSO variability.  

\subsection{Classification of Accretion-Driven Outbursts}
\label{sec:midir_burst}

This program was initially motivated by the goal of identifying large-amplitude, accretion-driven outbursts from YSOs in the nearest 500~pc.  Such outbursts, although rare, have long been thought to play an important role in mass accretion \citep{1996ARA&A..34..207H,2010ASPC..432..197D}.
%Our study reveals that the large-amplitude variability of YSOs encompasses a variety of behaviors beyond sustained outbursts. For this reason, it is important to discuss the relationship of our classification system to previous schemes. 
YSO outbursts were historically divided into luminous, decades-long FU~Ori outbursts and lower-amplitude EX~Lupi outbursts that lasted one year or less \citep{2022arXiv220311257F}. These two categories show distinct spectroscopic signatures that imply different modes of accretion \citep{2016ARA&A..54..135H}. EX~Lupi outbursts exhibit bright hydrogen emission lines indicative of magnetospheric accretion, i.e. accretion from the disks to the stars via magnetic field lines \citep{2010ApJ...719L..50A,2023ApJ...957..113W}. The near-IR spectra of FU-Ori type outbursts, in contrast, show no hydrogen lines and  CO overtones in deep absorption, suggesting the stars accrete directly through disks that extend to the stellar photospheres \citep{2018ApJ...861..145C,2016ARA&A..54..135H}. More recent work, however, has found that many long-duration bursts ($> 5$~yr) do show hydrogen line emission suggestive of magnetospheric accretion \citep{2021MNRAS.504..830G}.

Since 2004, 3-8~$\mu$m studies using multi-epoch data from Spitzer and WISE/NEOWISE identified outbursts from YSOs in nearby molecular clouds with durations of a decade or more, some from protostars that are not detected at shorter wavelengths \citep{2011A&A...526L...1C,2012ApJ...756...99F,2015ApJ...800L...5S,2019ApJ...872..183F,2022ApJ...924L..23Z}. The duration of these outbursts is similar to FU~Ori outbursts \citep{2025JKAS...58..209C}. Confirmation of the outburst type through spectra at visible or near-IR wavelengths (i.e. $< 2.5$~$\mu$m) is rare. This is due in part to the lack of a coordinated effort to obtain IR spectra of YSOs in nearby clouds, but for the most deeply embedded protostars, it is also due to our inability to obtain spectra in the $< 2.5$~$\mu$m range. 

\begin{figure}
    \centering
    \includegraphics[height=10 cm]{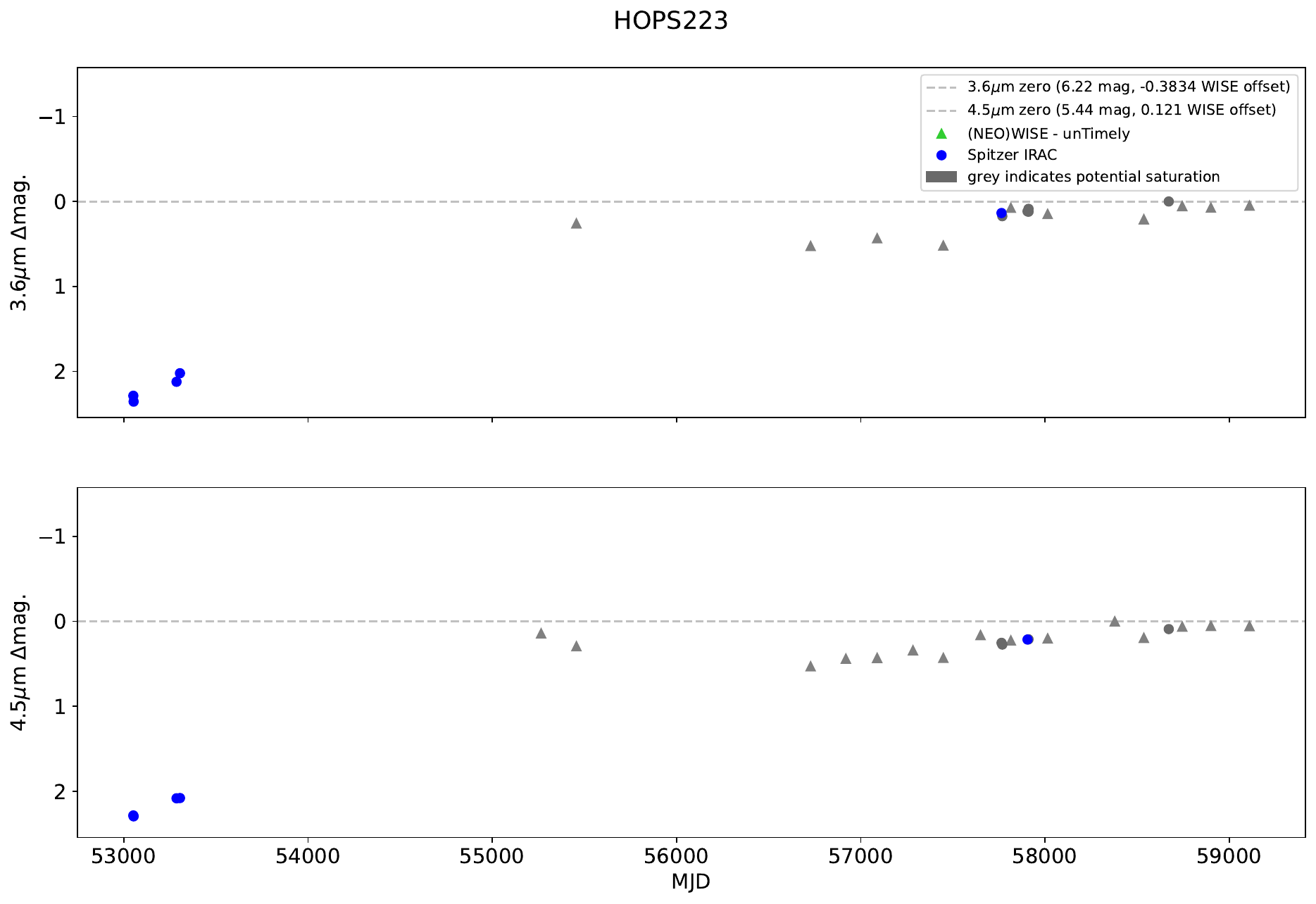}
    \caption{The light curve of the Orion protostar v2775~Ori, or HOPS~223. The light curves shows the ongoing burst from this protostar, which was first discovered by \citet{2011A&A...526L...1C} and studied by \citet{2012ApJ...756...99F}}..
    \label{fig:HOPS_223}
\end{figure}

Two protostellar outbursts for which there are mid-IR light curves and spectra are V2775~Ori and V1647~Ori. The outbursting protostar V2775~Ori (HOPS~223)  started its burst between 2005 and 2007 and has sustained the burst through the NEOWISE data \citep[Fig.~\ref{fig:HOPS_223},][]{2011A&A...526L...1C,2012ApJ...756...99F}. It appears to be in a face-on orientation where we can observe the protostar through the outflow cavity \citep{2016ApJS..224....5F,2023ApJ...944...49F}. The lack of hydrogen lines and the presence of deep CO and H$_2$O absorption features in the 1-2.5~$\mu$m spectrum of this protostar confirm that it is undergoing an FU~Ori type outburst that has lasted 18 years \citep[][]{2012ApJ...756...99F}. The protostar V1647~Ori (J05461313-0006045) underwent a burst with a similar mid-IR amplitude to V2775~Ori that lasted 16 years \citep[Fig.~\ref{fig:lc_comp_bursts},][]{2005ApJ...620L.107M,2024RNAAS...8...64K}. In contrast to V2775~Ori, V1647~Ori showed Br-$\gamma$ emission in its near-IR spectrum indicating magnetospheric (i.e. EX~Lupi type) accretion \citep{2009AJ....138.1137A}.  These two cases demonstrate that protostars showing similar 3-8~$\mu$m light curves in the sparsely sampled data available, can show distinct near-IR spectra. The similarities in the light curves of these two protostars may be in part due to the incomplete sampling. V1647~Ori may have had multiple episodes of high luminosity over the last 60 years, similar to YSOs classified as fluctuators \citep{2013ApJ...778..116N}. Given this uncertain landscape and our incomplete light curves, we avoid the classification of our YSOs into the standard categories of FU~Ori and EX~Lupi bursts based on their light curves.

\subsection{Bursts and Fades as Accretion-Driven Outbursts }
\label{sec:burst_fade}

We interpret the completed bursts, ongoing bursts, and fades as long-duration ($> 5$~yr), accretion-driven outbursts sampled at different points in their light curves. In this interpretation, the ongoing bursts are outbursts that were not completed by the final epoch of NEOWISE, and the fades are bursts that started before the ISOCAM observations in 1997. This is supported by the similar $\Delta t$ vs $\Delta m$ diagrams of these sources, with the highest amplitudes found at the longest timescales (Fig.~\ref{fig:dtvdmag}). We note that our classification of the light curves is provisional and dependent on incomplete sampling. Furthermore, accretion likely continues, although at a lower level, before and after the bursts \citep[e.g.][]{2022ApJ...924L..23Z}. Due to variations in this "quiescent" accretion rate, the pre- and post-burst levels may be different and variable.  

The  observed color variations support that the bursts and fades are parts of accretion-drive outbursts. \citet{2022ApJ...924L..23Z} found that during the $\Delta I2 \ge 2$~mag mid-IR outbursts of Class~0 protostars, the I1-I2 colors increased by $0.2-0.3$~mag as the outbursts peaked. They argued that the color changes were inconsistent with variations in the observed flux due to extinction, and that the outbursts must be due to changes in luminosity driven by variations in accretion. Other outbursts ascribed to accretion bursts show a range of W1-W2 color changes: small shifts to the red \citep{2021ApJ...920..132P,2023MNRAS.524.5548S}, little change in the colors \citep{2018ApJ...869..146H,2023ApJ...957....8W}, and finally shifts to the blue that are too small to be due to extinction \citep{2019AJ....158..240H}.

In Sec.~\ref{sec:colorplots}, we find that the observed changes in the $\Delta(I2-I1)/\Delta I2$ and $\Delta(W2-W1)/\Delta W2$ slopes during our bursts and fades are inconsistent with changes due to variations in the extinction. The bursts and fades show slopes clustered around zero for $\Delta I2$ or $\Delta W2$~ $\ge 0.75$~mag (Fig.~\ref{fig:slope_colormag}). The observed changes are  consistent with models of protostars and pre-ms stars with disks which show increases in luminosity are accompanied by small or positive (i.e. increasingly red) changes in the 3.6-4.5 colors \citep{2012AJ....144..192M,2022RNAAS...6....6H,2024AJ....167...82F}. We conclude that the observed color changes are consistent with those observed toward other outbursts and models of outbursts. 

Interpreting bursts and fades as parts of outbursts, our light curves include the beginnings and/or ends of outbursts toward 13 YSOs: four completed bursts, three ongoing bursts, and six fades. Since the completed bursts contain both the initial rise and the ending decay, we observe in total $7 \pm 2.6$ rises and $10 \pm 3.3$ decays (uncertainties are $\sqrt{N}$). These numbers are equal within the uncertainties.  Our numbers are consistent with a simple interpretation that we are observing outbursts where the rise and decay times are less than the 27 year span of our light curves.

Due to the regular sampling of the NEOWISE photometry, our decays are better sampled than our rises. Out of ten decays found in the completed bursts and fades, four completed bursts and one fade showed rapid declines that lasted $\le 2$~years, while the others show declines extending over more than a decade. In comparison, the rises are typically sampled over three, widely spaced, epochs:  ISO in 1997, Spitzer in 2004, and WISE  in 2010. Nevertheless, we find a 13-year rise between the ISOCAM and WISE data for J06473562-0622413. In addition, the rise to the maximum flux of J16271383-2443318 is partially sampled by NEOWISE, and lasts six years.  Overall, the rises are typically shorter, with 1/7 rises being longer than a decade and 5/10 decays being longer than a decade. \citet{2025ApJ...987...23C} also find longer durations for decays compared to rises. 
 
For the completed bursts, we can constrain the duration of the outbursts from the light curves. The duration of the J05461313-0006405 (V1647~Ori) outburst was 16 years, where the onset was measured in visible light data and the decline was detected by NEOWISE. The J05462569+0023416 burst was a span of 10-17~years, where the uncertainty comes from the poor constraints on the date of the rise. The duration of the J16271758-2405139 burst was 6-16 years where the uncertainty comes from both the times of the rise and that of the decline. In total, the completed bursts have durations between 6-17~years. For the ongoing bursts, we can only estimate lower limits to their durations. The duration of the J16264216-2431031 outburst is $\ge 10$~years where we take the initial time as the first detection of the YSO in outburst by WISE. J05362461-0622413 reached a maximum in 2010, after a slow rise that lasted $\ge 13$~years; thus this outburst also has a duration of $\ge 14$ years. J16271383-2443318 rose within 1 year to an initial maximum in 2013, implying a duration of $\ge 11 $ years. Finally, J18300772 rose between 1997 and 2004, giving it a duration of $\ge 20$~years. Thus, the ongoing bursts require durations greater than 13-20 years. 
%check year ranges J05461313-000640 - when did NEOWISE start
%J16264216-2431031  when was IRAC measurement 5-6 2013 
%J16271383-2443318 when was IRAC

Constraints on the durations can also be estimated for the fades using the time between their initial measurement and the point when they appear to have returned to their pre-burst luminosity. For J19025867-3707361, the fade has finished by 2004, consistent with a duration $\ge 7$~years. J05464312+0000525 and J32938708+311309 both return to their assumed pre-burst levels by 2013, implying durations of $\ge 16 $ years. For J03290473+3111349 and J16265196-2430396 the fades end by 2014, requiring a duration of a $\ge 17$ years. Finally, the fade of J05465358+0000061 ends in 2015, requiring a duration of $\ge 18$~ years. Thus, fades are consistent with outbursts with durations of $\ge 7$~years to $\ge 18$~years. In summary, when interpreting the light curves of completed bursts, ongoing bursts, and fades as observations of outbursts observed at different phases of their evolution, the observed outbursts have durations as low as 6 years, and in other cases, longer than 20~years. Most are consistent with durations between 10-20 years, with durations that vary from burst to burst.

\subsection{Fluctuating Protostars}
\label{sec:discuss_fluctuation}

In addition to the 13 YSOs showing bursts and fades, we find 26 YSOs (13 protostars and 13 pre-ms stars with disks) showing large-amplitude fluctuations. Unlike outbursts, where there is a sustained interval of high luminosity that exceeds 5~years, fluctuators show variations over a range of timescales, from dips in luminosity that last many years, to consecutive large peaks each lasting a few years, to rapid variations on timescales of a month (see Sec.~\ref{subsec:fluctuators}). The variety of behaviors included in this category suggest that this variability is driven by distinct processes operating on different timescales, or at least similar processes occurring at different radii in the disks of these YSOs. This is demonstrated by the $\Delta m_{4.5~\mu m}$ vs $\Delta t$ plot for fluctuators in Fig.~\ref{fig:dtvdmag}. Although the largest variations are apparent on $\Delta t > 1000$~days, the 90th percentile limit is comparatively constant across all timescales for fluctuators, with significantly higher values at $\Delta t < 100$~days than the bursts or fades.

The observed slopes in the changes of the photometry, i.e. $\Delta (I1-I2)/\Delta I2$ and $\Delta (W1-W2)/\Delta W2$  in Fig.~\ref{fig:slope_colormag}, are broader than those of the bursts and fades. Although the points with $\Delta I2$ or $\Delta W2$ values $\ge 1$~mag cluster around a slope of zero, similar to the bursts and fades, there is a broader distribution of slopes for the fluctuators. In this distribution, a number of points consistent with extinction. We conclude that the variations in the light curves are primarily due to variations in the mass accretion rate, as is the case for the bursts and fades, but that variations in the extinction contribute more to the fluctuators than bursts or fades.

Although we typically conceptualize accretion-driven, large-amplitude variability as episodes of high accretion rates punctuating long periods of a lower level, quiescent accretion, fluctuators show no clear quiescent level. Instead, most show continuous and/or repetitive variations over the 27 years covered in the light curves, and particularly over the 12 year period covered by NEOWISE, for which the light curves are sampled regularly twice a year. In this way, these sources do not fall into the typical categories of large-amplitude variables. They lack the sustained intervals of high luminosity characteristics of FU~Ori and V1647~Ori bursts as well as the mid-IR burst and fades in our sample and others \citep{2022ApJ...924L..23Z}. The level of variation in these bands is much larger than that observed in the W1 and W2 bands for outbursts of Ex-Lupi during the NEOWISE mission \citep{2023ApJ...957..113W}. 

We also find that there is significant variability during the bursts (Figs.~\ref{fig:lc_comp_bursts} and \ref{fig:lc_ongoing_bursts}). We require this variability to be less than half of the burst amplitude, or they are classified as fluctuators. The incomplete sampling of the light curves during the bursts, however, may omit variations that could result in the reclassification of YSOs as fluctuators. For example, we classify V1647~Ori  (J05461313-0006045) as a burst; however, at visible wavelengths it underwent a large decline in 2006 before returning to the previous level in 2008; \citet{2009ApJ...692L..67A} suggest that half of the decline was due to a decrease in the source luminosity and the other half was due to an increase in extinction from the formation of dust. This decline occurred during a gap in the Spitzer coverage. If our light curve did show a 3-8~$\mu$m decline to pre-burst levels, the first rise in 2003 and decline in 2006 would be similar to the shorter duration variations of fluctuators, while its sustained high luminosity between 2009 and 2017 would be classified as a burst.

Some of the fluctuators may also be undergoing bursts where the initial rise or final fade are not within our 27 year time frame. Supporting this picture, the variations in the fluctuators are comparable to the variability observed toward some of the ongoing bursts (Sec.~\ref{subsec:ongo_bursts}). Furthermore, the typical minimum to maximum variations of the fluctuators are between 1-0.75 ~mag, close to the $0.75$~mag cutoff for large-amplitude variability, suggesting that there could be many more fluctuators just below this threshold, with variations similar to those measured during bursts. The presence of fluctuations during bursts hints that states of elevated accretion may be unstable. In general, the nature of the fluctuators needs to be a focus of future studies. These must explore the relationship between the sampling of the light curve and classification, and they must explore the connection between bursts and fluctuations.

%Due to their numbers and complexity, the large-amplitude fluctuators bring a new challenge to our understanding of the variations in accretion. Understanding these variations, their relationships to bursts, and their overall importance to mass accretion should be considered a central problem in YSO variability.  

%This points to a more complex landscape for large-amplitude protostellar variability \citep{2023ASPC..534..355F,2017MNRAS.465.3039C,2021MNRAS.504..830G,2025arXiv250421237C}. 

%Times can also be estimated for the fades from their initial measurement to the point they are appear to  have returned to their post-burst luminosity.  For J05465358, J05464312, J32938708, returns by 2013 imply durations $\ge 16 years$. Note that J05465358 displayed a rapid drop that lated 2~years, while the others showed slow drops over 16 years.  For J03290473 the return by 2015 suggests a $\ge 18 years$, although this drop is noisy. The flux of FU~Ori has declined slowly since it's outburst in 1936. If some of the fades last longer than 27 years, we would expect the number of fades to be less than the number of outbursts.  In a survey of outbursts along the galactic plane, ?? find evidence for both slow rise and slow decline fades. Through a typical test, they find that the fades typically last longer ??.  

\section{Comparison with other Studies}

The NEOWISE and Spitzer light curves in this work were previously examined by other studies.  Spitzer studies used the photometry from the warm-mission YSOVAR program and the cryo-mission data obtained in 2004 \citep[Sec.~\ref{sec:Spitzer},][]{2014AJ....148...92R}. In the combined studies of the Ophiuchus and NGC~1333 regions \citep{2014AJ....148..122G,2015AJ....150..175R}, two large-amplitude variables in the NGC~1333 region, 032851.01+311818.5 and 032859.54+312146.7, are also found in our program as fluctuators (Table~\ref{tab:var_src}).

\citet{2021ApJ...920..132P} used the NEOWISE photometry to determine the lightcurves of 5400 YSOs in nearby star forming regions.  We note that the classification scheme is different than the one used here. They classify the light curves as linear, periodic, curved, or stochastic. The stochastic variables they subdivide into bursts, drops, or irregulars. In contrast to our definitions, they define bursts and drops as events of brightening or fading that last  a few epochs relative to a stable baseline, and they set a threshold in $\Delta W2$ of three times the standard deviation. Of our 39 variable YSOs, 37 are in their catalog, and 31 are classified as variables: one linear, seven curved, three bursts, and the remainder are irregulars. Fifteen of the 37 YSOs show variations of 0.75~mag or more in W2 in the NEOWISE photometry alone. Of these, we classify two as bursts, one as a fade, and 12 as fluctuators. \citet{2021ApJ...920..132P} classify these same YSOs as one burst, two curved, and 12 irregulars. Overall, the 12 fluctuators are more likely to be identified as large-amplitude variables with the NEOWISE data alone; these are classified as 1 curved and 11 irregulars. Thus, there is a close correspondence between the fluctuators and irregulars. One of the fluctuators is the protostar V371 Ser, or EC~53, which is tabulated as an eruptive variable by \citet{2021ApJ...920..132P}.

\begin{figure}
    \centering
    \includegraphics[height=5.5 cm]{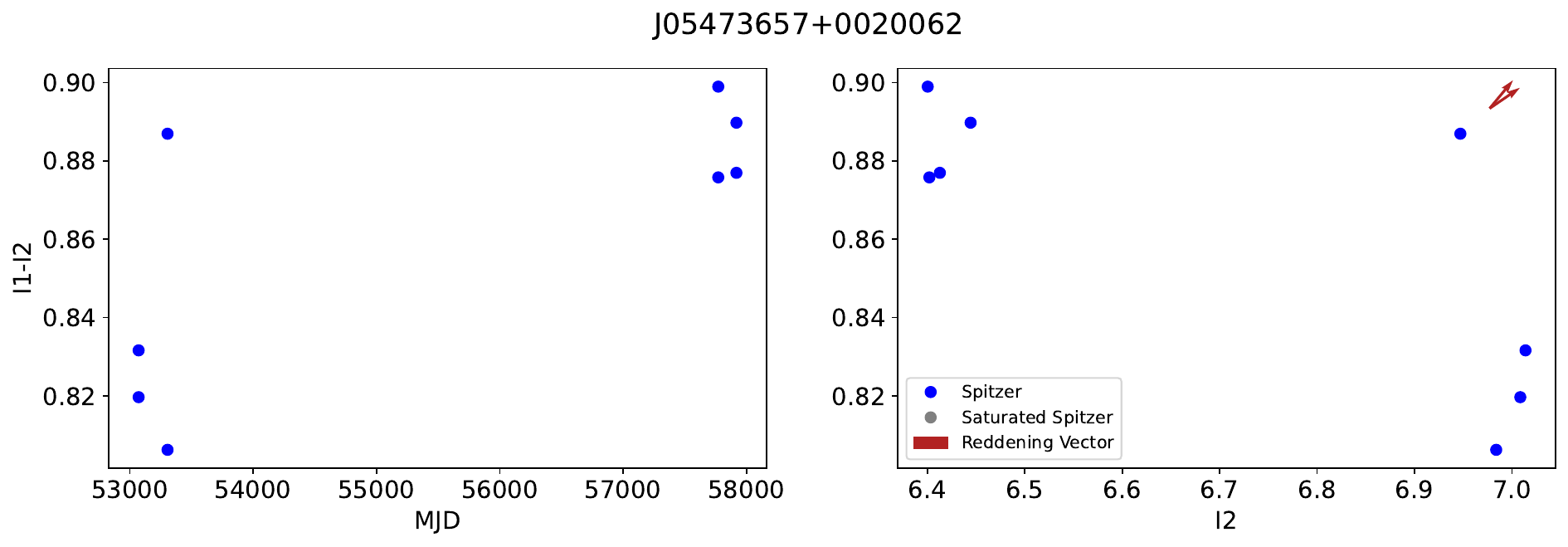}
    \caption{The CMD of the Orion protostar J05473657+0020062, or HOPS~364.}
    \label{fig:HOPS_364_color}
\end{figure}

Recently, \citet{2025JKAS...58..209C} published the Outbursting YSO Catalogue, or OYCAT. We find three sources in common with OYCAT. The completed burst and protostar J05461313-0006045, or V1647~Ori, is classified in OYCAT as a PVM (i.e.~a peculiar/v1647~Ori/MNor) eruptive class based on its spectrum and as an intermediate light curve (i.e duration between that an FU~Ori and EX~Lupi type source based on the light curve). The ongoing burst and protostar J05473657+0020062 (Appendix~A) is  classified in OYCAT as no burst detected (based on the light curve), but with an FUOr-like spectrum \citep{2018ApJ...861..145C}. In the ISOCAM and Spitzer data, we see the rise of this FU-Ori like outburst extending from 1997 to 2016 (Figs~\ref{fig:HOPS_364} and ~\ref{fig:HOPS_364_color}). 

Finally, the fluctuator and protostar J05460363-0014492 (HOPS~315) is classified in OYCAT as a PVM eruptive class with  an intermediate light curve class. J05460363-0014492 was recently found to have SiO gas lines and crystalline silicate features in absorption in its JWST MIRI MRS spectrum; these are likely seen in absorption against the emission generated around the mid-plane of a rapidly accreting disk \citep{2025Natur.643..649M}. It also shows Br-$\gamma$ line and CO overtones in emission \citep{2023MNRAS.521.5669C}. The light curve of this source shows that it was rising during the MIRI observations, implying an increase in the accretion rate \citep[Fig.~\ref{fig:lc_fluctuators},][]{2023MNRAS.521.5669C}.

\section{Summary}

We report a study combining ISOCAM 6.75~$\mu$m, Spitzer/IRAC 3.6 and 4.5~$\mu$m, and (NEO)WISE 3.5 and 4.5~$\mu$m photometry to construct 3-8~$\mu$m light curves extending up to 27 years. For each source, we create 3.6~$\mu$m light curves from the IRAC 3.6~$\mu$m and WISE 3.5~$\mu$m bands, and 4.5~$\mu$m light curves from the IRAC 4.5~$\mu$m, WISE 4.6~$\mu$m and ISOCAM 6.7~$\mu$m band. The IRAC photometry includes data from the cryo and warm-missions. We use IRAC 5.8~$\mu$m and 8~$\mu$m photometry from the cryo-mission to cross-calibrate changes in the ISOCAM photometry to the IRAC 4.5~$\mu$m data. With the inclusion of the ISOCAM photometry, the 4.5~$\mu$m light curves extend from 1997 to 2024. 

Out of the 221 YSOs with ISOCAM, Spitzer and (NEO)WISE photometry extending either from 1997 to 2020 or 1997 to 2024, we find 39 YSOs with large-amplitude variability, defined as exhibiting at least a factor of two variation in flux over the light curve (or a magnitude change of 0.75~mag). We classify the light curves of these large-amplitude variables into four categories: i.) completed bursts showing a rise of a factor of two or more in flux, a sustained high 3-8~$\mu$m flux for $\ge 5$~year, and then a decline to pre-rise levels, ii.) ongoing bursts showing a factor of two or more rise and then sustained high flux to the final epoch, iii.) fades showing a sustained decline of a factor of two or more in flux over the light curve, and iv.) fluctuators showing shorter timescale variations without a sustained, high-flux phase or sustained decline. From an analysis of the light curves, we report the following:  

\begin{enumerate}

\item{} We find completed bursts towards four YSOs (one protostar and three pre-ms stars with disks), ongoing bursts towards three YSOs (one protostar and two pre-ms stars with disks), and fades for six YSOs (three protostars and three pre-ms stars with disks). The remaining 26 YSOs with high-amplitude variations are classified as fluctuators, 13 of which are protostars and 13 pre-ms stars with disks. 
    
\item{} Thirteen of the 39 YSOs appear to sample YSO outbursts either in their entirety (the 4 completed bursts) or incompletely (the 9 ongoing bursts and fades). The amplitudes in the 4.5~$\mu$m light curves range from our limit of 0.75~mag up to 3.3~mag, and the durations may be as short as 6 years, or extend to more than 20 years, with all but one having durations $> 10$~yrs. The rises and declines of the outbursts can be either rapid (1-2~years) or slow (over a decade).  

\item{} The most common type of large-amplitude variable are fluctuators. This category contains a variety of behaviors: rapid variations over day to month scales, repetitive peaks separated by a few years, and dips extending several years. The maximum $\Delta m_{4.5}$ values of the fluctuators cluster between 0.75 and 1~mag, lower than the bursts and fades, but they show much higher amplitude variations than bursts and faces at time intervals of $< 100$~days.

\item{} Changes in the $I1-I2$ color with $I2$ and $W1-W2$ with $W2$ suggest that bursts and fades are primarily driven by changes in the luminosity of the central YSO due to episodes of rapid mass accretion. For fluctuators, the variations in fluxes also appear to be primarily driven by variations in the accretion rate, although with contributions from variations in extinction.

\item{} Significant variations in the 3-8~$\mu$m fluxes are present during outbursts. This suggests that high accretion rates may often be unstable and that there may be a connection between the fluctuators and outbursts. Given their prevalence in our sample, understanding the mechanism driving the fluctuators and their connection to outbursts is a central question in the study of YSO variability.

\end{enumerate}

\bibliography{main.bib}{}

@ARTICLE{2005ApJ...620L.107M,
       author = {{Muzerolle}, J. and {Megeath}, S.~T. and {Flaherty}, K.~M. and {Gordon}, K.~D. and {Rieke}, G.~H. and {Young}, E.~T. and {Lada}, C.~J.},
        title = "{The Outburst of V1647 Orionis Revealed by Spitzer}",
      journal = {\apjl},
         year = 2005,
        month = feb,
       volume = {620},
       number = {2},
        pages = {L107-L110},
          doi = {10.1086/428832},
archivePrefix = {arXiv},
       eprint = {astro-ph/0412272},
 primaryClass = {astro-ph},
       adsurl = {https://ui.adsabs.harvard.edu/abs/2005ApJ...620L.107M}
}

@ARTICLE{2023AJ....165...36M,
       author = {{Meisner}, Aaron M. and {Caselden}, Dan and {Schlafly}, Edward F. and {Kiwy}, Frank},
        title = "{unTimely: a Full-sky, Time-domain unWISE Catalog}",
      journal = {\aj},
         year = 2023,
        month = feb,
       volume = {165},
       number = {2},
          eid = {36},
        pages = {36},
          doi = {10.3847/1538-3881/aca2ab},
archivePrefix = {arXiv},
       eprint = {2209.14327},
 primaryClass = {astro-ph.IM},
       adsurl = {https://ui.adsabs.harvard.edu/abs/2023AJ....165...36M}
}

@ARTICLE{2016ARA&A..54..135H,
       author = {{Hartmann}, Lee and {Herczeg}, Gregory and {Calvet}, Nuria},
        title = "{Accretion onto Pre-Main-Sequence Stars}",
      journal = {\araa},
         year = 2016,
        month = sep,
       volume = {54},
        pages = {135-180},
          doi = {10.1146/annurev-astro-081915-023347},
       adsurl = {https://ui.adsabs.harvard.edu/abs/2016ARA&A..54..135H}
}

@ARTICLE{2022arXiv220311257F,
       author = {{Fischer}, William J. and {Hillenbrand}, Lynne A. and {Herczeg}, Gregory J. and {Johnstone}, Doug and {K{\'o}sp{\'a}l}, {\'A}gnes and {Dunham}, Michael M.},
        title = "{Accretion Variability as a Guide to Stellar Mass Assembly}",
      journal = {arXiv e-prints},
         year = 2022,
        month = mar,
          eid = {arXiv:2203.11257},
        pages = {arXiv:2203.11257},
          doi = {10.48550/arXiv.2203.11257},
archivePrefix = {arXiv},
       eprint = {2203.11257},
 primaryClass = {astro-ph.SR},
       adsurl = {https://ui.adsabs.harvard.edu/abs/2022arXiv220311257F}
}

@ARTICLE{2023ApJ...944...49F,
       author = {{Federman}, Samuel and {Megeath}, S. Thomas and {Tobin}, John J. and {Sheehan}, Patrick D. and {Pokhrel}, Riwaj and {Habel}, Nolan and {Stutz}, Amelia M. and {Fischer}, William J. and {Hartmann}, Lee and {Stanke}, Thomas and {Narang}, Mayank and {Osorio}, Mayra and {Atnagulov}, Prabhani and {Rahatgaonkar}, Rohan},
        title = "{300: An ACA 870 {\ensuremath{\mu}}m Continuum Survey of Orion Protostars and Their Evolution}",
      journal = {\apj},
         year = 2023,
        month = feb,
       volume = {944},
       number = {1},
          eid = {49},
        pages = {49},
          doi = {10.3847/1538-4357/ac9f4b},
archivePrefix = {arXiv},
       eprint = {2210.07925},
 primaryClass = {astro-ph.SR},
       adsurl = {https://ui.adsabs.harvard.edu/abs/2023ApJ...944...49F}
}

@ARTICLE{2017ApJ...840...69F,
       author = {{Fischer}, William J. and {Megeath}, S. Thomas and {Furlan}, Elise and {Ali}, Babar and {Stutz}, Amelia M. and {Tobin}, John J. and {Osorio}, Mayra and {Stanke}, Thomas and {Manoj}, P. and {Poteet}, Charles A. and {Booker}, Joseph J. and {Hartmann}, Lee and {Wilson}, Thomas L. and {Myers}, Philip C. and {Watson}, Dan M.},
        title = "{The Herschel Orion Protostar Survey: Luminosity and Envelope Evolution}",
      journal = {\apj},
         year = 2017,
        month = may,
       volume = {840},
       number = {2},
          eid = {69},
        pages = {69},
          doi = {10.3847/1538-4357/aa6d69},
archivePrefix = {arXiv},
       eprint = {1704.05847},
 primaryClass = {astro-ph.SR},
       adsurl = {https://ui.adsabs.harvard.edu/abs/2017ApJ...840...69F}
}

@ARTICLE{2022ApJ...924L..23Z,
       author = {{Zakri}, Wafa and {Megeath}, S.~T. and {Fischer}, William J. and {Gutermuth}, Robert and {Furlan}, Elise and {Hartmann}, Lee and {Karnath}, Nicole and {Osorio}, Mayra and {Safron}, Emily and {Stanke}, Thomas and {Stutz}, Amelia M. and {Tobin}, John J. and {Allen}, Thomas S. and {Federman}, Sam and {Habel}, Nolan and {Manoj}, P. and {Narang}, Mayank and {Pokhrel}, Riwaj and {Rebull}, Luisa and {Sheehan}, Patrick D. and {Watson}, Dan M.},
        title = "{The Rate, Amplitude, and Duration of Outbursts from Class 0 Protostars in Orion}",
      journal = {\apjl},
         year = 2022,
        month = jan,
       volume = {924},
       number = {2},
          eid = {L23},
        pages = {L23},
          doi = {10.3847/2041-8213/ac46ae},
archivePrefix = {arXiv},
       eprint = {2201.04647},
 primaryClass = {astro-ph.SR},
       adsurl = {https://ui.adsabs.harvard.edu/abs/2022ApJ...924L..23Z}
}

@ARTICLE{2021ApJ...920..132P,
       author = {{Park}, Wooseok and {Lee}, Jeong-Eun and {Contreras Pe{\~n}a}, Carlos and {Johnstone}, Doug and {Herczeg}, Gregory and {Lee}, Sieun and {Lee}, Seonjae and {Bhardwaj}, Anupam and {Moriarty-Schieven}, Gerald H.},
        title = "{Quantifying Variability of Young Stellar Objects in the Mid-infrared Over 6 Years with the Near-Earth Object Wide-field Infrared Survey Explorer}",
      journal = {\apj},
         year = 2021,
        month = oct,
       volume = {920},
       number = {2},
          eid = {132},
        pages = {132},
          doi = {10.3847/1538-4357/ac1745},
archivePrefix = {arXiv},
       eprint = {2107.10751},
 primaryClass = {astro-ph.SR},
       adsurl = {https://ui.adsabs.harvard.edu/abs/2021ApJ...920..132P}
}

@ARTICLE{2012ApJ...756...99F,
       author = {{Fischer}, William J. and {Megeath}, S. Thomas and {Tobin}, John J. and {Stutz}, Amelia M. and {Ali}, Babar and {Remming}, Ian and {Kounkel}, Marina and {Stanke}, Thomas and {Osorio}, Mayra and {Henning}, Thomas and {Manoj}, P. and {Wilson}, T.~L.},
        title = "{Multiwavelength Observations of V2775 Ori, an Outbursting Protostar in L 1641: Exploring the Edge of the FU Orionis Regime}",
      journal = {\apj},
         year = 2012,
        month = sep,
       volume = {756},
       number = {1},
          eid = {99},
        pages = {99},
          doi = {10.1088/0004-637X/756/1/99},
archivePrefix = {arXiv},
       eprint = {1207.2466},
 primaryClass = {astro-ph.SR},
       adsurl = {https://ui.adsabs.harvard.edu/abs/2012ApJ...756...99F}
}

@article{Vorobyov2006,
   author = {E. I. Vorobyov and Shantanu Basu},
   doi = {10.1086/507320},
   issn = {0004-637X},
   issue = {2},
   journal = {The Astrophysical Journal},
   month = {10},
   pages = {956-969},
   title = {The Burst Mode of Protostellar Accretion},
   volume = {650},
   year = {2006},
}

@ARTICLE{2012AJ....144..192M,
       author = {{Megeath}, S.~T. and {Gutermuth}, R. and {Muzerolle}, J. and {Kryukova}, E. and {Flaherty}, K. and {Hora}, J.~L. and {Allen}, L.~E. and {Hartmann}, L. and {Myers}, P.~C. and {Pipher}, J.~L. and {Stauffer}, J. and {Young}, E.~T. and {Fazio}, G.~G.},
        title = "{The Spitzer Space Telescope Survey of the Orion A and B Molecular Clouds. I. A Census of Dusty Young Stellar Objects and a Study of Their Mid-infrared Variability}",
      journal = {\aj},
         year = 2012,
        month = dec,
       volume = {144},
       number = {6},
          eid = {192},
        pages = {192},
          doi = {10.1088/0004-6256/144/6/192},
archivePrefix = {arXiv},
       eprint = {1209.3826},
 primaryClass = {astro-ph.GA},
       adsurl = {https://ui.adsabs.harvard.edu/abs/2012AJ....144..192M}
}

@ARTICLE{1996ARA&A..34..207H,
       author = {{Hartmann}, Lee and {Kenyon}, Scott J.},
        title = "{The FU Orionis Phenomenon}",
      journal = {\araa},
         year = 1996,
        month = jan,
       volume = {34},
        pages = {207-240},
          doi = {10.1146/annurev.astro.34.1.207},
       adsurl = {https://ui.adsabs.harvard.edu/abs/1996ARA&A..34..207H}
}

@INPROCEEDINGS{2014prpl.conf..387A,
       author = {{Audard}, M. and {{\'A}brah{\'a}m}, P. and {Dunham}, M.~M. and {Green}, J.~D. and {Grosso}, N. and {Hamaguchi}, K. and {Kastner}, J.~H. and {K{\'o}sp{\'a}l}, {\'A}. and {Lodato}, G. and {Romanova}, M.~M. and {Skinner}, S.~L. and {Vorobyov}, E.~I. and {Zhu}, Z.},
        title = "{Episodic Accretion in Young Stars}",
    booktitle = {Protostars and Planets VI},
         year = 2014,
       editor = {{Beuther}, Henrik and {Klessen}, Ralf S. and {Dullemond}, Cornelis P. and {Henning}, Thomas},
        month = jan,
        pages = {387-410},
          doi = {10.2458/azu_uapress_9780816531240-ch017},
archivePrefix = {arXiv},
       eprint = {1401.3368},
 primaryClass = {astro-ph.SR},
       adsurl = {https://ui.adsabs.harvard.edu/abs/2014prpl.conf..387A}
}

@ARTICLE{2004ApJ...606L.123B,
       author = {{Brice{\~n}o}, C. and {Vivas}, A.~K. and {Hern{\'a}ndez}, J. and {Calvet}, N. and {Hartmann}, L. and {Megeath}, T. and {Berlind}, P. and {Calkins}, M. and {Hoyer}, S.},
        title = "{MCNeil's Nebula in Orion: The Outburst History}",
      journal = {\apjl},
         year = 2004,
        month = may,
       volume = {606},
       number = {2},
        pages = {L123-L126},
          doi = {10.1086/421395},
archivePrefix = {arXiv},
       eprint = {astro-ph/0404012},
 primaryClass = {astro-ph},
       adsurl = {https://ui.adsabs.harvard.edu/abs/2004ApJ...606L.123B}
}

@ARTICLE{2016ApJS..224....5F,
       author = {{Furlan}, E. and {Fischer}, W.~J. and {Ali}, B. and {Stutz}, A.~M. and {Stanke}, T. and {Tobin}, J.~J. and {Megeath}, S.~T. and {Osorio}, M. and {Hartmann}, L. and {Calvet}, N. and {Poteet}, C.~A. and {Booker}, J. and {Manoj}, P. and {Watson}, D.~M. and {Allen}, L.},
        title = "{The Herschel Orion Protostar Survey: Spectral Energy Distributions and Fits Using a Grid of Protostellar Models}",
      journal = {\apjs},
         year = 2016,
        month = may,
       volume = {224},
       number = {1},
          eid = {5},
        pages = {5},
          doi = {10.3847/0067-0049/224/1/5},
archivePrefix = {arXiv},
       eprint = {1602.07314},
 primaryClass = {astro-ph.SR},
       adsurl = {https://ui.adsabs.harvard.edu/abs/2016ApJS..224....5F}
}

@ARTICLE{2023MNRAS.524.5548S,
       author = {{Siwak}, Micha{\l} and {Hillenbrand}, Lynne A. and {K{\'o}sp{\'a}l}, {\'A}gnes and {{\'A}brah{\'a}m}, P{\'e}ter and {Giannini}, Teresa and {De}, Kishalay and {Mo{\'o}r}, Attila and {Szil{\'a}gyi}, M{\'a}t{\'e} and {Jan{\'\i}k}, Jan and {Koen}, Chris and {Park}, Sunkyung and {Nagy}, Zs{\'o}fia and {Cruz-S{\'a}enz de Miera}, Fernando and {Fiorellino}, Eleonora and {Marton}, G{\'a}bor and {Kun}, M{\'a}ria and {Lucas}, Philip W. and {Udalski}, Andrzej and {Szab{\'o}}, Zs{\'o}fia Marianna},
        title = "{Gaia21bty: An EXor light curve exhibiting a FUor spectrum}",
      journal = {\mnras},
         year = 2023,
        month = oct,
       volume = {524},
       number = {4},
        pages = {5548-5565},
          doi = {10.1093/mnras/stad2135},
archivePrefix = {arXiv},
       eprint = {2307.08802},
 primaryClass = {astro-ph.SR},
       adsurl = {https://ui.adsabs.harvard.edu/abs/2023MNRAS.524.5548S}
}

@ARTICLE{2009ApJ...692L..67A,
       author = {{Aspin}, Colin and {Reipurth}, Bo and {Beck}, Tracy L. and {Aldering}, Greg and {Doering}, Ryan L. and {Hammel}, Heidi B. and {Lynch}, David K. and {Meixner}, Margaret and {Pecontal}, Emmanuel and {Russell}, Ray W. and {Sitko}, Michael L. and {Thomas}, Rollin C. and {U}, Vivian},
        title = "{V1647 Orionis: Reinvigorated Accretion and the Re-Appearance of McNeil's Nebula}",
      journal = {\apjl},
         year = 2009,
        month = feb,
       volume = {692},
       number = {2},
        pages = {L67-L71},
          doi = {10.1088/0004-637X/692/2/L67},
archivePrefix = {arXiv},
       eprint = {0901.1338},
 primaryClass = {astro-ph.SR},
       adsurl = {https://ui.adsabs.harvard.edu/abs/2009ApJ...692L..67A}
}

@ARTICLE{2009AJ....138.1137A,
       author = {{Aspin}, Colin and {Reipurth}, Bo},
        title = "{V1647 Orionis: Optical Photometric and Spectroscopic Monitoring Through the 2003-2006 Outburst}",
      journal = {\aj},
         year = 2009,
        month = oct,
       volume = {138},
       number = {4},
        pages = {1137-1158},
          doi = {10.1088/0004-6256/138/4/1137},
archivePrefix = {arXiv},
       eprint = {0907.4402},
 primaryClass = {astro-ph.SR},
       adsurl = {https://ui.adsabs.harvard.edu/abs/2009AJ....138.1137A}
}

@MISC{2015nwis.rept....1C,
       author = {{Cutri}, R.~M. and {Mainzer}, A. and {Conrow}, T. and {Masci}, F. and {Bauer}, J. and {Dailey}, J. and {Kirkpatrick}, J.~D. and {Fajardo-Acosta}, S. and {Gelino}, C. and {Grillmair}, C. and {Wheelock}, S.~L. and {Yan}, L. and {Harbut}, M. and {Beck}, R. and {Wittman}, M. and {Wright}, E.~L. and {Masiero}, J. and {Grav}, T. and {Sonnett}, S. and {Nugent}, C. and {Kramer}, E. and {Stevenson}, R. and {Eisenhardt}, P.~R.~M. and {Fabinsky}, B. and {Tholen}, D. and {Papin}, M. and {Fowler}, J. and {McCallon}, H.},
        title = "{Explanatory Supplement to the NEOWISE Data Release Products}",
 howpublished = {Explanatory Supplement to the NEOWISE Data Release Products},
         year = 2015,
        month = mar,
        pages = {1},
       adsurl = {https://ui.adsabs.harvard.edu/abs/2015nwis.rept....1C}
}

@INPROCEEDINGS{2023ASPC..534..355F,
       author = {{Fischer}, W.~J. and {Hillenbrand}, L.~A. and {Herczeg}, G.~J. and {Johnstone}, D. and {Kospal}, A. and {Dunham}, M.~M.},
        title = "{Accretion Variability as a Guide to Stellar Mass Assembly}",
    booktitle = {Protostars and Planets VII},
         year = 2023,
       editor = {{Inutsuka}, S. and {Aikawa}, Y. and {Muto}, T. and {Tomida}, K. and {Tamura}, M.},
       series = {Astronomical Society of the Pacific Conference Series},
       volume = {534},
        month = jul,
        pages = {355},
          doi = {10.48550/arXiv.2203.11257},
archivePrefix = {arXiv},
       eprint = {2203.11257},
 primaryClass = {astro-ph.SR},
       adsurl = {https://ui.adsabs.harvard.edu/abs/2023ASPC..534..355F}
}

@ARTICLE{2017ApJ...844..109F,
       author = {{Forbrich}, Jan and {Reid}, Mark J. and {Menten}, Karl M. and {Rivilla}, Victor M. and {Wolk}, Scott J. and {Rau}, Urvashi and {Chandler}, Claire J.},
        title = "{Extreme Radio Flares and Associated X-Ray Variability from Young Stellar Objects in the Orion Nebula Cluster}",
      journal = {\apj},
         year = 2017,
        month = aug,
       volume = {844},
       number = {2},
          eid = {109},
        pages = {109},
          doi = {10.3847/1538-4357/aa7aa4},
archivePrefix = {arXiv},
       eprint = {1706.05562},
 primaryClass = {astro-ph.HE},
       adsurl = {https://ui.adsabs.harvard.edu/abs/2017ApJ...844..109F}
}

@ARTICLE{2020A&A...638L...4G,
       author = {{Grosso}, Nicolas and {Hamaguchi}, Kenji and {Principe}, David A. and {Kastner}, Joel H.},
        title = "{Evidence for magnetic activity at starbirth: a powerful X-ray flare from the Class 0 protostar HOPS 383}",
      journal = {\aap},
         year = 2020,
        month = jun,
       volume = {638},
          eid = {L4},
        pages = {L4},
          doi = {10.1051/0004-6361/202038185},
archivePrefix = {arXiv},
       eprint = {2006.02676},
 primaryClass = {astro-ph.HE},
       adsurl = {https://ui.adsabs.harvard.edu/abs/2020A&A...638L...4G}
}

@ARTICLE{2015AJ....150..118P,
       author = {{Poppenhaeger}, K. and {Cody}, A.~M. and {Covey}, K.~R. and {G{\"u}nther}, H.~M. and {Hillenbrand}, L.~A. and {Plavchan}, P. and {Rebull}, L.~M. and {Stauffer}, J.~R. and {Wolk}, S.~J. and {Espaillat}, C. and {Forbrich}, J. and {Gutermuth}, R.~A. and {Hora}, J.~L. and {Morales-Calder{\'o}n}, M. and {Song}, Inseok},
        title = "{YSOVAR: Mid-infrared Variability of Young Stellar Objects and Their Disks in the Cluster IRAS 20050+2720}",
      journal = {\aj},
         year = 2015,
        month = oct,
       volume = {150},
       number = {4},
          eid = {118},
        pages = {118},
          doi = {10.1088/0004-6256/150/4/118},
archivePrefix = {arXiv},
       eprint = {1507.04325},
 primaryClass = {astro-ph.SR},
       adsurl = {https://ui.adsabs.harvard.edu/abs/2015AJ....150..118P}
}

@ARTICLE{2014AJ....148...92R,
       author = {{Rebull}, L.~M. and {Cody}, A.~M. and {Covey}, K.~R. and {G{\"u}nther}, H.~M. and {Hillenbrand}, L.~A. and {Plavchan}, P. and {Poppenhaeger}, K. and {Stauffer}, J.~R. and {Wolk}, S.~J. and {Gutermuth}, R. and {Morales-Calder{\'o}n}, M. and {Song}, I. and {Barrado}, D. and {Bayo}, A. and {James}, D. and {Hora}, J.~L. and {Vrba}, F.~J. and {Alves de Oliveira}, C. and {Bouvier}, J. and {Carey}, S.~J. and {Carpenter}, J.~M. and {Favata}, F. and {Flaherty}, K. and {Forbrich}, J. and {Hernandez}, J. and {McCaughrean}, M.~J. and {Megeath}, S.~T. and {Micela}, G. and {Smith}, H.~A. and {Terebey}, S. and {Turner}, N. and {Allen}, L. and {Ardila}, D. and {Bouy}, H. and {Guieu}, S.},
        title = "{Young Stellar Object VARiability (YSOVAR): Long Timescale Variations in the Mid-infrared}",
      journal = {\aj},
         year = 2014,
        month = nov,
       volume = {148},
       number = {5},
          eid = {92},
        pages = {92},
          doi = {10.1088/0004-6256/148/5/92},
archivePrefix = {arXiv},
       eprint = {1408.6756},
 primaryClass = {astro-ph.SR},
       adsurl = {https://ui.adsabs.harvard.edu/abs/2014AJ....148...92R}
}

@ARTICLE{2016ApJ...833..104F,
       author = {{Flaherty}, Kevin M. and {DeMarchi}, Lindsay and {Muzerolle}, James and {Balog}, Zoltan and {Herbst}, William and {Megeath}, S. Thomas and {Furlan}, Elise and {Gutermuth}, Robert},
        title = "{Spitzer Observations of Long-term Infrared Variability among Young Stellar Objects in Chamaeleon I}",
      journal = {\apj},
         year = 2016,
        month = dec,
       volume = {833},
       number = {1},
          eid = {104},
        pages = {104},
          doi = {10.3847/1538-4357/833/1/104},
archivePrefix = {arXiv},
       eprint = {1609.09100},
 primaryClass = {astro-ph.SR},
       adsurl = {https://ui.adsabs.harvard.edu/abs/2016ApJ...833..104F}
}

@ARTICLE{2014AJ....147...83S,
       author = {{Stauffer}, John and {Cody}, Ann Marie and {Baglin}, Annie and {Alencar}, Silvia and {Rebull}, Luisa and {Hillenbrand}, Lynne A. and {Venuti}, Laura and {Turner}, Neal J. and {Carpenter}, John and {Plavchan}, Peter and {Findeisen}, Krzysztof and {Carey}, Sean and {Terebey}, Susan and {Morales-Calder{\'o}n}, Mar{\'\i}a and {Bouvier}, Jerome and {Micela}, Giusi and {Flaccomio}, Ettore and {Song}, Inseok and {Gutermuth}, Rob and {Hartmann}, Lee and {Calvet}, Nuria and {Whitney}, Barbara and {Barrado}, David and {Vrba}, Frederick J. and {Covey}, Kevin and {Herbst}, William and {Furesz}, Gabor and {Aigrain}, Suzanne and {Favata}, Fabio},
        title = "{CSI 2264: Characterizing Accretion-burst Dominated Light Curves for Young Stars in NGC 2264}",
      journal = {\aj},
         year = 2014,
        month = apr,
       volume = {147},
       number = {4},
          eid = {83},
        pages = {83},
          doi = {10.1088/0004-6256/147/4/83},
archivePrefix = {arXiv},
       eprint = {1401.6600},
 primaryClass = {astro-ph.SR},
       adsurl = {https://ui.adsabs.harvard.edu/abs/2014AJ....147...83S}
}

@ARTICLE{2013AJ....145...66F,
       author = {{Flaherty}, K.~M. and {Muzerolle}, J. and {Rieke}, G. and {Gutermuth}, R. and {Balog}, Z. and {Herbst}, W. and {Megeath}, S.~T.},
        title = "{Kinks and Dents in Protoplanetary Disks: Rapid Infrared Variability as Evidence for Large Structural Perturbations}",
      journal = {\aj},
         year = 2013,
        month = mar,
       volume = {145},
       number = {3},
          eid = {66},
        pages = {66},
          doi = {10.1088/0004-6256/145/3/66},
archivePrefix = {arXiv},
       eprint = {1212.1731},
 primaryClass = {astro-ph.SR},
       adsurl = {https://ui.adsabs.harvard.edu/abs/2013AJ....145...66F}
}

@ARTICLE{2011ApJ...733...50M,
       author = {{Morales-Calder{\'o}n}, M. and {Stauffer}, J.~R. and {Hillenbrand}, L.~A. and {Gutermuth}, R. and {Song}, I. and {Rebull}, L.~M. and {Plavchan}, P. and {Carpenter}, J.~M. and {Whitney}, B.~A. and {Covey}, K. and {Alves de Oliveira}, C. and {Winston}, E. and {McCaughrean}, M.~J. and {Bouvier}, J. and {Guieu}, S. and {Vrba}, F.~J. and {Holtzman}, J. and {Marchis}, F. and {Hora}, J.~L. and {Wasserman}, L.~H. and {Terebey}, S. and {Megeath}, T. and {Guinan}, E. and {Forbrich}, J. and {Hu{\'e}lamo}, N. and {Riviere-Marichalar}, P. and {Barrado}, D. and {Stapelfeldt}, K. and {Hern{\'a}ndez}, J. and {Allen}, L.~E. and {Ardila}, D.~R. and {Bayo}, A. and {Favata}, F. and {James}, D. and {Werner}, M. and {Wood}, K.},
        title = "{Ysovar: The First Sensitive, Wide-area, Mid-infrared Photometric Monitoring of the Orion Nebula Cluster}",
      journal = {\apj},
         year = 2011,
        month = may,
       volume = {733},
       number = {1},
          eid = {50},
        pages = {50},
          doi = {10.1088/0004-637X/733/1/50},
archivePrefix = {arXiv},
       eprint = {1103.5238},
 primaryClass = {astro-ph.SR},
       adsurl = {https://ui.adsabs.harvard.edu/abs/2011ApJ...733...50M}
}

@ARTICLE{2010ApJ...710..470D,
       author = {{Dunham}, Michael M. and {Evans}, Neal J., II and {Terebey}, Susan and {Dullemond}, Cornelis P. and {Young}, Chadwick H.},
        title = "{Evolutionary Signatures in the Formation of Low-Mass Protostars. II. Toward Reconciling Models and Observations}",
      journal = {\apj},
         year = 2010,
        month = feb,
       volume = {710},
       number = {1},
        pages = {470-502},
          doi = {10.1088/0004-637X/710/1/470},
archivePrefix = {arXiv},
       eprint = {0912.5229},
 primaryClass = {astro-ph.GA},
       adsurl = {https://ui.adsabs.harvard.edu/abs/2010ApJ...710..470D}
}

@ARTICLE{2011ApJ...736...53O,
       author = {{Offner}, Stella S.~R. and {McKee}, Christopher F.},
        title = "{The Protostellar Luminosity Function}",
      journal = {\apj},
         year = 2011,
        month = jul,
       volume = {736},
       number = {1},
          eid = {53},
        pages = {53},
          doi = {10.1088/0004-637X/736/1/53},
archivePrefix = {arXiv},
       eprint = {1105.0671},
 primaryClass = {astro-ph.SR},
       adsurl = {https://ui.adsabs.harvard.edu/abs/2011ApJ...736...53O}
}

@ARTICLE{2019ApJ...872..183F,
       author = {{Fischer}, William J. and {Safron}, Emily and {Megeath}, S. Thomas},
        title = "{Constraining the Rate of Protostellar Accretion Outbursts in the Orion Molecular Clouds}",
      journal = {\apj},
         year = 2019,
        month = feb,
       volume = {872},
       number = {2},
          eid = {183},
        pages = {183},
          doi = {10.3847/1538-4357/ab01dc},
archivePrefix = {arXiv},
       eprint = {1901.08062},
 primaryClass = {astro-ph.SR},
       adsurl = {https://ui.adsabs.harvard.edu/abs/2019ApJ...872..183F}
}

@ARTICLE{2024ApJ...966..215M,
       author = {{Mairs}, Steve and {Lee}, Seonjae and {Johnstone}, Doug and {Broughton}, Colton and {Lee}, Jeong-Eun and {Herczeg}, Gregory J. and {Bell}, Graham S. and {Chen}, Zhiwei and {Contreras-Pe{\~n}a}, Carlos and {Francis}, Logan and {Hatchell}, Jennifer and {Kim}, Mi-Ryang and {Liu}, Sheng-Yuan and {Park}, Geumsook and {Qiu}, Keping and {Wang}, Yao-Te and {Zhang}, Xu and {JCMT Transient Team}},
        title = "{The JCMT Transient Survey: Six Year Summary of 450/850 {\ensuremath{\mu}}m Protostellar Variability and Calibration Pipeline Version 2.0}",
      journal = {\apj},
         year = 2024,
        month = may,
       volume = {966},
       number = {2},
          eid = {215},
        pages = {215},
          doi = {10.3847/1538-4357/ad35b6},
archivePrefix = {arXiv},
       eprint = {2401.03549},
 primaryClass = {astro-ph.IM},
       adsurl = {https://ui.adsabs.harvard.edu/abs/2024ApJ...966..215M}
}

@ARTICLE{1987PASP...99..191S,
       author = {{Stetson}, Peter B.},
        title = "{DAOPHOT: A Computer Program for Crowded-Field Stellar Photometry}",
      journal = {\pasp},
         year = 1987,
        month = mar,
       volume = {99},
        pages = {191},
          doi = {10.1086/131977},
       adsurl = {https://ui.adsabs.harvard.edu/abs/1987PASP...99..191S}
}

@ARTICLE{2009ApJS..181..321E,
       author = {{Evans}, Neal J., II and {Dunham}, Michael M. and {J{\o}rgensen}, Jes K. and {Enoch}, Melissa L. and {Mer{\'\i}n}, Bruno and {van Dishoeck}, Ewine F. and {Alcal{\'a}}, Juan M. and {Myers}, Philip C. and {Stapelfeldt}, Karl R. and {Huard}, Tracy L. and {Allen}, Lori E. and {Harvey}, Paul M. and {van Kempen}, Tim and {Blake}, Geoffrey A. and {Koerner}, David W. and {Mundy}, Lee G. and {Padgett}, Deborah L. and {Sargent}, Anneila I.},
        title = "{The Spitzer c2d Legacy Results: Star-Formation Rates and Efficiencies; Evolution and Lifetimes}",
      journal = {\apjs},
         year = 2009,
        month = apr,
       volume = {181},
       number = {2},
        pages = {321-350},
          doi = {10.1088/0067-0049/181/2/321},
       adsurl = {https://ui.adsabs.harvard.edu/abs/2009ApJS..181..321E}
}

@ARTICLE{2009ApJS..184...18G,
       author = {{Gutermuth}, R.~A. and {Megeath}, S.~T. and {Myers}, P.~C. and {Allen}, L.~E. and {Pipher}, J.~L. and {Fazio}, G.~G.},
        title = "{A Spitzer Survey of Young Stellar Clusters Within One Kiloparsec of the Sun: Cluster Core Extraction and Basic Structural Analysis}",
      journal = {\apjs},
         year = 2009,
        month = sep,
       volume = {184},
       number = {1},
        pages = {18-83},
          doi = {10.1088/0067-0049/184/1/18},
       adsurl = {https://ui.adsabs.harvard.edu/abs/2009ApJS..184...18G}
}

@INPROCEEDINGS{2000ESASP.445..201O,
       author = {{Olofsson}, G. and {Kaas}, A.~A. and {Bontemps}, S. and {Nordh}, L. and {Huldtgren}, M. and {Abergel}, A. and {Andr{\'e}}, P. and {Boulanger}, F. and {Burgdorf}, M. and {Cabrit}, S. and {Casali}, M.~M. and {Cesarsky}, C.~J. and {Copet}, E. and {Davies}, J. and {Falgarone}, E. and {Montmerle}, T. and {Perault}, M. and {Persi}, P. and {Prusti}, T. and {Puget}, J.~L. and {Sibille}, F.},
        title = "{ISOCAM Observations of Nearby Star Formation Regions}",
    booktitle = {Star Formation from the Small to the Large Scale},
         year = 2000,
       editor = {{Favata}, F. and {Kaas}, A. and {Wilson}, A.},
       series = {ESA Special Publication},
       volume = {445},
        month = jun,
        pages = {201},
       adsurl = {https://ui.adsabs.harvard.edu/abs/2000ESASP.445..201O}
}

@ARTICLE{2004A&A...421..623K,
       author = {{Kaas}, A.~A. and {Olofsson}, G. and {Bontemps}, S. and {Andr{\'e}}, P. and {Nordh}, L. and {Huldtgren}, M. and {Prusti}, T. and {Persi}, P. and {Delgado}, A.~J. and {Motte}, F. and {Abergel}, A. and {Boulanger}, F. and {Burgdorf}, M. and {Casali}, M.~M. and {Cesarsky}, C.~J. and {Davies}, J. and {Falgarone}, E. and {Montmerle}, T. and {Perault}, M. and {Puget}, J.~L. and {Sibille}, F.},
        title = "{The young stellar population in the Serpens Cloud Core:  An ISOCAM survey}",
      journal = {\aap},
         year = 2004,
        month = jul,
       volume = {421},
        pages = {623-642},
          doi = {10.1051/0004-6361:20035775},
       adsurl = {https://ui.adsabs.harvard.edu/abs/2004A&A...421..623K}
}

@ARTICLE{2020ApJ...896...60P,
       author = {{Pokhrel}, Riwaj and {Gutermuth}, Robert A. and {Betti}, Sarah K. and {Offner}, Stella S.~R. and {Myers}, Philip C. and {Megeath}, S. Thomas and {Sokol}, Alyssa D. and {Ali}, Babar and {Allen}, Lori and {Allen}, Thomas S. and {Dunham}, Michael M. and {Fischer}, William J. and {Henning}, Thomas and {Heyer}, Mark and {Hora}, Joseph L. and {Pipher}, Judith L. and {Tobin}, John J. and {Wolk}, Scott J.},
        title = "{Star-Gas Surface Density Correlations in 12 Nearby Molecular Clouds. I. Data Collection and Star-sampled Analysis}",
      journal = {\apj},
         year = 2020,
        month = jun,
       volume = {896},
       number = {1},
          eid = {60},
        pages = {60},
          doi = {10.3847/1538-4357/ab92a2},
archivePrefix = {arXiv},
       eprint = {2005.05466},
 primaryClass = {astro-ph.GA},
       adsurl = {https://ui.adsabs.harvard.edu/abs/2020ApJ...896...60P}
}

@ARTICLE{2016AJ....151....5M,
       author = {{Megeath}, S.~T. and {Gutermuth}, R. and {Muzerolle}, J. and {Kryukova}, E. and {Hora}, J.~L. and {Allen}, L.~E. and {Flaherty}, K. and {Hartmann}, L. and {Myers}, P.~C. and {Pipher}, J.~L. and {Stauffer}, J. and {Young}, E.~T. and {Fazio}, G.~G.},
        title = "{The Spitzer Space Telescope Survey of the Orion A and B Molecular Clouds. II. The Spatial Distribution and Demographics of Dusty Young Stellar Objects}",
      journal = {\aj},
         year = 2016,
        month = jan,
       volume = {151},
       number = {1},
          eid = {5},
        pages = {5},
          doi = {10.3847/0004-6256/151/1/5},
archivePrefix = {arXiv},
       eprint = {1511.01202},
 primaryClass = {astro-ph.GA},
       adsurl = {https://ui.adsabs.harvard.edu/abs/2016AJ....151....5M}
}

@ARTICLE{2005PASP..117..978R,
       author = {{Reach}, William T. and {Megeath}, S.~T. and {Cohen}, Martin and {Hora}, J. and {Carey}, Sean and {Surace}, Jason and {Willner}, S.~P. and {Barmby}, P. and {Wilson}, Gillian and {Glaccum}, William and {Lowrance}, Patrick and {Marengo}, Massimo and {Fazio}, Giovanni G.},
        title = "{Absolute Calibration of the Infrared Array Camera on the Spitzer Space Telescope}",
      journal = {\pasp},
         year = 2005,
        month = sep,
       volume = {117},
       number = {835},
        pages = {978-990},
          doi = {10.1086/432670},
archivePrefix = {arXiv},
       eprint = {astro-ph/0507139},
 primaryClass = {astro-ph},
       adsurl = {https://ui.adsabs.harvard.edu/abs/2005PASP..117..978R}
}

@ARTICLE{2012AJ....144...31K,
       author = {{Kryukova}, E. and {Megeath}, S.~T. and {Gutermuth}, R.~A. and {Pipher}, J. and {Allen}, T.~S. and {Allen}, L.~E. and {Myers}, P.~C. and {Muzerolle}, J.},
        title = "{Luminosity Functions of Spitzer-identified Protostars in Nine Nearby Molecular Clouds}",
      journal = {\aj},
         year = 2012,
        month = aug,
       volume = {144},
       number = {2},
          eid = {31},
        pages = {31},
          doi = {10.1088/0004-6256/144/2/31},
archivePrefix = {arXiv},
       eprint = {1204.1535},
 primaryClass = {astro-ph.SR},
       adsurl = {https://ui.adsabs.harvard.edu/abs/2012AJ....144...31K}
}

@ARTICLE{2024RNAAS...8...64K,
       author = {{Kulkarni}, Chinmay S. and {Behling}, Thomas and {Burns-Watson}, Nathanael and {Jones}, Jason and {Robbins}, Tyler and {Megeath}, S. Thomas and {Federman}, Samuel and {Fischer}, William J. and {Gutermuth}, Robert and {Pokhrel}, Riwaj and {Zakri}, Wafa and {Oliveira}, Savio B.},
        title = "{The Infrared Variability of V1647 Ori over 26 yr with Spitzer, (NEO)WISE, and ISO}",
      journal = {Research Notes of the American Astronomical Society},
         year = 2024,
        month = mar,
       volume = {8},
       number = {3},
          eid = {64},
        pages = {64},
          doi = {10.3847/2515-5172/ad2f40},
       adsurl = {https://ui.adsabs.harvard.edu/abs/2024RNAAS...8...64K}
}

@ARTICLE{2004IAUC.8284....1M,
       author = {{McNeil}, J.~W. and {Reipurth}, B. and {Meech}, K.},
        title = "{IRAS 05436-0007}",
      journal = {\iaucirc},
         year = 2004,
        month = feb,
       volume = {8284},
        pages = {1},
       adsurl = {https://ui.adsabs.harvard.edu/abs/2004IAUC.8284....1M}
}

@ARTICLE{2004IAUC.8354....1S,
       author = {{Samus}, N.~N.},
        title = "{V1647 Orionis = IRAS 05436-0007}",
      journal = {\iaucirc},
         year = 2004,
        month = jun,
       volume = {8354},
        pages = {1},
       adsurl = {https://ui.adsabs.harvard.edu/abs/2004IAUC.8354....1S}
}

@ARTICLE{2014ApJ...791..131K,
       author = {{Koenig}, X.~P. and {Leisawitz}, D.~T.},
        title = "{A Classification Scheme for Young Stellar Objects Using the Wide-field Infrared Survey Explorer AllWISE Catalog: Revealing Low-density Star Formation in the Outer Galaxy}",
      journal = {\apj},
         year = 2014,
        month = aug,
       volume = {791},
       number = {2},
          eid = {131},
        pages = {131},
          doi = {10.1088/0004-637X/791/2/131},
archivePrefix = {arXiv},
       eprint = {1407.2262},
 primaryClass = {astro-ph.GA},
       adsurl = {https://ui.adsabs.harvard.edu/abs/2014ApJ...791..131K}
}

@INPROCEEDINGS{2010ASPC..432..197D,
       author = {{Dunham}, M.~M. and {Evans}, N.~J., II and {Terebey}, S. and {Dullemond}, C.~P. and {Young}, C.~H.},
        title = "{Evolutionary Models of the Formation of Protostars out of Low-Mass, Dense Cores: Towards Reconciling Models and Observations}",
    booktitle = {New Horizons in Astronomy: Frank N. Bash Symposium 2009},
         year = 2010,
       editor = {{Stanford}, L.~M. and {Green}, J.~D. and {Hao}, L. and {Mao}, Y.},
       series = {Astronomical Society of the Pacific Conference Series},
       volume = {432},
        month = oct,
        pages = {197},
       adsurl = {https://ui.adsabs.harvard.edu/abs/2010ASPC..432..197D}
}

@ARTICLE{2004ApJS..154...10F,
       author = {{Fazio}, G.~G. and {Hora}, J.~L. and {Allen}, L.~E. and {Ashby}, M.~L.~N. and {Barmby}, P. and {Deutsch}, L.~K. and {Huang}, J. -S. and {Kleiner}, S. and {Marengo}, M. and {Megeath}, S.~T. and {Melnick}, G.~J. and {Pahre}, M.~A. and {Patten}, B.~M. and {Polizotti}, J. and {Smith}, H.~A. and {Taylor}, R.~S. and {Wang}, Z. and {Willner}, S.~P. and {Hoffmann}, W.~F. and {Pipher}, J.~L. and {Forrest}, W.~J. and {McMurty}, C.~W. and {McCreight}, C.~R. and {McKelvey}, M.~E. and {McMurray}, R.~E. and {Koch}, D.~G. and {Moseley}, S.~H. and {Arendt}, R.~G. and {Mentzell}, J.~E. and {Marx}, C.~T. and {Losch}, P. and {Mayman}, P. and {Eichhorn}, W. and {Krebs}, D. and {Jhabvala}, M. and {Gezari}, D.~Y. and {Fixsen}, D.~J. and {Flores}, J. and {Shakoorzadeh}, K. and {Jungo}, R. and {Hakun}, C. and {Workman}, L. and {Karpati}, G. and {Kichak}, R. and {Whitley}, R. and {Mann}, S. and {Tollestrup}, E.~V. and {Eisenhardt}, P. and {Stern}, D. and {Gorjian}, V. and {Bhattacharya}, B. and {Carey}, S. and {Nelson}, B.~O. and {Glaccum}, W.~J. and {Lacy}, M. and {Lowrance}, P.~J. and {Laine}, S. and {Reach}, W.~T. and {Stauffer}, J.~A. and {Surace}, J.~A. and {Wilson}, G. and {Wright}, E.~L. and {Hoffman}, A. and {Domingo}, G. and {Cohen}, M.},
        title = "{The Infrared Array Camera (IRAC) for the Spitzer Space Telescope}",
      journal = {\apjs},
         year = 2004,
        month = sep,
       volume = {154},
       number = {1},
        pages = {10-17},
          doi = {10.1086/422843},
archivePrefix = {arXiv},
       eprint = {astro-ph/0405616},
 primaryClass = {astro-ph},
       adsurl = {https://ui.adsabs.harvard.edu/abs/2004ApJS..154...10F}
}

@ARTICLE{2013MNRAS.430.2910S,
       author = {{Scholz}, Alexander and {Froebrich}, Dirk and {Wood}, Kenneth},
        title = "{A systematic survey for eruptive young stellar objects using mid-infrared photometry}",
      journal = {\mnras},
         year = 2013,
        month = apr,
       volume = {430},
       number = {4},
        pages = {2910-2922},
          doi = {10.1093/mnras/stt091},
archivePrefix = {arXiv},
       eprint = {1301.3152},
 primaryClass = {astro-ph.SR},
       adsurl = {https://ui.adsabs.harvard.edu/abs/2013MNRAS.430.2910S}
}

@ARTICLE{2015ApJ...800L...5S,
       author = {{Safron}, Emily J. and {Fischer}, William J. and {Megeath}, S. Thomas and {Furlan}, Elise and {Stutz}, Amelia M. and {Stanke}, Thomas and {Billot}, Nicolas and {Rebull}, Luisa M. and {Tobin}, John J. and {Ali}, Babar and {Allen}, Lori E. and {Booker}, Joseph and {Watson}, Dan M. and {Wilson}, T.~L.},
        title = "{Hops 383: an Outbursting Class 0 Protostar in Orion}",
      journal = {\apjl},
         year = 2015,
        month = feb,
       volume = {800},
       number = {1},
          eid = {L5},
        pages = {L5},
          doi = {10.1088/2041-8205/800/1/L5},
archivePrefix = {arXiv},
       eprint = {1501.00492},
 primaryClass = {astro-ph.SR},
       adsurl = {https://ui.adsabs.harvard.edu/abs/2015ApJ...800L...5S}
}

@ARTICLE{2011A&A...526L...1C,
       author = {{Caratti o Garatti}, A. and {Garcia Lopez}, R. and {Scholz}, A. and {Giannini}, T. and {Eisl{\"o}ffel}, J. and {Nisini}, B. and {Massi}, F. and {Antoniucci}, S. and {Ray}, T.~P.},
        title = "{The outburst of an embedded low-mass YSO in L1641}",
      journal = {\aap},
         year = 2011,
        month = feb,
       volume = {526},
          eid = {L1},
        pages = {L1},
          doi = {10.1051/0004-6361/201016146},
archivePrefix = {arXiv},
       eprint = {1012.0281},
 primaryClass = {astro-ph.SR},
       adsurl = {https://ui.adsabs.harvard.edu/abs/2011A&A...526L...1C}
}

@INPROCEEDINGS{2016IAUS..314..209T,
       author = {{Terebey}, S. and {Cody}, A.~M. and {Rebull}, L.~M. and {Stauffer}, J.~R.},
        title = "{Mid-infrared Variability and Accretion in NGC 2264 Protostars}",
    booktitle = {Young Stars \& Planets Near the Sun},
         year = 2016,
       editor = {{Kastner}, J.~H. and {Stelzer}, B. and {Metchev}, S.~A.},
       series = {IAU Symposium},
       volume = {314},
        month = jan,
        pages = {209-210},
          doi = {10.1017/S1743921315006353},
       adsurl = {https://ui.adsabs.harvard.edu/abs/2016IAUS..314..209T}
}

@ARTICLE{2007ApJ...663.1069F,
       author = {{Flaherty}, K.~M. and {Pipher}, J.~L. and {Megeath}, S.~T. and {Winston}, E.~M. and {Gutermuth}, R.~A. and {Muzerolle}, J. and {Allen}, L.~E. and {Fazio}, G.~G.},
        title = "{Infrared Extinction toward Nearby Star-forming Regions}",
      journal = {\apj},
         year = 2007,
        month = jul,
       volume = {663},
       number = {2},
        pages = {1069-1082},
          doi = {10.1086/518411},
archivePrefix = {arXiv},
       eprint = {astro-ph/0703777},
 primaryClass = {astro-ph},
       adsurl = {https://ui.adsabs.harvard.edu/abs/2007ApJ...663.1069F}
}

@ARTICLE{2024AJ....167...82F,
       author = {{Fischer}, William J. and {Battersby}, Cara and {Johnstone}, Doug and {Lee}, Rachel and {Sewi{\l}o}, Marta and {Beuther}, Henrik and {Hasegawa}, Yasuhiro and {Ginsburg}, Adam and {Pontoppidan}, Klaus},
        title = "{Far-infrared Luminosity Bursts Trace Mass Accretion onto Protostars}",
      journal = {\aj},
         year = 2024,
        month = feb,
       volume = {167},
       number = {2},
          eid = {82},
        pages = {82},
          doi = {10.3847/1538-3881/ad188b},
archivePrefix = {arXiv},
       eprint = {2310.12867},
 primaryClass = {astro-ph.SR},
       adsurl = {https://ui.adsabs.harvard.edu/abs/2024AJ....167...82F}
}

@ARTICLE{2019AJ....158..240H,
       author = {{Hillenbrand}, Lynne A. and {Reipurth}, Bo and {Connelley}, Michael and {Cutri}, Roc M. and {Isaacson}, Howard},
        title = "{Gaia 19ajj: A Young Star Brightening Due to Enhanced Accretion and Reduced Extinction}",
      journal = {\aj},
         year = 2019,
        month = dec,
       volume = {158},
       number = {6},
          eid = {240},
        pages = {240},
          doi = {10.3847/1538-3881/ab4e16},
archivePrefix = {arXiv},
       eprint = {1910.05790},
 primaryClass = {astro-ph.SR},
       adsurl = {https://ui.adsabs.harvard.edu/abs/2019AJ....158..240H}
}

@ARTICLE{2023ApJ...957..113W,
       author = {{Wang}, Mu-Tian and {Herczeg}, Gregory J. and {Liu}, Hui-Gen and {Fang}, Min and {Johnstone}, Doug and {Lee}, Ho-Gyu and {Walter}, Frederick M. and {Hambsch}, Franz-Josef and {Contreras Pe{\~n}a}, Carlos and {Lee}, Jeong-Eun and {Millward}, Mervyn and {Pearce}, Andrew and {Monard}, Berto and {Zhou}, Lihang},
        title = "{The Accretion History of EX Lup: A Century of Bursts, Outbursts, and Quiescence}",
      journal = {\apj},
         year = 2023,
        month = nov,
       volume = {957},
       number = {2},
          eid = {113},
        pages = {113},
          doi = {10.3847/1538-4357/acf2f4},
archivePrefix = {arXiv},
       eprint = {2308.11895},
 primaryClass = {astro-ph.SR},
       adsurl = {https://ui.adsabs.harvard.edu/abs/2023ApJ...957..113W}
}

@ARTICLE{2023ApJ...957....8W,
       author = {{Wang}, Tinggui and {Li}, Jiaxun and {N. Mace}, Gregory and {Ji}, Tuo and {Jiang}, Ning and {Zhu}, Qingfeng and {Fang}, Min},
        title = "{A Gigantic Mid-infrared Outburst in an Embedded Class I Young Stellar Object J064722.95+031644.6}",
      journal = {\apj},
         year = 2023,
        month = nov,
       volume = {957},
       number = {1},
          eid = {8},
        pages = {8},
          doi = {10.3847/1538-4357/acf92e},
archivePrefix = {arXiv},
       eprint = {2309.11016},
 primaryClass = {astro-ph.SR},
       adsurl = {https://ui.adsabs.harvard.edu/abs/2023ApJ...957....8W}
}

@ARTICLE{2018ApJ...869..146H,
       author = {{Hillenbrand}, Lynne A. and {Contreras Pe{\~n}a}, Carlos and {Morrell}, Sam and {Naylor}, Tim and {Kuhn}, Michael A. and {Cutri}, Roc M. and {Rebull}, Luisa M. and {Hodgkin}, Simon and {Froebrich}, Dirk and {Mainzer}, Amy K.},
        title = "{Gaia 17bpi: An FU Ori-type Outburst}",
      journal = {\apj},
         year = 2018,
        month = dec,
       volume = {869},
       number = {2},
          eid = {146},
        pages = {146},
          doi = {10.3847/1538-4357/aaf414},
archivePrefix = {arXiv},
       eprint = {1812.06640},
 primaryClass = {astro-ph.SR},
       adsurl = {https://ui.adsabs.harvard.edu/abs/2018ApJ...869..146H}
}

@ARTICLE{2022RNAAS...6....6H,
       author = {{Hillenbrand}, Lynne A. and {Rodriguez}, Antonio C.},
        title = "{Expected FU Ori Outburst Amplitudes from the Optical to the Mid-infrared}",
      journal = {Research Notes of the American Astronomical Society},
         year = 2022,
        month = jan,
       volume = {6},
       number = {1},
          eid = {6},
        pages = {6},
          doi = {10.3847/2515-5172/ac4807},
archivePrefix = {arXiv},
       eprint = {2201.01012},
 primaryClass = {astro-ph.SR},
       adsurl = {https://ui.adsabs.harvard.edu/abs/2022RNAAS...6....6H}
}

@ARTICLE{2018ApJ...861..145C,
       author = {{Connelley}, Michael S. and {Reipurth}, Bo},
        title = "{A Near-infrared Spectroscopic Survey of FU Orionis Objects}",
      journal = {\apj},
         year = 2018,
        month = jul,
       volume = {861},
       number = {2},
          eid = {145},
        pages = {145},
          doi = {10.3847/1538-4357/aaba7b},
archivePrefix = {arXiv},
       eprint = {1806.08880},
 primaryClass = {astro-ph.SR},
       adsurl = {https://ui.adsabs.harvard.edu/abs/2018ApJ...861..145C}
}

@ARTICLE{2010ApJ...719L..50A,
       author = {{Aspin}, Colin and {Reipurth}, Bo and {Herczeg}, Gregory J. and {Capak}, Peter},
        title = "{The 2008 Extreme Outburst of the Young Eruptive Variable Star EX Lupi}",
      journal = {\apjl},
         year = 2010,
        month = aug,
       volume = {719},
       number = {1},
        pages = {L50-L55},
          doi = {10.1088/2041-8205/719/1/L50},
archivePrefix = {arXiv},
       eprint = {1007.4178},
 primaryClass = {astro-ph.SR},
       adsurl = {https://ui.adsabs.harvard.edu/abs/2010ApJ...719L..50A}
}

@ARTICLE{2015ApJS..220...11D,
       author = {{Dunham}, Michael M. and {Allen}, Lori E. and {Evans}, II, Neal J. and {Broekhoven-Fiene}, Hannah and {Cieza}, Lucas A. and {Di Francesco}, James and {Gutermuth}, Robert A. and {Harvey}, Paul M. and {Hatchell}, Jennifer and {Heiderman}, Amanda and {Huard}, Tracy L. and {Johnstone}, Doug and {Kirk}, Jason M. and {Matthews}, Brenda C. and {Miller}, Jennifer F. and {Peterson}, Dawn E. and {Young}, Kaisa E.},
        title = "{Young Stellar Objects in the Gould Belt}",
      journal = {\apjs},
         year = 2015,
        month = sep,
       volume = {220},
       number = {1},
          eid = {11},
        pages = {11},
          doi = {10.1088/0067-0049/220/1/11},
archivePrefix = {arXiv},
       eprint = {1508.03199},
 primaryClass = {astro-ph.GA},
       adsurl = {https://ui.adsabs.harvard.edu/abs/2015ApJS..220...11D}
}

@ARTICLE{2008ApJ...674..336G,
       author = {{Gutermuth}, R.~A. and {Myers}, P.~C. and {Megeath}, S.~T. and {Allen}, L.~E. and {Pipher}, J.~L. and {Muzerolle}, J. and {Porras}, A. and {Winston}, E. and {Fazio}, G.},
        title = "{Spitzer Observations of NGC 1333: A Study of Structure and Evolution in a Nearby Embedded Cluster}",
      journal = {\apj},
         year = 2008,
        month = feb,
       volume = {674},
       number = {1},
        pages = {336-356},
          doi = {10.1086/524722},
archivePrefix = {arXiv},
       eprint = {0710.1860},
 primaryClass = {astro-ph},
       adsurl = {https://ui.adsabs.harvard.edu/abs/2008ApJ...674..336G}
}

@ARTICLE{2023ApJS..266...32P,
       author = {{Pokhrel}, Riwaj and {Megeath}, S. Thomas and {Gutermuth}, Robert A. and {Furlan}, Elise and {Fischer}, William J. and {Federman}, Samuel and {Tobin}, John J. and {Stutz}, Amelia M. and {Hartmann}, Lee and {Osorio}, Mayra and {Watson}, Dan M. and {Stanke}, Thomas and {Manoj}, P. and {Narang}, Mayank and {Atnagulov}, Prabhani and {Habel}, Nolan and {Zakri}, Wafa},
        title = "{Extension of HOPS out to 500 pc (eHOPS). I. Identification and Modeling of Protostars in the Aquila Molecular Clouds}",
      journal = {\apjs},
         year = 2023,
        month = jun,
       volume = {266},
       number = {2},
          eid = {32},
        pages = {32},
          doi = {10.3847/1538-4365/acbfac},
archivePrefix = {arXiv},
       eprint = {2209.12090},
 primaryClass = {astro-ph.GA},
       adsurl = {https://ui.adsabs.harvard.edu/abs/2023ApJS..266...32P}
}

@ARTICLE{2009ApJ...692..973E,
       author = {{Enoch}, Melissa L. and {Evans}, II, Neal J. and {Sargent}, Anneila I. and {Glenn}, Jason},
        title = "{Properties of the Youngest Protostars in Perseus, Serpens, and Ophiuchus}",
      journal = {\apj},
         year = 2009,
        month = feb,
       volume = {692},
       number = {2},
        pages = {973-997},
          doi = {10.1088/0004-637X/692/2/973},
archivePrefix = {arXiv},
       eprint = {0809.4012},
 primaryClass = {astro-ph},
       adsurl = {https://ui.adsabs.harvard.edu/abs/2009ApJ...692..973E}
}

@ARTICLE{2001A&A...372..173B,
       author = {{Bontemps}, S. and {Andr{\'e}}, P. and {Kaas}, A.~A. and {Nordh}, L. and {Olofsson}, G. and {Huldtgren}, M. and {Abergel}, A. and {Blommaert}, J. and {Boulanger}, F. and {Burgdorf}, M. and {Cesarsky}, C.~J. and {Cesarsky}, D. and {Copet}, E. and {Davies}, J. and {Falgarone}, E. and {Lagache}, G. and {Montmerle}, T. and {P{\'e}rault}, M. and {Persi}, P. and {Prusti}, T. and {Puget}, J.~L. and {Sibille}, F.},
        title = "{ISOCAM observations of the rho Ophiuchi cloud: Luminosity and mass functions of the pre-main sequence embedded cluster}",
      journal = {\aap},
         year = 2001,
        month = jun,
       volume = {372},
        pages = {173-194},
          doi = {10.1051/0004-6361:20010474},
archivePrefix = {arXiv},
       eprint = {astro-ph/0103373},
 primaryClass = {astro-ph},
       adsurl = {https://ui.adsabs.harvard.edu/abs/2001A&A...372..173B}
}

@ARTICLE{1999A&A...350..883O,
       author = {{Olofsson}, G. and {Huldtgren}, M. and {Kaas}, A.~A. and {Bontemps}, S. and {Nordh}, L. and {Abergel}, A. and {Andr{\'e}}, P. and {Boulanger}, F. and {Burgdorf}, M. and {Casali}, M.~M. and {Cesarsky}, C.~J. and {Davies}, J. and {Falgarone}, E. and {Montmerle}, T. and {Perault}, M. and {Persi}, P. and {Prusti}, T. and {Puget}, J.~L. and {Sibille}, F.},
        title = "{ISOCAM observations of the RCrA star formation region}",
      journal = {\aap},
         year = 1999,
        month = oct,
       volume = {350},
        pages = {883-890},
       adsurl = {https://ui.adsabs.harvard.edu/abs/1999A&A...350..883O}
}

@ARTICLE{2000A&A...357..219P,
       author = {{Persi}, P. and {Marenzi}, A.~R. and {Olofsson}, G. and {Kaas}, A.~A. and {Nordh}, L. and {Huldtgren}, M. and {Abergel}, A. and {Andr{\'e}}, P. and {Bontemps}, S. and {Boulanger}, F. and {Burggdorf}, M. and {Casali}, M.~M. and {Cesarsky}, C.~J. and {Copet}, E. and {Davies}, J. and {Falgarone}, E. and {Montmerle}, T. and {Perault}, M. and {Prusti}, T. and {Puget}, J.~L. and {Sibille}, F.},
        title = "{ISOCAM observations of the Chamaeleon I dark cloud}",
      journal = {\aap},
         year = 2000,
        month = may,
       volume = {357},
        pages = {219-224},
       adsurl = {https://ui.adsabs.harvard.edu/abs/2000A&A...357..219P}
}

@MISC{2012ivoa.rept.1015R,
       author = {{Rodrigo}, Carlos and {Solano}, Enrique and {Bayo}, Amelia},
        title = "{SVO Filter Profile Service Version 1.0}",
 howpublished = {IVOA Working Draft 15 October 2012},
         year = 2012,
        month = oct,
        pages = {1015},
          doi = {10.5479/ADS/bib/2012ivoa.rept.1015R},
       adsurl = {https://ui.adsabs.harvard.edu/abs/2012ivoa.rept.1015R}
}

@ARTICLE{1996A&A...315L..32C,
       author = {{Cesarsky}, C.~J. and {Abergel}, A. and {Agnese}, P. and {Altieri}, B. and {Augueres}, J.~L. and {Aussel}, H. and {Biviano}, A. and {Blommaert}, J. and {Bonnal}, J.~F. and {Bortoletto}, F. and {Boulade}, O. and {Boulanger}, F. and {Cazes}, S. and {Cesarsky}, D.~A. and {Chedin}, A. and {Claret}, A. and {Combes}, M. and {Cretolle}, J. and {Davies}, J.~K. and {Desert}, F.~X. and {Elbaz}, D. and {Engelmann}, J.~J. and {Epstein}, G. and {Franceschini}, A. and {Gallais}, P. and {Gastaud}, R. and {Gorisse}, M. and {Guest}, S. and {Hawarden}, T. and {Imbault}, D. and {Kleczewski}, M. and {Lacombe}, F. and {Landriu}, D. and {Lapegue}, J. and {Lena}, P. and {Longair}, M.~S. and {Mandolesi}, R. and {Metcalfe}, L. and {Mosquet}, N. and {Nordh}, L. and {Okumura}, K. and {Ott}, S. and {Perault}, M. and {Perrier}, F. and {Persi}, P. and {Puget}, P. and {Purkins}, T. and {Rio}, Y. and {Robert}, T. and {Rouan}, D. and {Roy}, A. and {Saint-Pe}, O. and {Sam Lone}, J. and {Sargent}, A. and {Sauvage}, M. and {Sibille}, F. and {Siebenmorgen}, R. and {Sirou}, F. and {Soufflot}, A. and {Starck}, J.~L. and {Tiphene}, D. and {Tran}, D. and {Ventura}, G. and {Vigroux}, L. and {Vivares}, F. and {Wade}, R.},
        title = "{ISOCAM in flight.}",
      journal = {\aap},
         year = 1996,
        month = nov,
       volume = {315},
        pages = {L32-L37},
       adsurl = {https://ui.adsabs.harvard.edu/abs/1996A&A...315L..32C}
}

@PROCEEDINGS{2003ESASP1262.....B,
        title = "{The ISO Handbook Volume II: CAM - The ISO Camera (v 2.0)}",
    booktitle = {ESA Special Publication},
         year = 2003,
       editor = {{Blommaert}, Joris and {Siebenmorgen}, Ralf and {Coulais}, Alain and {Metcalfe}, Leo and {Miville-Desch{\^e}nes}, Marc-Antoine and {Okumura}, Koryo and {Ott}, Stephan and {Pollack}, Andy and {Sauvage}, Marc and {Starck}, Jean-Luc},
       series = {ESA Special Publication},
       volume = {1262},
        month = nov,
       adsurl = {https://ui.adsabs.harvard.edu/abs/2003ESASP1262.....B}
}

@ARTICLE{2010AJ....140.1868W,
       author = {{Wright}, Edward L. and {Eisenhardt}, Peter R.~M. and {Mainzer}, Amy K. and {Ressler}, Michael E. and {Cutri}, Roc M. and {Jarrett}, Thomas and {Kirkpatrick}, J. Davy and {Padgett}, Deborah and {McMillan}, Robert S. and {Skrutskie}, Michael and {Stanford}, S.~A. and {Cohen}, Martin and {Walker}, Russell G. and {Mather}, John C. and {Leisawitz}, David and {Gautier}, III, Thomas N. and {McLean}, Ian and {Benford}, Dominic and {Lonsdale}, Carol J. and {Blain}, Andrew and {Mendez}, Bryan and {Irace}, William R. and {Duval}, Valerie and {Liu}, Fengchuan and {Royer}, Don and {Heinrichsen}, Ingolf and {Howard}, Joan and {Shannon}, Mark and {Kendall}, Martha and {Walsh}, Amy L. and {Larsen}, Mark and {Cardon}, Joel G. and {Schick}, Scott and {Schwalm}, Mark and {Abid}, Mohamed and {Fabinsky}, Beth and {Naes}, Larry and {Tsai}, Chao-Wei},
        title = "{The Wide-field Infrared Survey Explorer (WISE): Mission Description and Initial On-orbit Performance}",
      journal = {\aj},
         year = 2010,
        month = dec,
       volume = {140},
       number = {6},
        pages = {1868-1881},
          doi = {10.1088/0004-6256/140/6/1868},
archivePrefix = {arXiv},
       eprint = {1008.0031},
 primaryClass = {astro-ph.IM},
       adsurl = {https://ui.adsabs.harvard.edu/abs/2010AJ....140.1868W}
}

@ARTICLE{2014ApJ...792...30M,
       author = {{Mainzer}, A. and {Bauer}, J. and {Cutri}, R.~M. and {Grav}, T. and {Masiero}, J. and {Beck}, R. and {Clarkson}, P. and {Conrow}, T. and {Dailey}, J. and {Eisenhardt}, P. and {Fabinsky}, B. and {Fajardo-Acosta}, S. and {Fowler}, J. and {Gelino}, C. and {Grillmair}, C. and {Heinrichsen}, I. and {Kendall}, M. and {Kirkpatrick}, J. Davy and {Liu}, F. and {Masci}, F. and {McCallon}, H. and {Nugent}, C.~R. and {Papin}, M. and {Rice}, E. and {Royer}, D. and {Ryan}, T. and {Sevilla}, P. and {Sonnett}, S. and {Stevenson}, R. and {Thompson}, D.~B. and {Wheelock}, S. and {Wiemer}, D. and {Wittman}, M. and {Wright}, E. and {Yan}, L.},
        title = "{Initial Performance of the NEOWISE Reactivation Mission}",
      journal = {\apj},
         year = 2014,
        month = sep,
       volume = {792},
       number = {1},
          eid = {30},
        pages = {30},
          doi = {10.1088/0004-637X/792/1/30},
archivePrefix = {arXiv},
       eprint = {1406.6025},
 primaryClass = {astro-ph.EP},
       adsurl = {https://ui.adsabs.harvard.edu/abs/2014ApJ...792...30M}
}

@inproceedings{10.1117/12.857814,
author = {William A. Mahoney and Lisa J. Garcia and Joseph Hunt Jr. and Douglas B. McElroy and Vince G. Mannings and David S. Mittman and JoAnn C. O'Linger and Marc Sarrel and Elena Scire},
title = {{Spitzer warm mission transition and operations}},
volume = {7737},
booktitle = {Observatory Operations: Strategies, Processes, and Systems III},
editor = {David R. Silva and Alison B. Peck and B. Thomas Soifer},
organization = {International Society for Optics and Photonics},
publisher = {SPIE},
pages = {77371W},
year = {2010},
doi = {10.1117/12.857814},
URL = {https://doi.org/10.1117/12.857814}
}

@ARTICLE{2024ApJ...962L..16N,
       author = {{Narang}, Mayank and {Manoj}, P. and {Tyagi}, Himanshu and {Watson}, Dan M. and {Megeath}, S. Thomas and {Federman}, Samuel and {Rubinstein}, Adam E. and {Gutermuth}, Robert and {Caratti o Garatti}, Alessio and {Beuther}, Henrik and et al.},
        title = "{Discovery of a Collimated Jet from the Low-luminosity Protostar IRAS 16253‑2429 in a Quiescent Accretion Phase with the JWST}",
      journal = {\apjl},
         year = 2024,
        month = feb,
       volume = {962},
       number = {1},
          eid = {L16},
        pages = {L16},
          doi = {10.3847/2041-8213/ad1de3},
archivePrefix = {arXiv},
       eprint = {2310.14061},
 primaryClass = {astro-ph.SR},
       adsurl = {https://ui.adsabs.harvard.edu/abs/2024ApJ...962L..16N}
}

@ARTICLE{2025MNRAS.541.4025H,
       author = {{Hartmann}, Lee and {Tobin}, John J. and {Sheehan}, Patrick and {Kounkel}, Marina and {Zhao}, Claire},
        title = "{On the protostellar mass{\textendash}luminosity relation}",
      journal = {\mnras},
         year = 2025,
        month = aug,
       volume = {541},
       number = {4},
        pages = {4025-4030},
          doi = {10.1093/mnras/staf1220},
archivePrefix = {arXiv},
       eprint = {2507.18728},
 primaryClass = {astro-ph.SR},
       adsurl = {https://ui.adsabs.harvard.edu/abs/2025MNRAS.541.4025H}
}

@ARTICLE{2012ApJ...747...52D,
       author = {{Dunham}, Michael M. and {Vorobyov}, Eduard I.},
        title = "{Resolving the Luminosity Problem in Low-mass Star Formation}",
      journal = {\apj},
         year = 2012,
        month = mar,
       volume = {747},
       number = {1},
          eid = {52},
        pages = {52},
          doi = {10.1088/0004-637X/747/1/52},
archivePrefix = {arXiv},
       eprint = {1112.4789},
 primaryClass = {astro-ph.GA},
       adsurl = {https://ui.adsabs.harvard.edu/abs/2012ApJ...747...52D}
}

@ARTICLE{2011BASI...39..387M,
       author = {{Mahabal}, A.~A. and {Djorgovski}, S.~G. and {Drake}, A.~J. and {Donalek}, C. and {Graham}, M.~J. and {Williams}, R.~D. and {Chen}, Y. and {Moghaddam}, B. and {Turmon}, M. and {Beshore}, E. and et al.},
        title = "{Discovery, classification, and scientific exploration of transient events from the Catalina Real-time Transient Survey}",
      journal = {Bulletin of the Astronomical Society of India},
         year = 2011,
        month = sep,
       volume = {39},
       number = {3},
        pages = {387-408},
          doi = {10.48550/arXiv.1111.0313},
archivePrefix = {arXiv},
       eprint = {1111.0313},
 primaryClass = {astro-ph.IM},
       adsurl = {https://ui.adsabs.harvard.edu/abs/2011BASI...39..387M}
}

@ARTICLE{2015ApJ...798...89F,
       author = {{Findeisen}, Krzysztof and {Cody}, Ann Marie and {Hillenbrand}, Lynne},
        title = "{Simulated Performance of Timescale Metrics for Aperiodic Light Curves}",
      journal = {\apj},
         year = 2015,
        month = jan,
       volume = {798},
       number = {2},
          eid = {89},
        pages = {89},
          doi = {10.1088/0004-637X/798/2/89},
archivePrefix = {arXiv},
       eprint = {1410.7882},
 primaryClass = {astro-ph.IM},
       adsurl = {https://ui.adsabs.harvard.edu/abs/2015ApJ...798...89F}
}

@ARTICLE{2025MNRAS.543.1133R,
       author = {{Ryan}, Benjamin W. and {Stokes-Geddes}, Holly and {Froebrich}, Dirk},
        title = "{A survey for variable young stars with small telescopes {\textendash} X. Comparing stochastic YSO light curves}",
      journal = {\mnras},
         year = 2025,
        month = oct,
       volume = {543},
       number = {2},
        pages = {1133-1145},
          doi = {10.1093/mnras/staf1495},
archivePrefix = {arXiv},
       eprint = {2509.07710},
 primaryClass = {astro-ph.SR},
       adsurl = {https://ui.adsabs.harvard.edu/abs/2025MNRAS.543.1133R}
}

@ARTICLE{2014ApJ...782...51A,
       author = {{Antoniucci}, S. and {Giannini}, T. and {Li Causi}, G. and {Lorenzetti}, D.},
        title = "{On the Mid-infrared Variability of Candidate Eruptive Variables (EXors): A Comparison between Spitzer and WISE Data}",
      journal = {\apj},
         year = 2014,
        month = feb,
       volume = {782},
       number = {1},
          eid = {51},
        pages = {51},
          doi = {10.1088/0004-637X/782/1/51},
archivePrefix = {arXiv},
       eprint = {1401.1970},
 primaryClass = {astro-ph.SR},
       adsurl = {https://ui.adsabs.harvard.edu/abs/2014ApJ...782...51A}
}

@ARTICLE{2015AJ....150..175R,
       author = {{Rebull}, L.~M. and {Stauffer}, J.~R. and {Cody}, A.~M. and {G{\"u}nther}, H.~M. and {Hillenbrand}, L.~A. and {Poppenhaeger}, K. and {Wolk}, S.~J. and {Hora}, J. and {Hernandez}, J. and {Bayo}, A. and et al.},
        title = "{YSOVAR: Mid-infrared Variability in NGC 1333}",
      journal = {\aj},
         year = 2015,
        month = dec,
       volume = {150},
       number = {6},
          eid = {175},
        pages = {175},
          doi = {10.1088/0004-6256/150/6/175},
archivePrefix = {arXiv},
       eprint = {1508.04705},
 primaryClass = {astro-ph.SR},
       adsurl = {https://ui.adsabs.harvard.edu/abs/2015AJ....150..175R}
}

@ARTICLE{2013ApJ...778..116N,
       author = {{Ninan}, J.~P. and {Ojha}, D.~K. and {Bhatt}, B.~C. and {Ghosh}, S.~K. and {Mohan}, V. and {Mallick}, K.~K. and {Tamura}, M. and {Henning}, Th.},
        title = "{Reappearance of McNeil's Nebula (V1647 Orionis) and its Outburst Environment}",
      journal = {\apj},
         year = 2013,
        month = dec,
       volume = {778},
       number = {2},
          eid = {116},
        pages = {116},
          doi = {10.1088/0004-637X/778/2/116},
archivePrefix = {arXiv},
       eprint = {1309.4967},
 primaryClass = {astro-ph.SR},
       adsurl = {https://ui.adsabs.harvard.edu/abs/2013ApJ...778..116N}
}

@ARTICLE{2009ApJ...690..496C,
       author = {{Chapman}, Nicholas L. and {Mundy}, Lee G. and {Lai}, Shih-Ping and {Evans}, II, Neal J.},
        title = "{The Mid-Infrared Extinction Law in the Ophiuchus, Perseus, and Serpens Molecular Clouds}",
      journal = {\apj},
         year = 2009,
        month = jan,
       volume = {690},
       number = {1},
        pages = {496-511},
          doi = {10.1088/0004-637X/690/1/496},
archivePrefix = {arXiv},
       eprint = {0809.1106},
 primaryClass = {astro-ph},
       adsurl = {https://ui.adsabs.harvard.edu/abs/2009ApJ...690..496C}
}

@ARTICLE{2025JKAS...58..209C,
       author = {{Contreras Pe{\~n}a}, Carlos and {Lee}, Jeong-Eun and {Herczeg}, Gregory and {Johnstone}, Doug and {{\'A}brah{\'a}m}, P{\'e}ter and {Antoniucci}, Simone and {Audard}, Marc and {Ashraf}, Mizna and {Baek}, Giseon and {Garatti}, Alessio Caratti o. and et al.},
        title = "{The Outbursting YSOs Catalogue (OYCAT)}",
      journal = {Journal of Korean Astronomical Society},
         year = 2025,
        month = sep,
       volume = {58},
        pages = {209-230},
          doi = {10.5303/JKAS.2025.58.2.209},
archivePrefix = {arXiv},
       eprint = {2509.24876},
 primaryClass = {astro-ph.SR},
       adsurl = {https://ui.adsabs.harvard.edu/abs/2025JKAS...58..209C}
}

@ARTICLE{2025Natur.643..649M,
       author = {{McClure}, M.~K. and {van't Hoff}, Merel and {Francis}, Logan and {Bergin}, Edwin and {Rocha}, Will R.~M. and {Sturm}, J.~A. and {Harsono}, Daniel and {van Dishoeck}, Ewine F. and {Black}, John H. and {Noble}, J.~A. and et al.},
        title = "{Refractory solid condensation detected in an embedded protoplanetary disk}",
      journal = {\nat},
         year = 2025,
        month = jul,
       volume = {643},
       number = {8072},
        pages = {649-653},
          doi = {10.1038/s41586-025-09163-z},
       adsurl = {https://ui.adsabs.harvard.edu/abs/2025Natur.643..649M}
}

@ARTICLE{2021MNRAS.504..830G,
       author = {{Guo}, Zhen and {Lucas}, P.~W. and {Contreras Pe{\~n}a}, C. and {Smith}, L.~C. and {Morris}, C. and {Kurtev}, R.~G. and {Borissova}, J. and {Alonso-Garc{\'\i}a}, J. and {Minniti}, D. and {Chen{\'e}}, A.-N. and et al.},
        title = "{Analysis of physical processes in eruptive YSOs with near-infrared spectra and multiwavelength light curves}",
      journal = {\mnras},
         year = 2021,
        month = jun,
       volume = {504},
       number = {1},
        pages = {830-856},
          doi = {10.1093/mnras/stab882},
archivePrefix = {arXiv},
       eprint = {2103.13335},
 primaryClass = {astro-ph.SR},
       adsurl = {https://ui.adsabs.harvard.edu/abs/2021MNRAS.504..830G}
}

@ARTICLE{2023MNRAS.521.5669C,
       author = {{Contreras Pe{\~n}a}, Carlos and {Herczeg}, Gregory J. and {Ashraf}, Mizna and {Jose}, Jessy and {Lee}, Ho-Gyu and {Johnstone}, Doug and {Lee}, Jeong-Eun and {Zhou}, Xing-yu and {Liu}, Hanpu and {Yoon}, Sung-Yong},
        title = "{Photometric and spectroscopic monitoring of YSOs in nearby star-forming regions - I. Eruptive YSOs}",
      journal = {\mnras},
         year = 2023,
        month = jun,
       volume = {521},
       number = {4},
        pages = {5669-5685},
          doi = {10.1093/mnras/stad820},
archivePrefix = {arXiv},
       eprint = {2303.08390},
 primaryClass = {astro-ph.SR},
       adsurl = {https://ui.adsabs.harvard.edu/abs/2023MNRAS.521.5669C}
}

@ARTICLE{2014AJ....148..122G,
       author = {{G{\"u}nther}, H.~M. and {Cody}, A.~M. and {Covey}, K.~R. and {Hillenbrand}, L.~A. and {Plavchan}, P. and {Poppenhaeger}, K. and {Rebull}, L.~M. and {Stauffer}, J.~R. and {Wolk}, S.~J. and {Allen}, L. and et al.},
        title = "{YSOVAR: Mid-infrared Variability in the Star-forming Region Lynds 1688}",
      journal = {\aj},
         year = 2014,
        month = dec,
       volume = {148},
       number = {6},
          eid = {122},
        pages = {122},
          doi = {10.1088/0004-6256/148/6/122},
archivePrefix = {arXiv},
       eprint = {1408.3063},
 primaryClass = {astro-ph.SR},
       adsurl = {https://ui.adsabs.harvard.edu/abs/2014AJ....148..122G}
}

@ARTICLE{2025ApJ...987...23C,
       author = {{Contreras Pe{\~n}a}, Carlos and {Lee}, Jeong-Eun and {Lee}, Ho-Gyu and {Herczeg}, Gregory and {Johnstone}, Doug and {Liu}, Hanpu and {Lucas}, Philip W. and {Guo}, Zhen and {Kuhn}, Michael A. and {Smith}, Leigh C. and et al.},
        title = "{``Oh FUors, Where Art Thou?'': A Search for Long-lasting Young Stellar Object Outbursts Hiding in Infrared Surveys}",
      journal = {\apj},
         year = 2025,
        month = jul,
       volume = {987},
       number = {1},
          eid = {23},
        pages = {23},
          doi = {10.3847/1538-4357/add25f},
archivePrefix = {arXiv},
       eprint = {2504.21237},
 primaryClass = {astro-ph.SR},
       adsurl = {https://ui.adsabs.harvard.edu/abs/2025ApJ...987...23C}
}

@ARTICLE{2024ApJ...962...38L,
       author = {{Lee}, Sieun and {Lee}, Jeong-Eun and {Contreras Pe{\~n}a}, Carlos and {Johnstone}, Doug and {Herczeg}, Gregory and {Lee}, Seonjae},
        title = "{Mid-infrared Variability of Young Stellar Objects on Timescales of Days to Years}",
      journal = {\apj},
         year = 2024,
        month = feb,
       volume = {962},
       number = {1},
          eid = {38},
        pages = {38},
          doi = {10.3847/1538-4357/ad14f8},
archivePrefix = {arXiv},
       eprint = {2312.05753},
 primaryClass = {astro-ph.SR},
       adsurl = {https://ui.adsabs.harvard.edu/abs/2024ApJ...962...38L}
}
\bibliographystyle{aasjournal}

\begin{acknowledgements}
In this paper, we remember the contributions of co-author WJF, who tragically passed away at the height of his abilities during the work on this paper. This paper uses observations from the Spitzer Space Telescope, operated by JPL/Caltech under a contract with NASA, and the Wide-field Infrared Survey Explorer, a joint project of the University of California, Los Angeles, and JPL/Caltech, funded by NASA. It is also based on observations with ISO, an ESA project with instruments funded by ESA Member States (especially the PI countries: France, Germany, the Netherlands and the United Kingdom) and with the participation of ISAS and NASA. It makes use of the NASA/IPAC Infrared Science Archive, operated by JPL/Caltech under a contract with NASA. S.T.M. and R.A.G. were supported by the NASA ADAP grant 80NSSC19K0591, and S.T.M. was supported by the NASA ADAP grant 80NSSC20K0454. R.P. was supported by the NASA ADAP grant 80NSSC18K1564. N.B-W's work was supported by the summer REU program at the University of Toledo. This research has made use of the SVO Filter Profile Service "Carlos Rodrigo", funded by MCIN/AEI/10.13039/501100011033/ through grant PID2023-146210NB-I00
\end{acknowledgements}

\newpage
\section{Appendix A}

\begin{figure}
    \centering
    \includegraphics[height=10 cm]{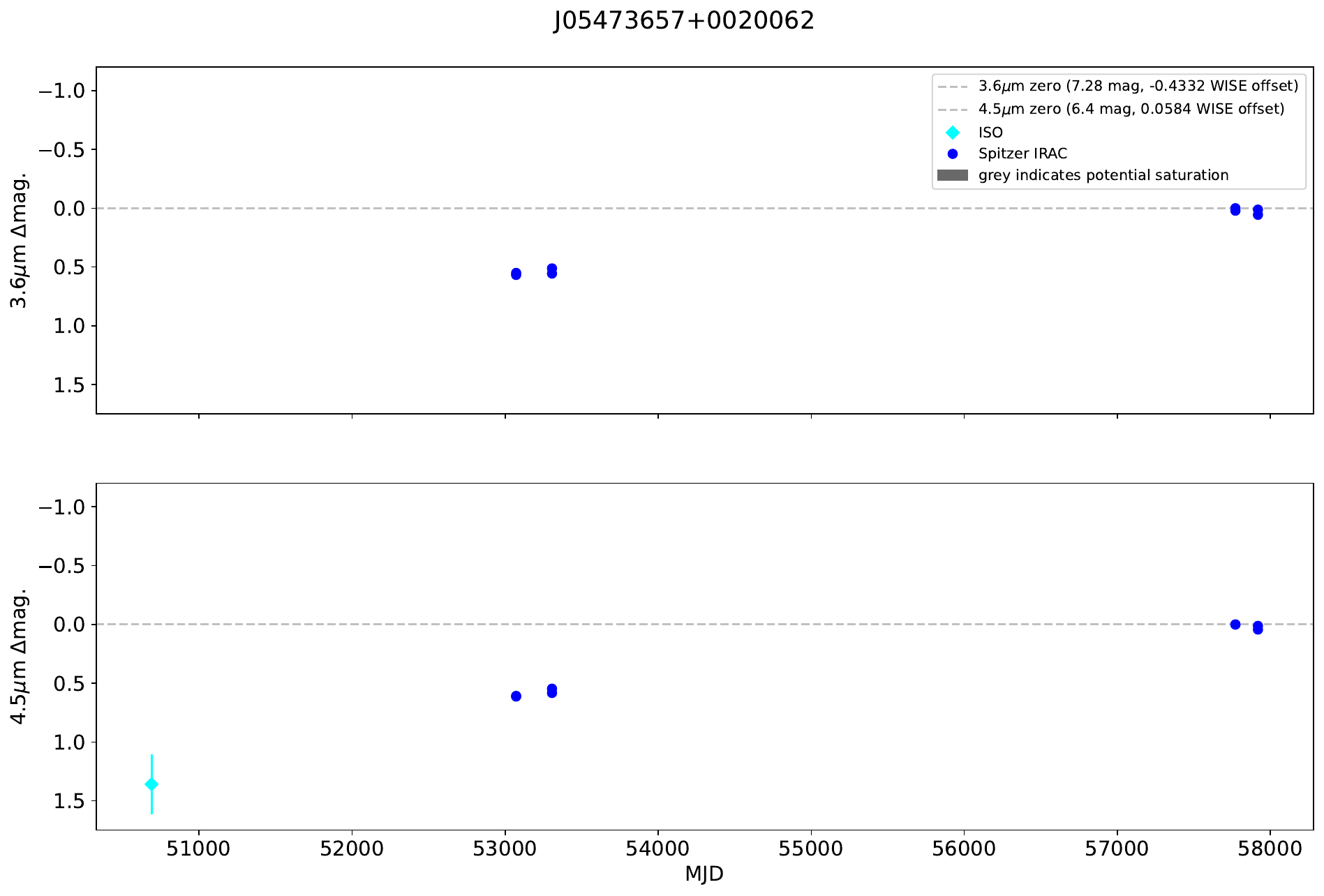}
    \caption{The light curves of the Orion protostar J05473657+0020062, or HOPS~364.}
    \label{fig:HOPS_364}
\end{figure}

In Figure.~\ref{fig:HOPS_364}, we show the light curve of J05473657+0020062 (HOPS~364), which shows a gradual burst in the ISO and Spitzer photometry. Due to extended nebulosity coincident with this protostar, there are no WISE or NEOWISE detections. For this reason, we do not include this source in our analysis. This protostar shows an FU~Ori like spectrum in the near-IR \citep{2018ApJ...861..145C}.

We also show the light curve of v2775~Ori (HOPS 223), which also shows an FU~Ori like spectrum \citep{2012ApJ...756...99F}. This is not included in the main text due to the lack of an ISO observation, and is put here for comparison with the other light curves. 

%In Fig.~\ref{fig:iso_plots}, we present the differences between the IRAC and ISOCAM photometry for each of the clouds, as described in Sec.~\ref{sec:ISOCAM}. In each of these, the median offset has been subtracted; the median offsets are given in Table.~\ref{table:median}. This table also gives the MADs for each cloud.  

% \figsetstart

\begin{figure*}[htp]
    \centering
    \includegraphics[width=14cm]{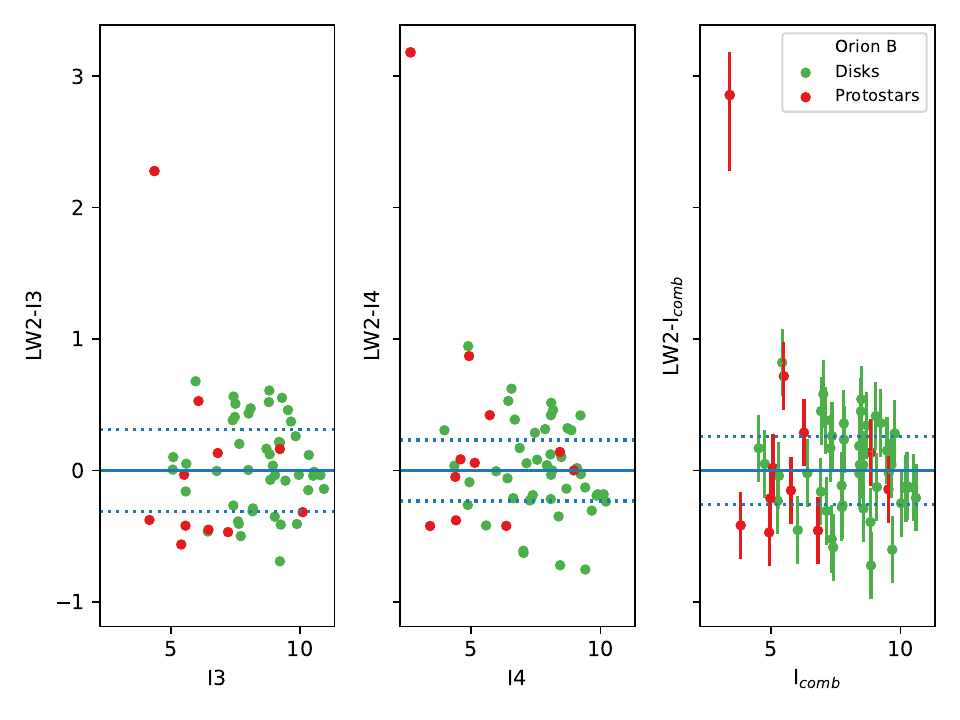}
    \caption{Difference between ISO magnitude (6.7 $\mu$m) and Spitzer magnitude versus Spitzer magnitude for YSOs in the Orion~B field. Leftmost uses the 5.8$\mu$m magnitude for Spitzer, center: 8$\mu$m, and right: the two combined.}
    \label{fig:iso_plots}
\end{figure*}

\addtocounter{figure}{-1}

\begin{figure*}[htp]
    \centering
    \includegraphics[width=14cm]{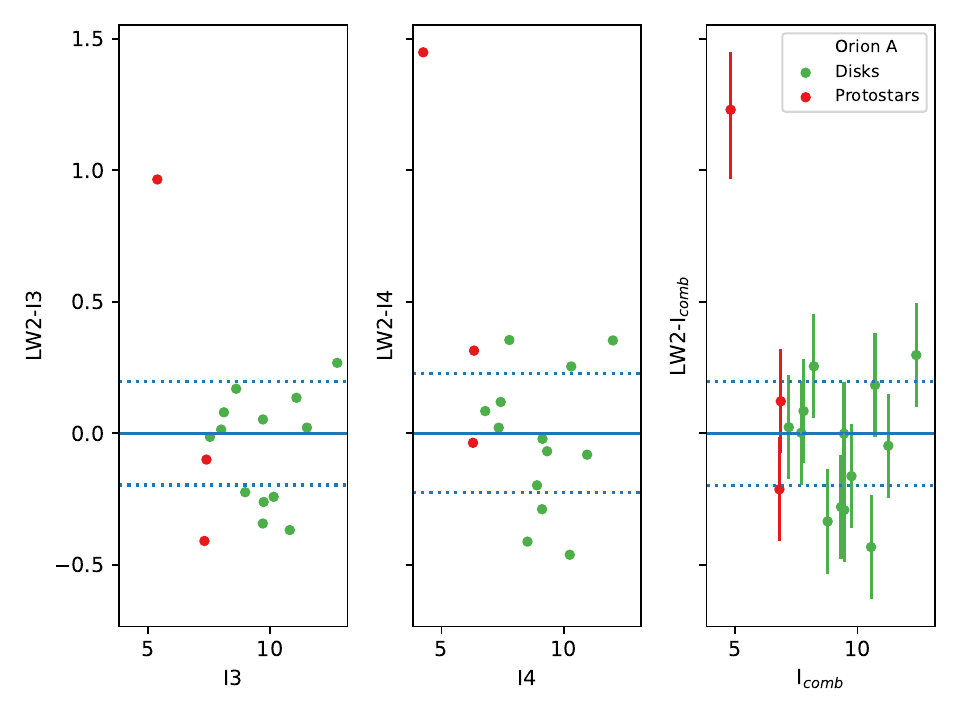}
         \caption{Orion~A}
%    \label{fig:iso_plots}
\end{figure*}

\addtocounter{figure}{-1}

\begin{figure*}[htp]
    \centering
    \includegraphics[width=14cm]{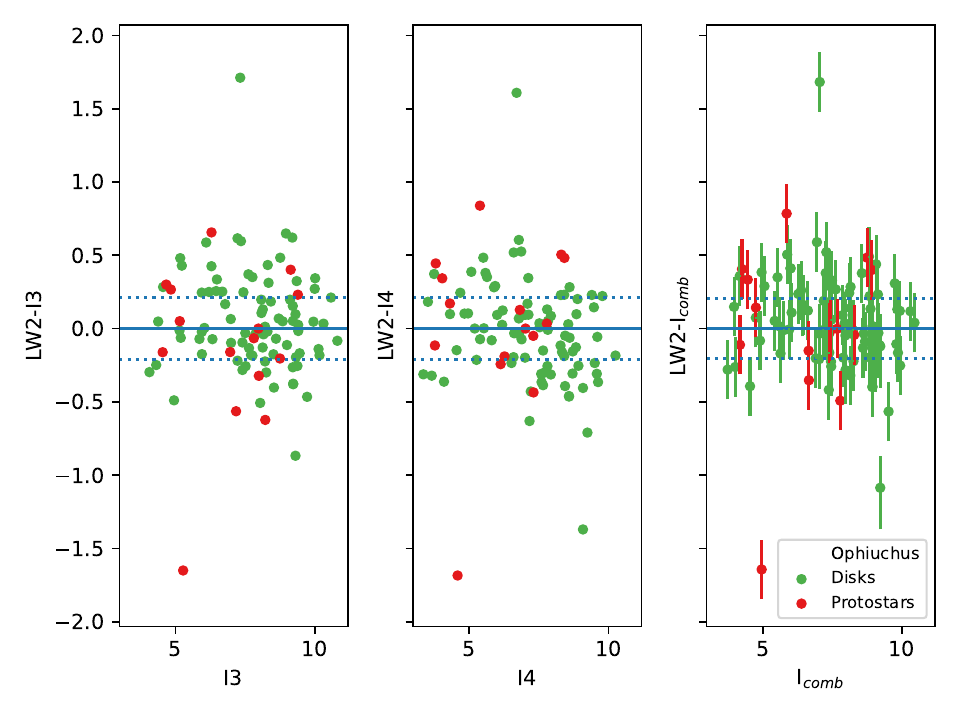}
    \caption{Ophiuchus}
%   \label{fig:iso_oph_app}
\end{figure*}

\addtocounter{figure}{-1}

\begin{figure*}[htp]
    \centering
    \includegraphics[width=14cm]{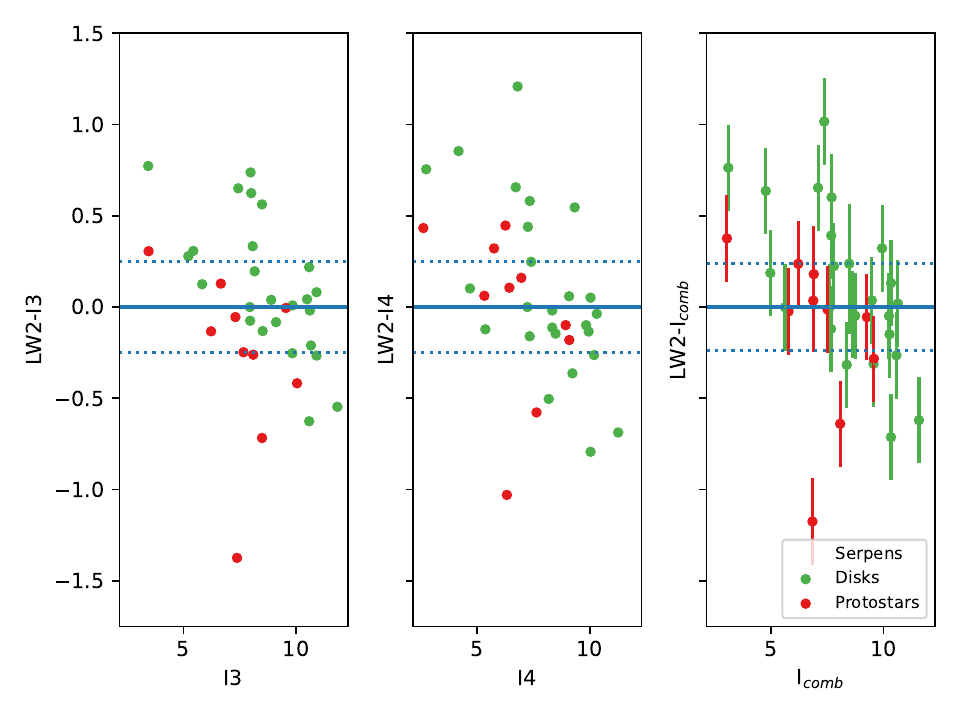}
    \caption{Serpens}
%    \label{fig:iso_serpens}
\end{figure*}

\addtocounter{figure}{-1}

\begin{figure*}[htp]
    \centering
    \includegraphics[width=14cm]{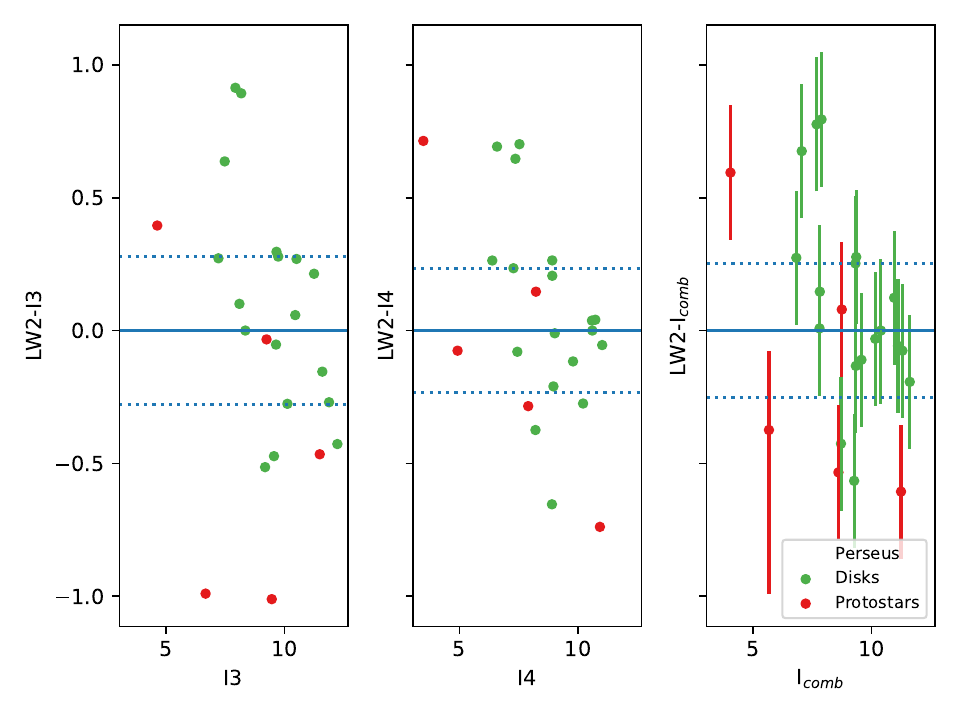}
    \caption{NGC~1333}
%    \label{fig:iso_1333}
\end{figure*}

\addtocounter{figure}{-1}

\begin{figure*}[htp]
    \centering
    \includegraphics[width=14cm]{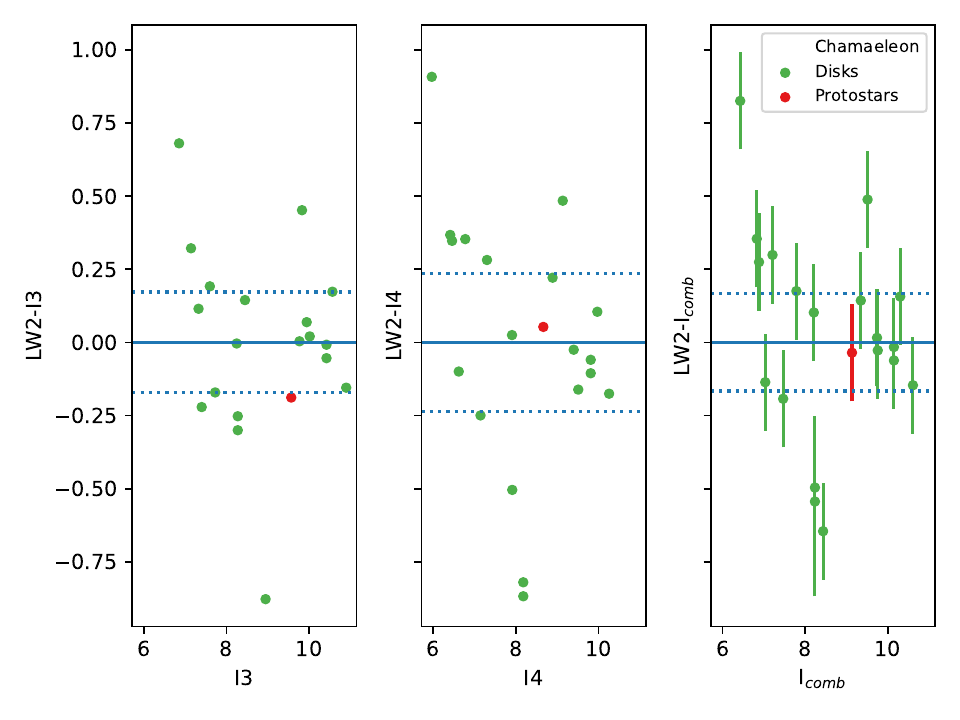}
    \caption{Chameleon}
 %   \label{fig:iso_cham}
\end{figure*}

\addtocounter{figure}{-1}

\begin{figure*}[htp]
    \centering
    \includegraphics[width=14cm]{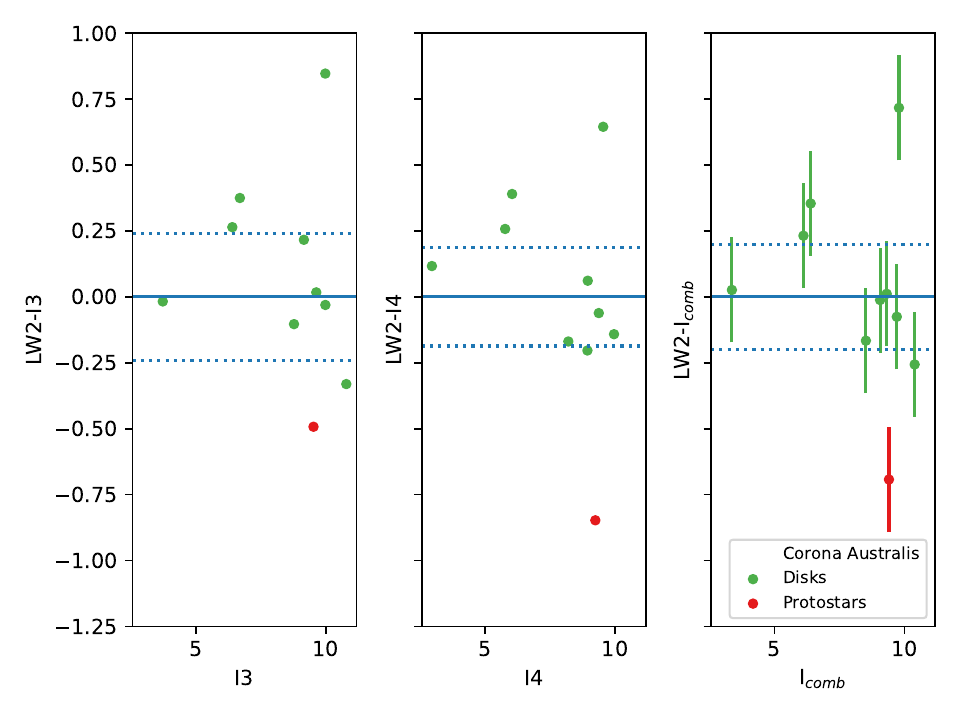}
    \caption{R~Corona Australis }
    \label{fig:iso_rcra}
\end{figure*}

.%Should we include the actual values used to filter these on?The unTimely catalog \citet{2023AJ....165...36M} presents an alternative reduction of both the WISE and NEOWISE data, including original mission data alongside the reactivation releases.

%\section{Appendix~B}

%Here we provide the complete light curves for the large-amplitude variables. These include those for completed bursts in Fig.~\ref{fig:lc_comp_bursts_appendix}, those for ongoing bursts in Fig.~\ref{fig:lc_ongoing_bursts_appendix}, those for fades in Fig.~\ref{fig:lc_fades_appendix}, and those for fluctuators in Fig.~\ref{fig:lc_fluctuators_appendix}.

%J03285841+3122176 X
%J05362461-0622413 X
%J05461313-0006045 X
%J05462569+0023416 X
%J05473657+0020062 X
%J16264216-2431031 X
%J16271383-2443318 X
%J16271758-2405139 X
%J18295225+0115476 X
%J18300772+0112044 X

\begin{figure*}
    \centering
    \includegraphics [height=10.5cm]{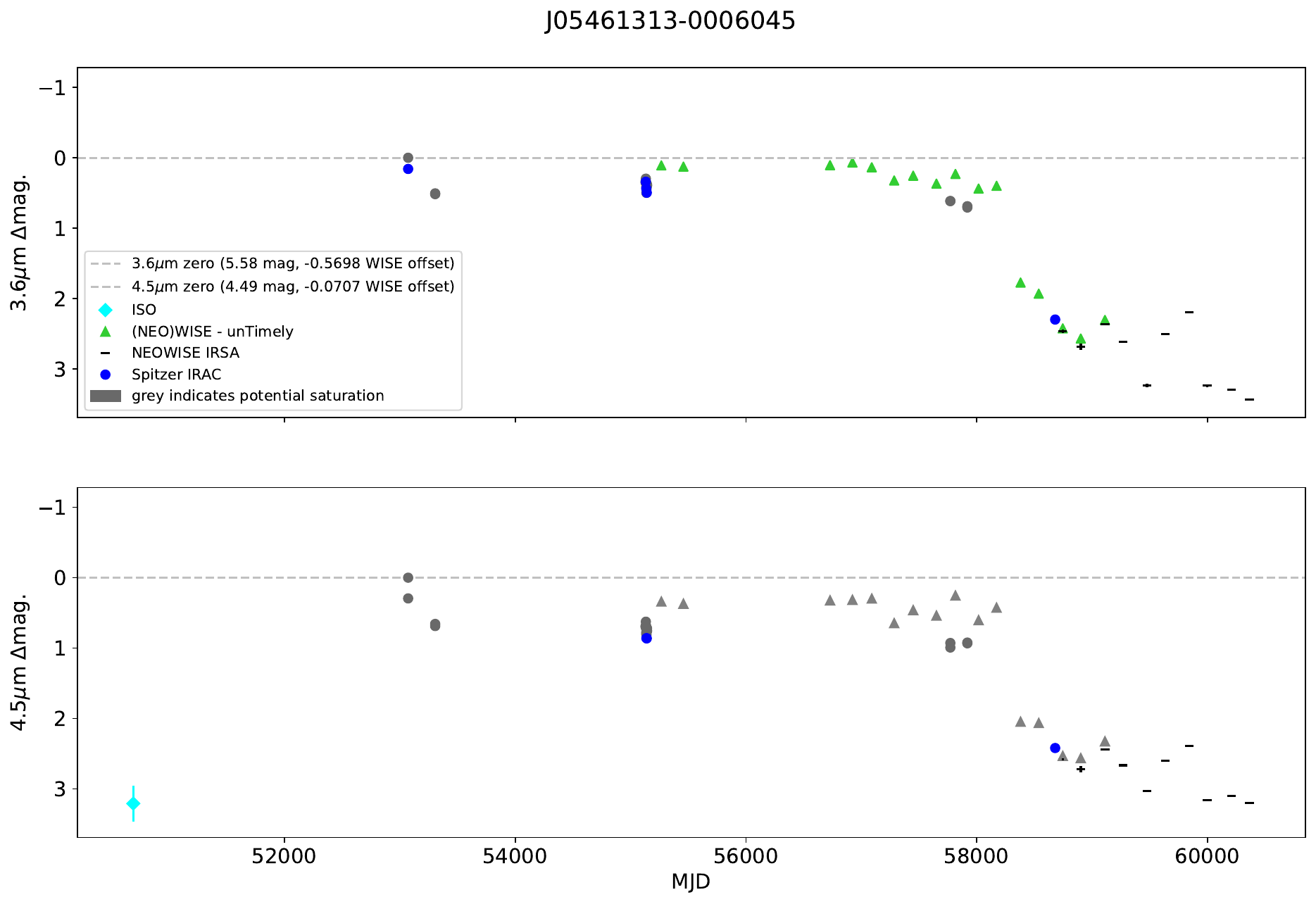}
    \includegraphics [height=10.5cm]{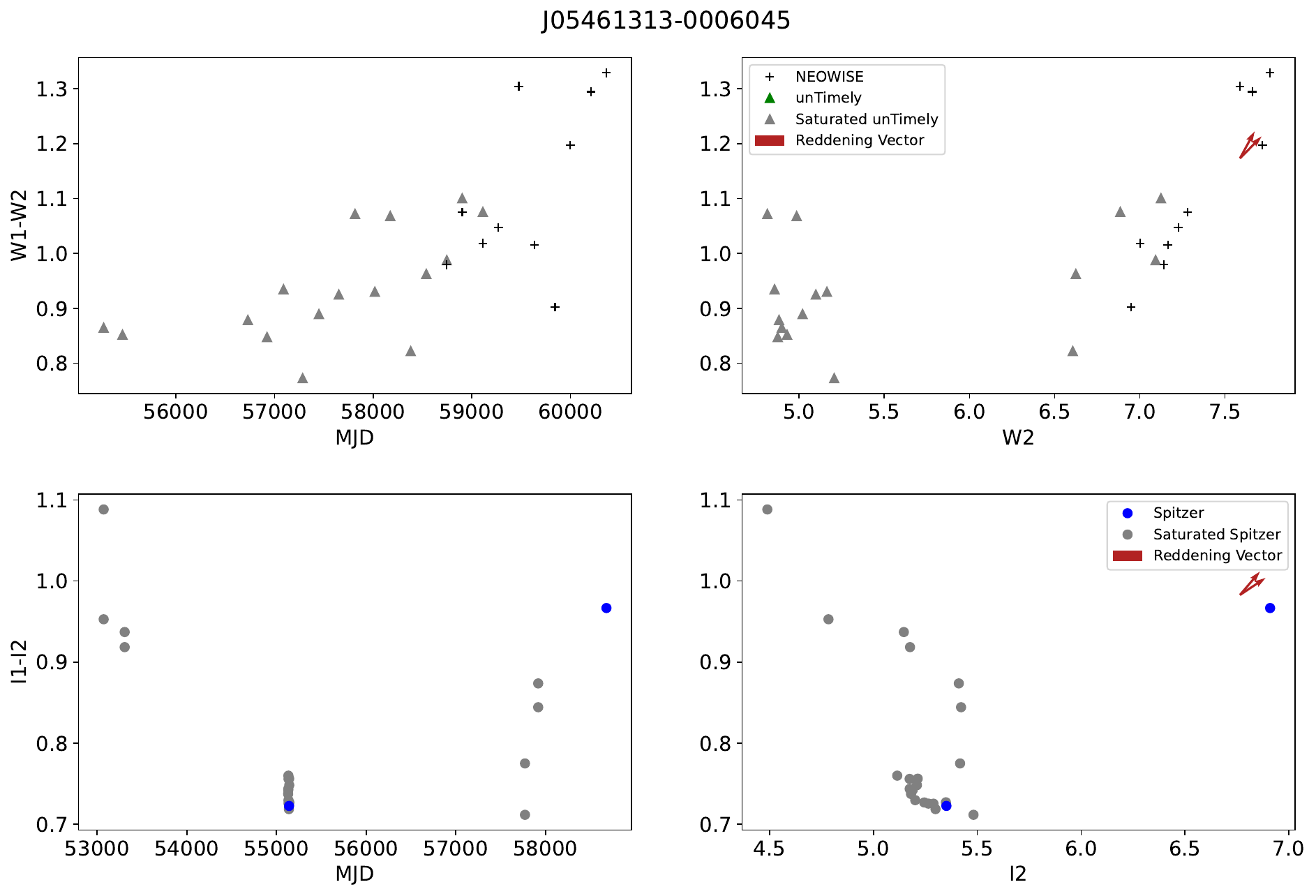}
    \caption{Top two panels: 3.6 and 4.5~$\mu$m light curves of the ongoing bursts. The magnitudes corresponding to a zero $\Delta$mag in the light curves and the offsets applied to the WISE photometry are given in the legend. Bottom four panels: clockwise from upper left, the W1-W2 vs MJD, W1-W2 vs 4.5~$\mu m$ magnitude, I1-I2 vs 4.5~$\mu m$ magnitude, and I1-I2 vs MJD diagrams.}
    %The light curves, color curves, and CMD diagrams of completed bursts in our sample. These are the protostar J05461313-0006405 in Orion B (also known as V1647~Ori or HOPS~388), the pre-main sequence star with disk J05462569+0023416 in Orion B, and the pre-main sequence star with disk J16271758-2405139 in Ophiuchus.} 
    \label{fig:lc_comp_bursts_appendix}
\end{figure*}

\addtocounter{figure}{-1}

\begin{figure*}
    \centering
    \includegraphics[height=10.5cm]{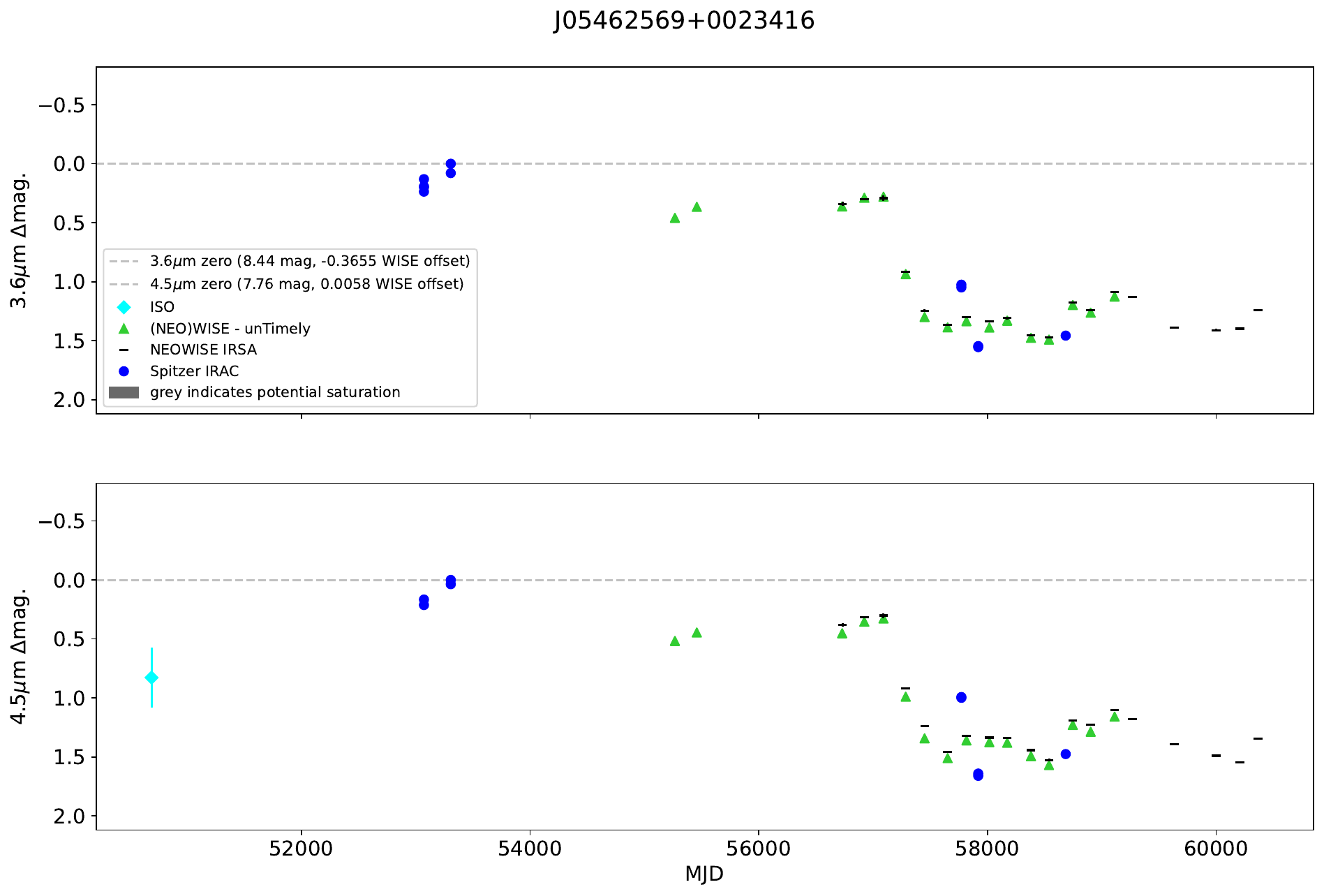}
     \includegraphics[height=10.5cm]{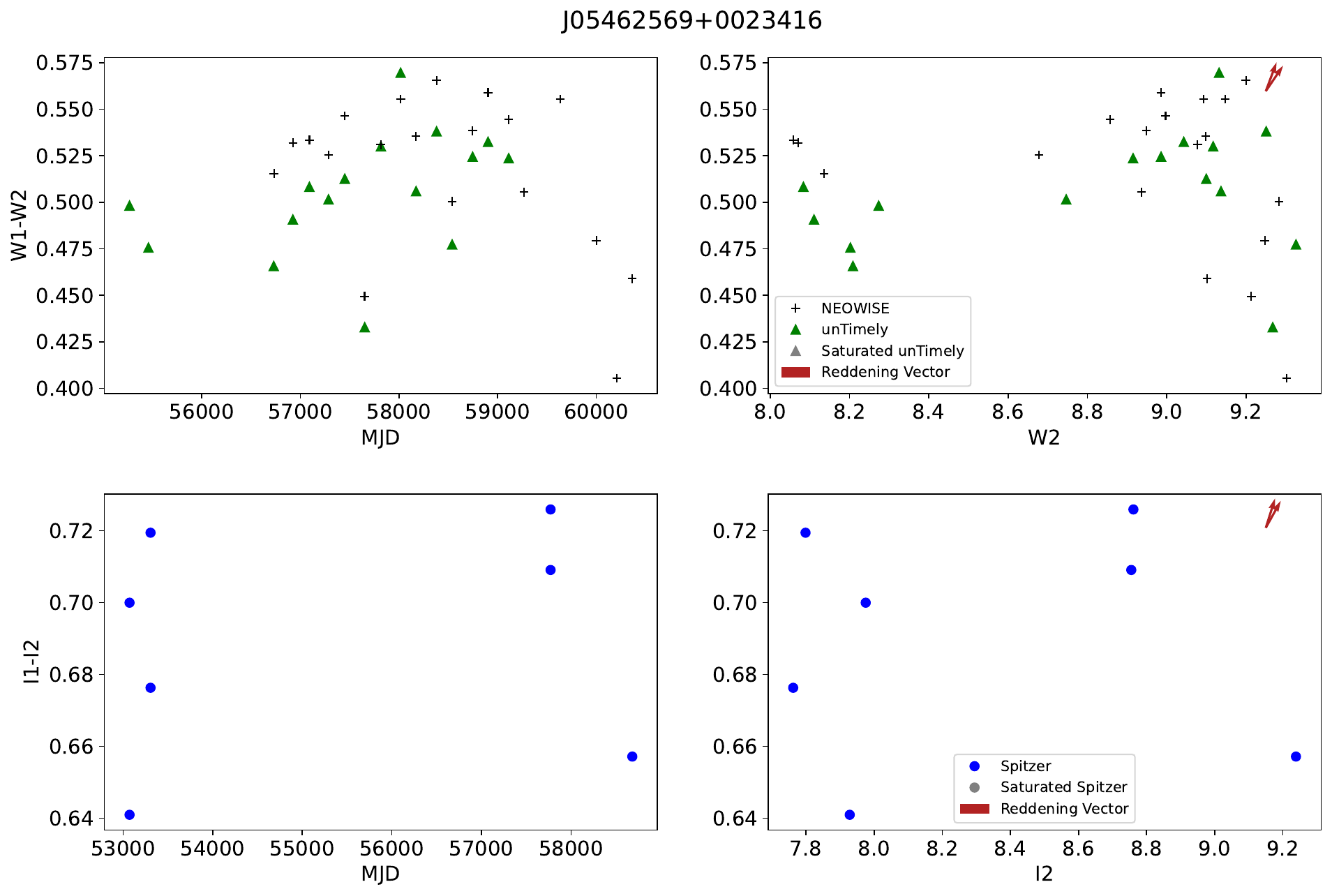}
    \caption{See above.}
\end{figure*}

\addtocounter{figure}{-1}

\begin{figure*}
    \centering
    \includegraphics[height=10.5cm]{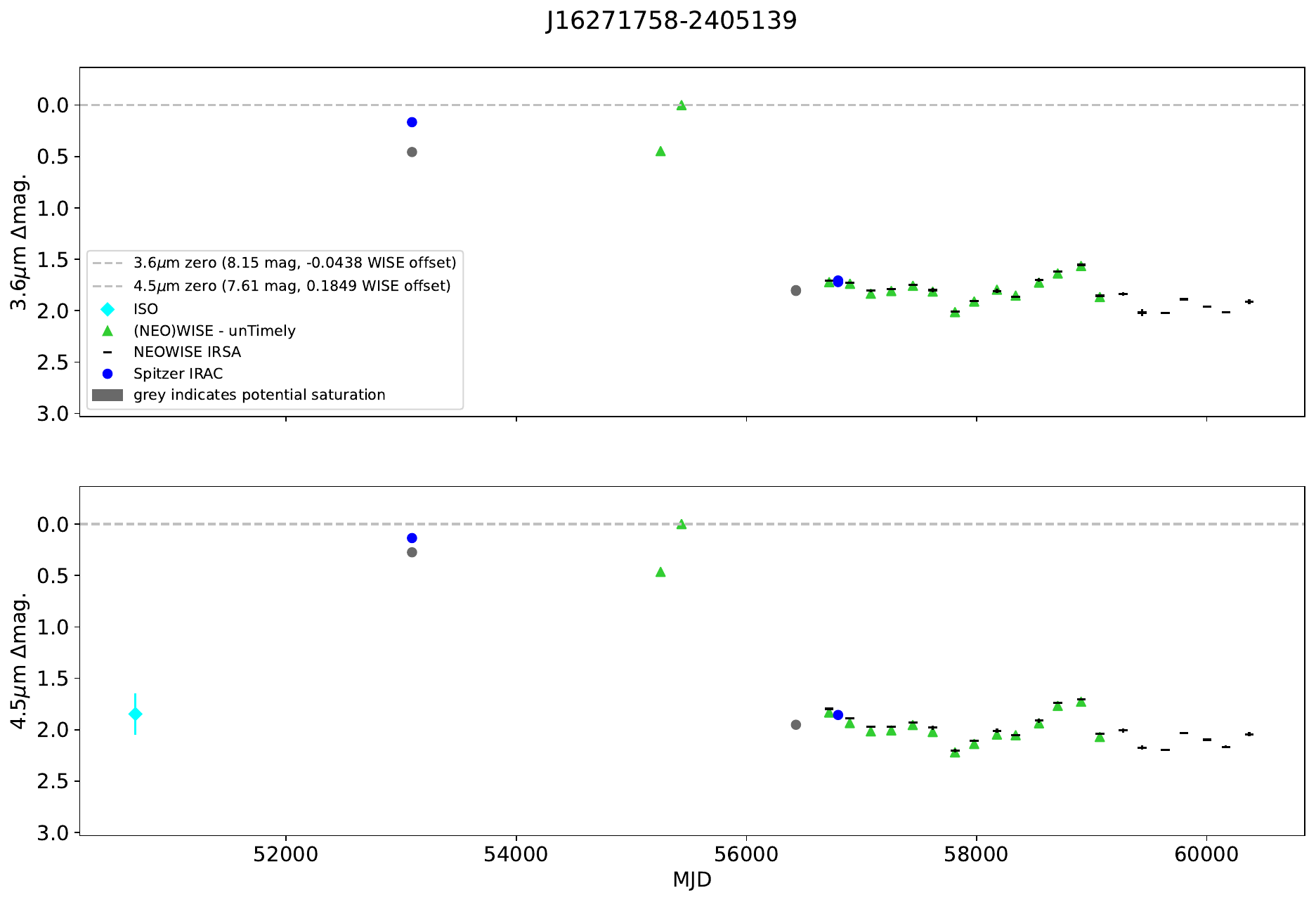}
    \includegraphics[height=10.5cm]{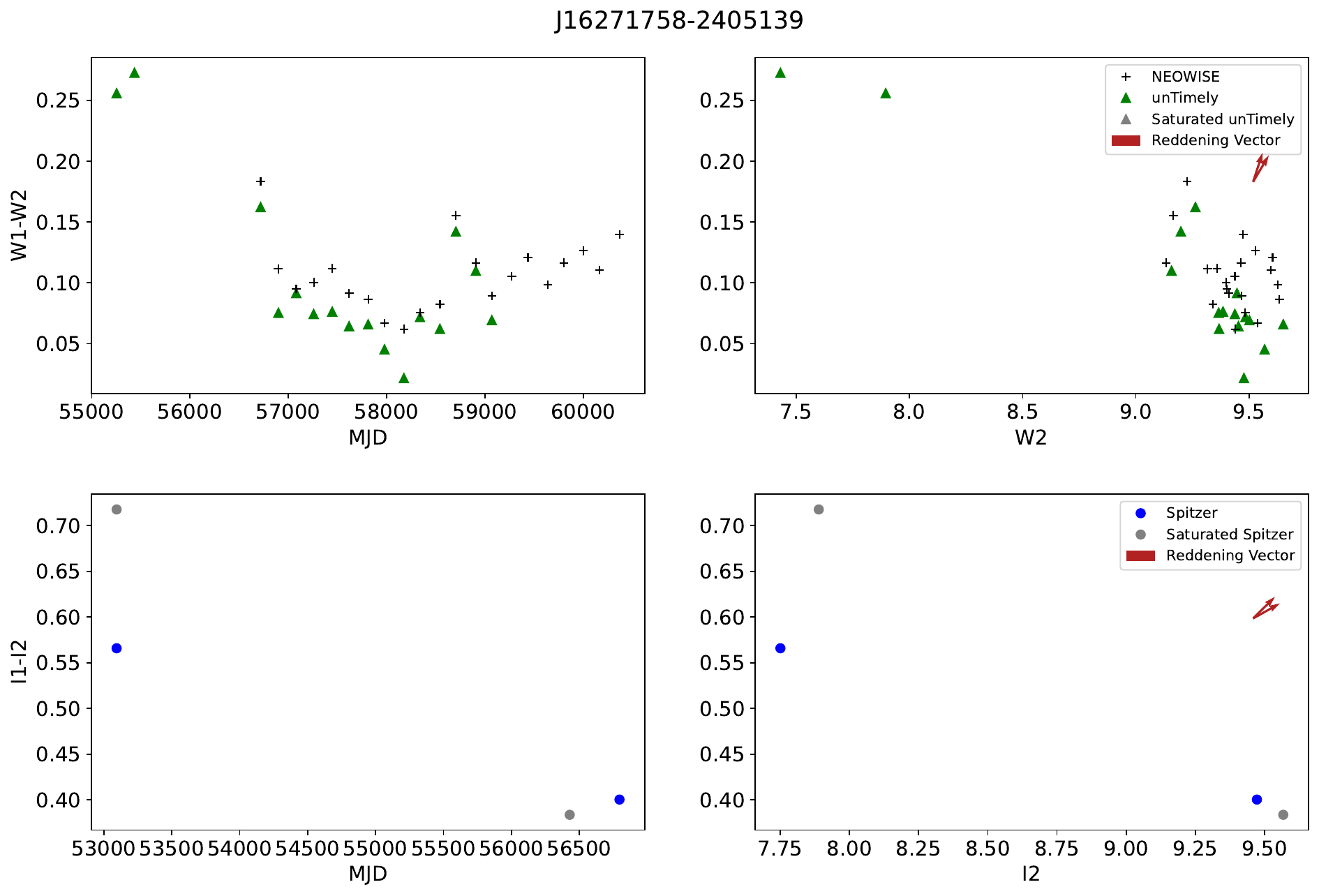}
    \caption{See above.}
\end{figure*}

% Added from ongoing bursts after last NEOWISE epoch added
\begin{figure*}
    \centering
    \includegraphics[width=15cm, height=10.5cm]{J16264216-2431031_DLC_v3.pdf}
    \includegraphics[width=15cm, height=10.5cm]{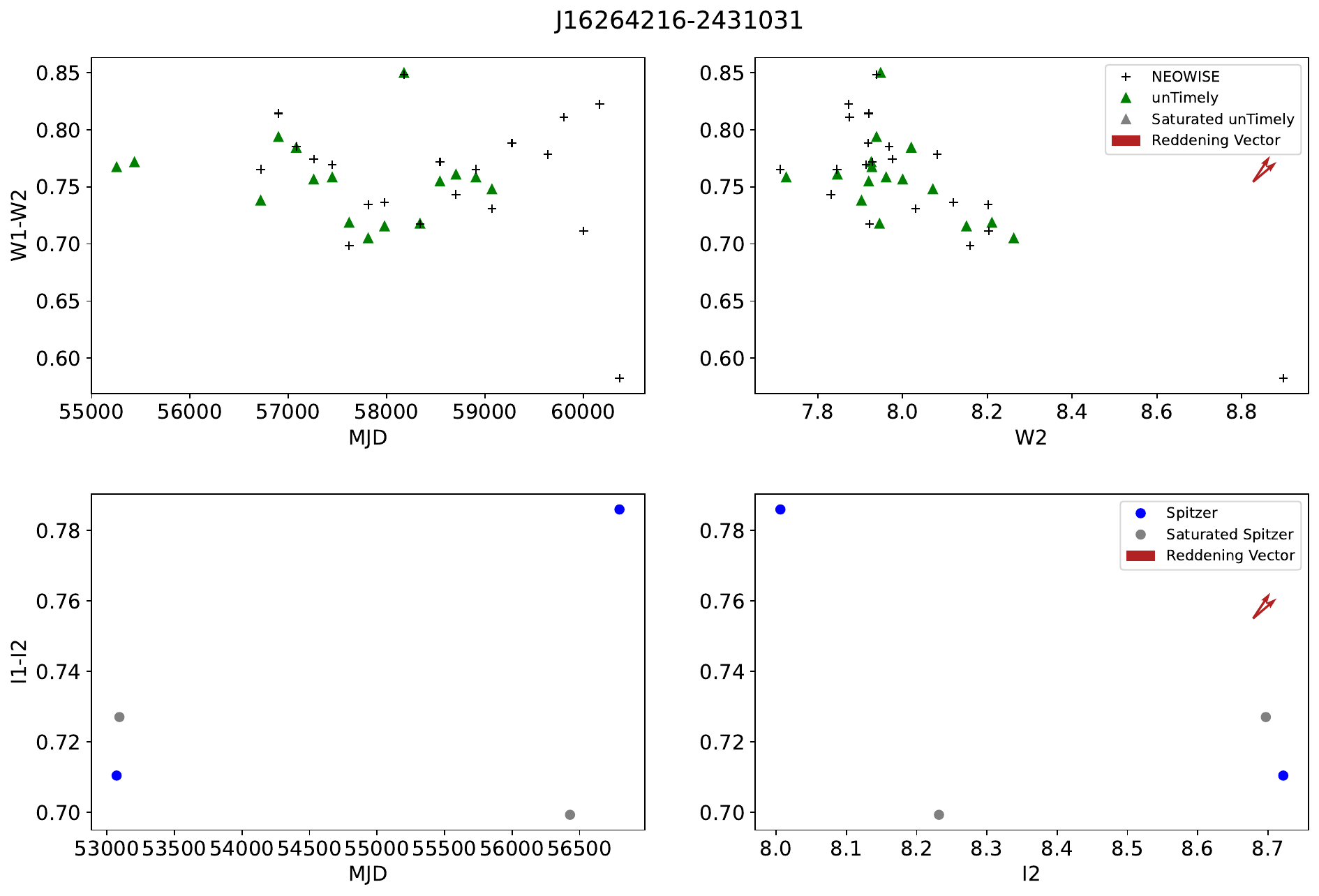}
    \caption{See above}
\end{figure*}

% \begin{figure*}
%    \centering
%    \includegraphics[width=15cm, height=10.5cm]{bursts/orion_B/J05460463+0004582_DLC.pdf}
%    \caption{The ongoing burst J05460463+0004582, or EM* LkHA 298. This is a disk %(??) in the Orion B cloud.}
%    \label{fig:EM_LkHA_298}
% \end{figure*}

% \begin{figure*}
%     \centering
%     \includegraphics[width=15cm, height=10.5cm]{bursts/rho_oph/J16255617-2420485_DLC.pdf}
%     \caption{Burst of ISO Oph 6 in the Ophiuchus region.}
%     \label{fig:ISO_Oph_6}
% \end{figure*}

\begin{figure*}
    \centering
    \includegraphics[ height=10 cm]{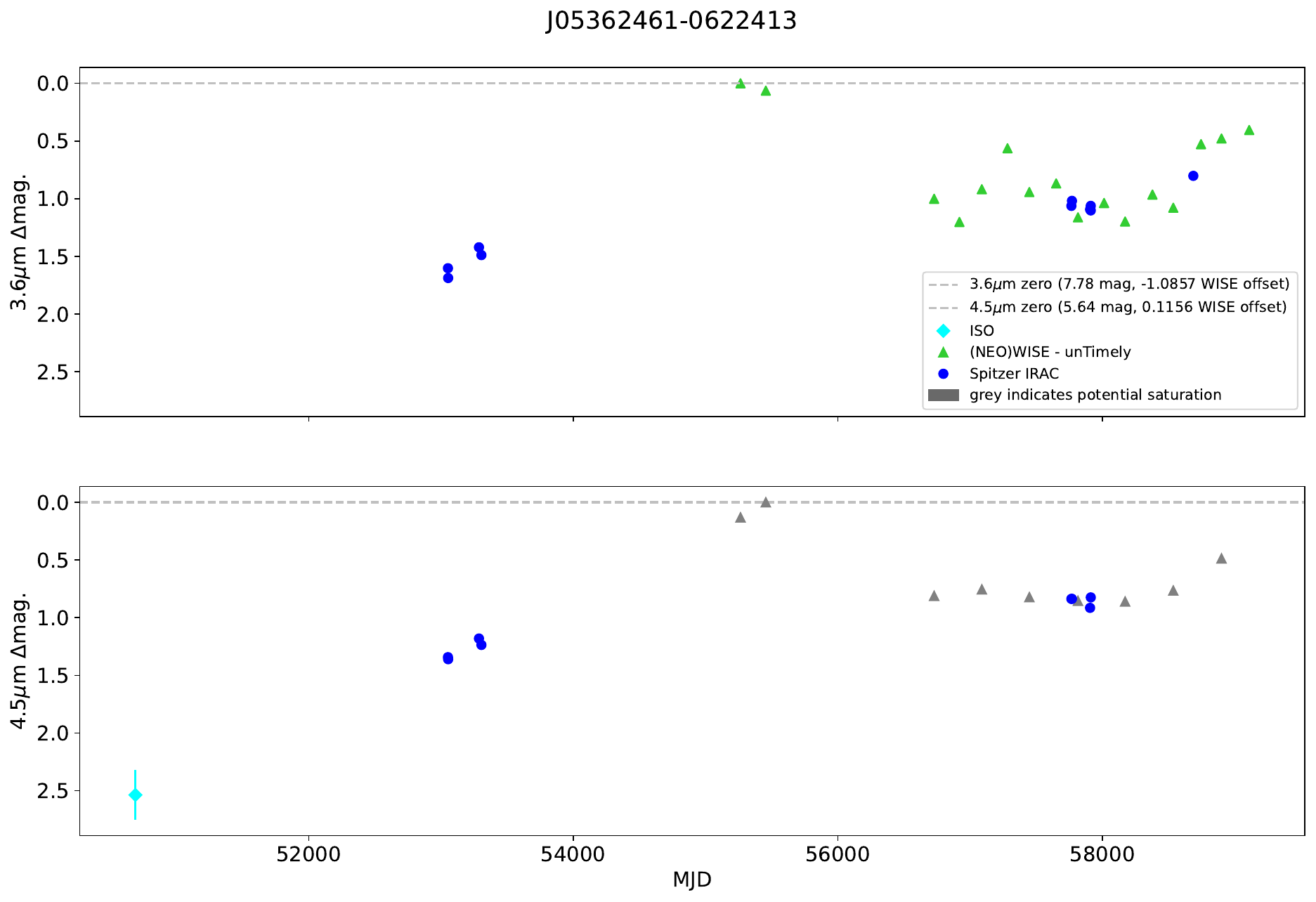}
    \includegraphics[ height=10 cm]{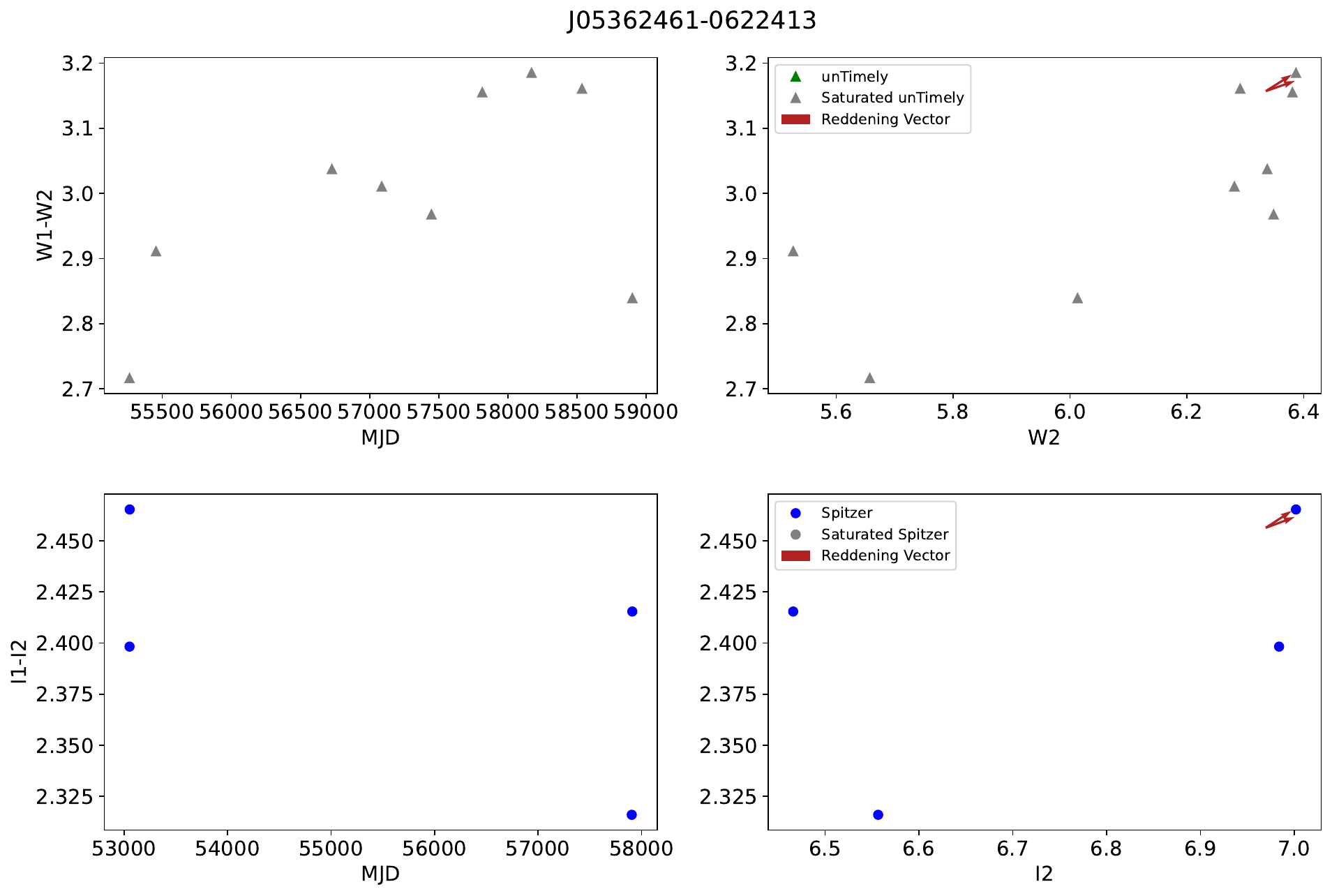}
    \caption{Top two panels: 3.6 and 4.5~$\mu$m light curves of the ongoing bursts. The magnitudes corresponding to a zero $\Delta$mag in the light curves and the offsets applied to the WISE photometry are given in the legend. Bottom four panels: clockwise from upper left, the W1-W2 vs MJD, W1-W2 vs 4.5~$\mu m$ magnitude, I1-I2 vs 4.5~$\mu m$ magnitude, and I1-I2 vs MJD diagrams.}
    %The light curves, color curves, and CMD diagrams of the ongoing bursts in our sample. These are the protostar (J05362461-0622413 (HOPS~178) in the Orion A cloud, the pre-main sequence stars with disks J16264216-2431031 (WL~7 or ISO~Oph~59) and J16271383-2443318 (ISO~Oph~117) in the Ophiuchus cloud, and the pre-main sequence star with disk J18300772+0112044(GMM2009~Serpens~84) in the Serpens region.}
    \label{fig:lc_ongoing_bursts_appendix}
\end{figure*}

\addtocounter{figure}{-1}

% moved to completed bursts section
% \begin{figure*}
%     \centering
%     \includegraphics[width=15cm, height=10.5cm]{J16264216-2431031_DLC_v3.pdf}
%     \includegraphics[width=15cm, height=10.5cm]{J16264216-2431031_ColP_v3.pdf}
%     \caption{See above}
% \end{figure*}

\addtocounter{figure}{-1}

\begin{figure*}
    \centering
    \includegraphics[width=15cm, height=10.5cm]{J16271383-2443318_DLC_v3.pdf}
    \includegraphics[width=15cm, height=10.5cm]{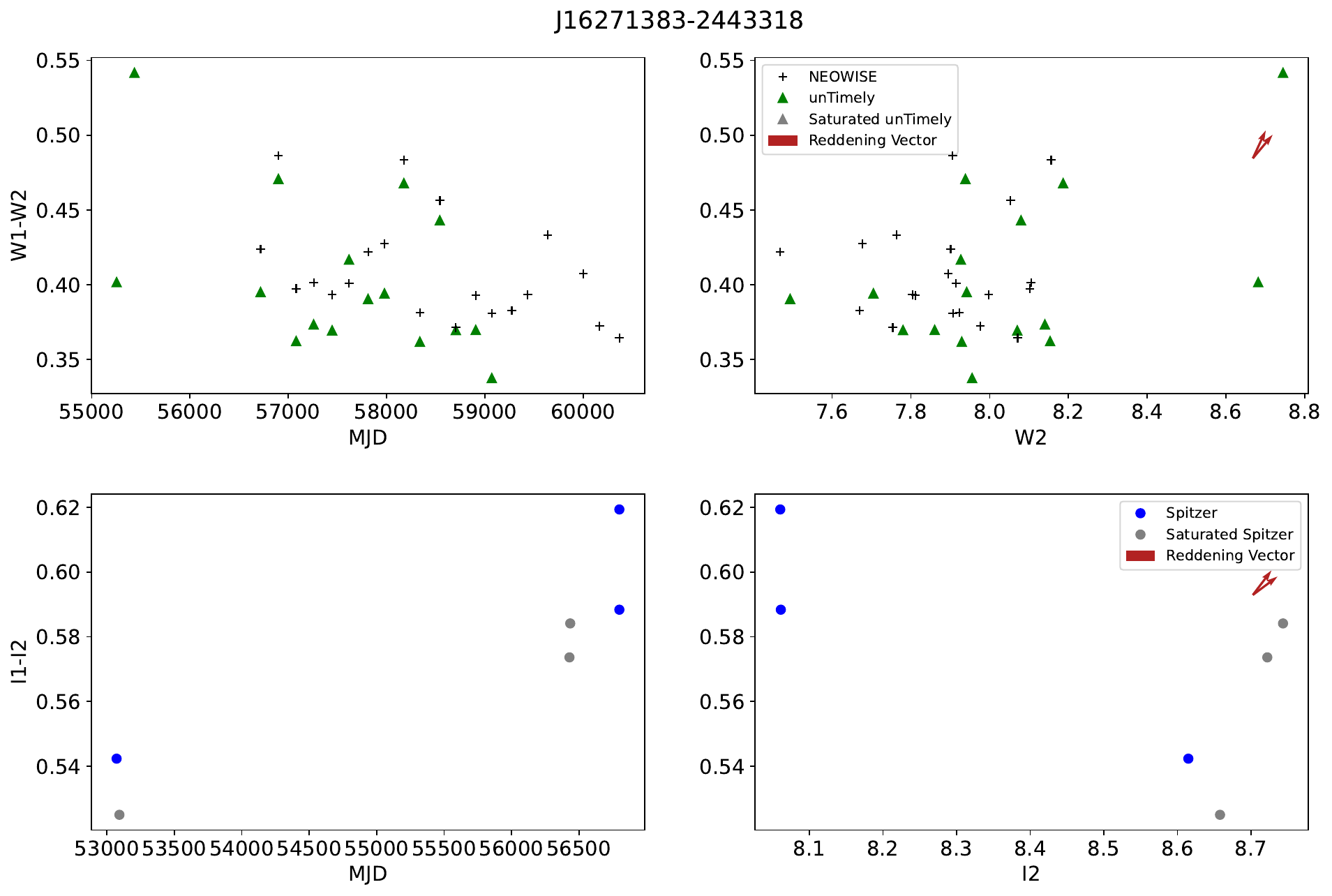}
    \caption{See above}
\end{figure*}

\addtocounter{figure}{-1}

\begin{figure*}
    \centering
    \includegraphics[width=15cm, height=10.5cm]{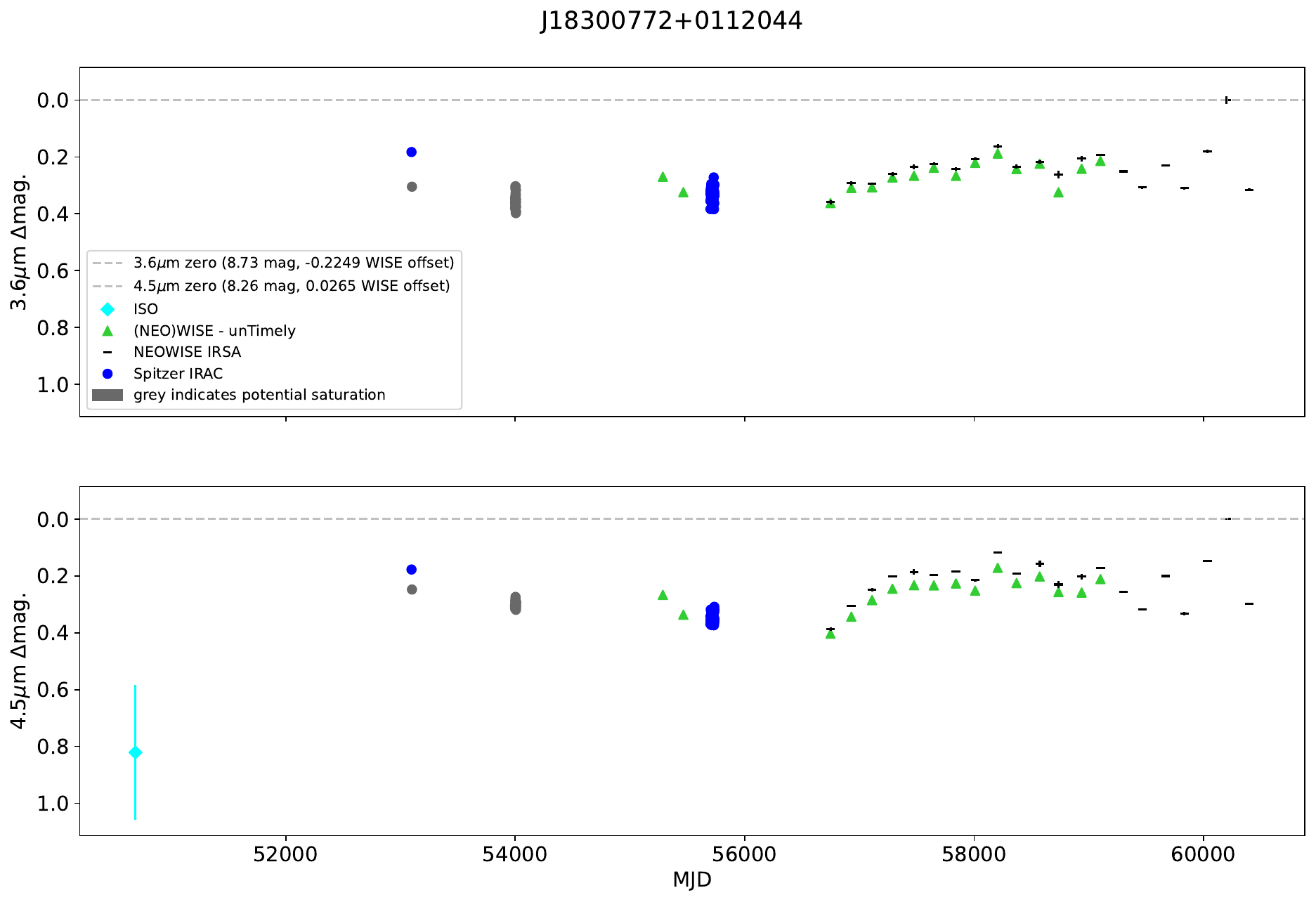}
    \includegraphics[width=15cm, height=10.5cm]{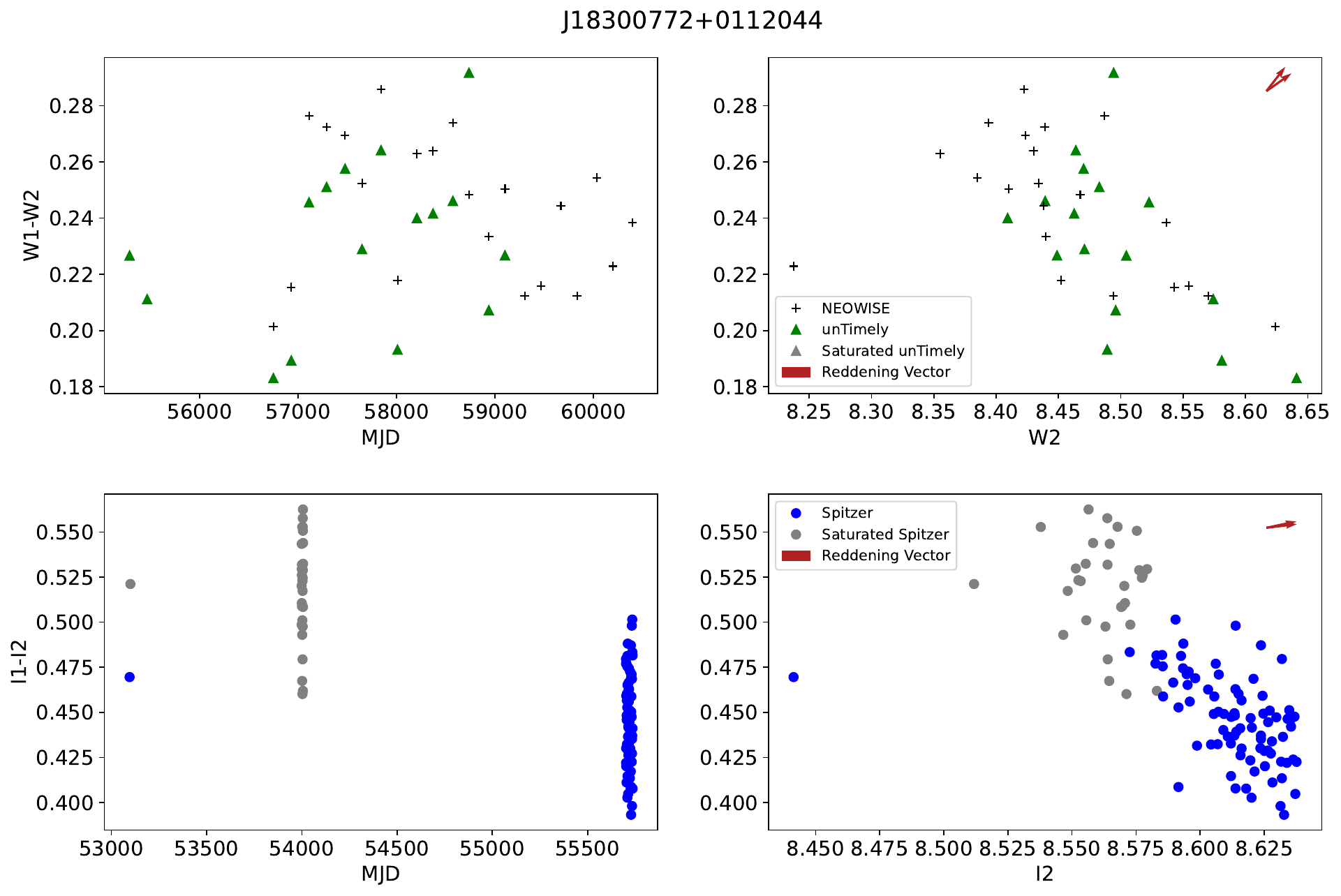}
    \caption{See above}
\end{figure*}

% J18300772+0112044 X
% J03283708+3113309 X
% J03290473+3111349 X
% J05464312+0000525 X
% J05465358+0000061 X
% J16265196-2430396 X
% J19025867-3707361 

\begin{figure*}
    \centering
    \includegraphics[width=15cm, height=10.5cm]{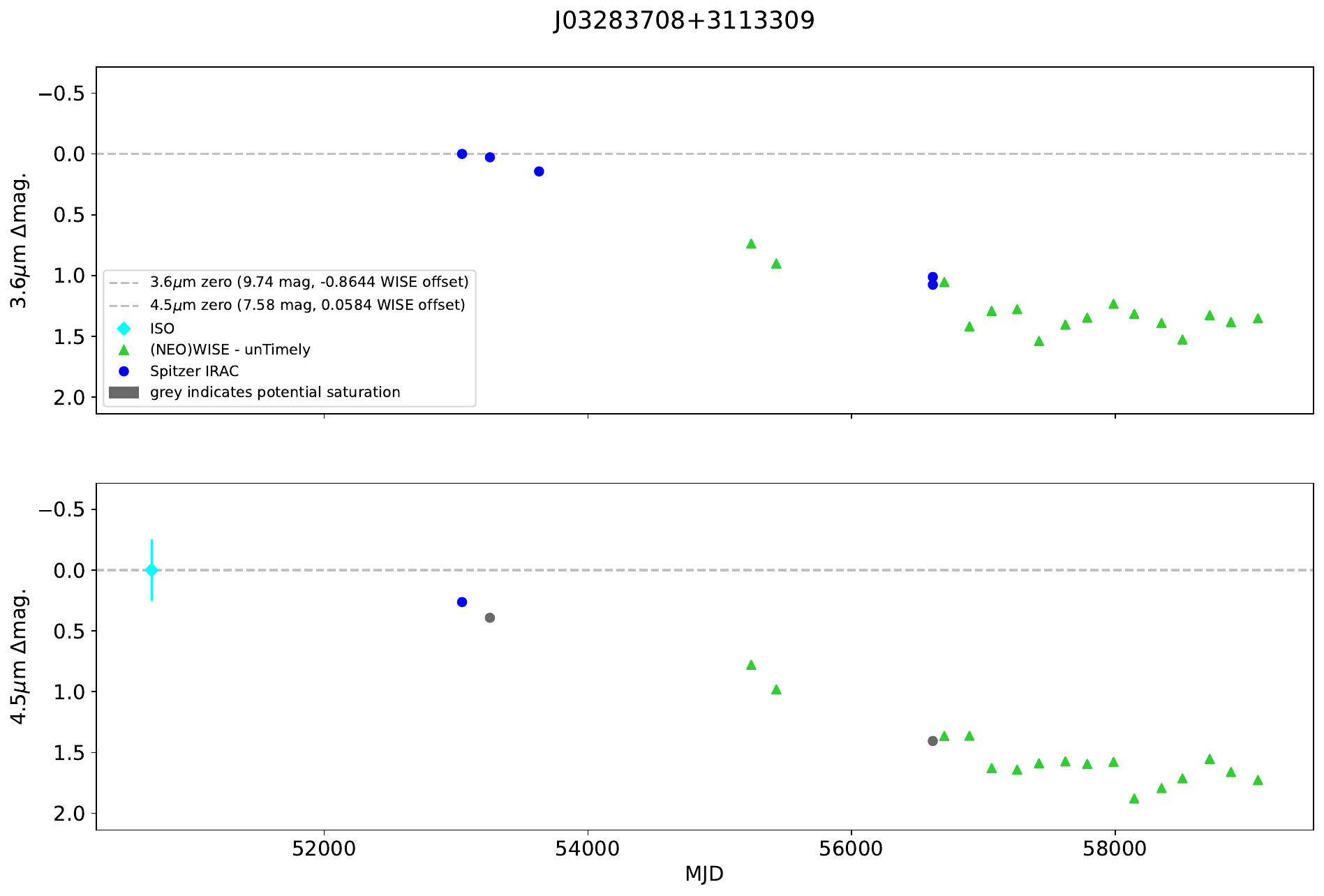}
    \includegraphics[width=15cm, height=10.5cm]{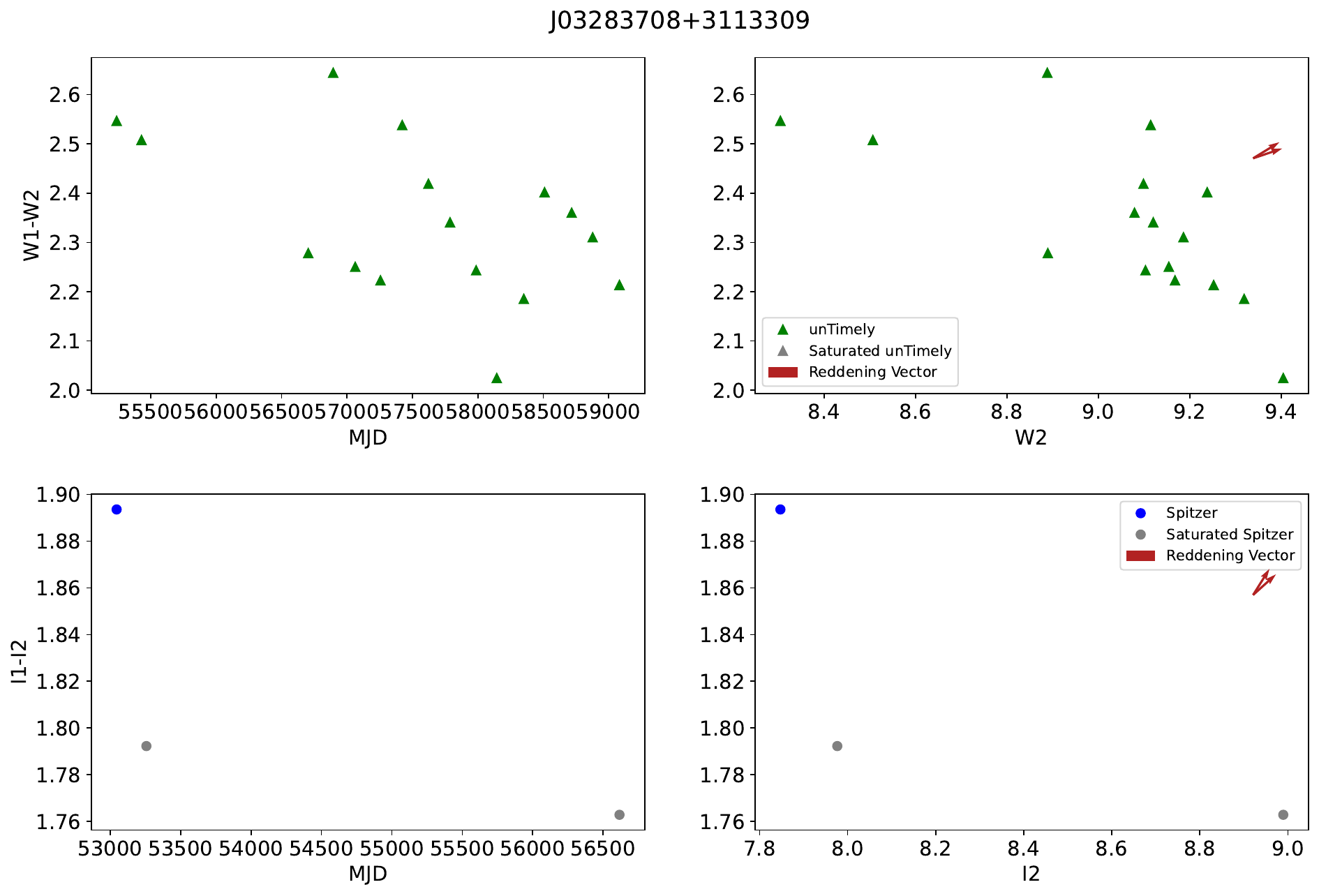}
    \caption{Top two panels: 3.6 and 4.5~$\mu$m light curves of the fading YSOs. The magnitudes corresponding to a zero $\Delta$mag in the light curves and the offsets applied to the WISE photometry are given in the legend. Bottom four panels: clockwise from upper left, the W1-W2 vs MJD, W1-W2 vs 4.5~$\mu m$ magnitude, I1-I2 vs 4.5~$\mu m$ magnitude, and I1-I2 vs MJD diagrams.}
    %The light curves, color curves, and CMD diagrams of the fading YSOs in our sample. These are the protostar J03283708+3113309 and the pre-ms star with disk J03290473+3111349 in the NGC 1333 cluster, the protostar J05464312+0000525 (HOPS~363) and pre-ms star with disk J05465358+0000061 in the Orion~B cloud, and pre-ms star with disk J16265196-2430396 (ISO~Oph~75) in the Ophiuchus cloud, and the protostar J19025867-3707361 in Corona Australis cluster.}
    \label{fig:lc_fades_appendix}
\end{figure*}
\addtocounter{figure}{-1}

\begin{figure*}
    \centering
    \includegraphics[width=15cm, height=10.5cm]{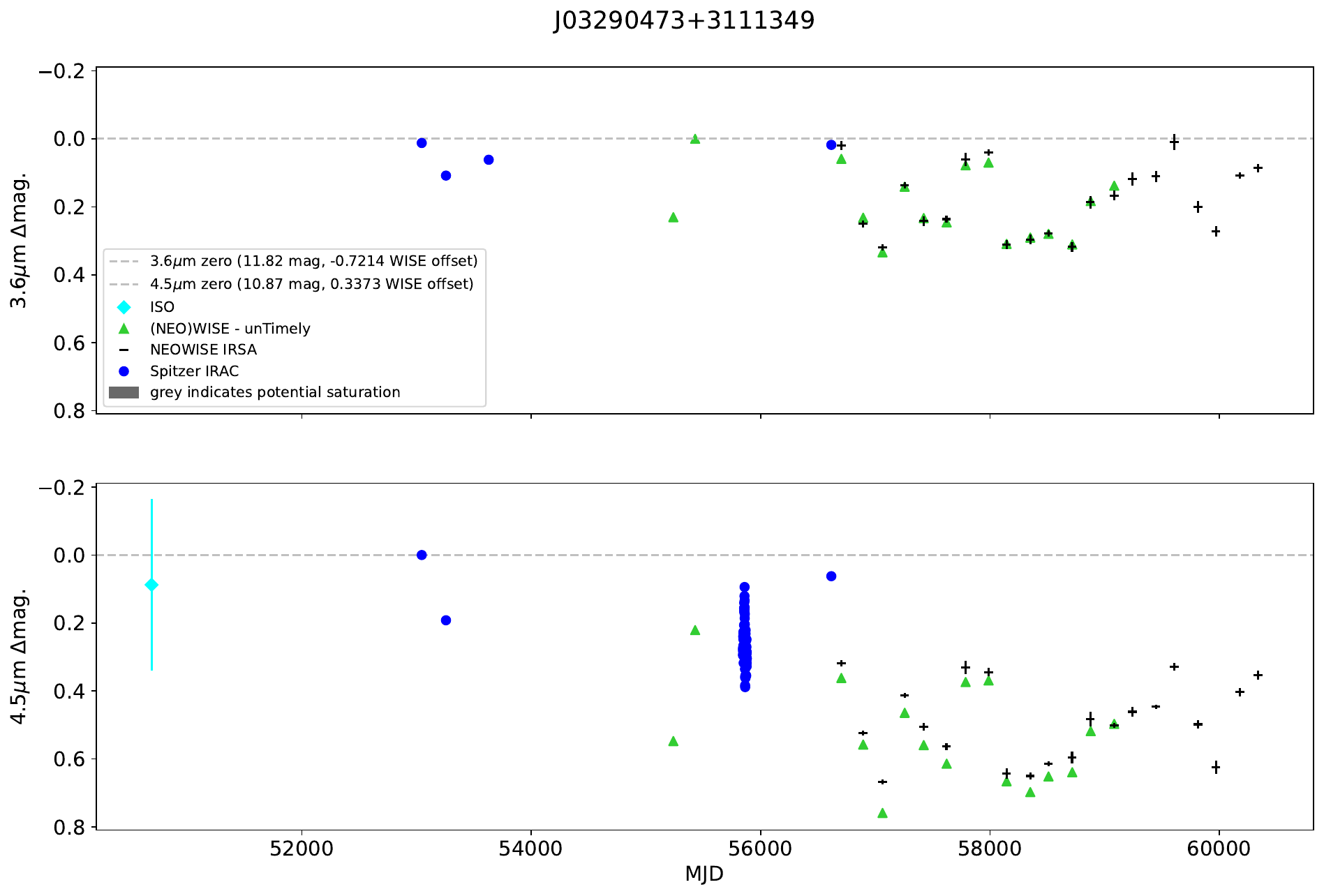}
    \includegraphics[width=15cm, height=10.5cm]{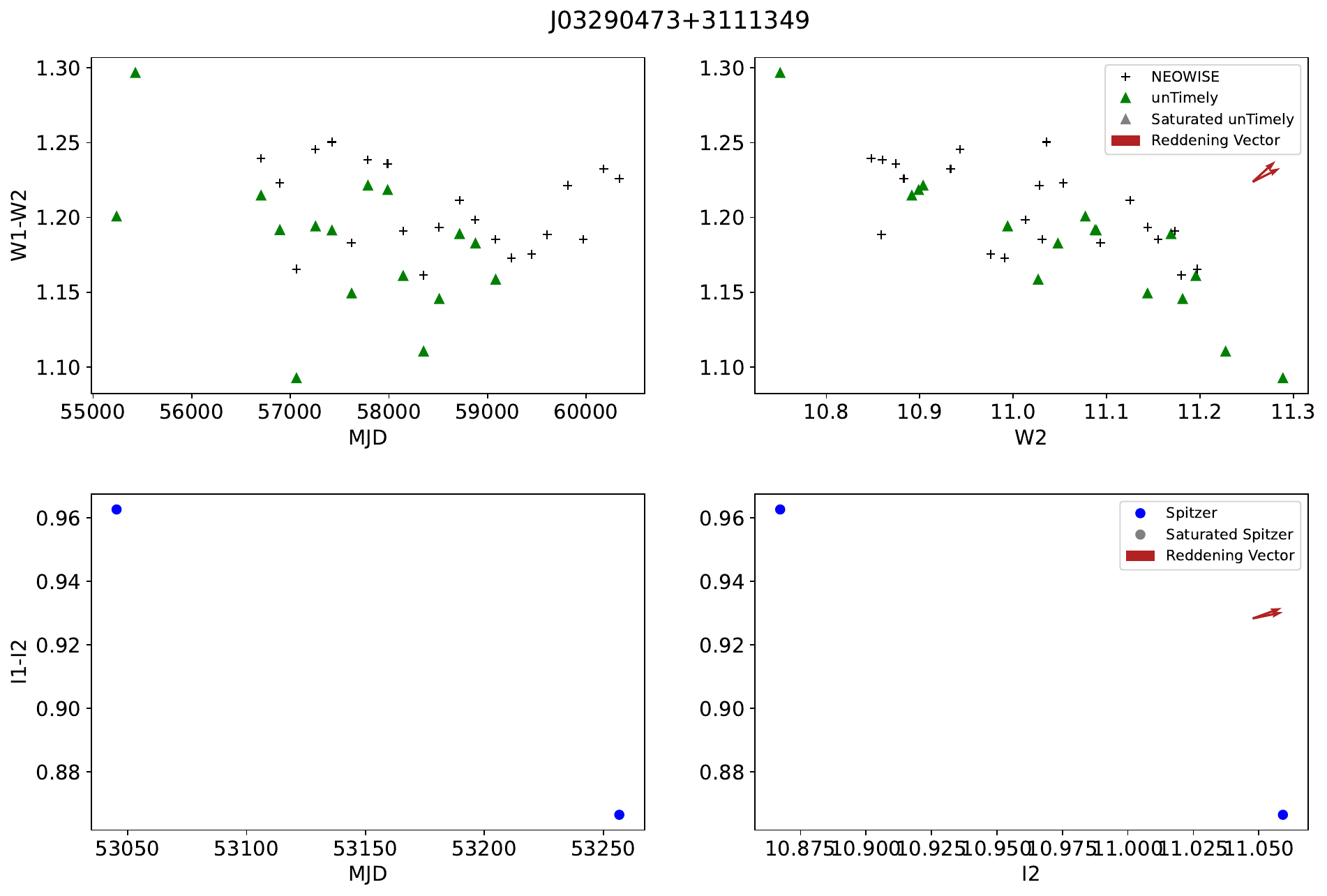}
    \caption{See above.}
\end{figure*}
\addtocounter{figure}{-1}

\begin{figure*}
    \centering
    \includegraphics[width=15cm, height=10.5cm]{J05465358+0000061_DLC_v3.pdf}
    \includegraphics[width=15cm, height=10.5cm]{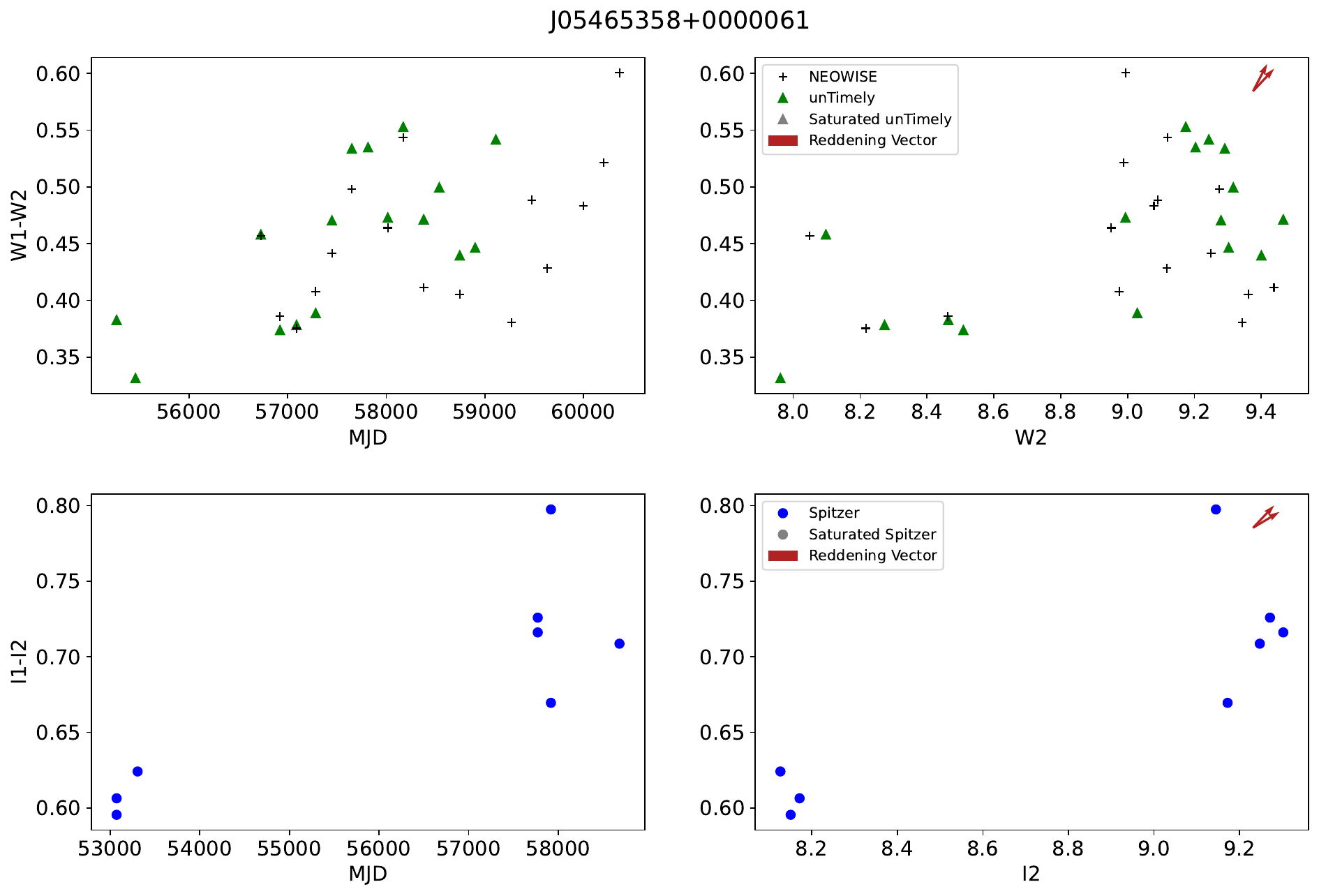}
    \caption{See above.}
\end{figure*}
\addtocounter{figure}{-1}

\begin{figure*}
    \centering
    \includegraphics[width=15cm, height=10.5cm]{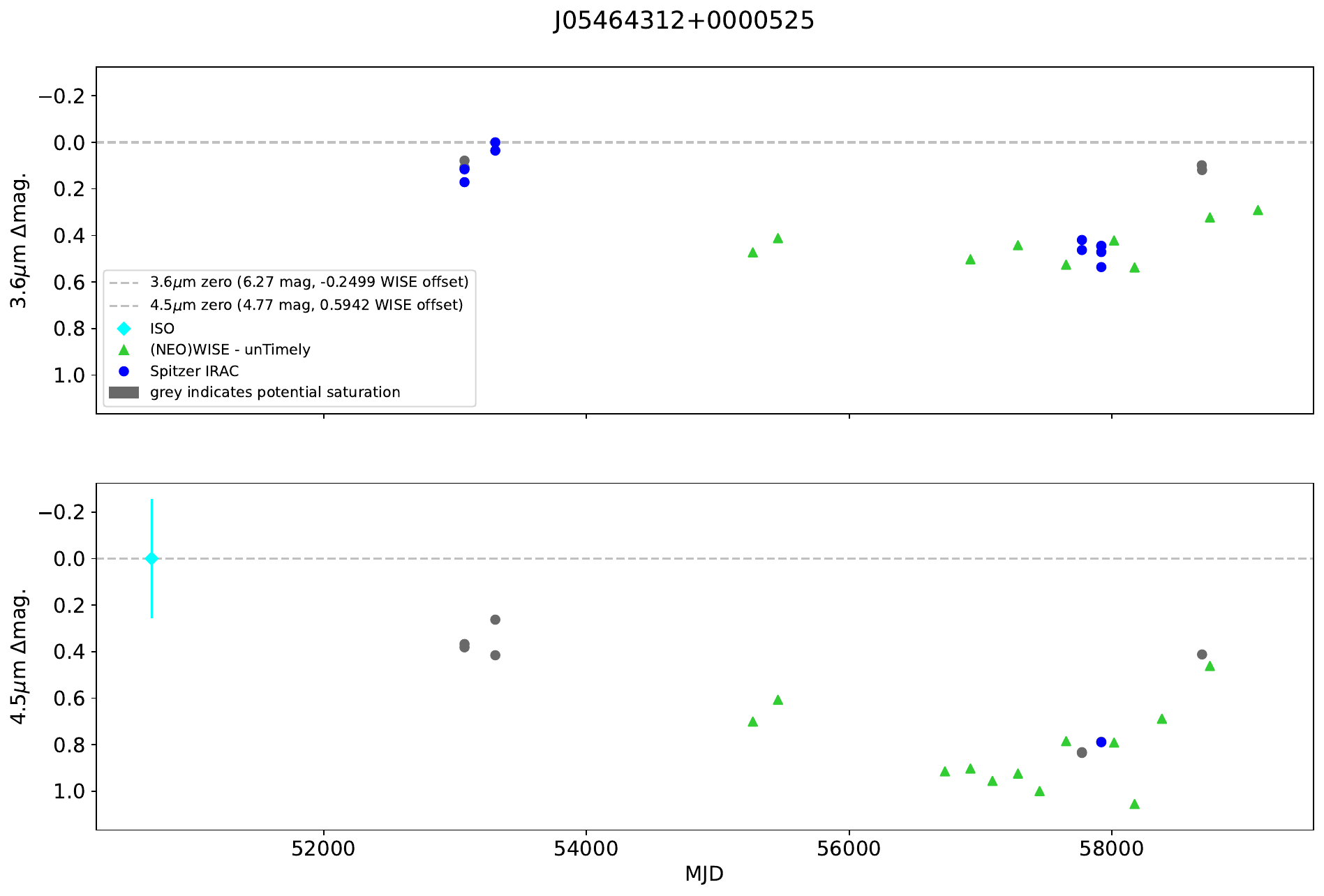}
    \includegraphics[width=15cm, height=10.5cm]{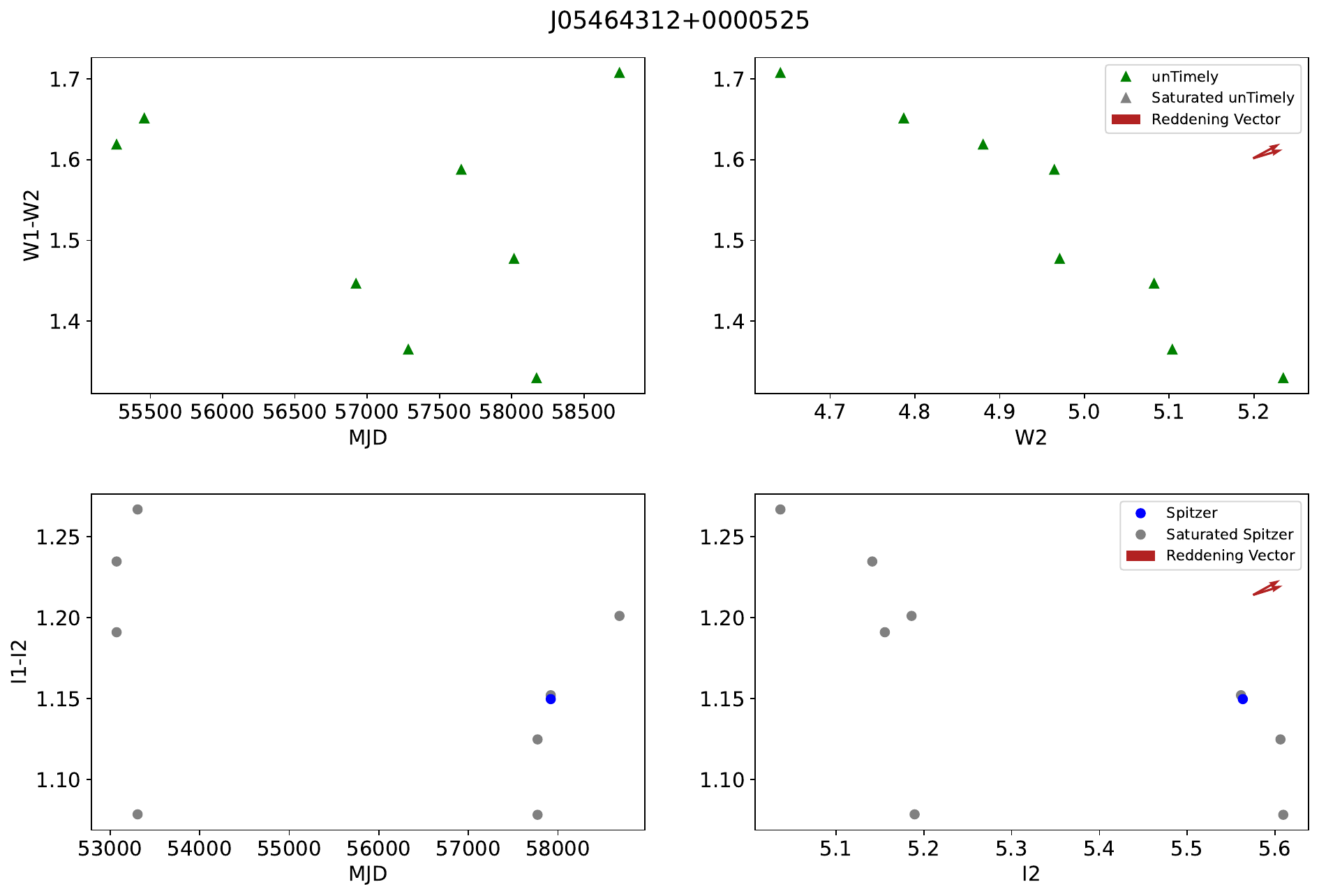}
    \caption{See above.}
\end{figure*}

\addtocounter{figure}{-1}

\begin{figure*}
    \centering
    \includegraphics[width=15cm, height=10.5cm]{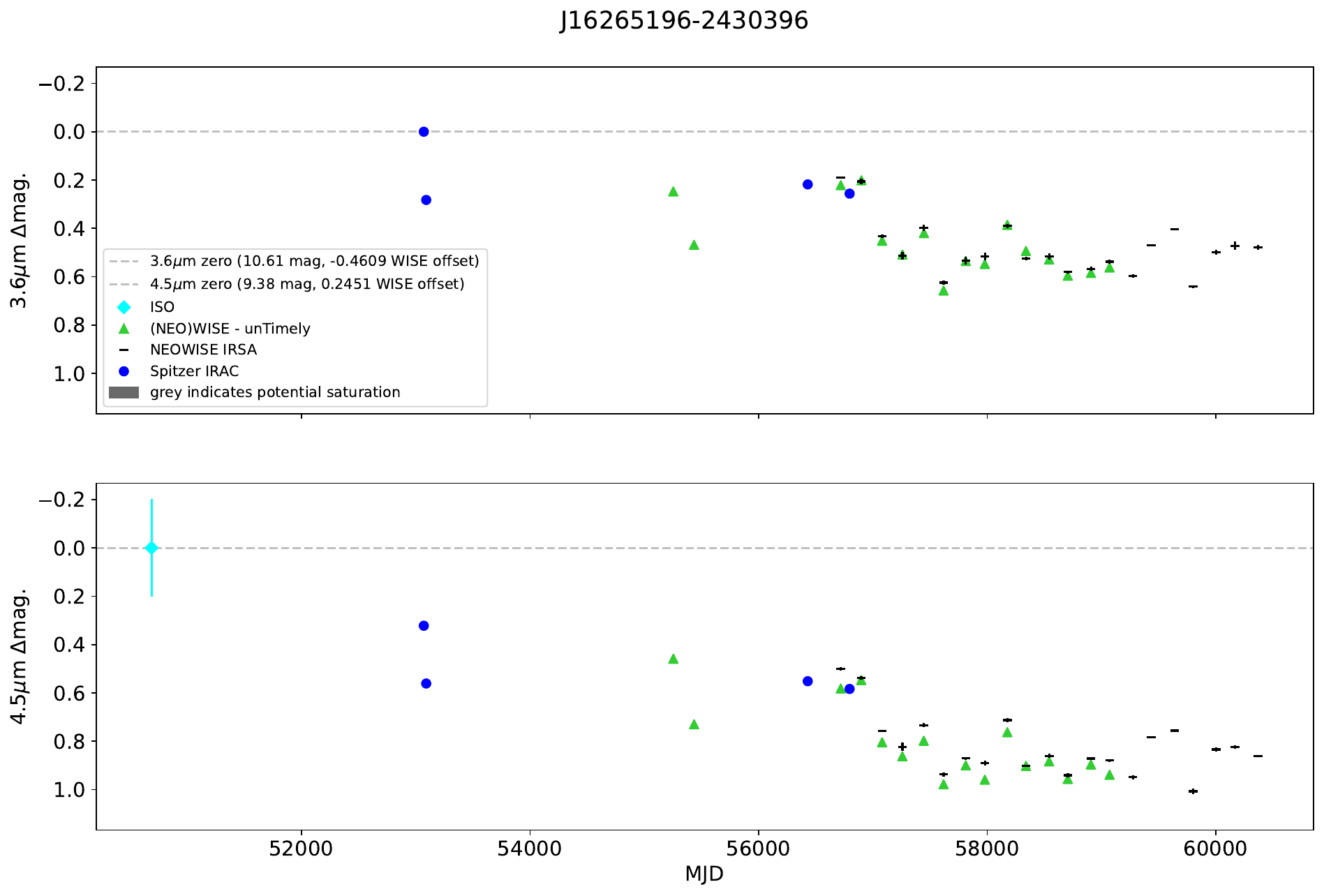}
    \includegraphics[width=15cm, height=10.5cm]{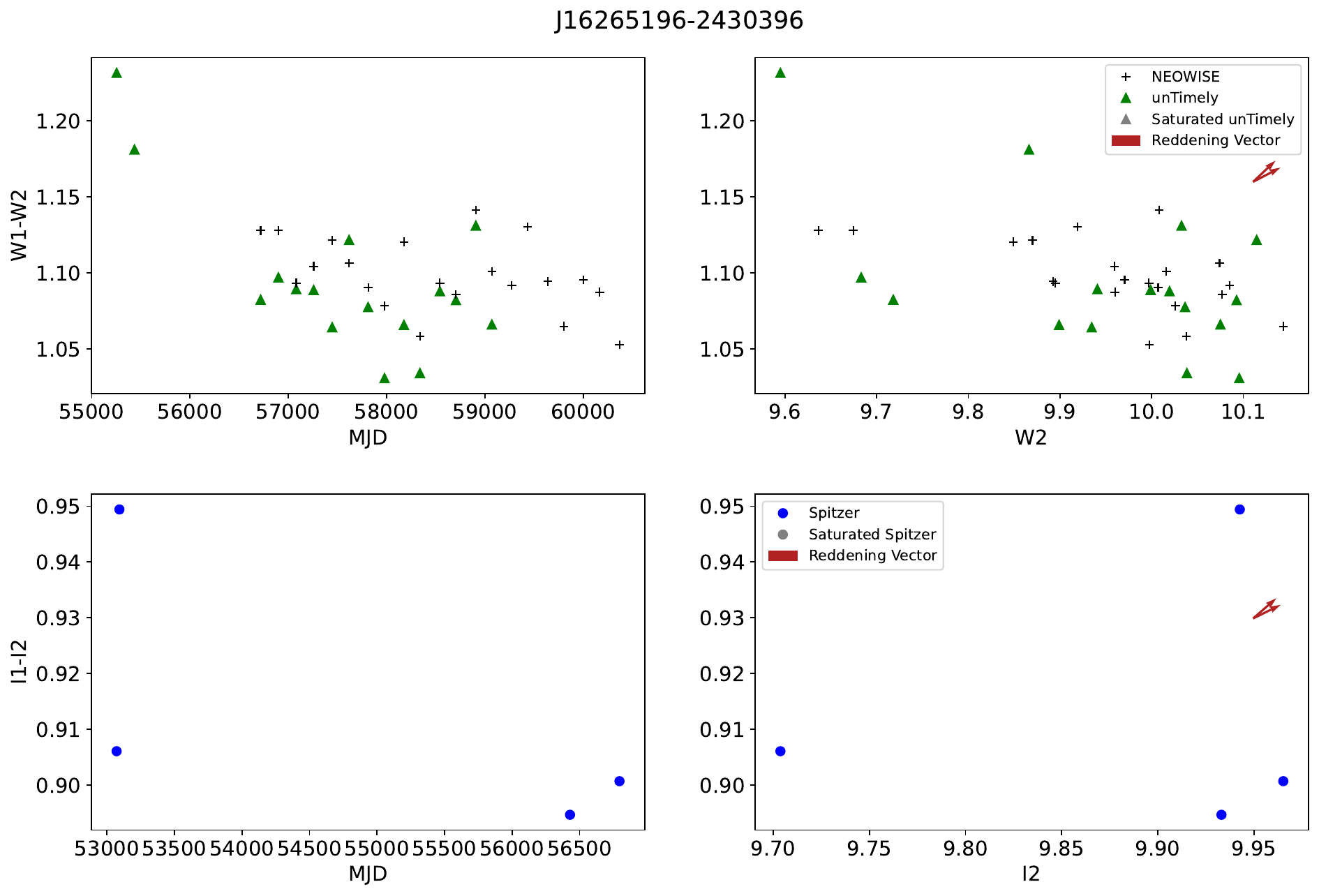}
    \caption{See above.}
\end{figure*}
\addtocounter{figure}{-1}

\begin{figure*}
    \centering
    \includegraphics[width=15cm, height=10.5cm]{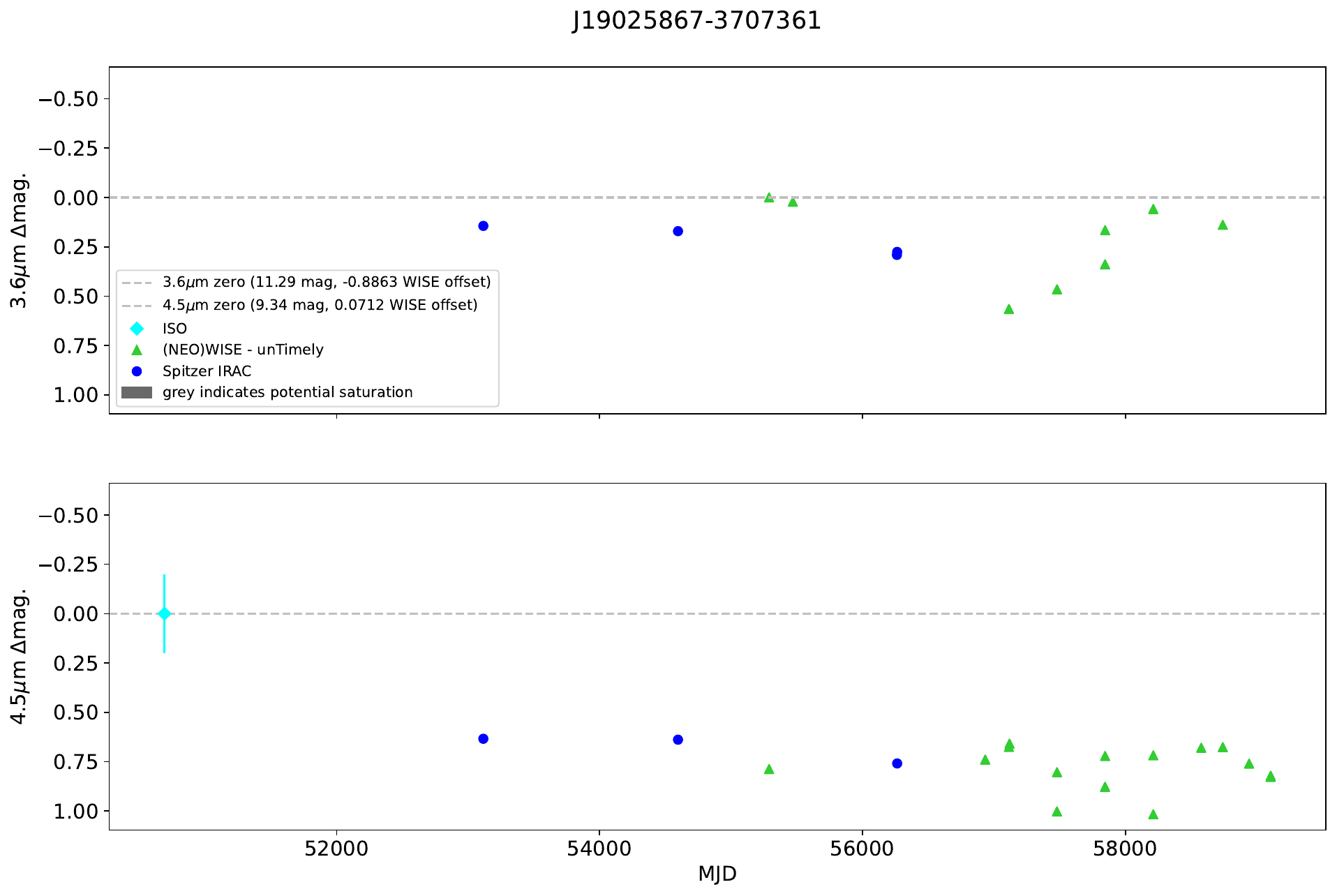}
    \includegraphics[width=15cm, height=10.5cm]{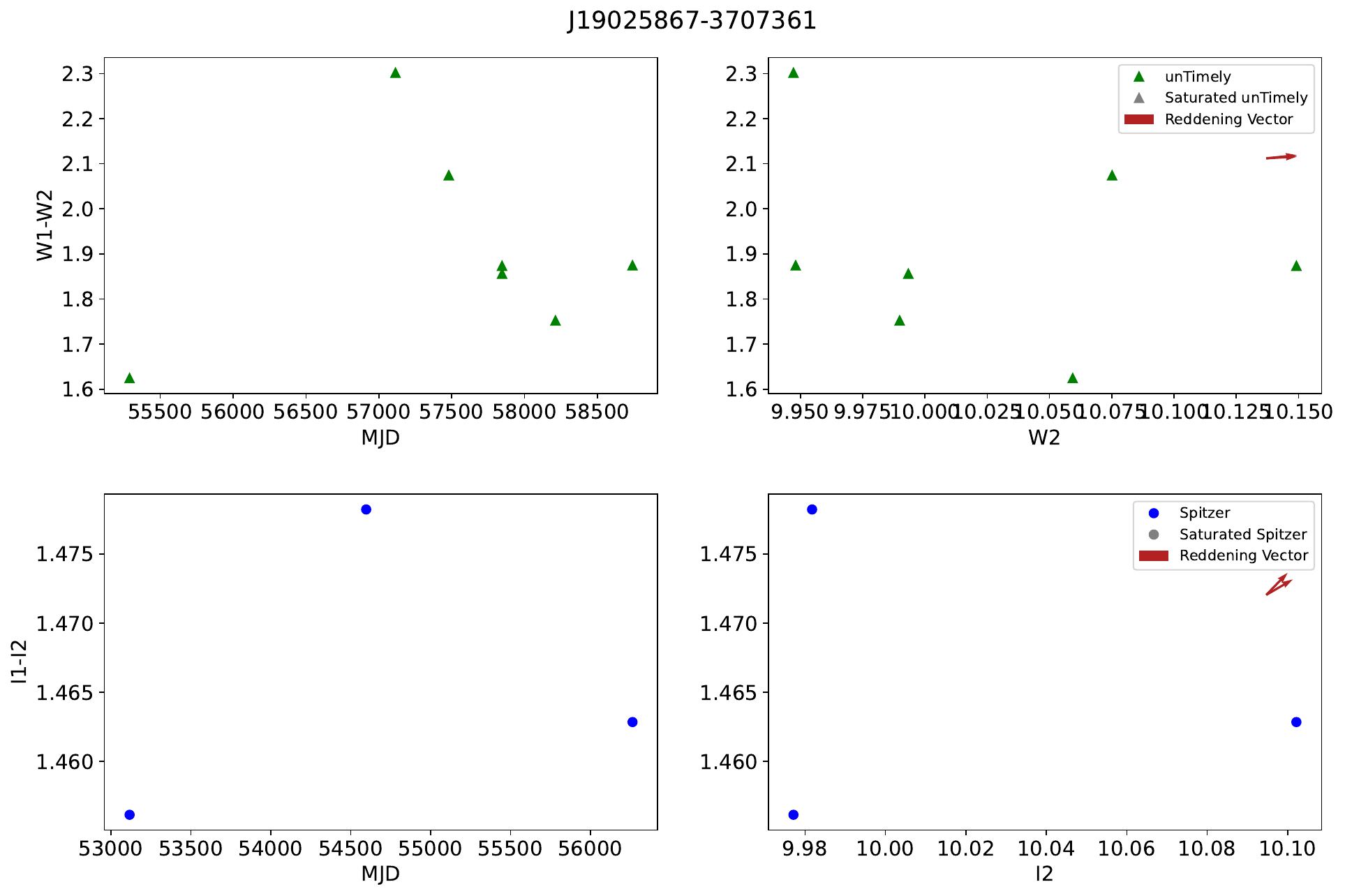}
     \caption{See above.}
\end{figure*}

% J03283452+3107055 XX
% J03284529+3105419 XX
% J03285102+3118184 XX
% J03285120+3119549 XX
% J03285216+3122453 XX
% J03285955+3121467 XX
% J03292042+3118343 XX
% J05405172-0226486 XX
% J05413972-0202241 XX
% J05415555-0223405 XX
% J05455630+0007085 XX
% J05460363-0014492 XX
% J05460477-0014163 XX
% J05464741+0012594 XX
% J05471062+0021141 XX
% J11094191-7634585 XX
% J11095438-7631114 XX
% J16262755-2441538 XX
% J16264048-2427146 XX
% J16264421-2434487 XX
% J16272180-2429536 XX
% J16274162-2446450 XX
% J16274709-2445353 XX
% J18295117+0116404 XX
% J18295954+0111583 XX
% J19001555-3657578 X
%X
%X

\begin{figure*}
    \centering
    \includegraphics[width=15cm, height=10.5cm]{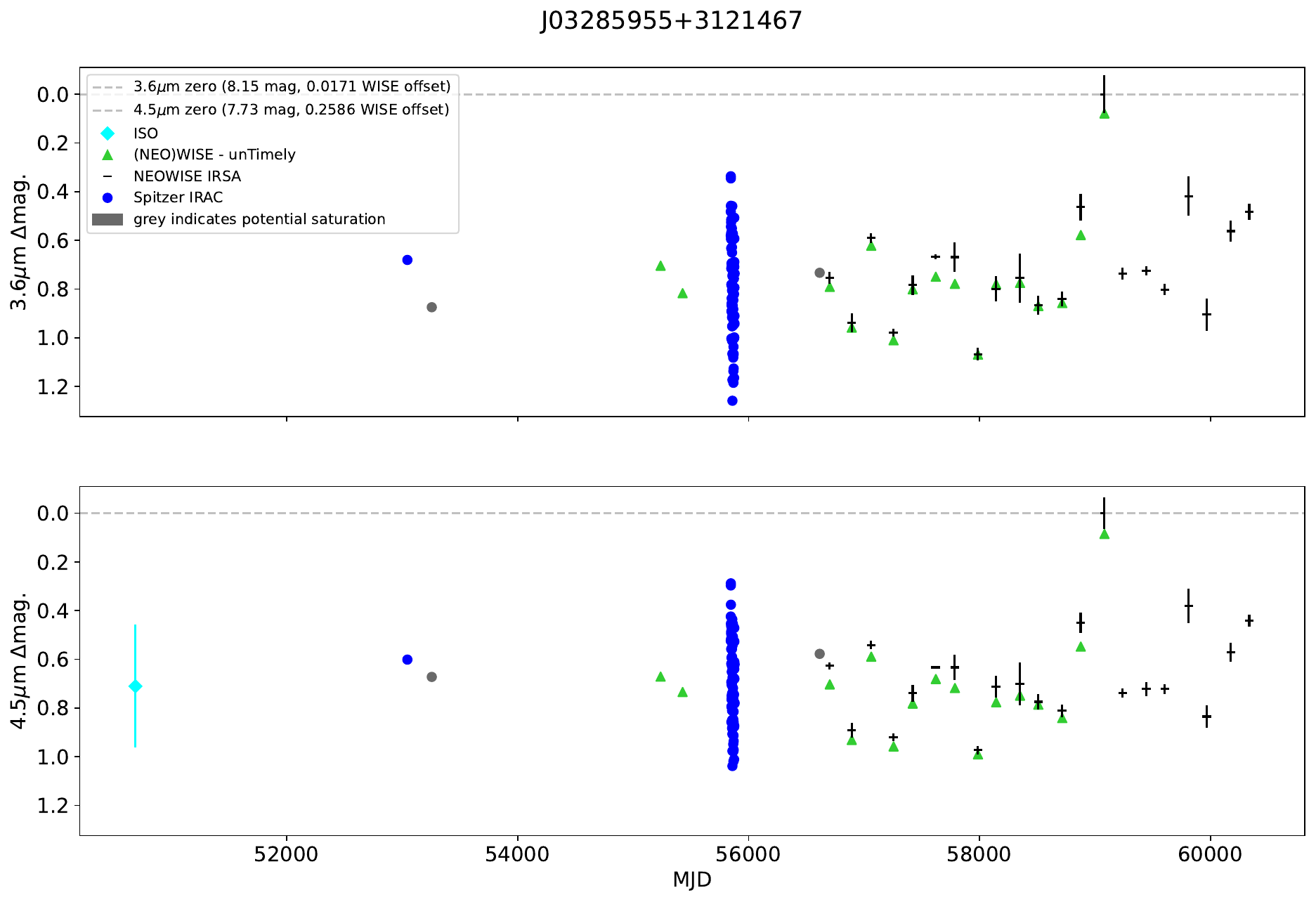}
    \includegraphics[width=15cm, height=10.5cm]{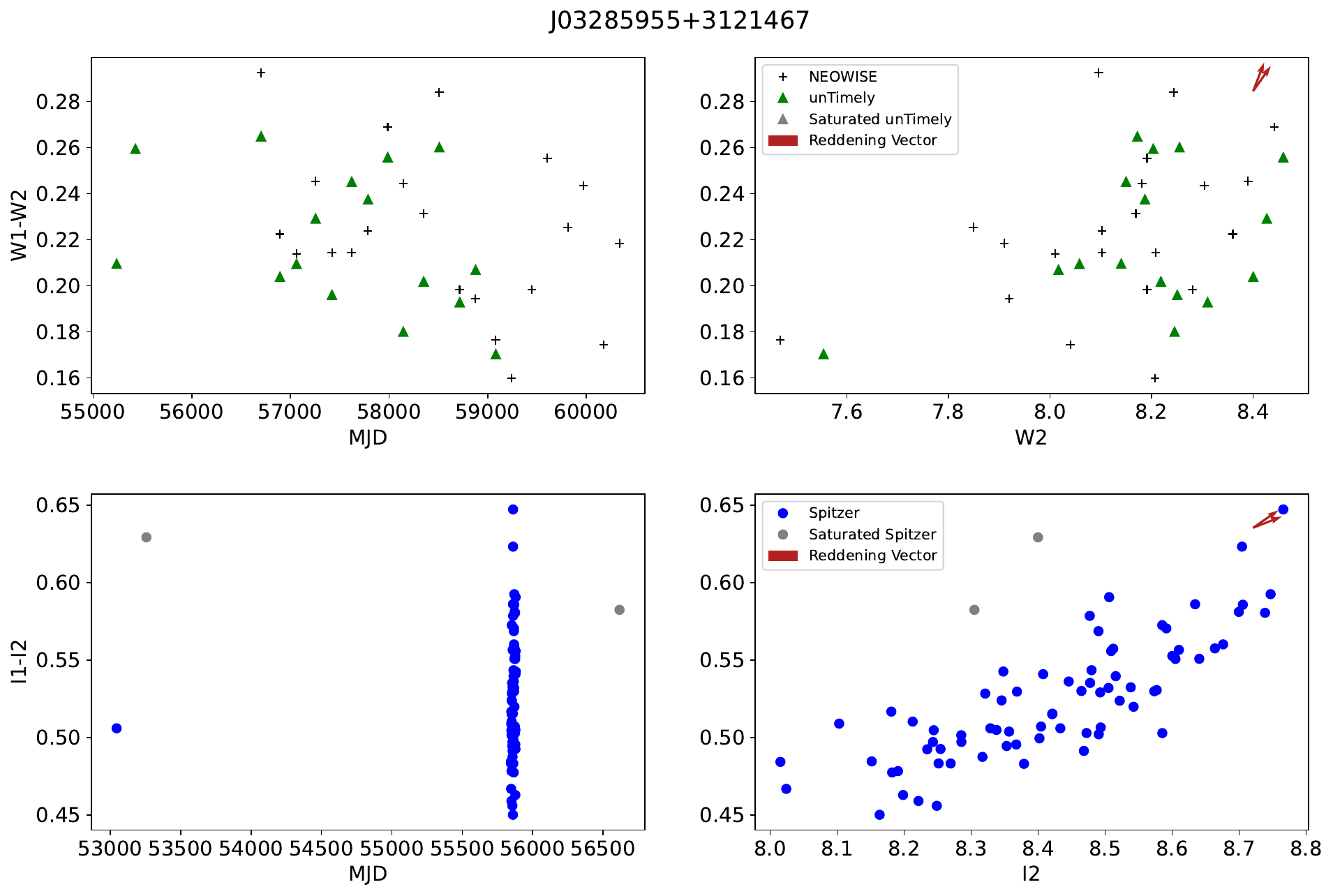}
    \caption{Top two panels: 3.6 and 4.5~$\mu$m light curves of the fluctuating YSOs. The magnitudes corresponding to a zero $\Delta$mag in the light curves and the offsets applied to the WISE photometry are given in the legend. Bottom four panels: clockwise from upper left, the W1-W2 vs MJD, W1-W2 vs 4.5~$\mu m$ magnitude, I1-I2 vs 4.5~$\mu m$ magnitude, and I1-I2 vs MJD diagrams.}
    \label{fig:lc_fluctuators_appendix}
\end{figure*}
% J03285955+3121467 X
\addtocounter{figure}{-1}

\begin{figure*}
    \centering
    \includegraphics[width=15cm, height=10.5cm]{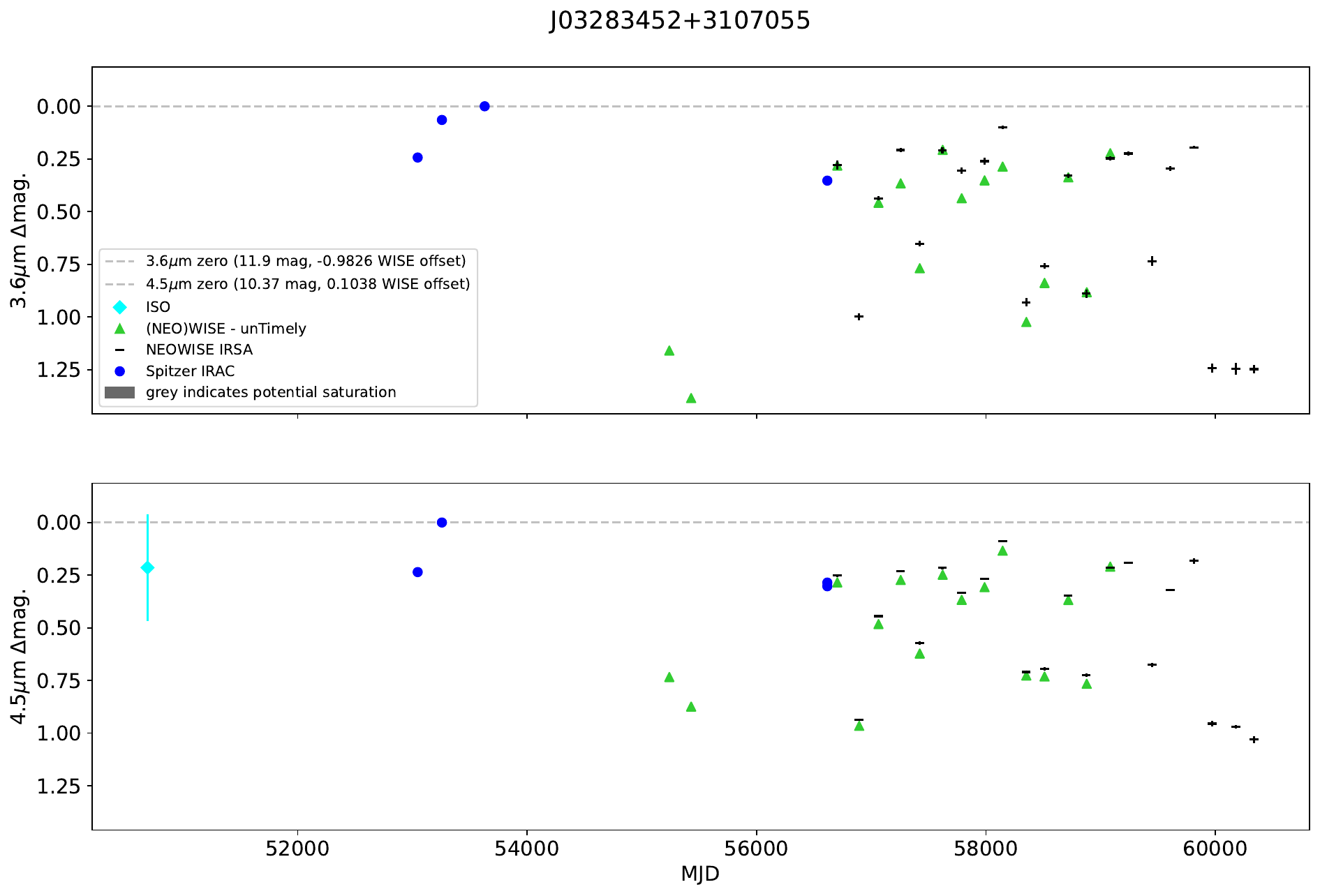}
     \includegraphics[width=15cm, height=10.5cm]{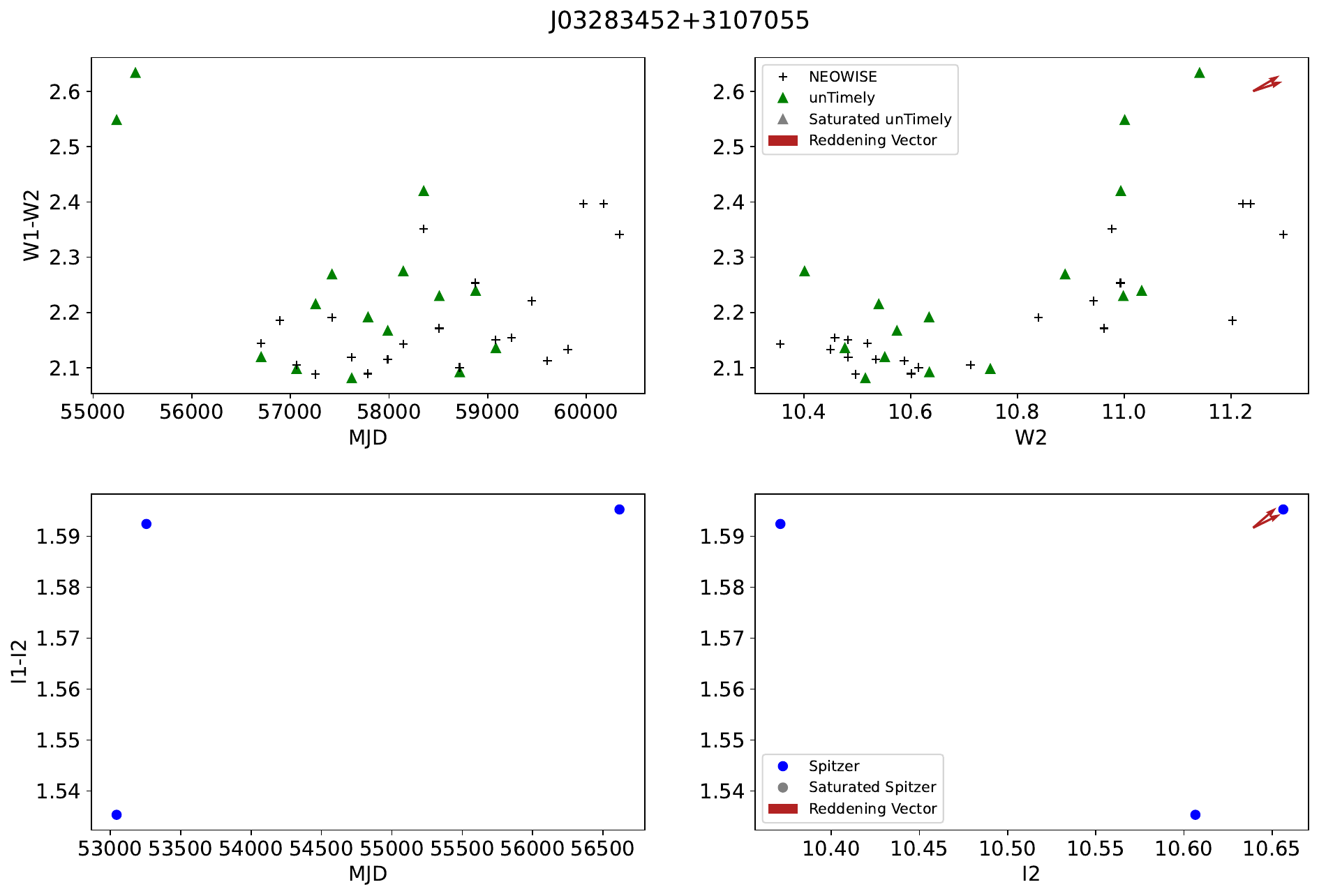}
    \caption{See above.}
\end{figure*}
\addtocounter{figure}{-1}

\begin{figure*}
    \centering
    \includegraphics[width=15cm, height=10.5cm]{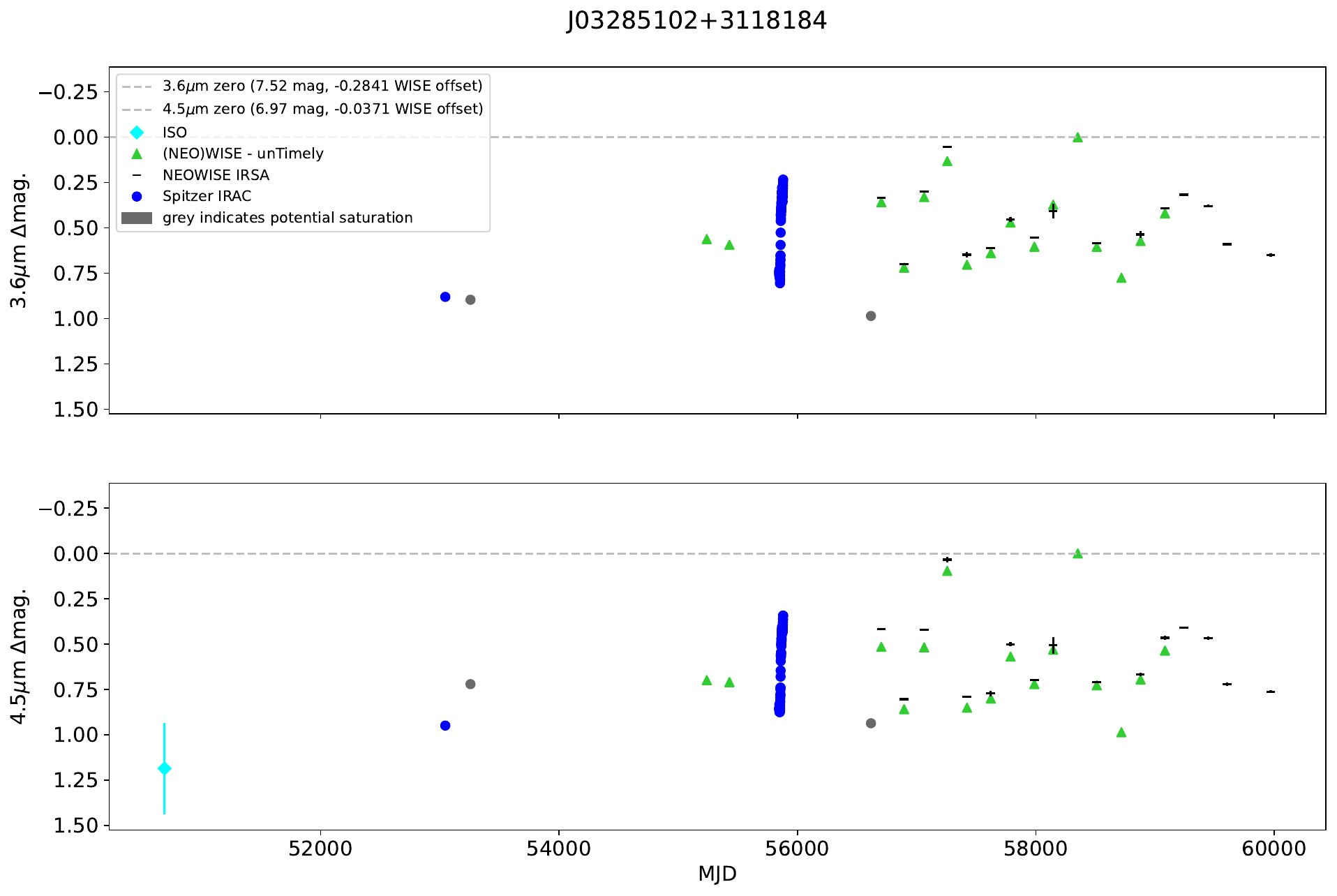}
    \includegraphics[width=15cm, height=10.5cm]{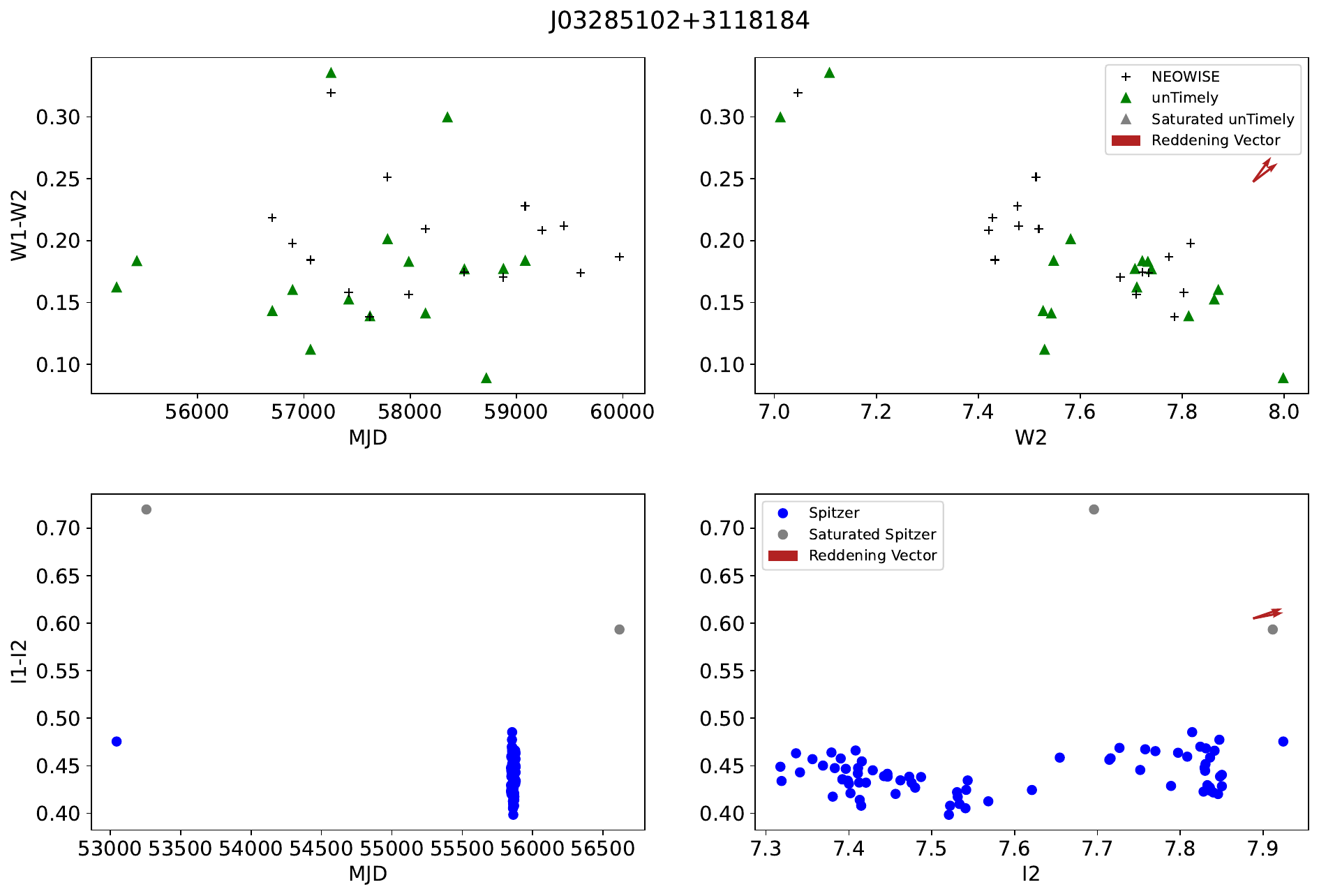}
    \caption{See above.}
    \label{fig:J03285102+3118184}
\end{figure*}
\addtocounter{figure}{-1}

\begin{figure*}
    \centering
    \includegraphics[width=15cm, height=10.5cm]{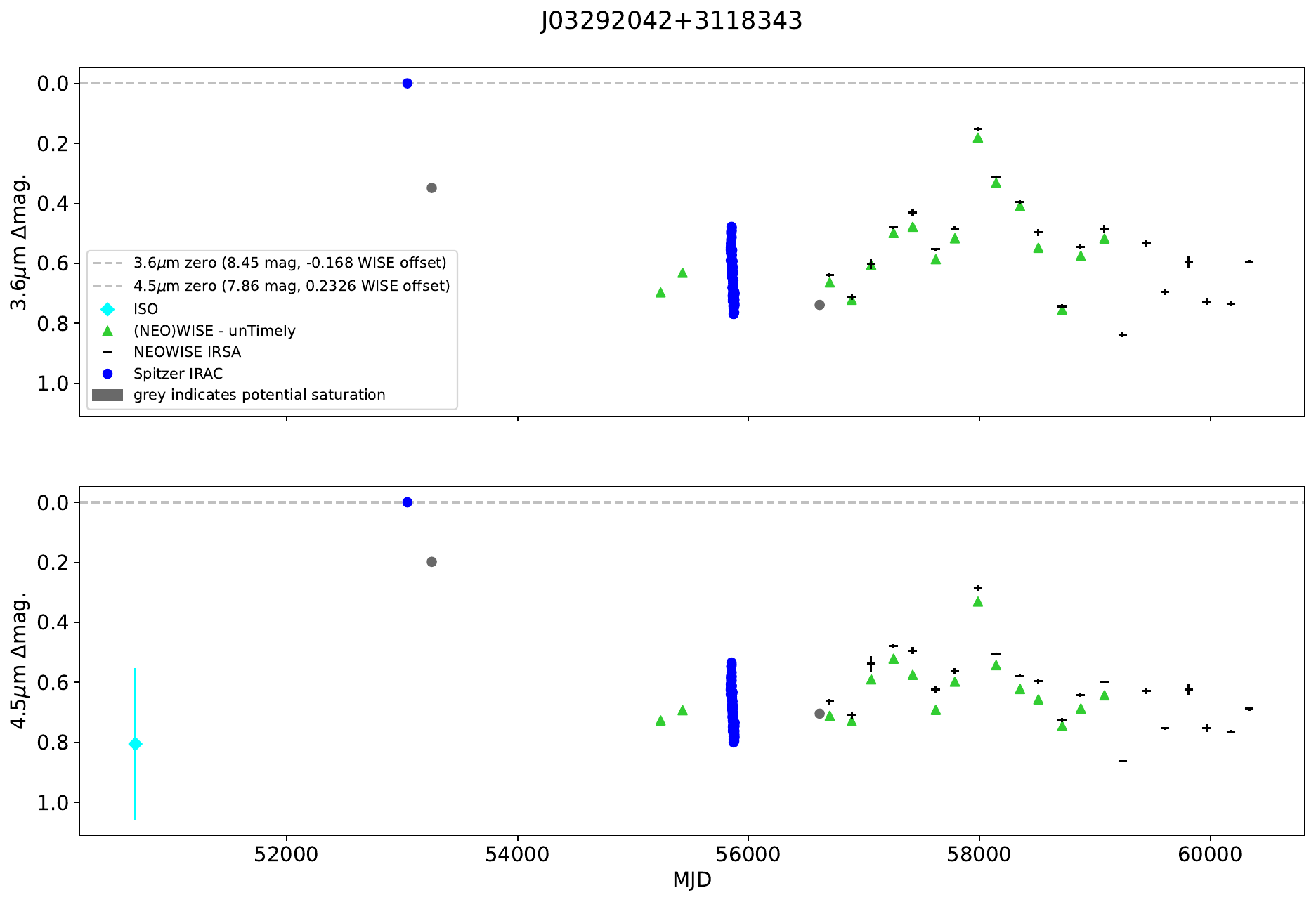}
    \includegraphics[width=15cm, height=10.5cm]{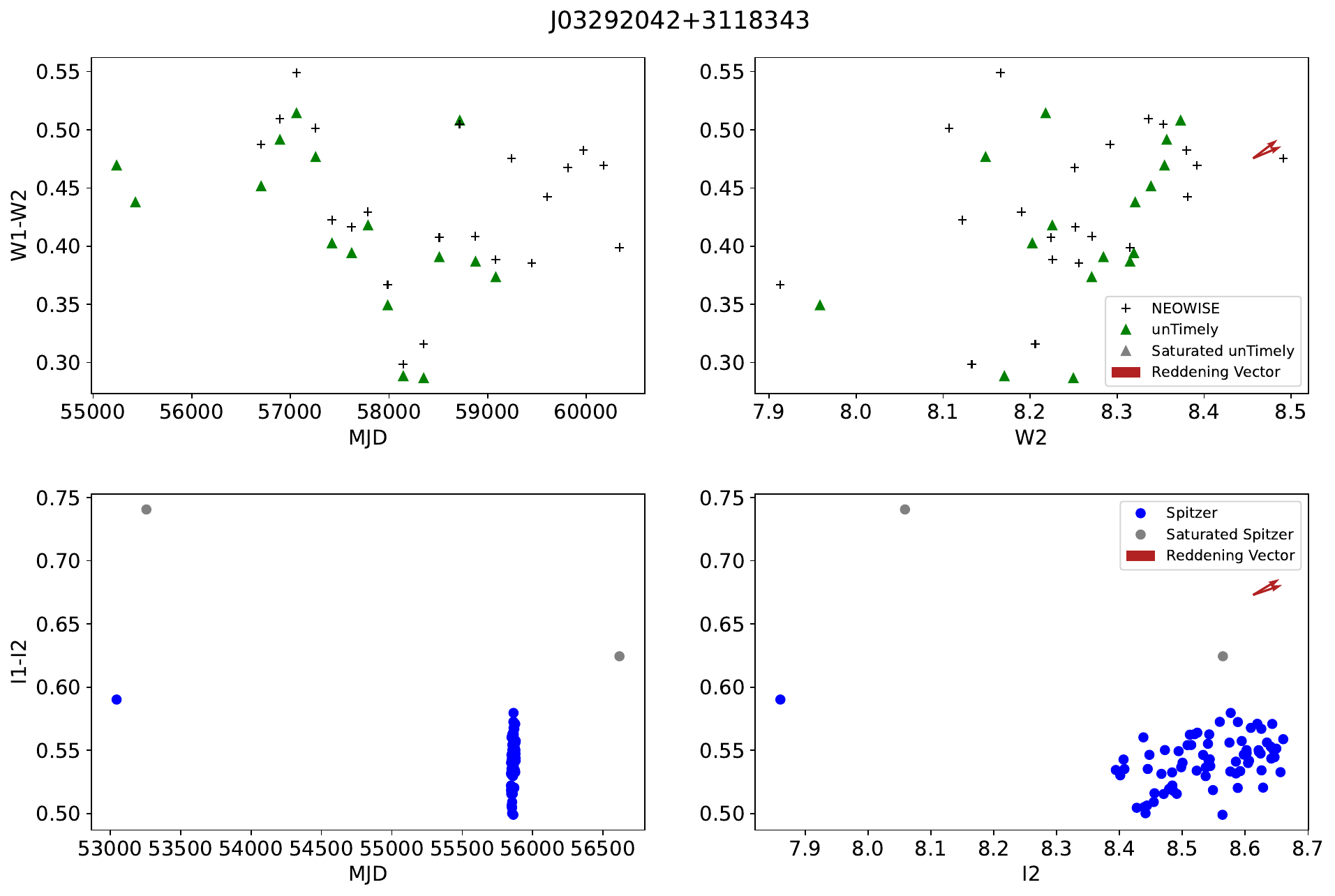}
    \caption{See above.}
    \label{fig:HH_17}
\end{figure*}
\addtocounter{figure}{-1}

\begin{figure*}
    \centering
    \includegraphics[width=15cm, height=10.5cm]{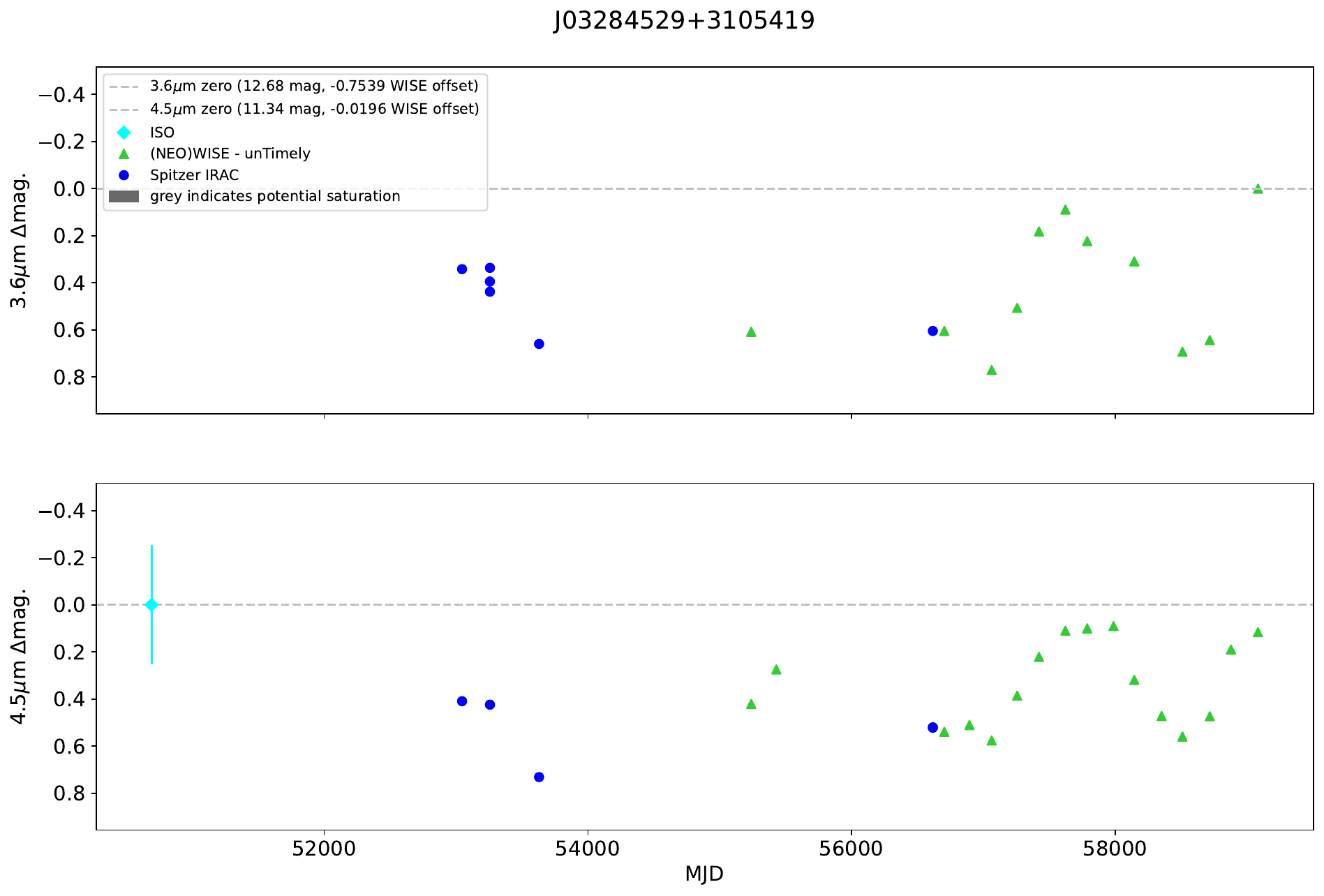}
    \includegraphics[width=15cm, height=10.5cm]{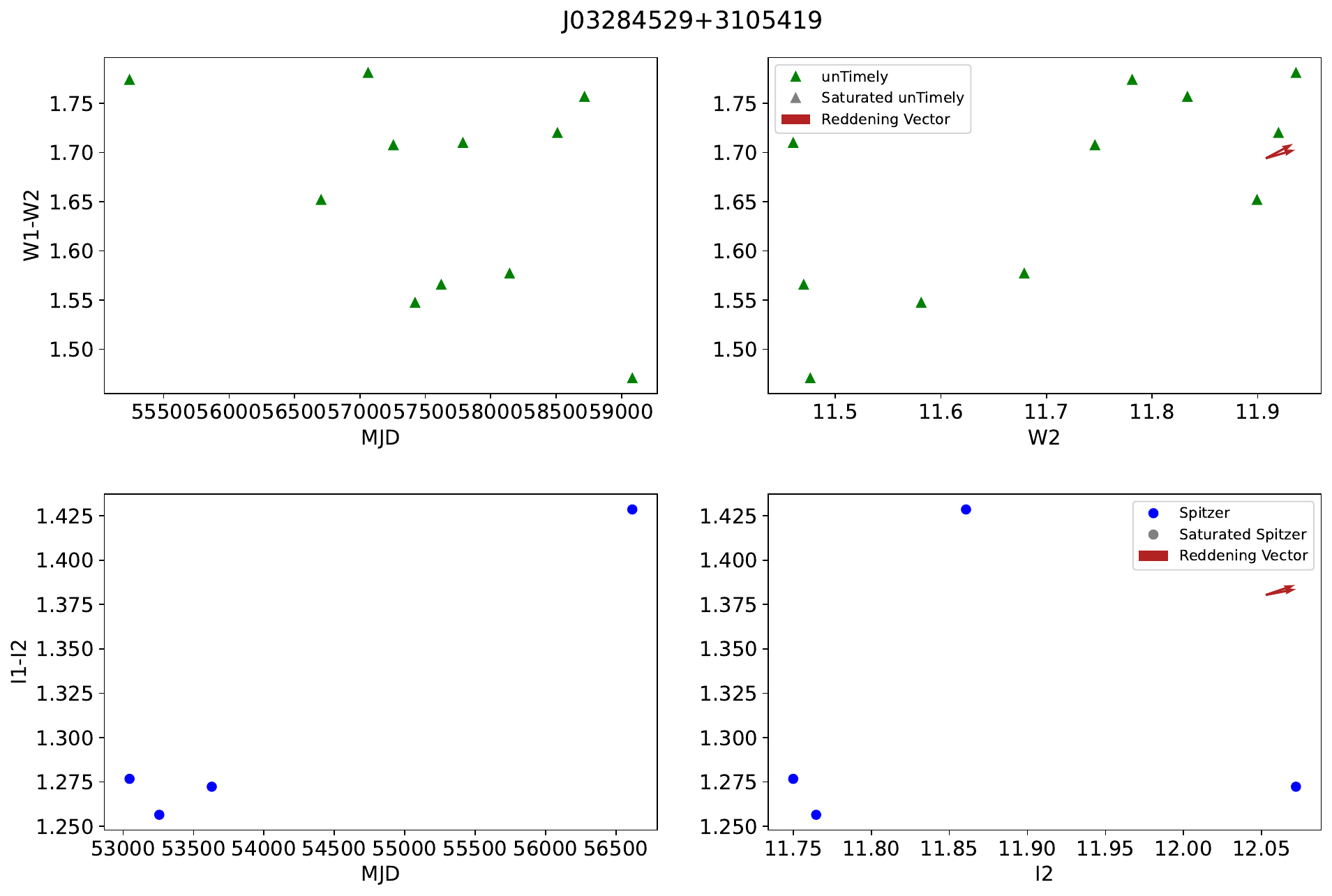}
    \caption{See above.}
    \label{fig:IRAS_03256+3055}
\end{figure*}
\addtocounter{figure}{-1}

\begin{figure*}
    \centering
    \includegraphics[width=15cm, height=10.5cm]{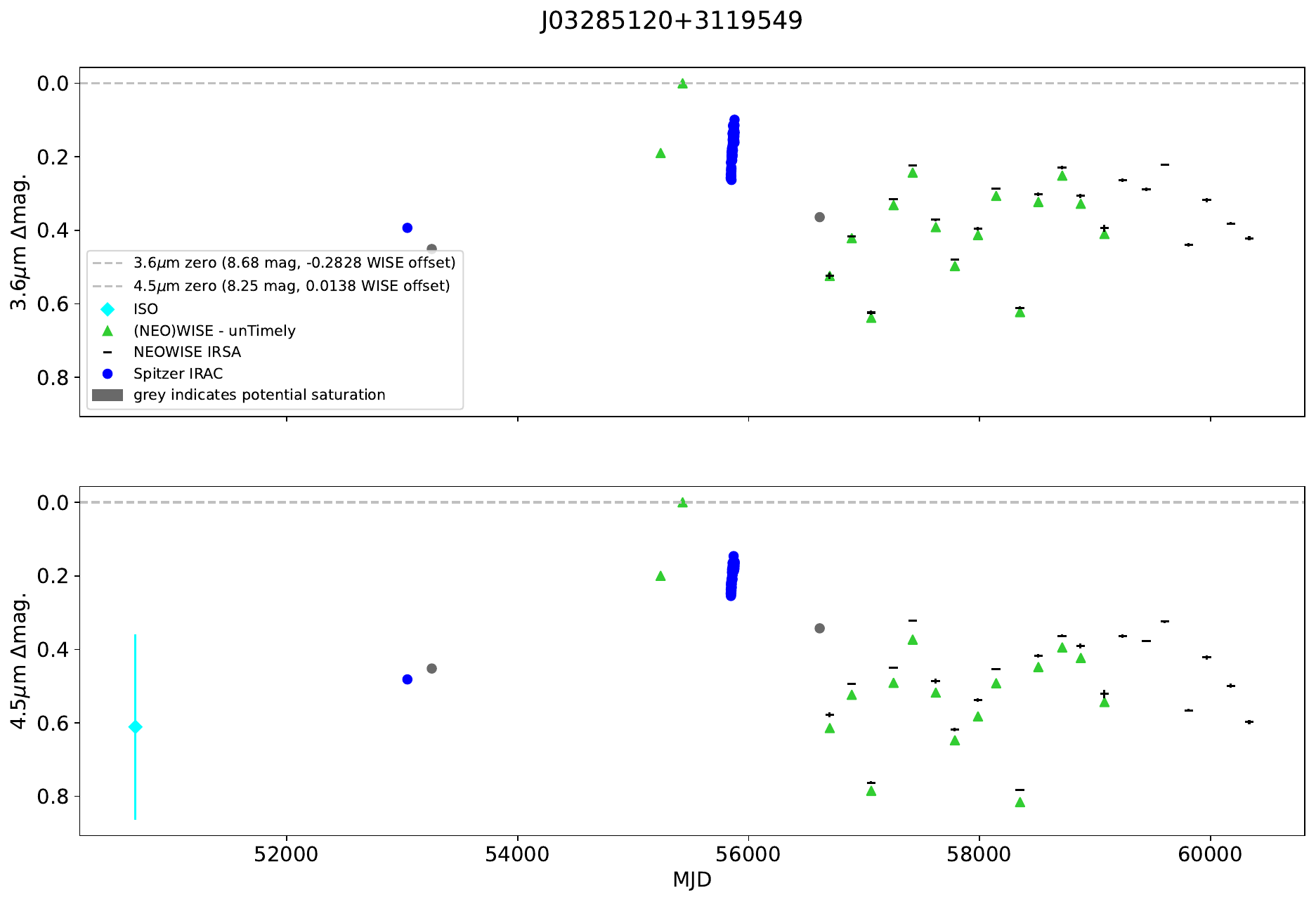}
    \includegraphics[width=15cm, height=10.5cm]{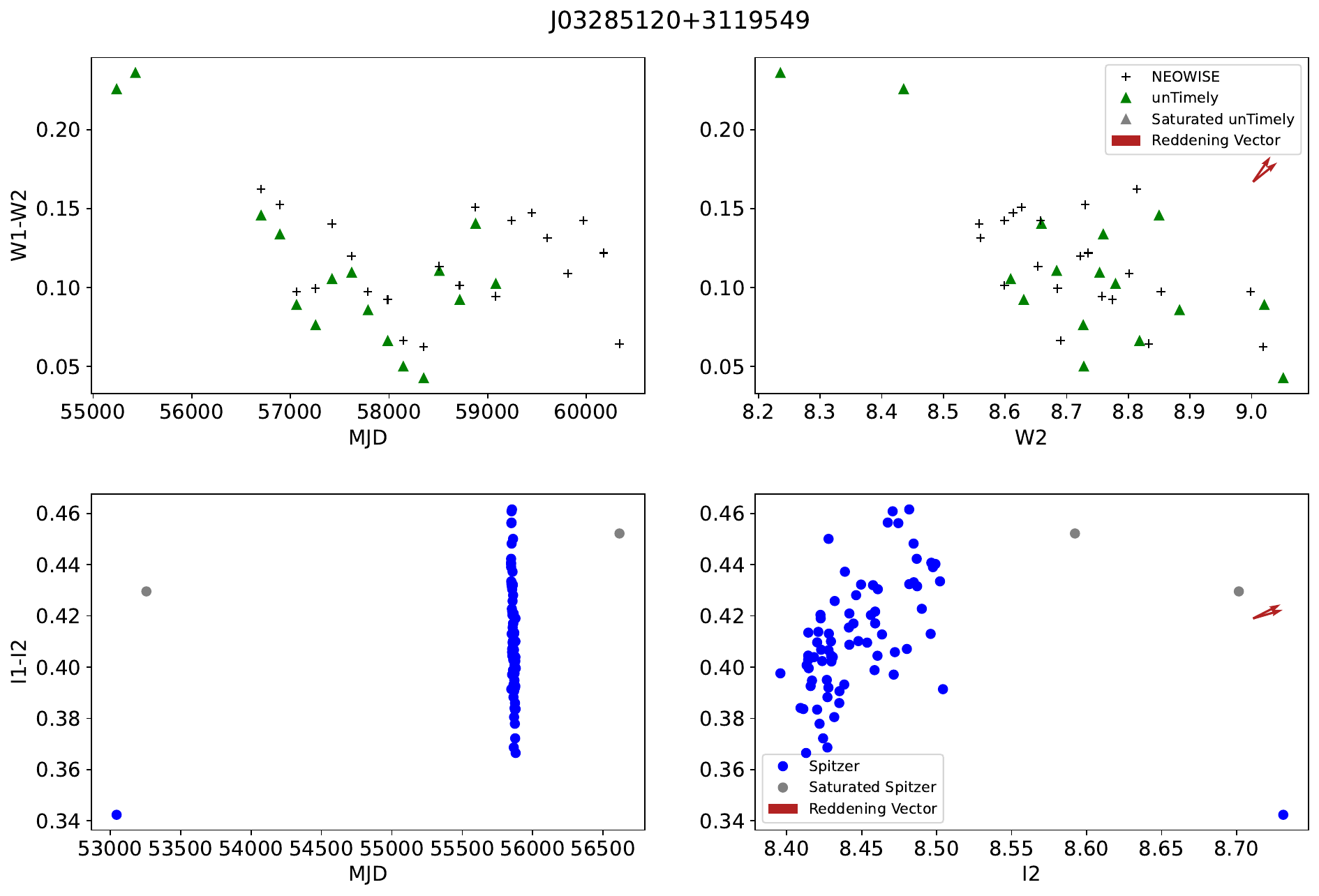}
    \caption{See above.}
    \label{fig:GMM2008_52}
\end{figure*}
\addtocounter{figure}{-1}

\begin{figure*}
    \centering
    \includegraphics[width=15cm, height=10.5cm]{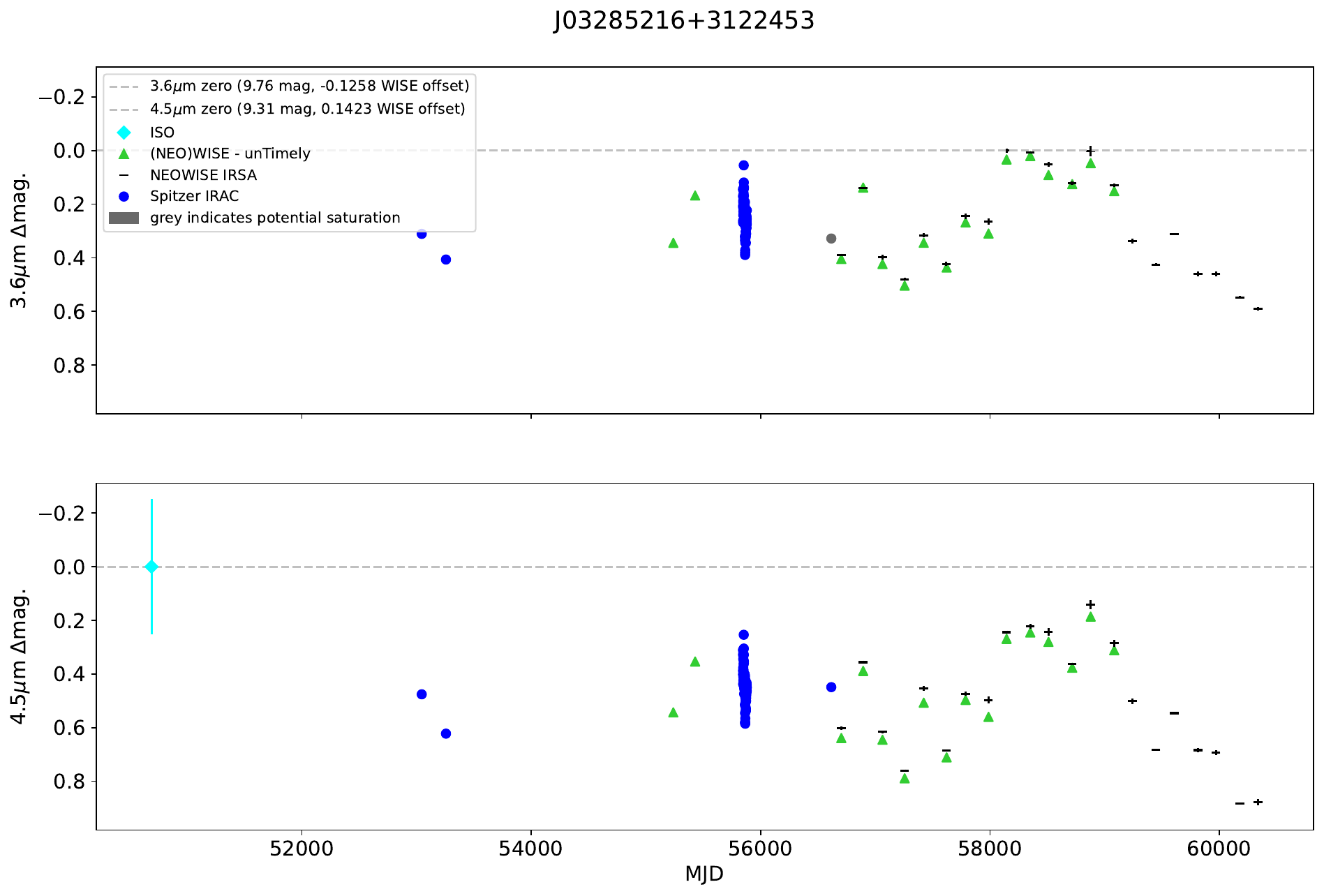}
    \includegraphics[width=15cm, height=10.5cm]{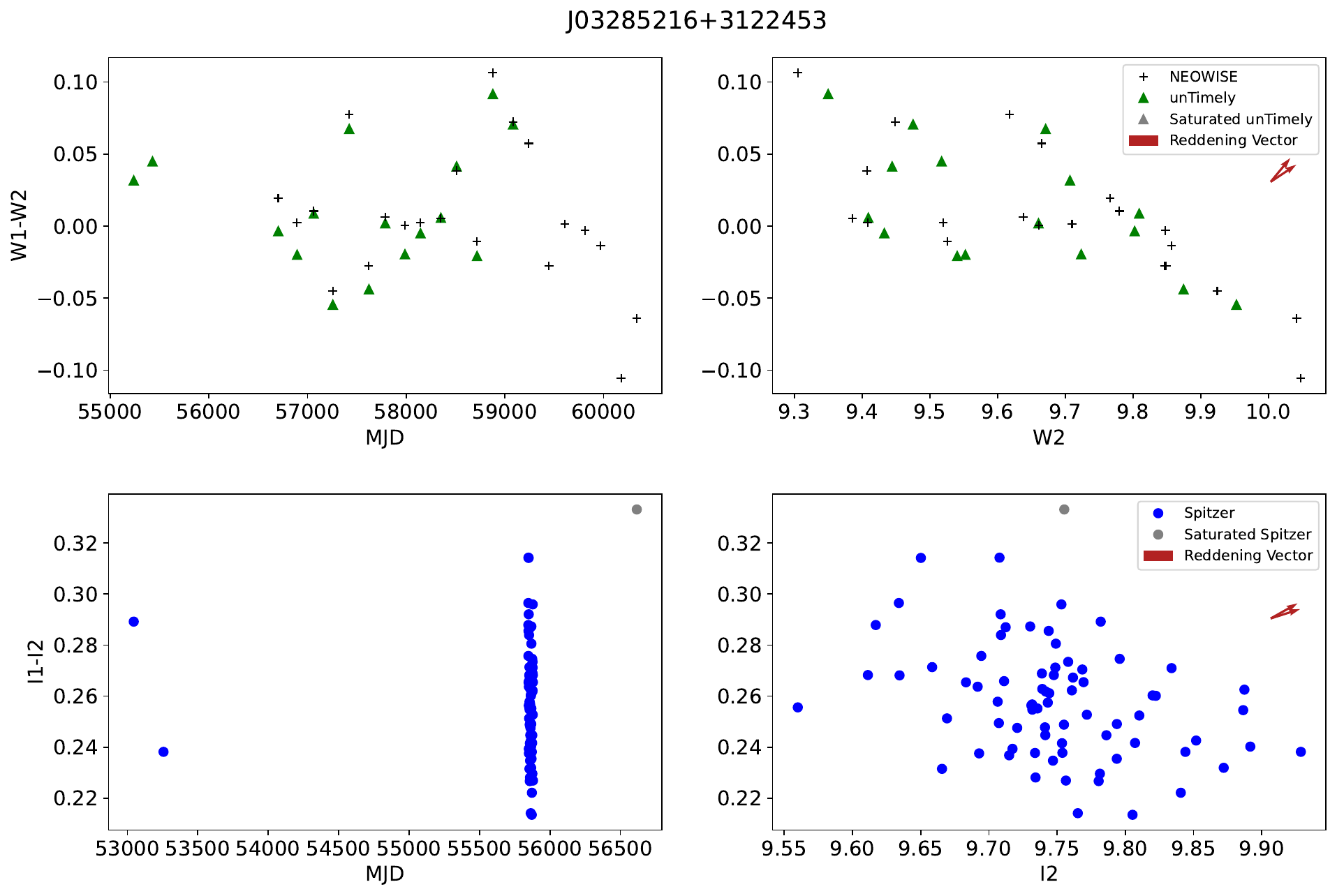}
    \caption{See above.}
    \label{fig:GMM2008_54}
\end{figure*}
% J03285216+3122453 X
\addtocounter{figure}{-1}

\begin{figure*}
    \centering
    \includegraphics[width=15cm, height=10.5cm]{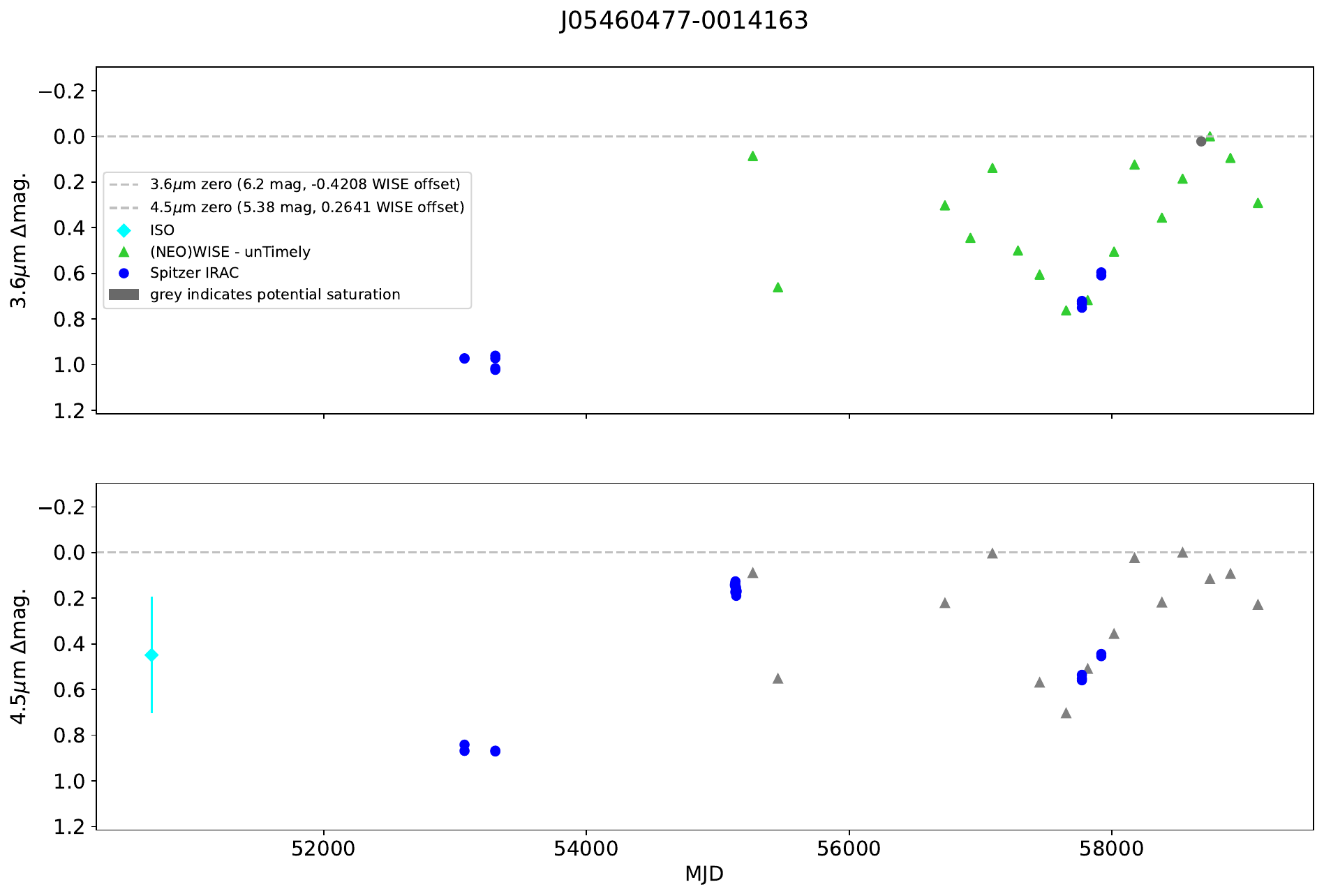}
    \includegraphics[width=15cm, height=10.5cm]{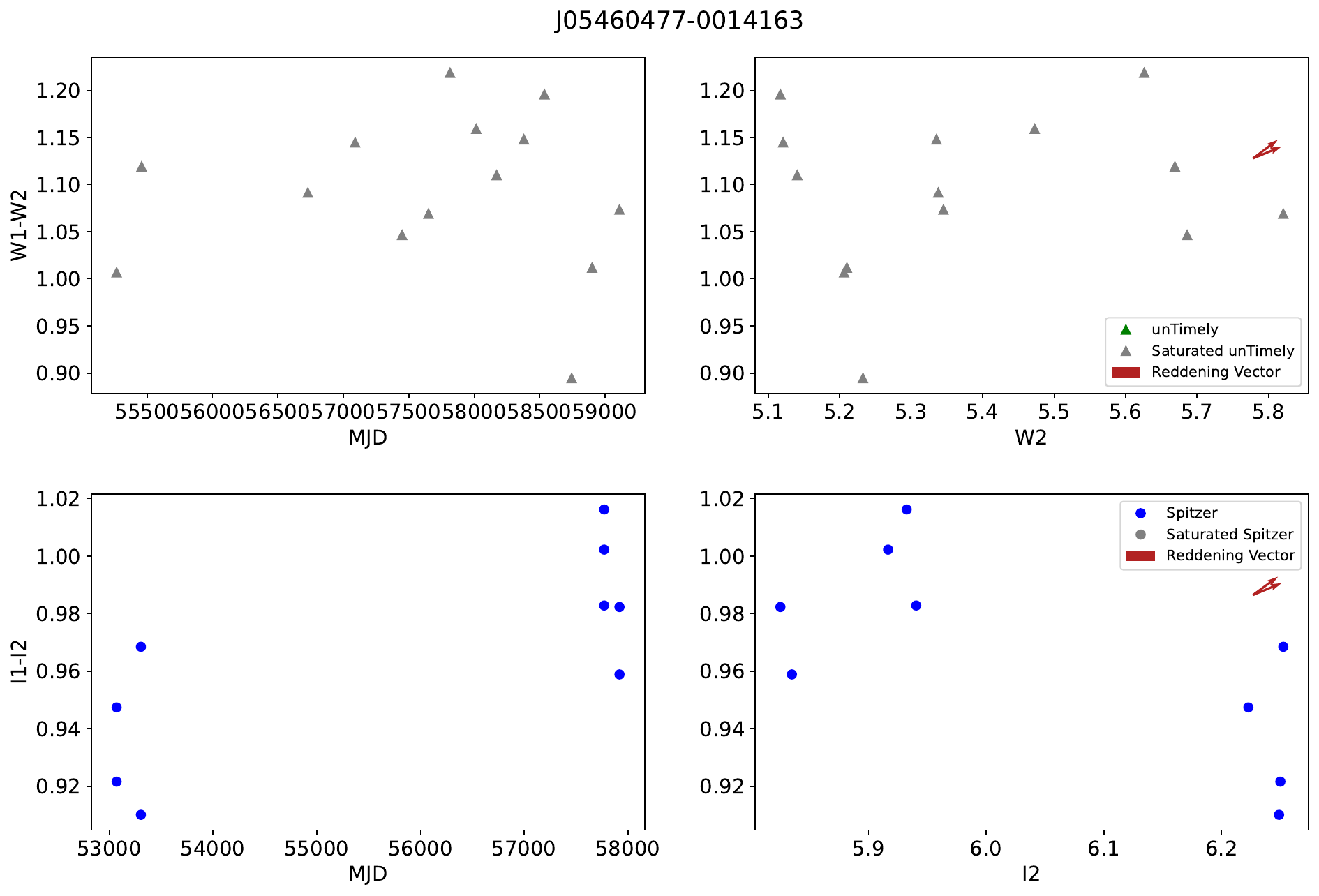}
    \caption{See above.}
\end{figure*}
\addtocounter{figure}{-1}

\begin{figure*}
    \centering
    \includegraphics[width=15cm, height=10.5cm]{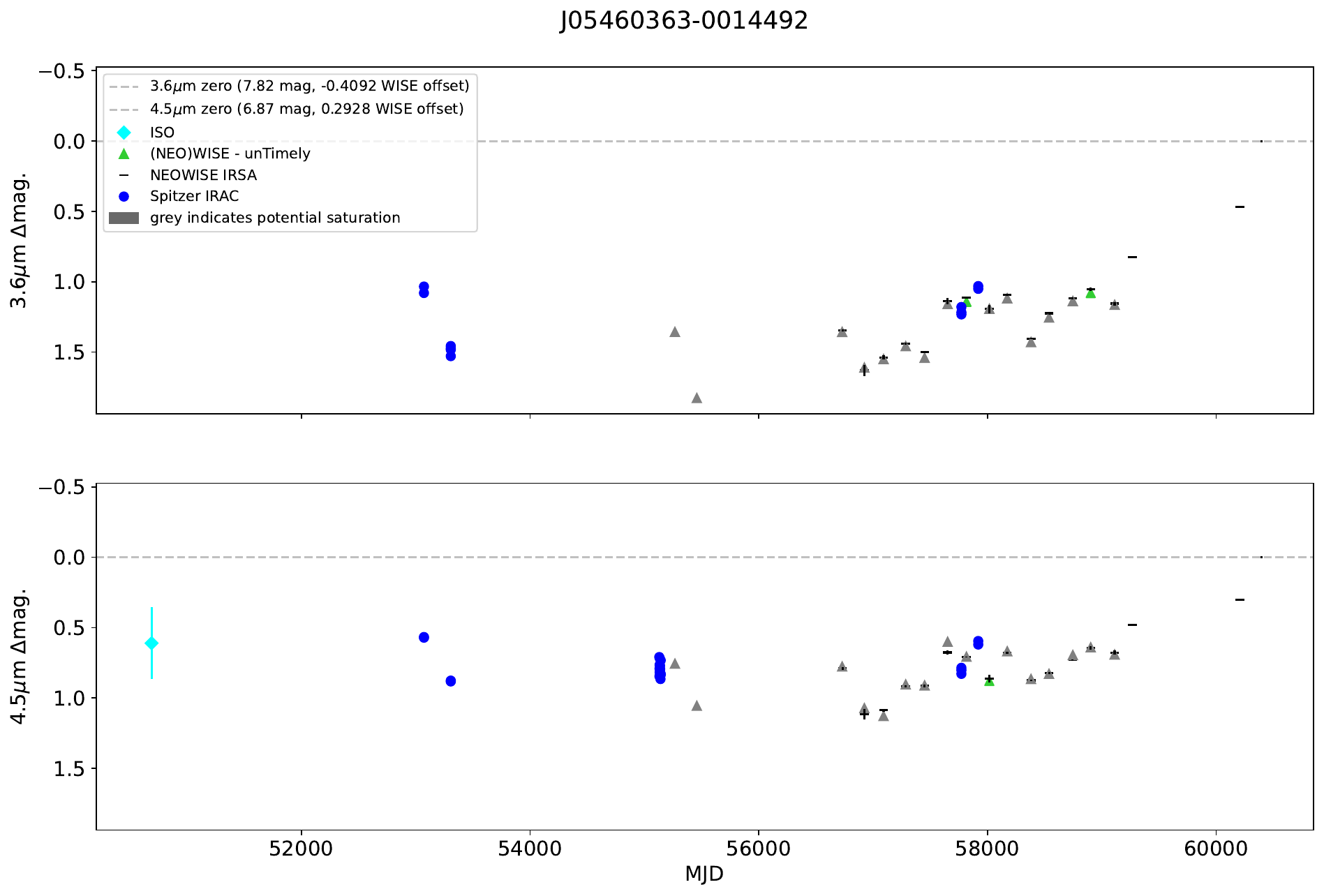}
    \includegraphics[width=15cm, height=10.5cm]{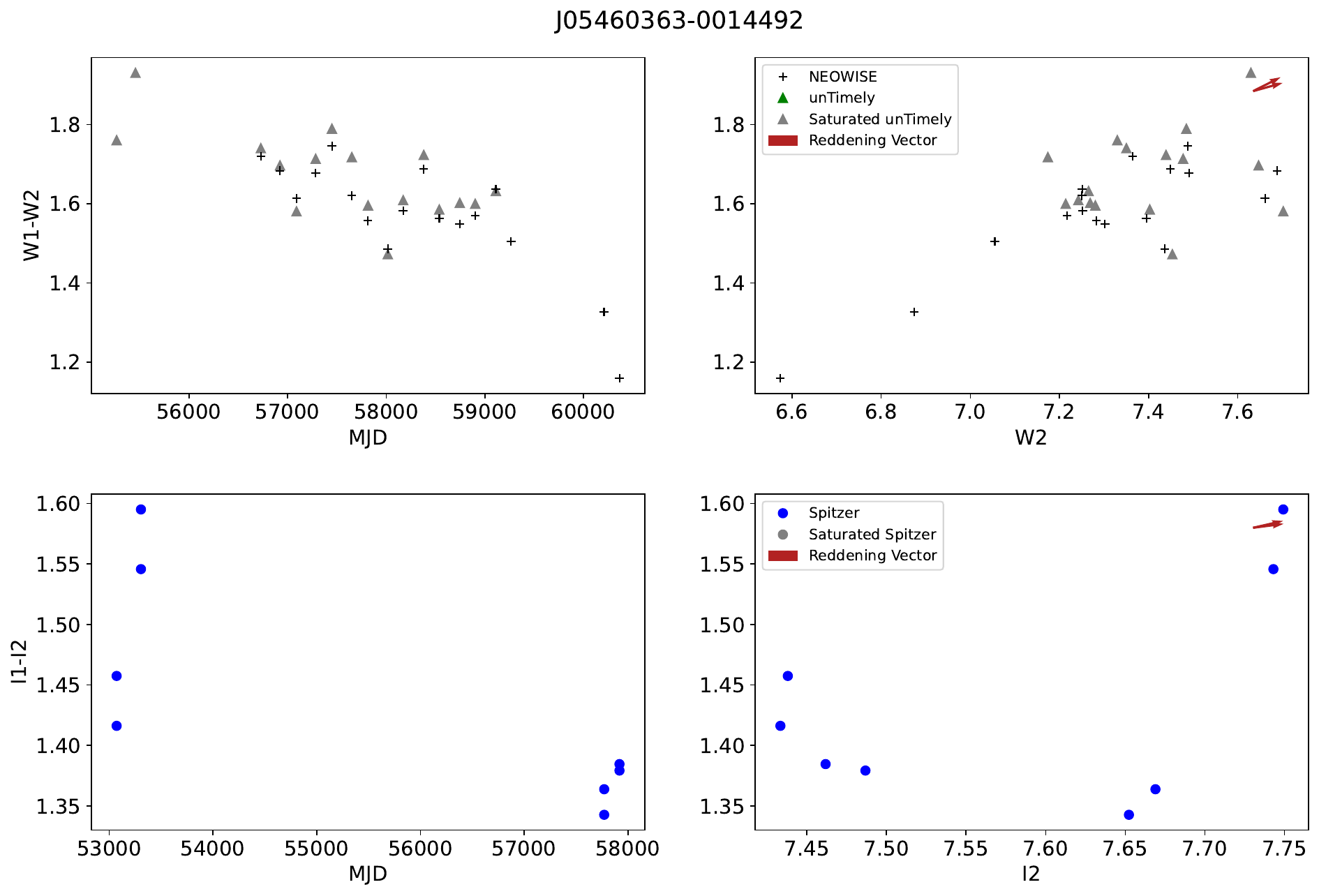}
    \caption{See above.}
\end{figure*}
\addtocounter{figure}{-1}

\begin{figure*}
    \centering
    \includegraphics[width=15cm, height=10.5cm]{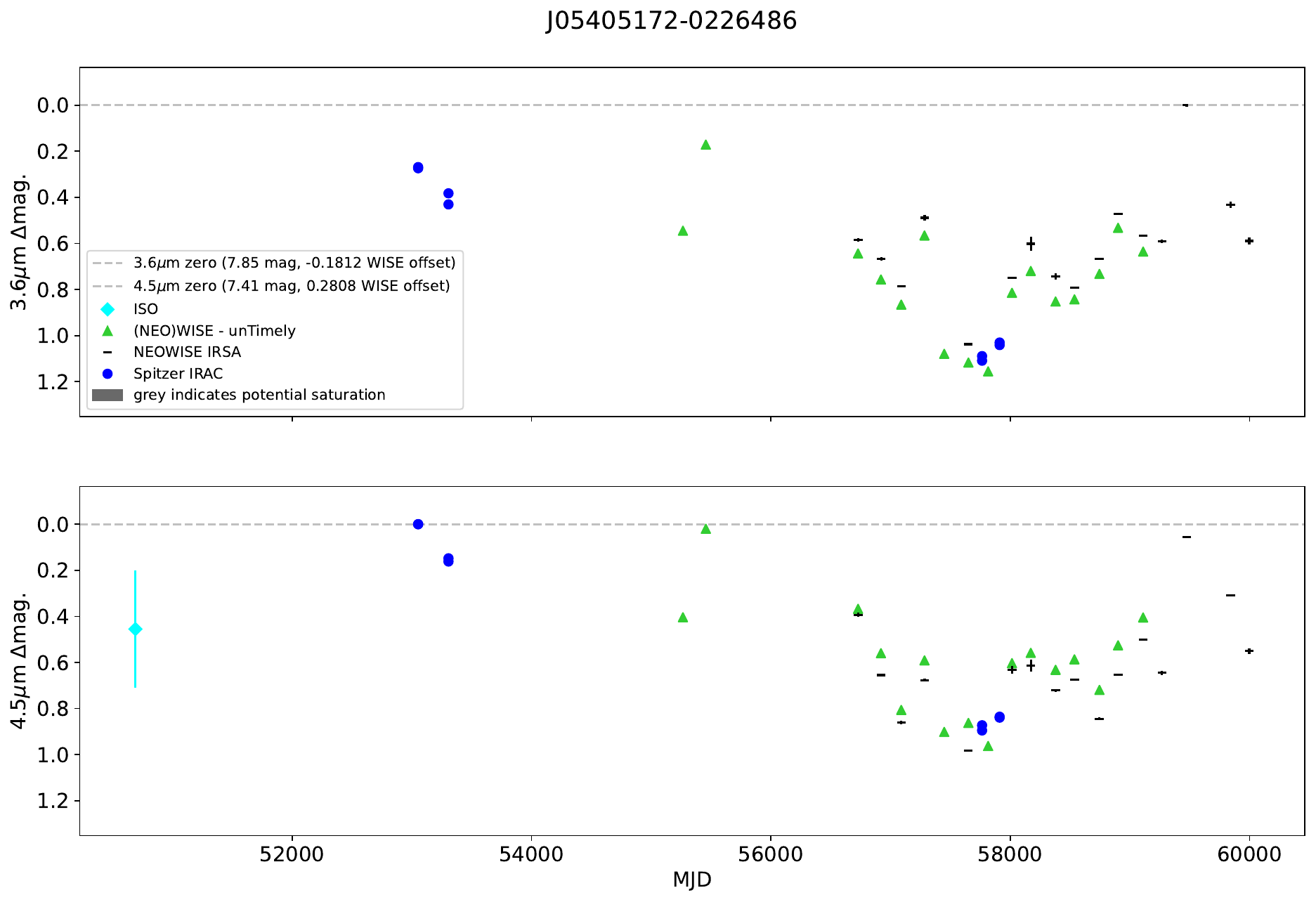}
    \includegraphics[width=15cm, height=10.5cm]{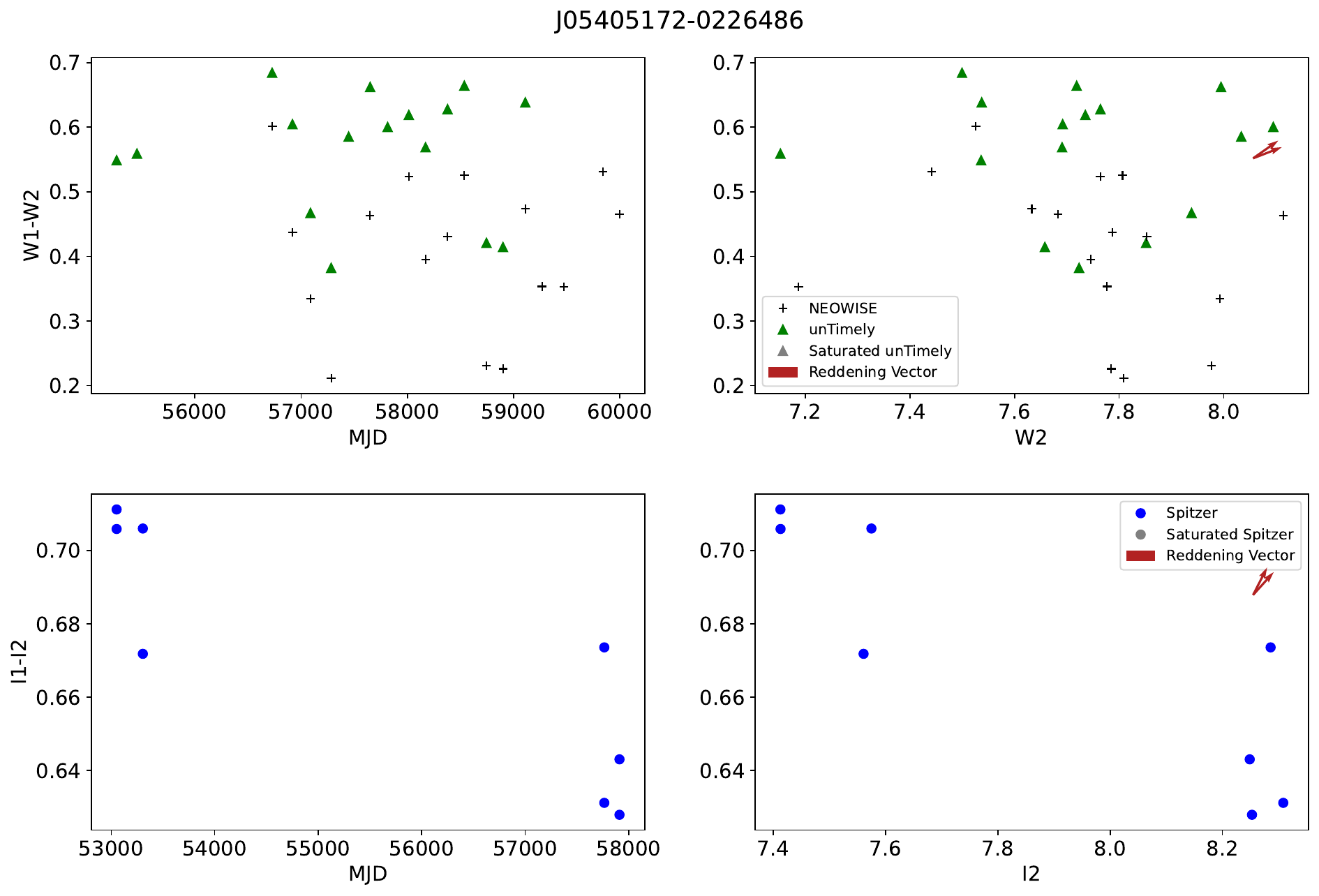}
    \caption{See above.}
    \label{fig:HOPS_294}
\end{figure*}
% J05405172-0226486 X
\addtocounter{figure}{-1}

\begin{figure*}
    \centering
    \includegraphics[width=15cm, height=10.5cm]{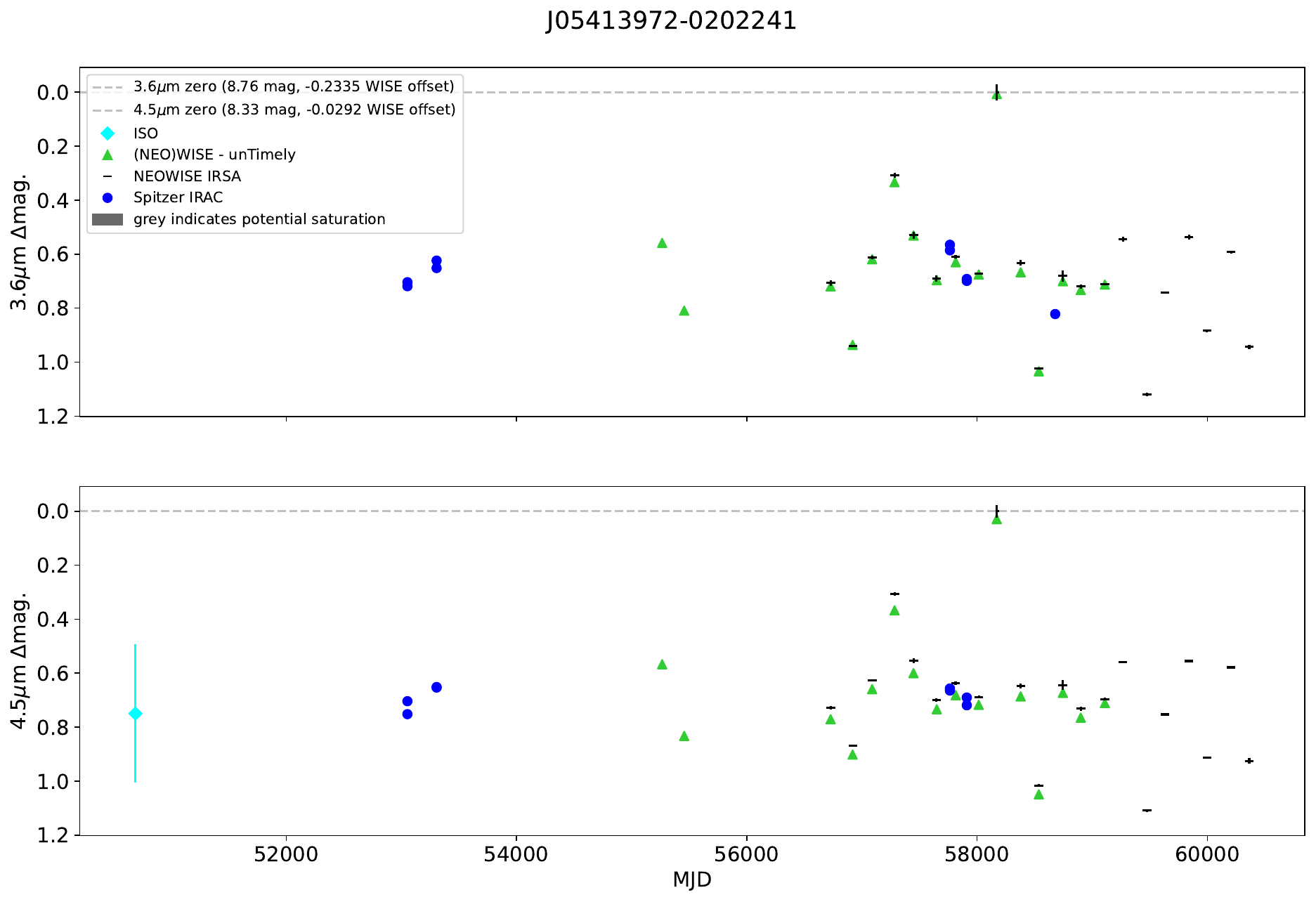}
    \includegraphics[width=15cm, height=10.5cm]{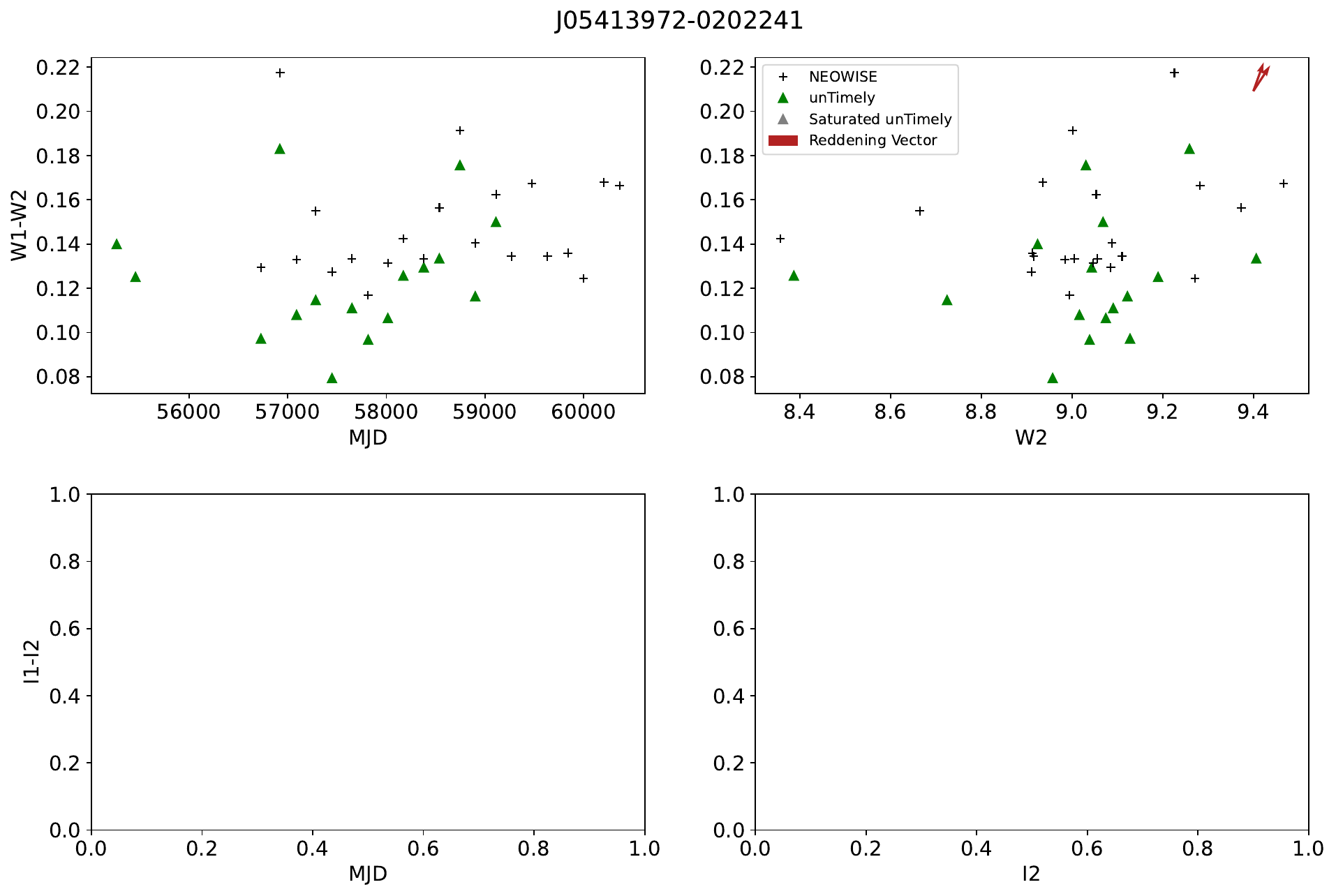}
    \caption{See above.}
    \label{fig:MGM2012_2873}
\end{figure*}
% J05413972-0202241 X
\addtocounter{figure}{-1}

\begin{figure*}
    \centering
    \includegraphics[width=15cm, height=10.5cm]{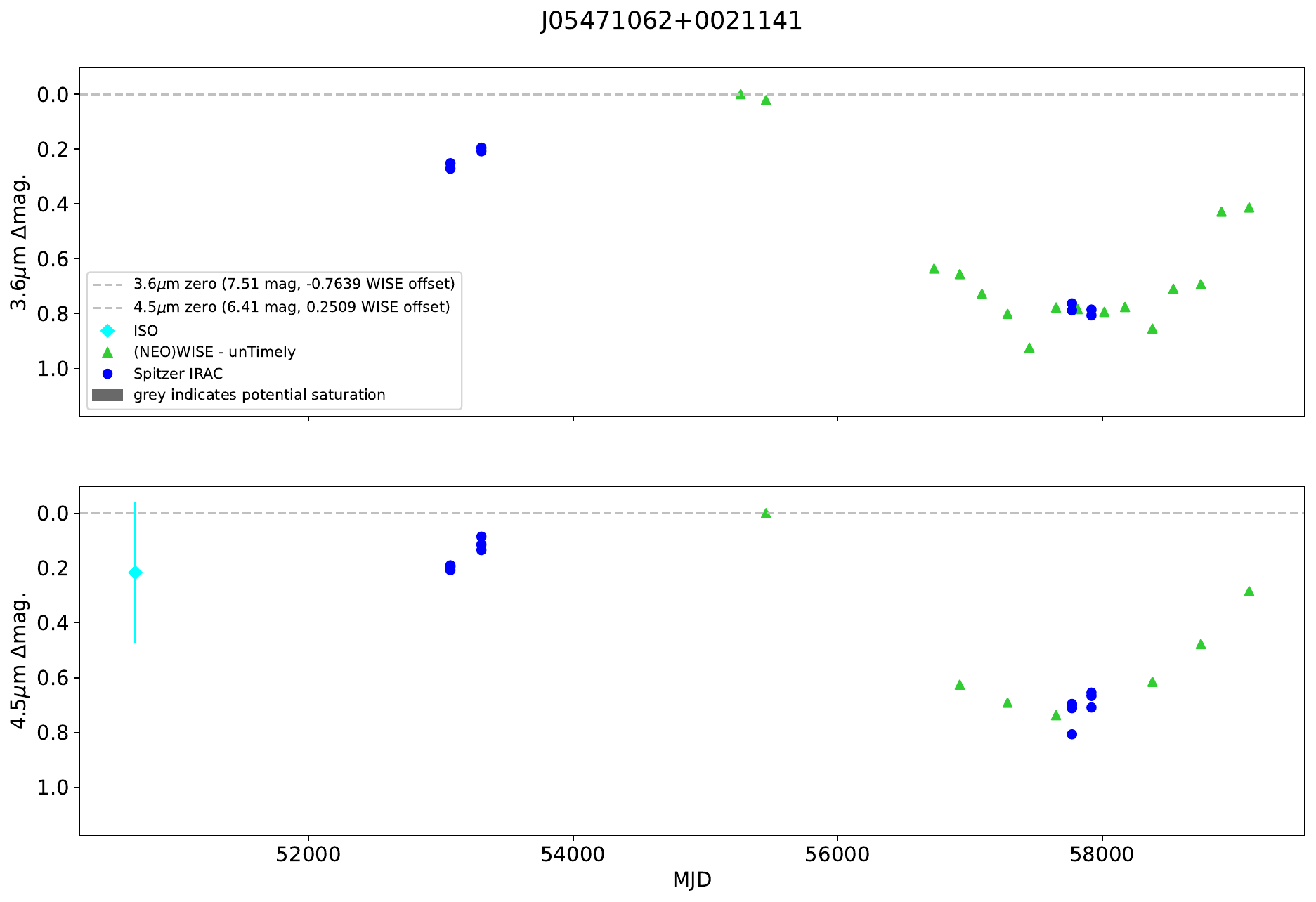}
    \includegraphics[width=15cm, height=10.5cm]{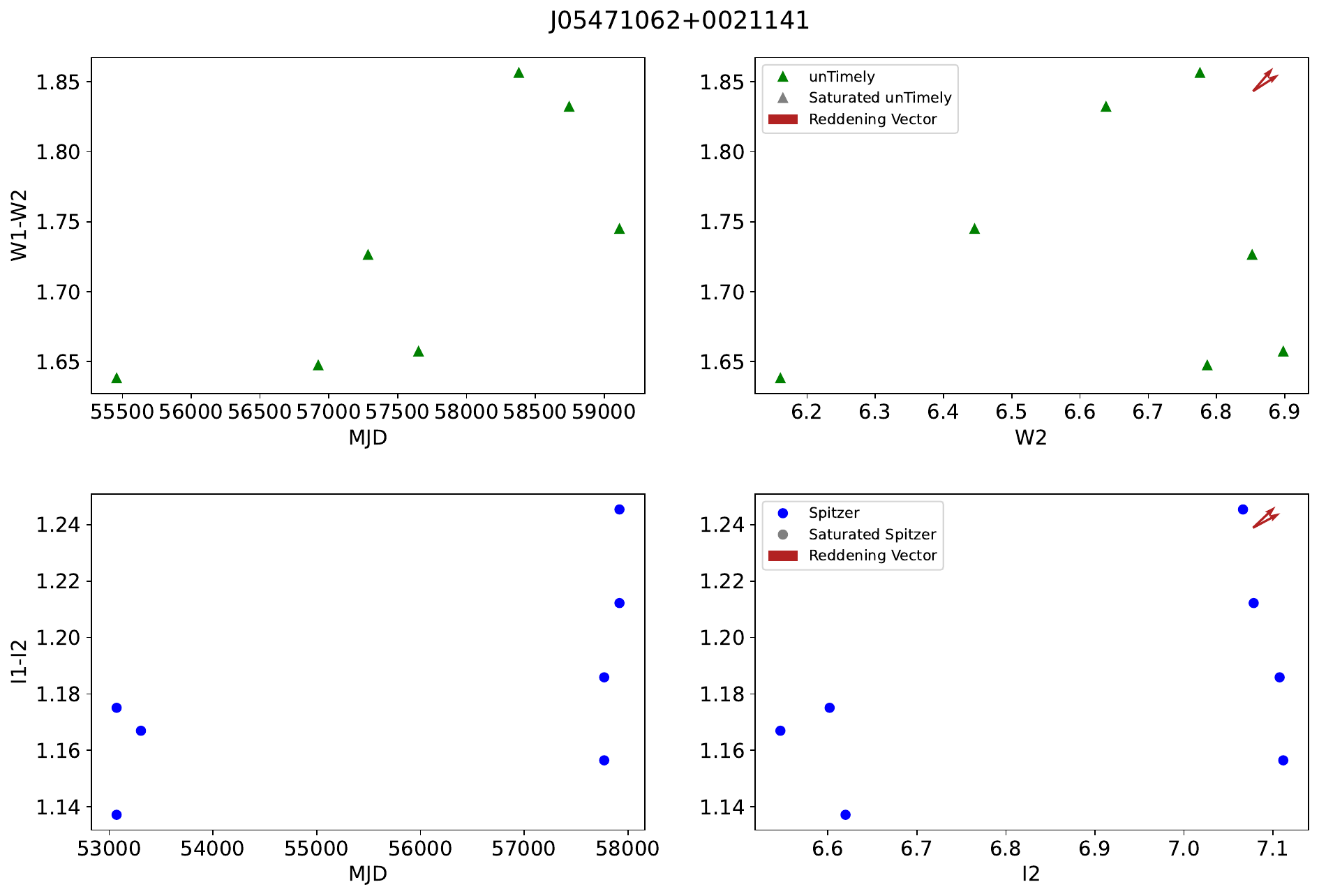}
    \caption{See above.}
\end{figure*}
% J05471062+0021141 X
\addtocounter{figure}{-1}

\begin{figure*}
    \centering
    \includegraphics[width=15cm, height=10.5cm]{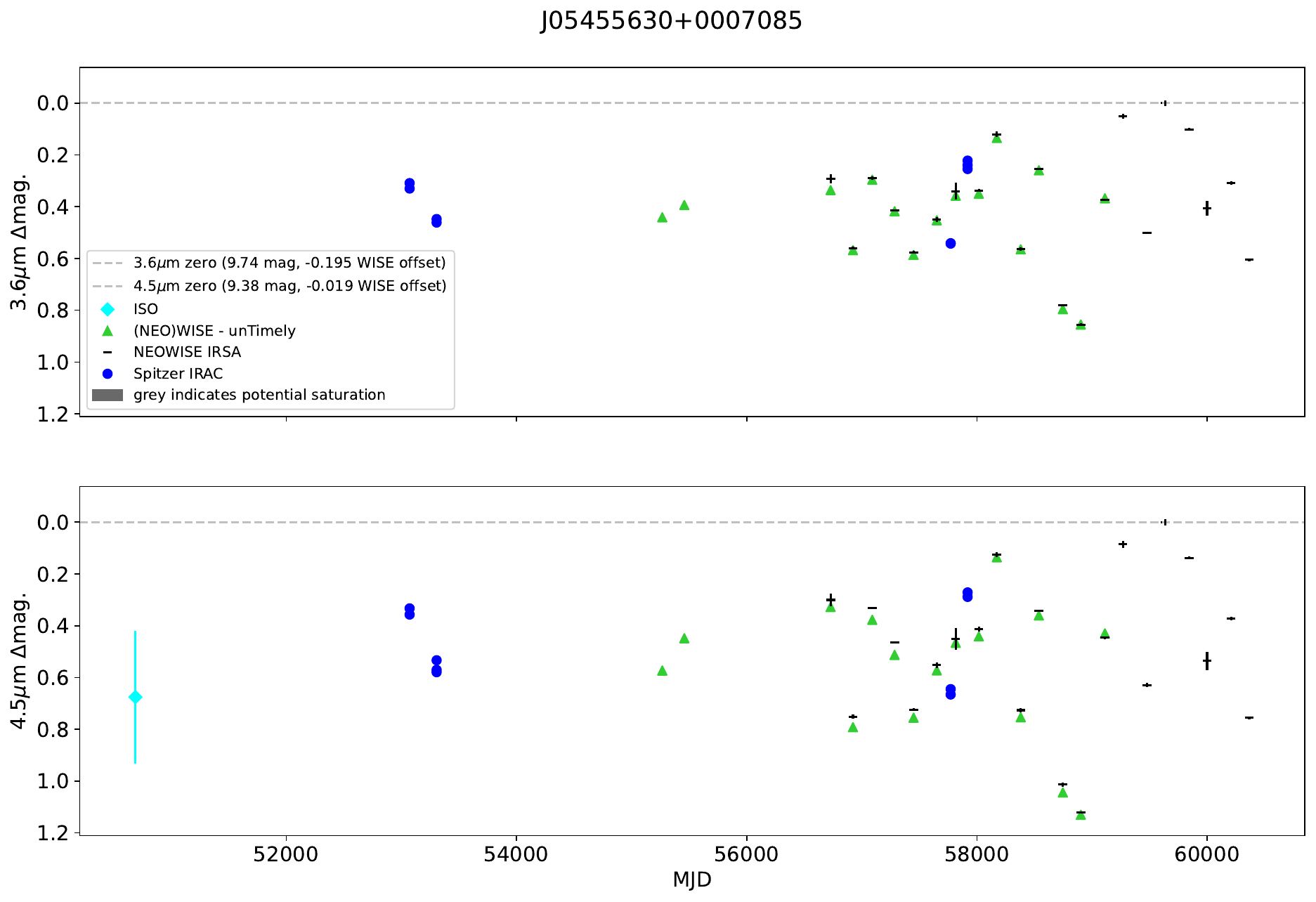}
    \includegraphics[width=15cm, height=10.5cm]{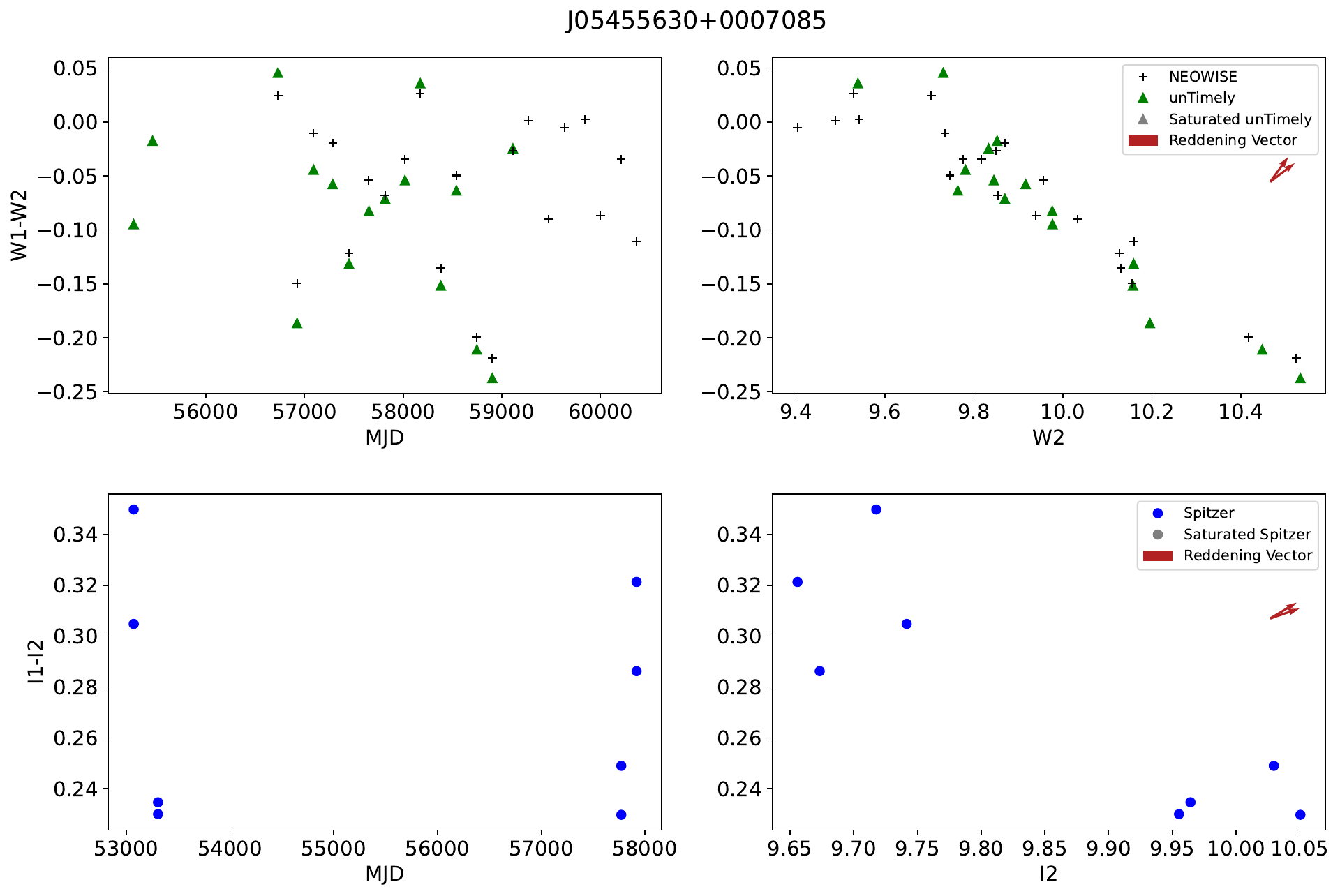}
    \caption{See above.}
    \label{fig:MGM2012_3264}
\end{figure*}
% J05455630+0007085 X
\addtocounter{figure}{-1}

\begin{figure*}
    \centering
    \includegraphics[width=15cm, height=10.5cm]{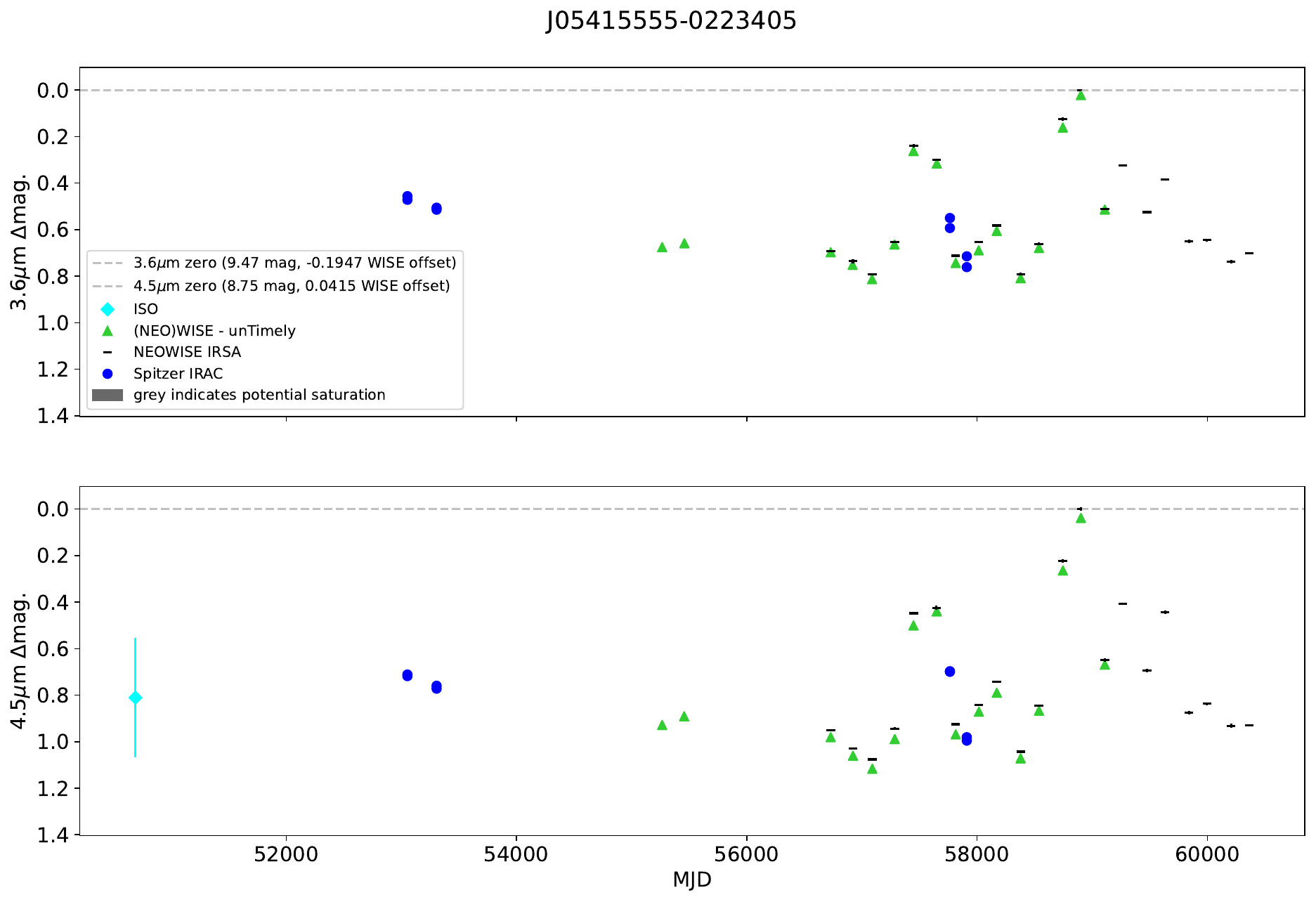}
    \includegraphics[width=15cm, height=10.5cm]{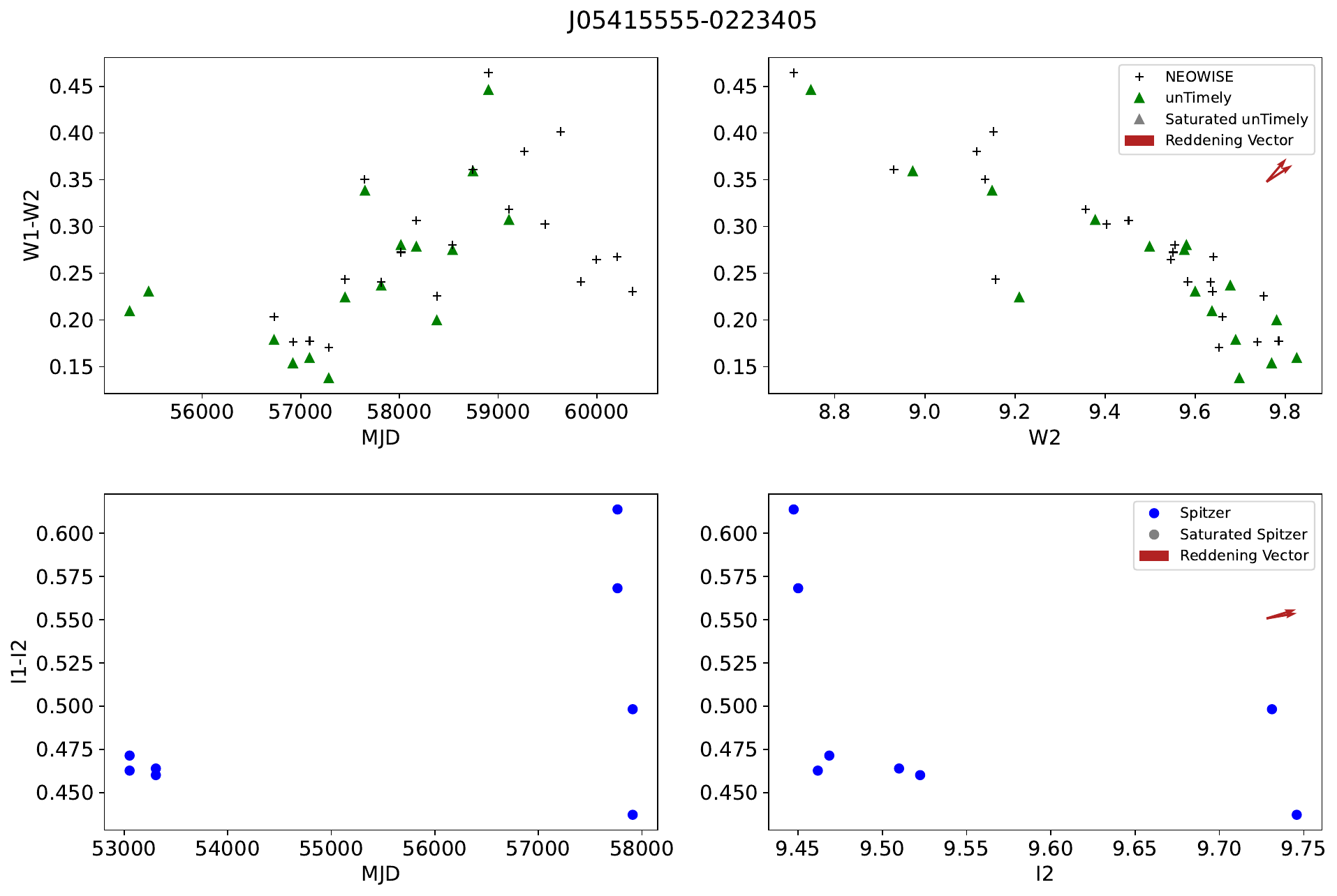}
    \caption{See above.}
    \label{fig:MGM2012_2829}
\end{figure*}
% J05415555-0223405 X
\addtocounter{figure}{-1}

\begin{figure*}
    \centering
    \includegraphics[width=15cm, height=10.5cm]{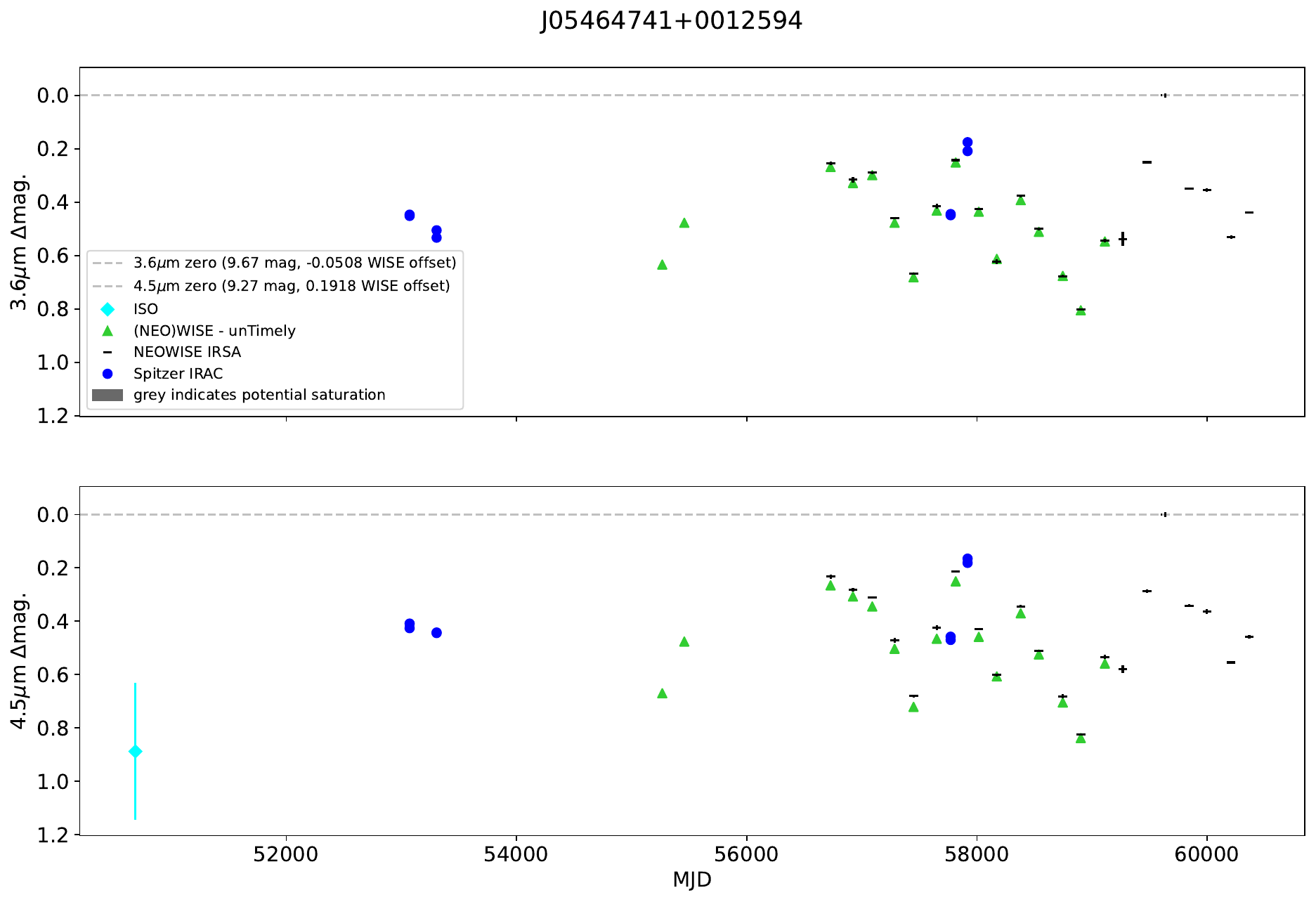}
    \includegraphics[width=15cm, height=10.5cm]{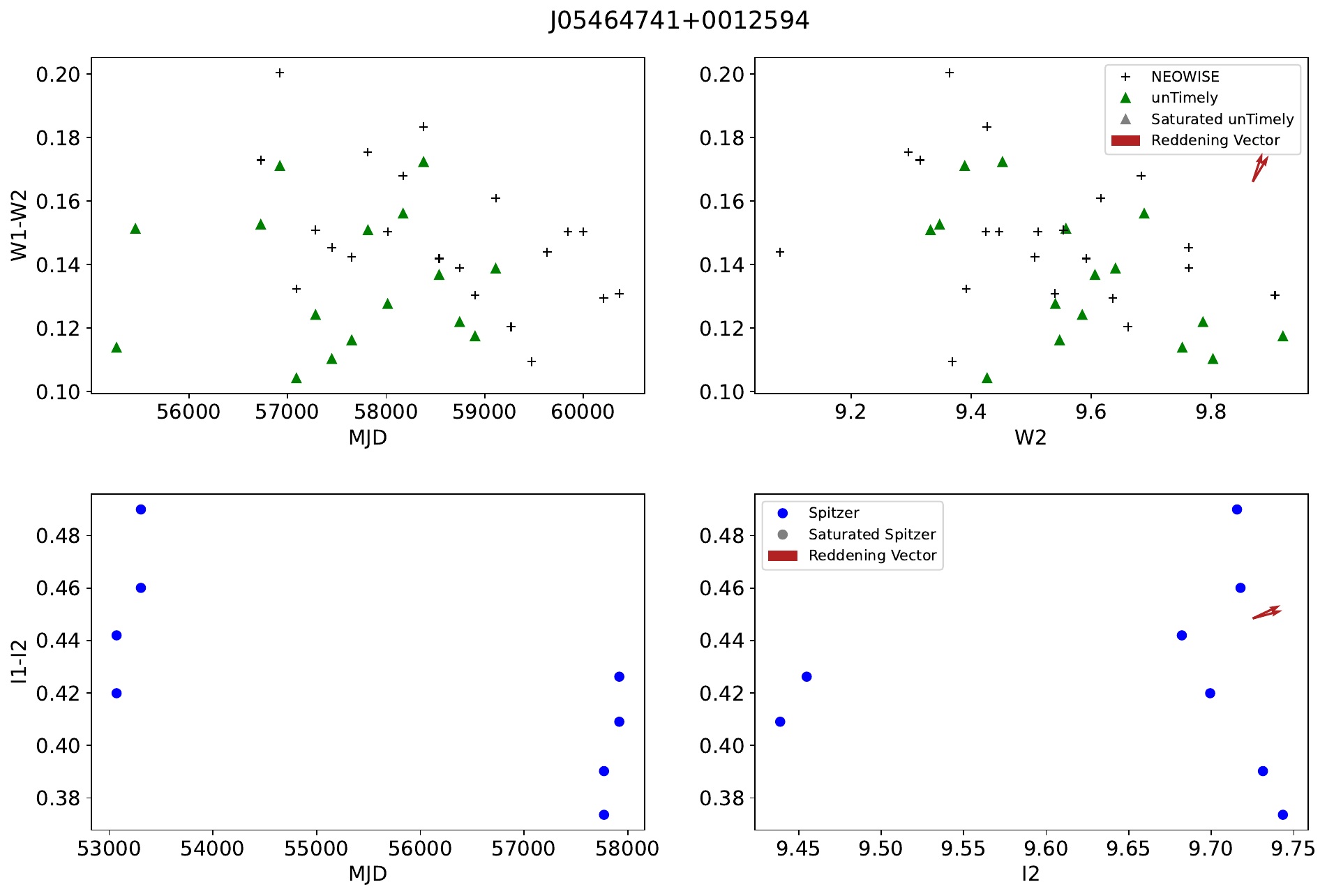}
    \caption{See above.}
\end{figure*}
% J05464741+0012594 X
\addtocounter{figure}{-1}

\begin{figure*}
    \centering
    \includegraphics[width=15cm, height=10.5cm]{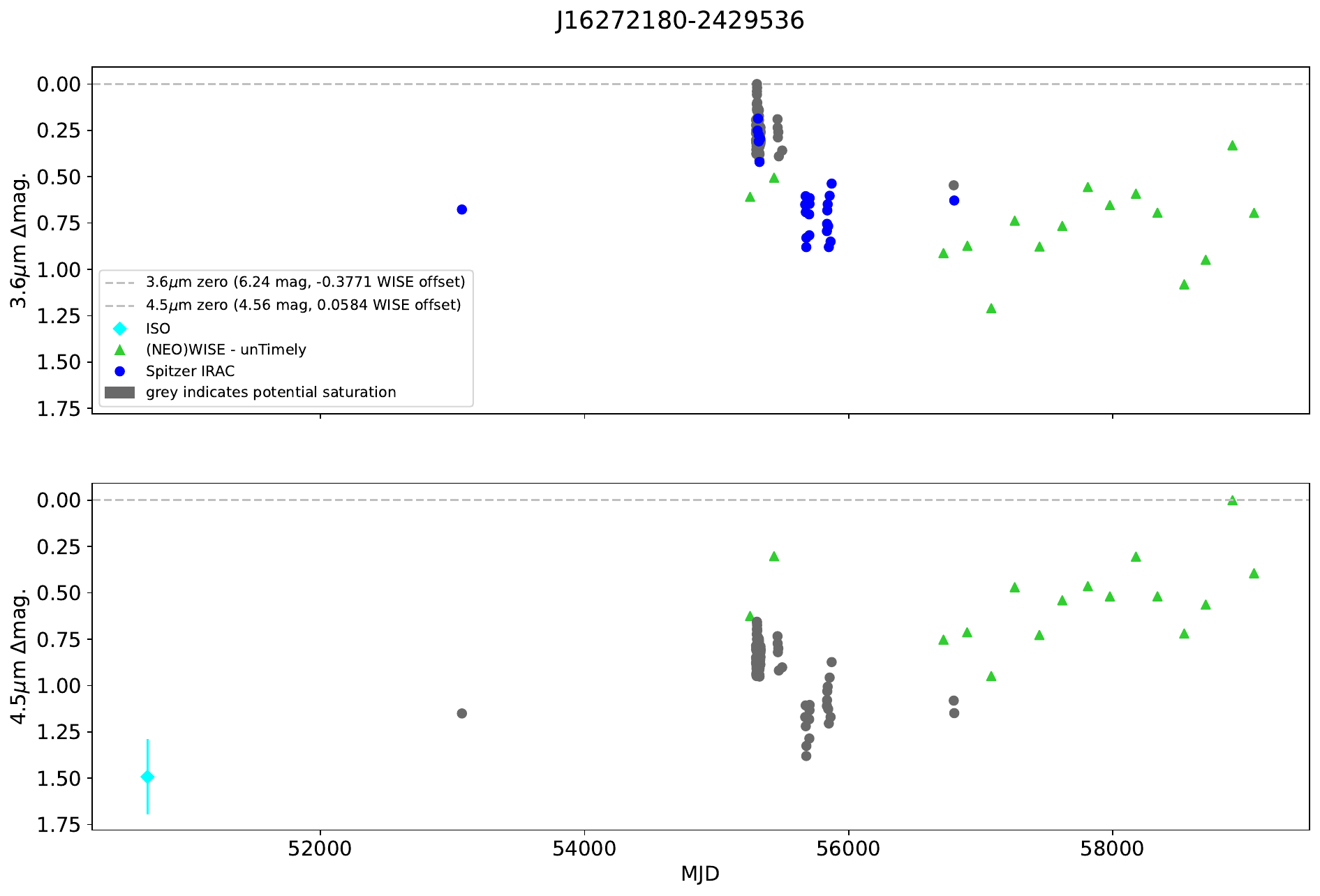}
    \includegraphics[width=15cm, height=10.5cm]{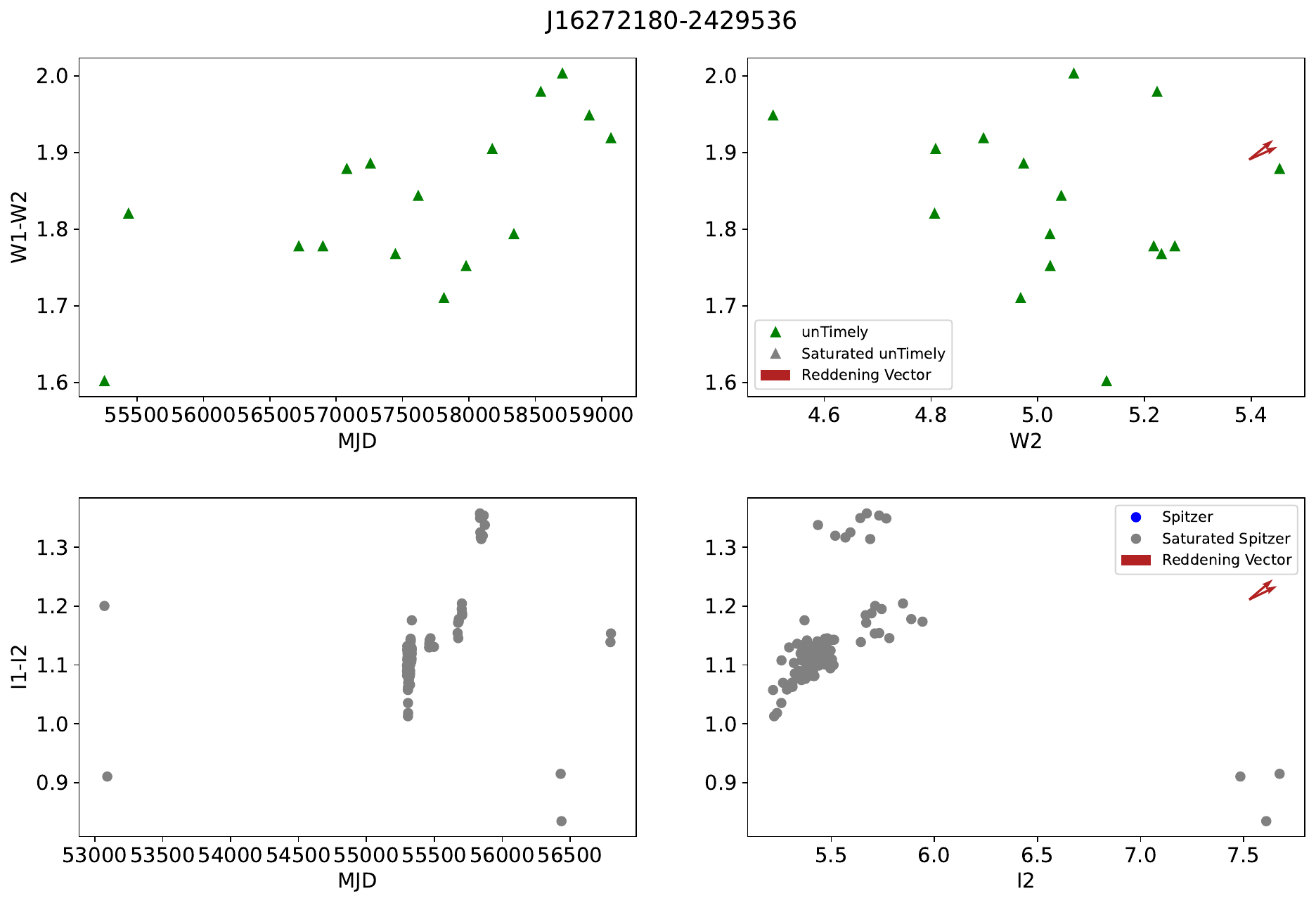}
    \caption{See above.}
\end{figure*}
\addtocounter{figure}{-1}

\begin{figure*}
    \centering
    \includegraphics[width=15cm, height=10.5cm]{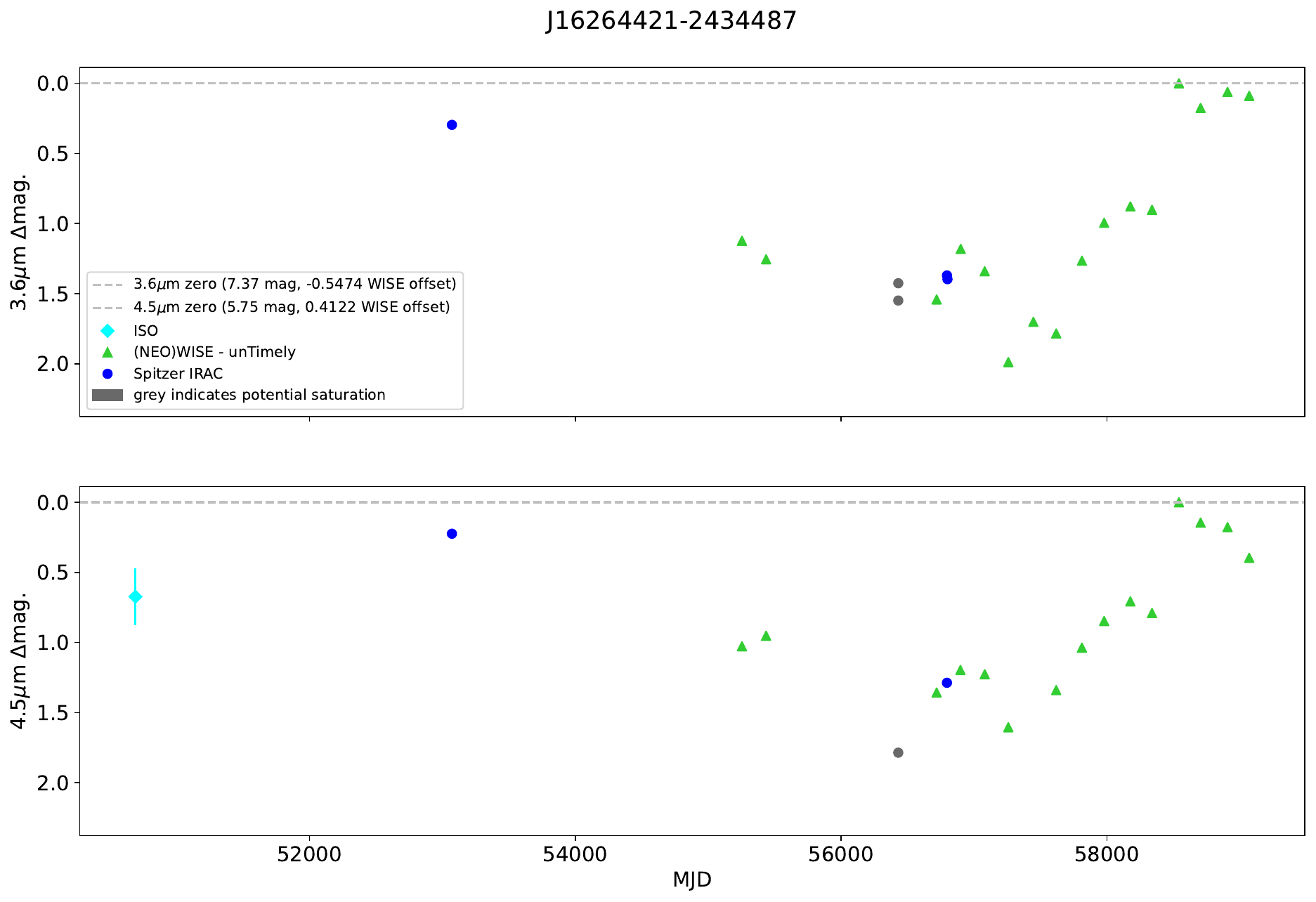}
    \includegraphics[width=15cm, height=10.5cm]{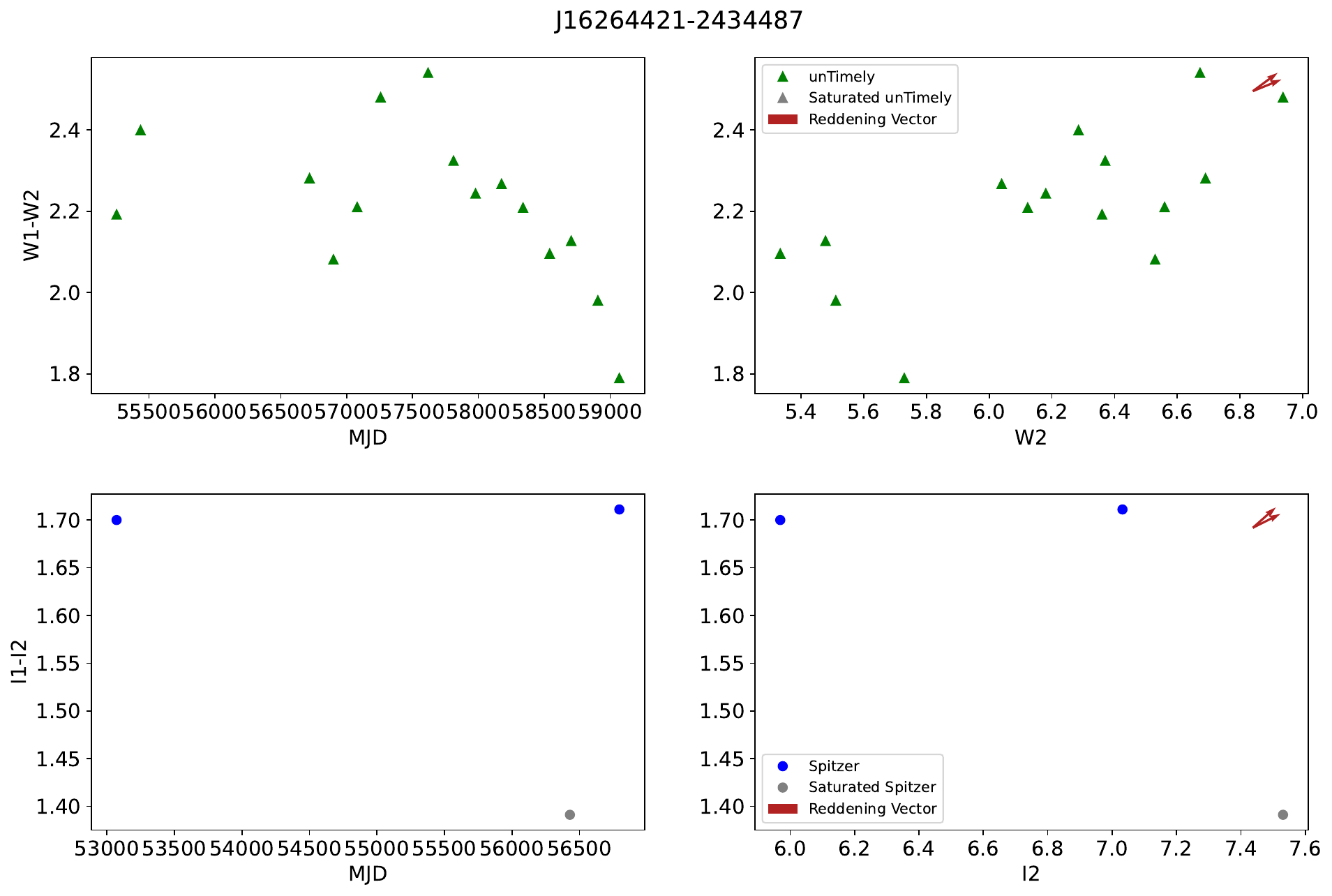}
    \caption{See above.}
\end{figure*}
\addtocounter{figure}{-1}

\begin{figure*}
    \centering
    \includegraphics[width=15cm, height=10.5cm]{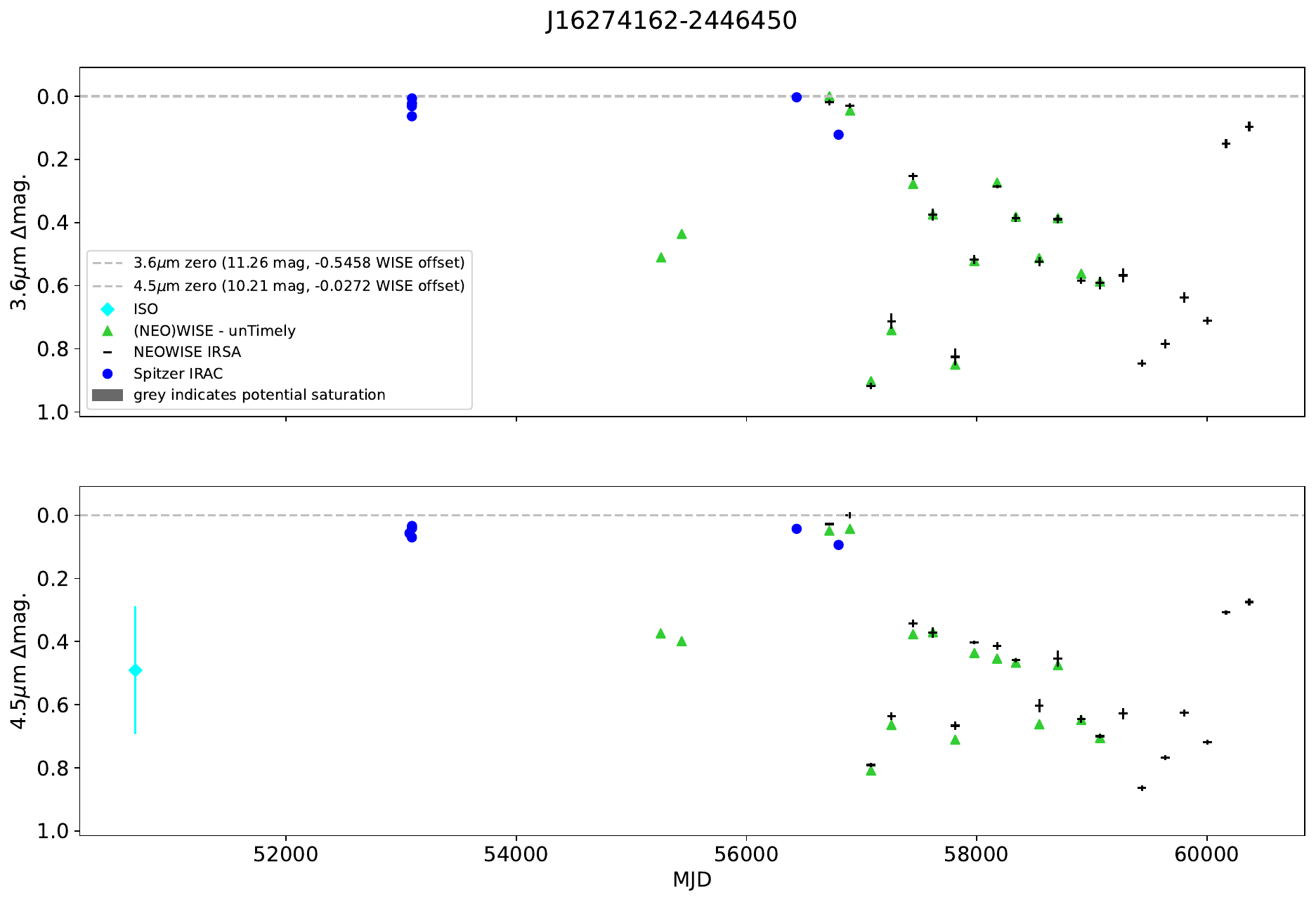}
    \includegraphics[width=15cm, height=10.5cm]{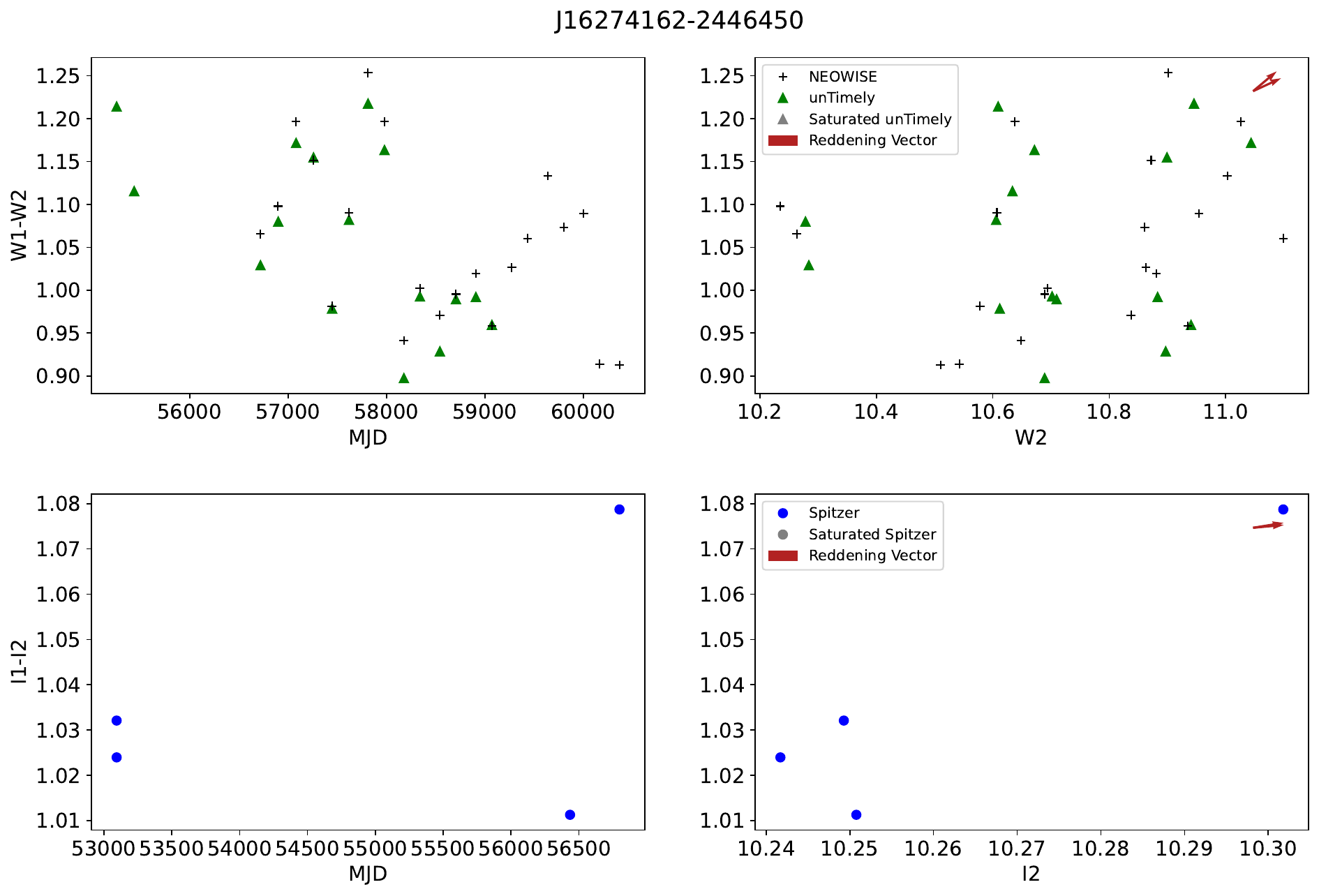}
    \caption{See above.}
\end{figure*}
\addtocounter{figure}{-1}

\begin{figure*}
    \centering
    \includegraphics[width=15cm, height=10.5cm]{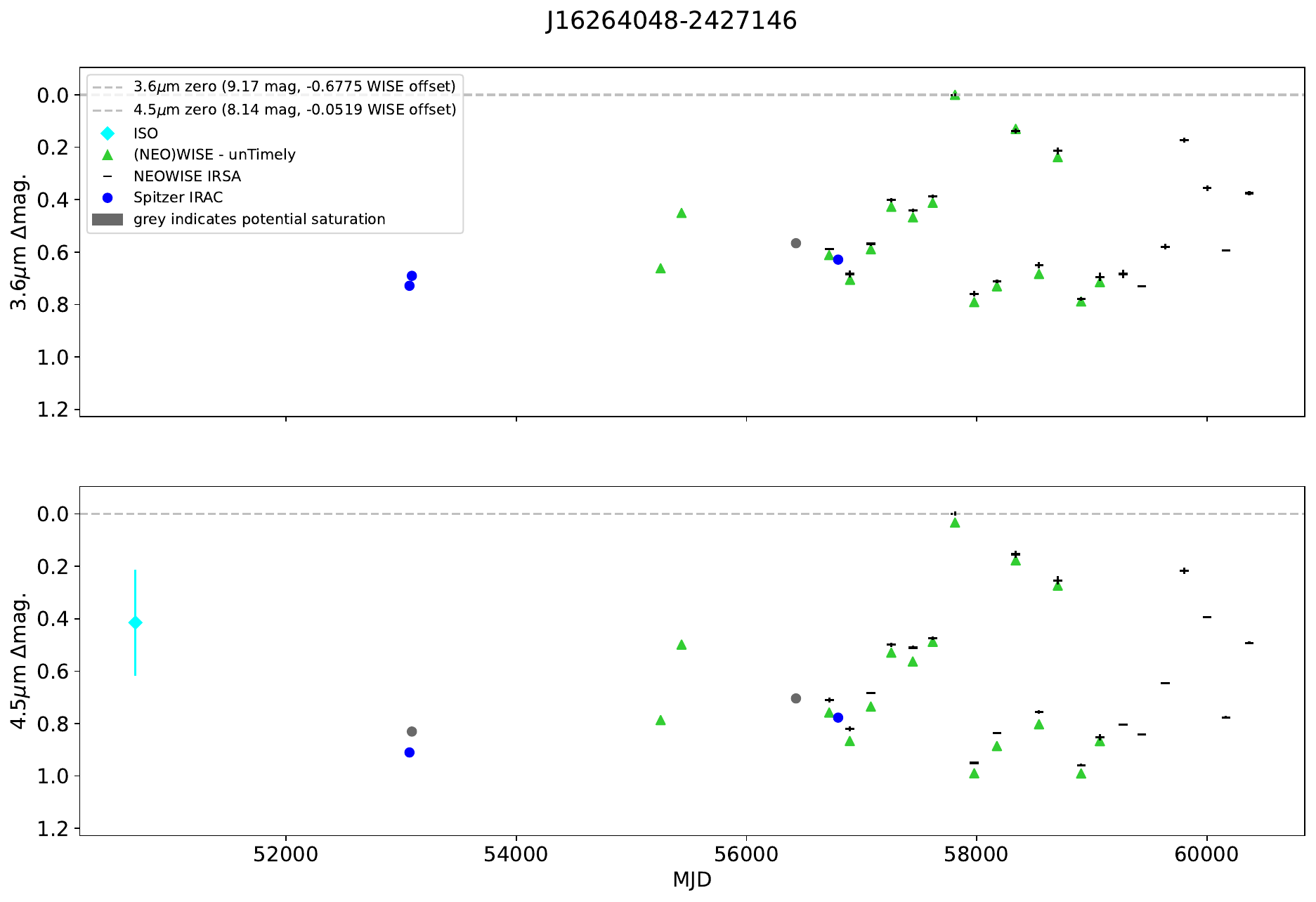}
    \includegraphics[width=15cm, height=10.5cm]{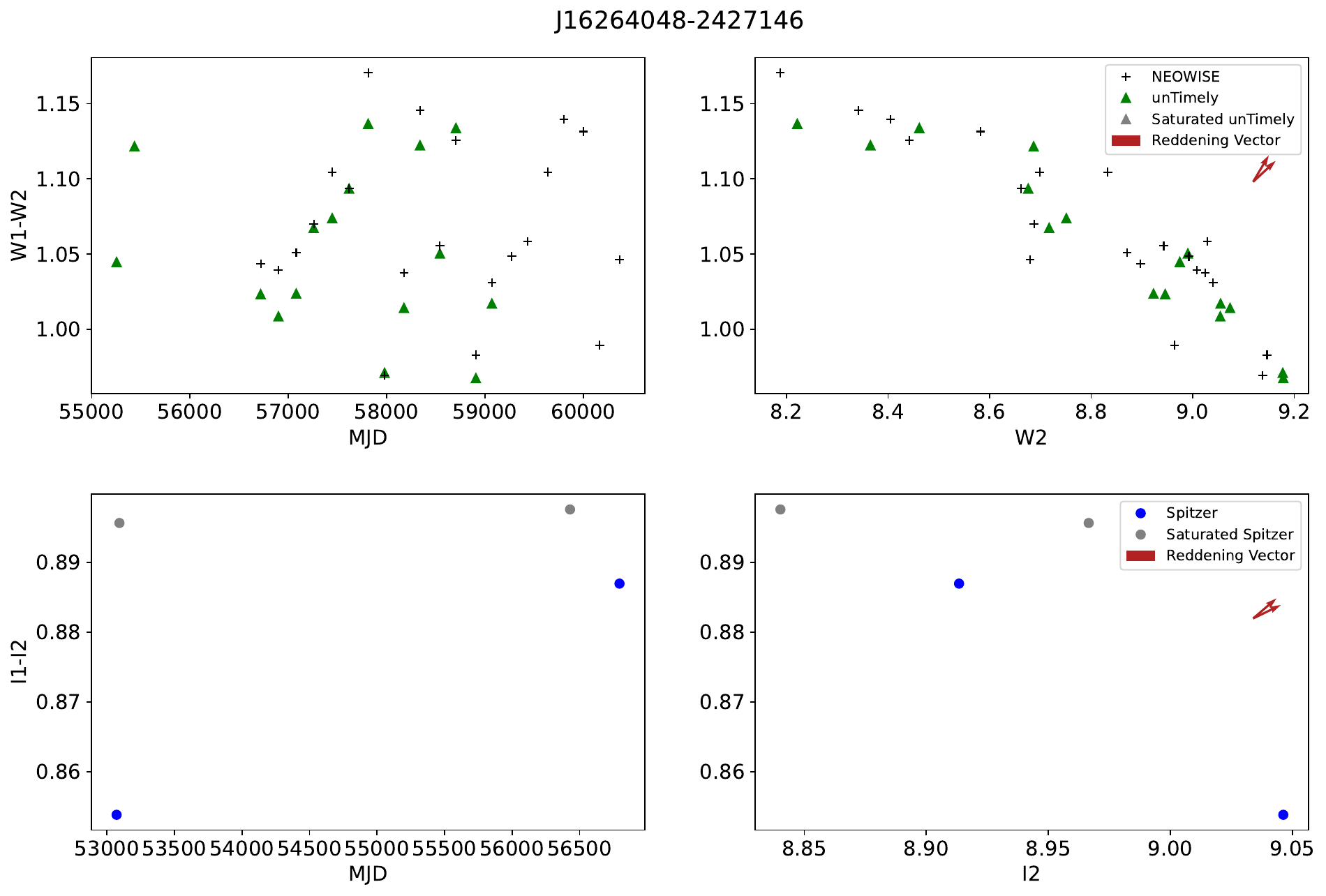}
    \caption{See above.}
\end{figure*}
\addtocounter{figure}{-1}

\begin{figure*}
    \centering
    \includegraphics[width=15cm, height=10.5cm]{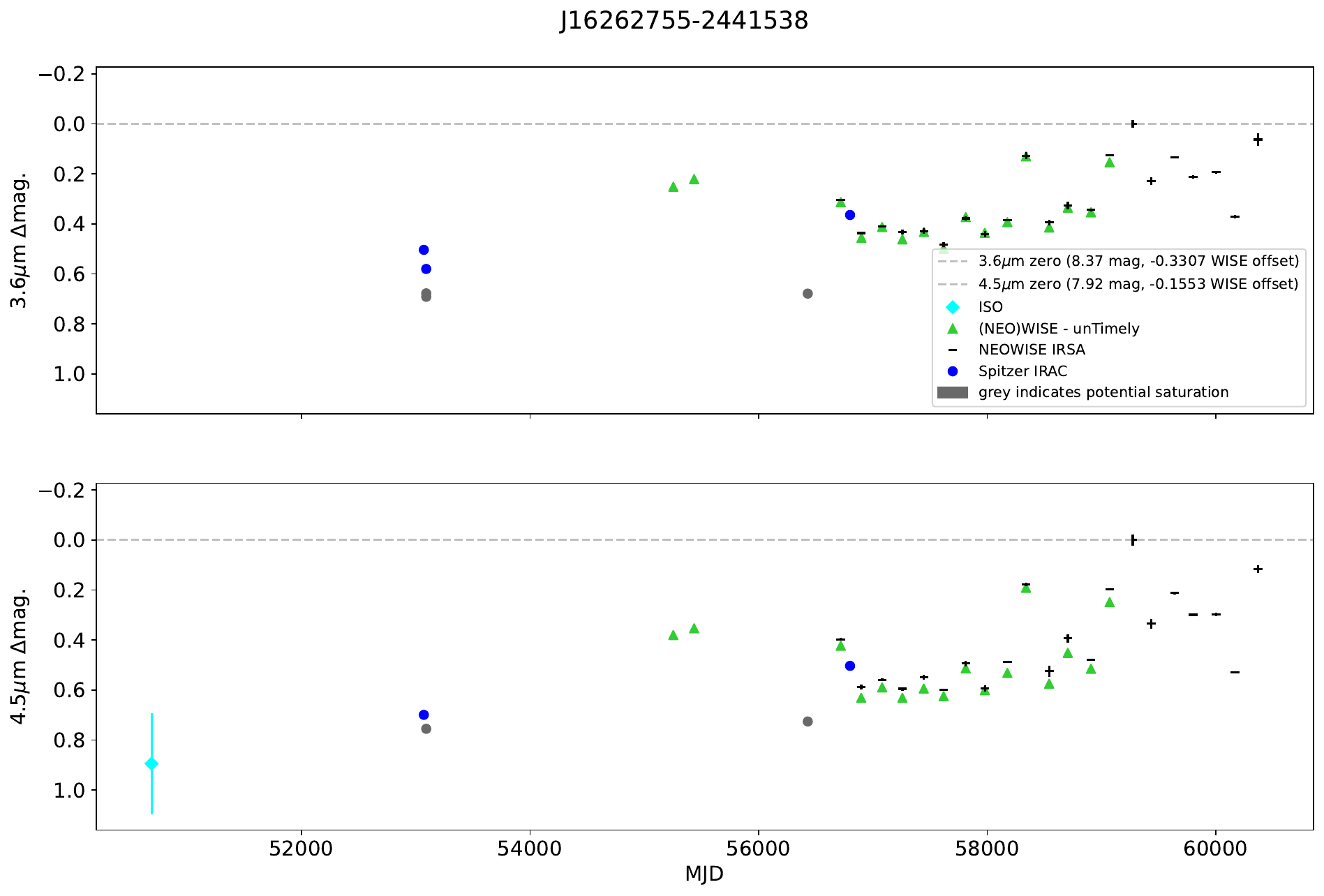}
    \includegraphics[width=15cm, height=10.5cm]{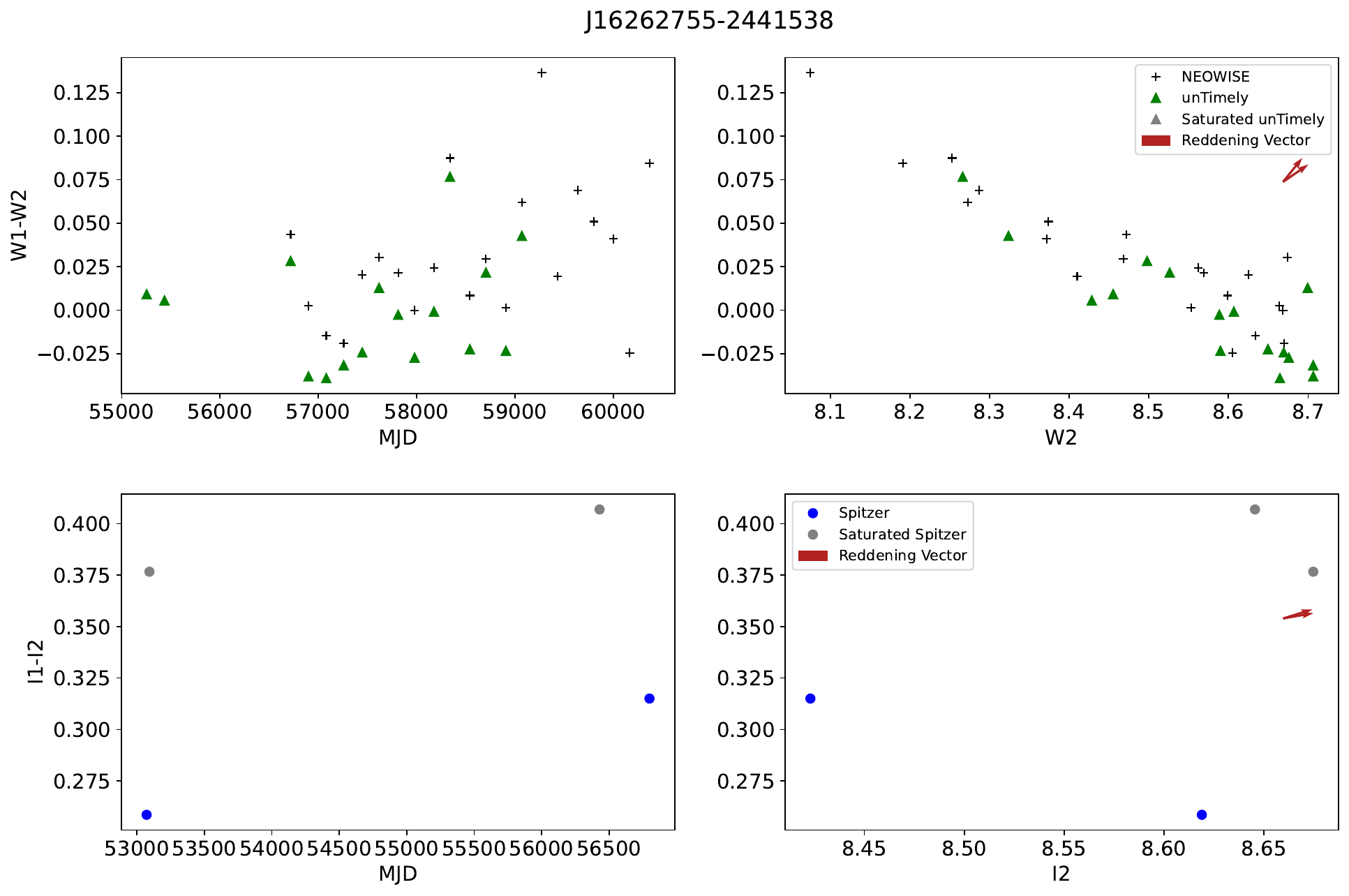}
    \caption{See above.}
\end{figure*}
\addtocounter{figure}{-1}

\begin{figure*}
    \centering
    \includegraphics[width=15cm, height=10.5cm]{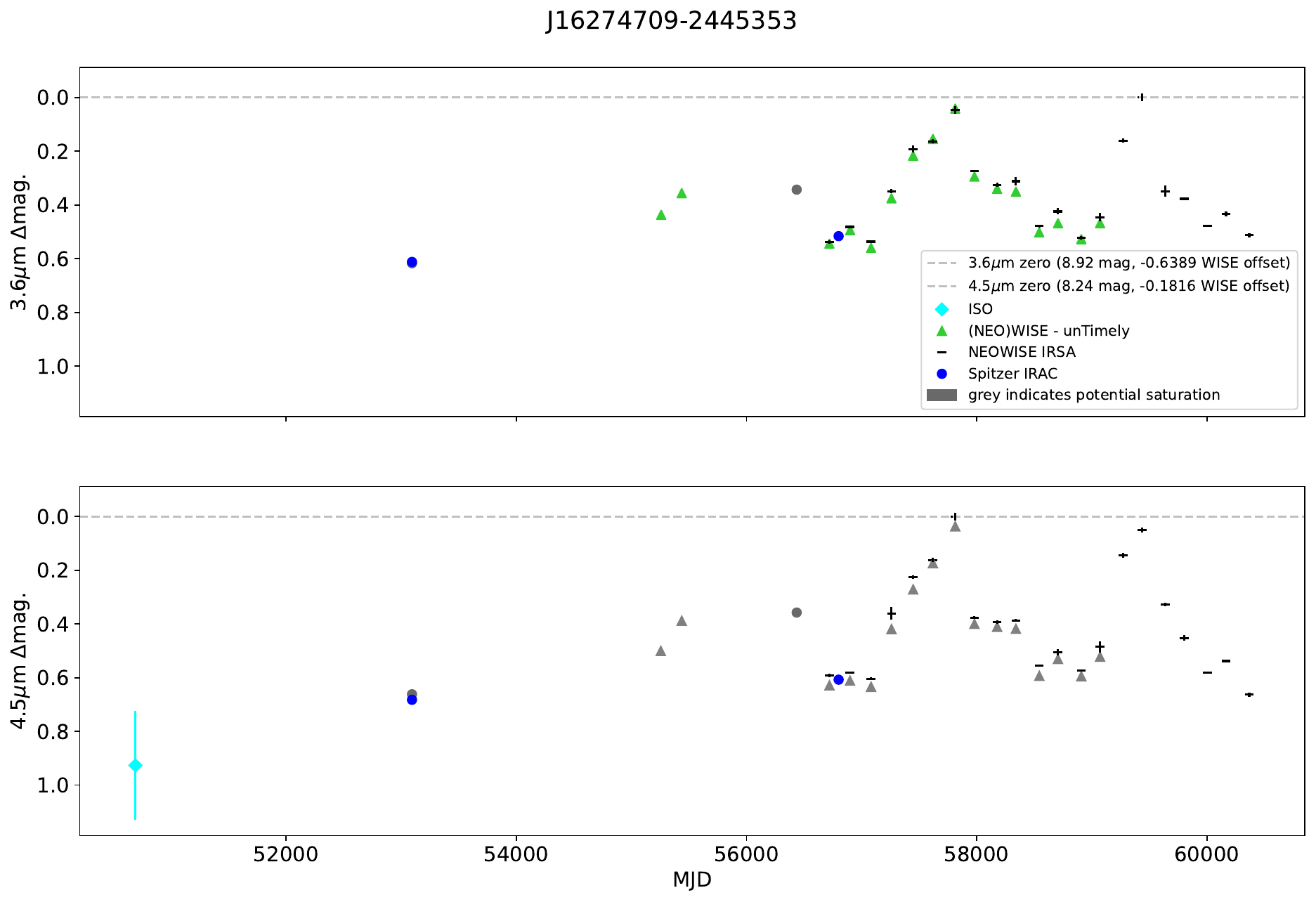}
    \includegraphics[width=15cm, height=10.5cm]{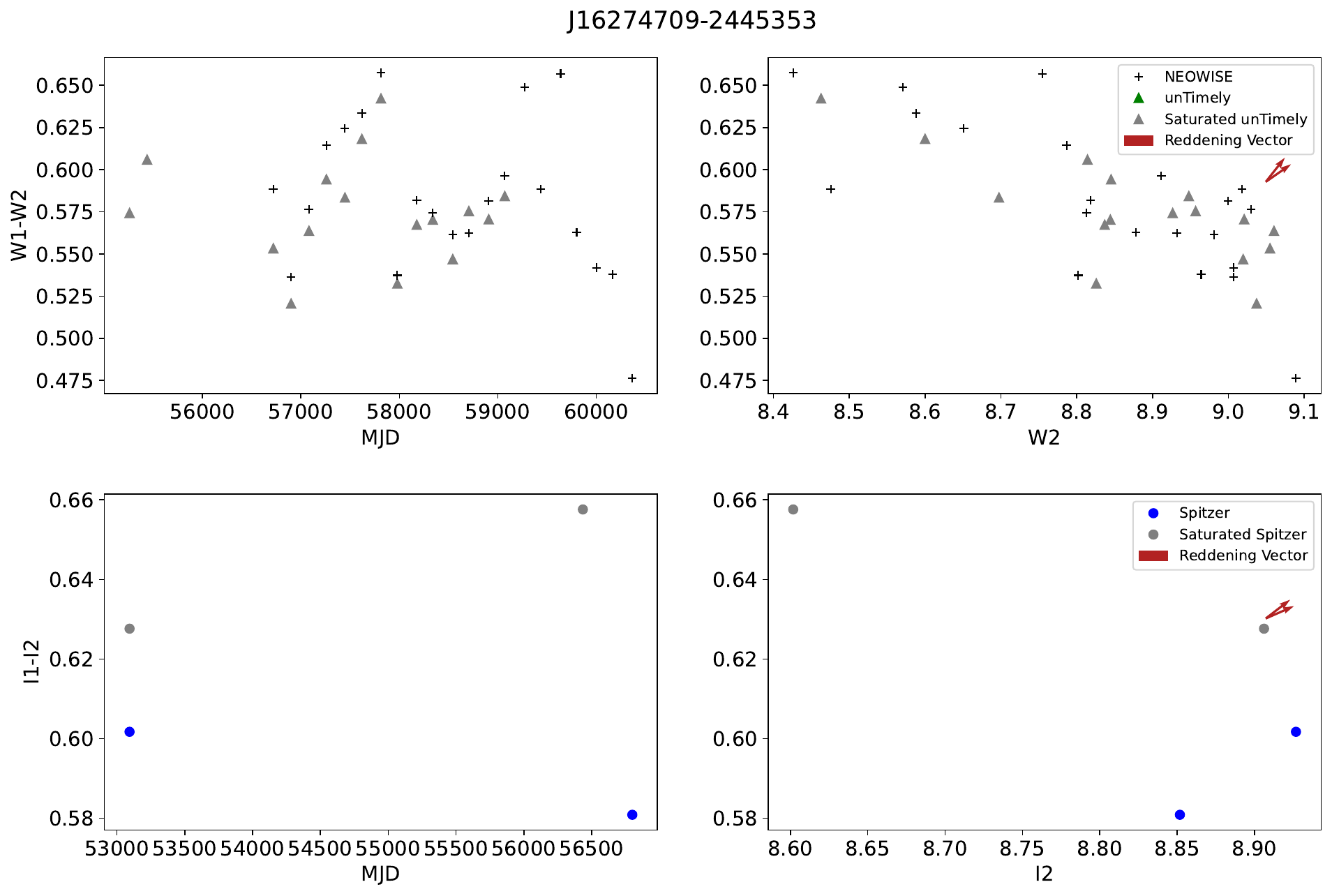}
    \caption{See above.}
\end{figure*}
\addtocounter{figure}{-1}

\begin{figure*}
    \centering
    \includegraphics[width=15cm, height=10.5cm]{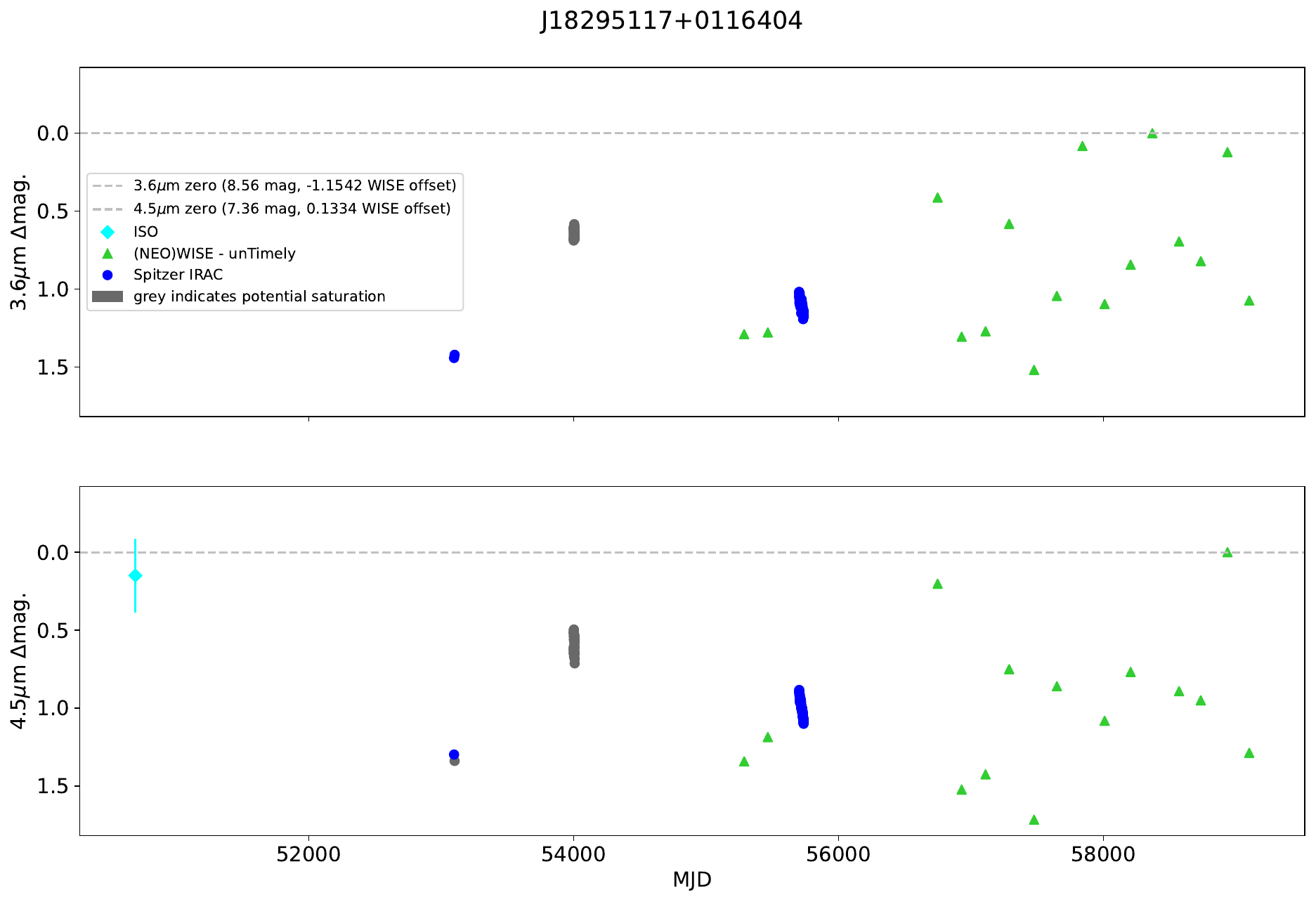}
    \includegraphics[width=15cm, height=10.5cm]{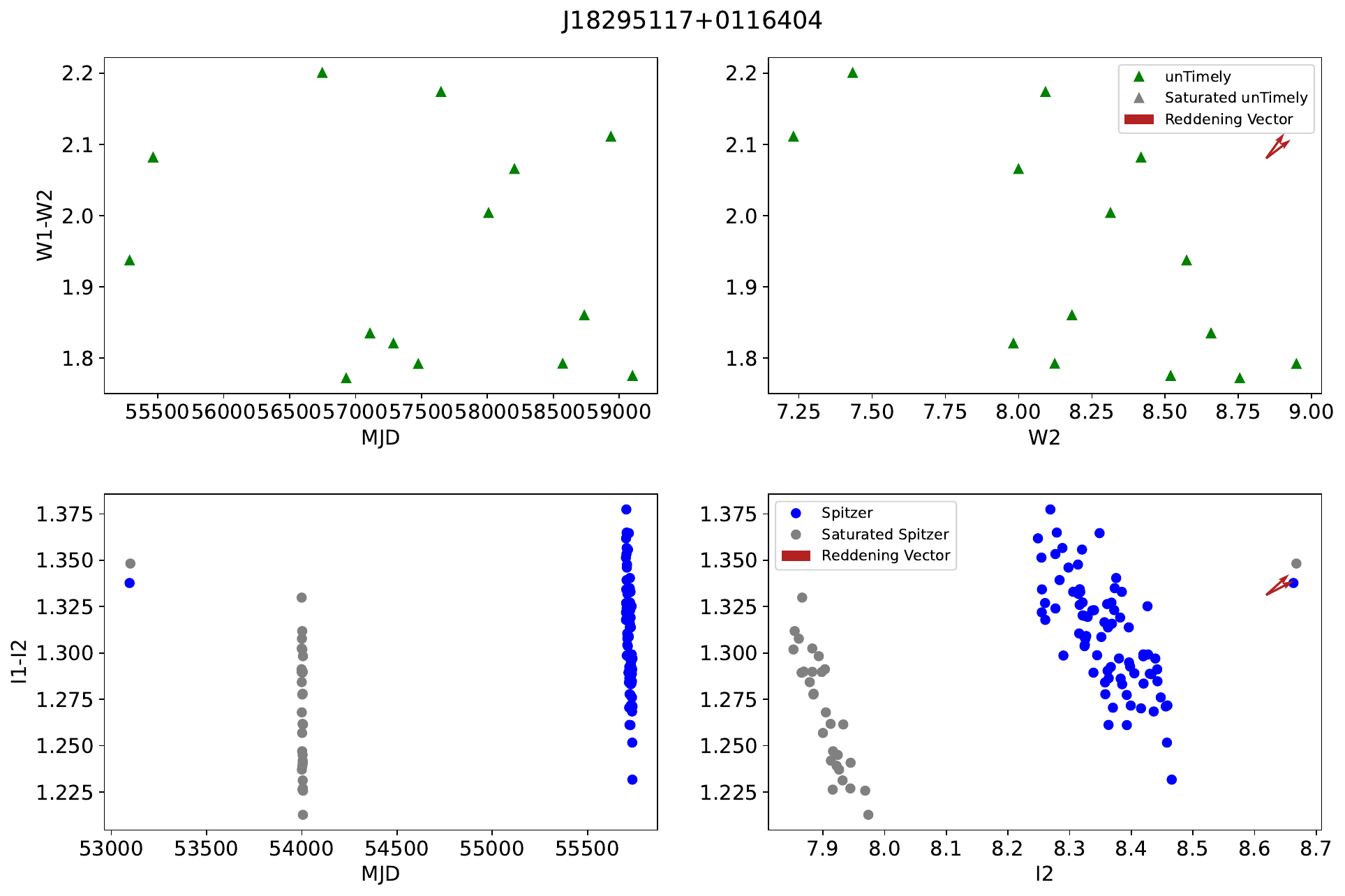}
    \caption{See above.}
\end{figure*}
% J18295117+0116404 X
\addtocounter{figure}{-1}

\begin{figure*}
    \centering
    \includegraphics[width=15cm, height=10.5cm]{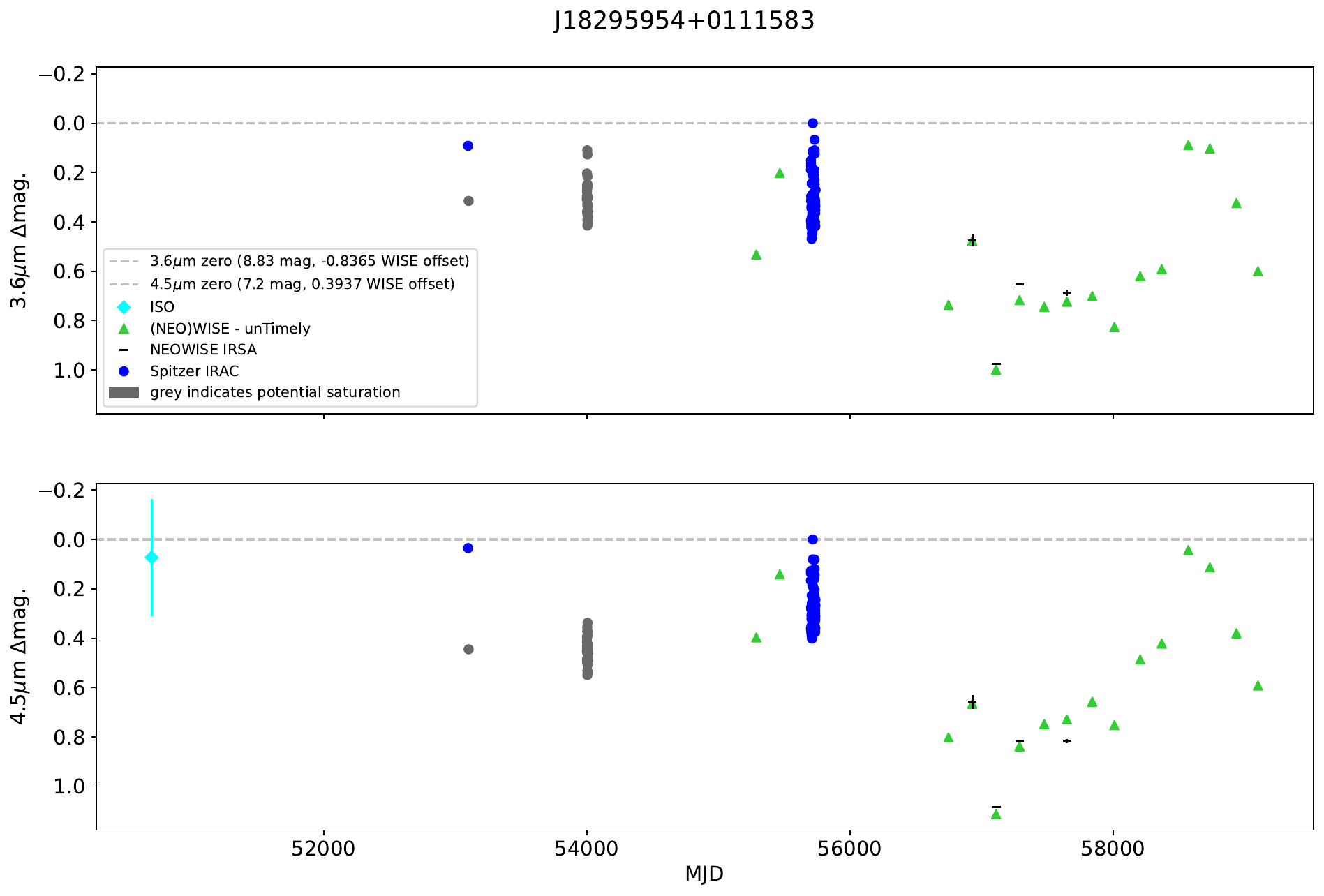}
    \includegraphics[width=15cm, height=10.5cm]{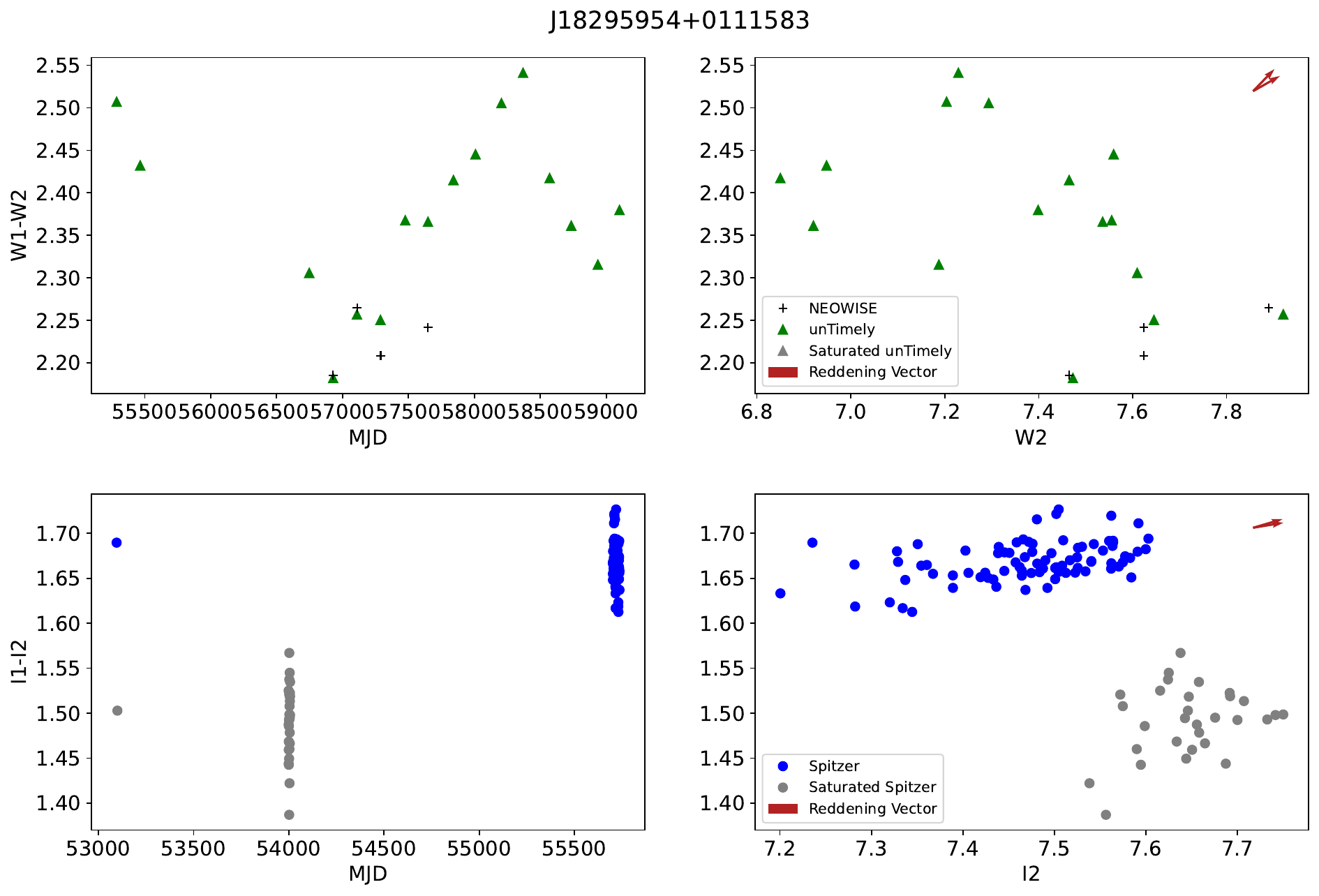}
    \caption{See above.}
\end{figure*}
% J18295954+0111583 X
\addtocounter{figure}{-1}

\begin{figure*}
    \centering
    \includegraphics[width=15cm, height=10.5cm]{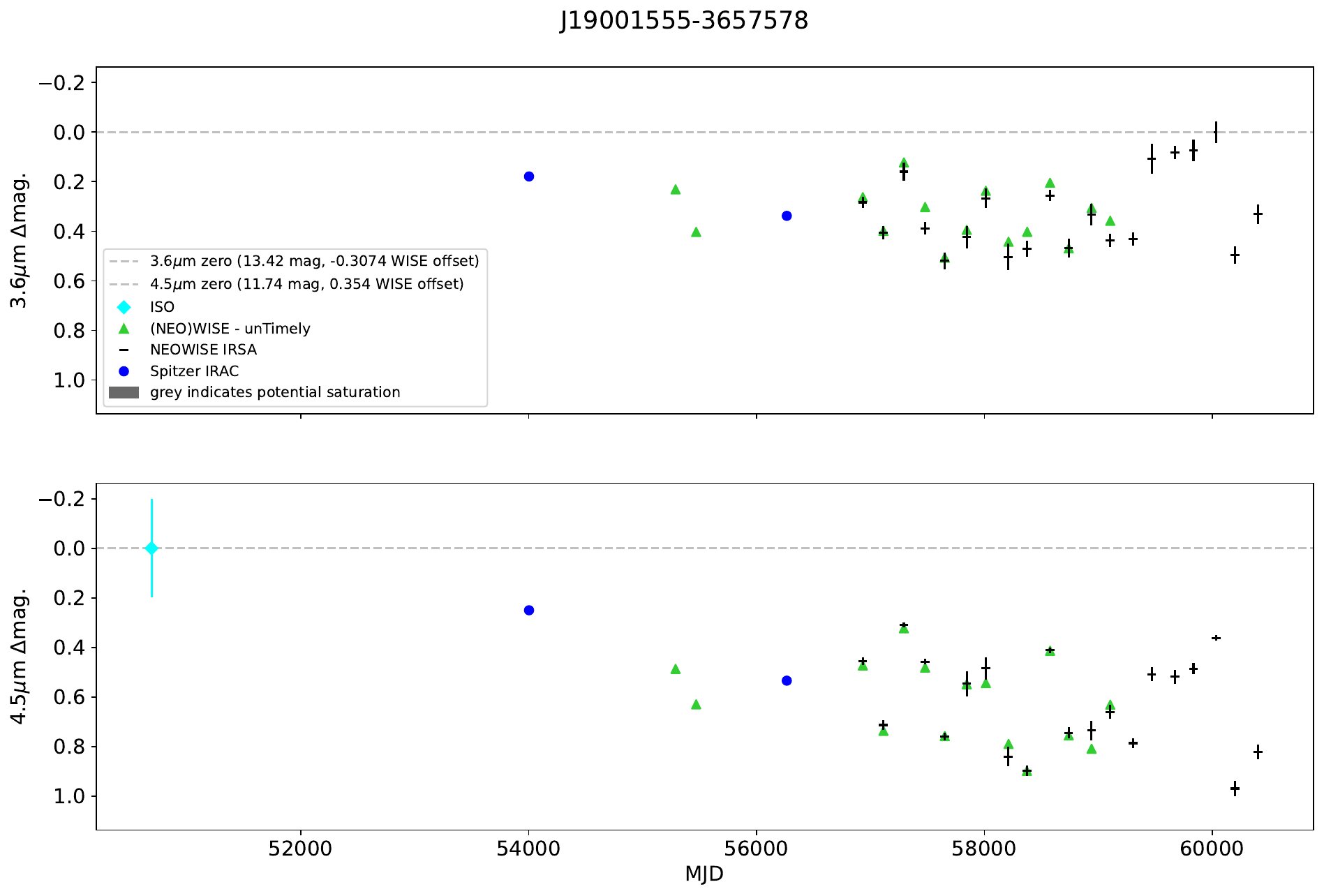}
     \includegraphics[width=15cm, height=10.5cm]{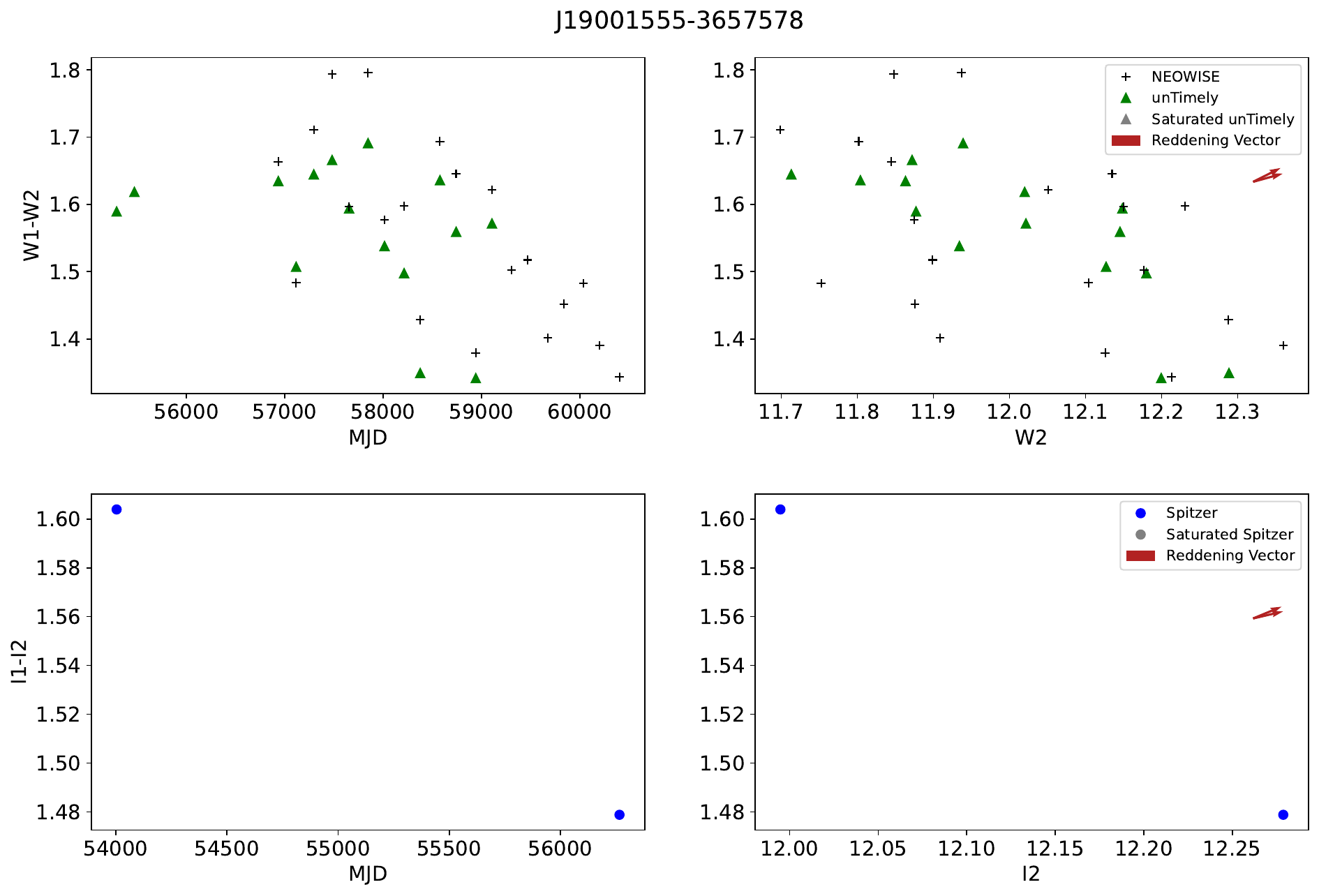}
    \caption{See above.}
    % J19001555-3657578 X
\end{figure*}

\begin{figure*}
    \centering
    \includegraphics[width=15cm, height=10.5cm]{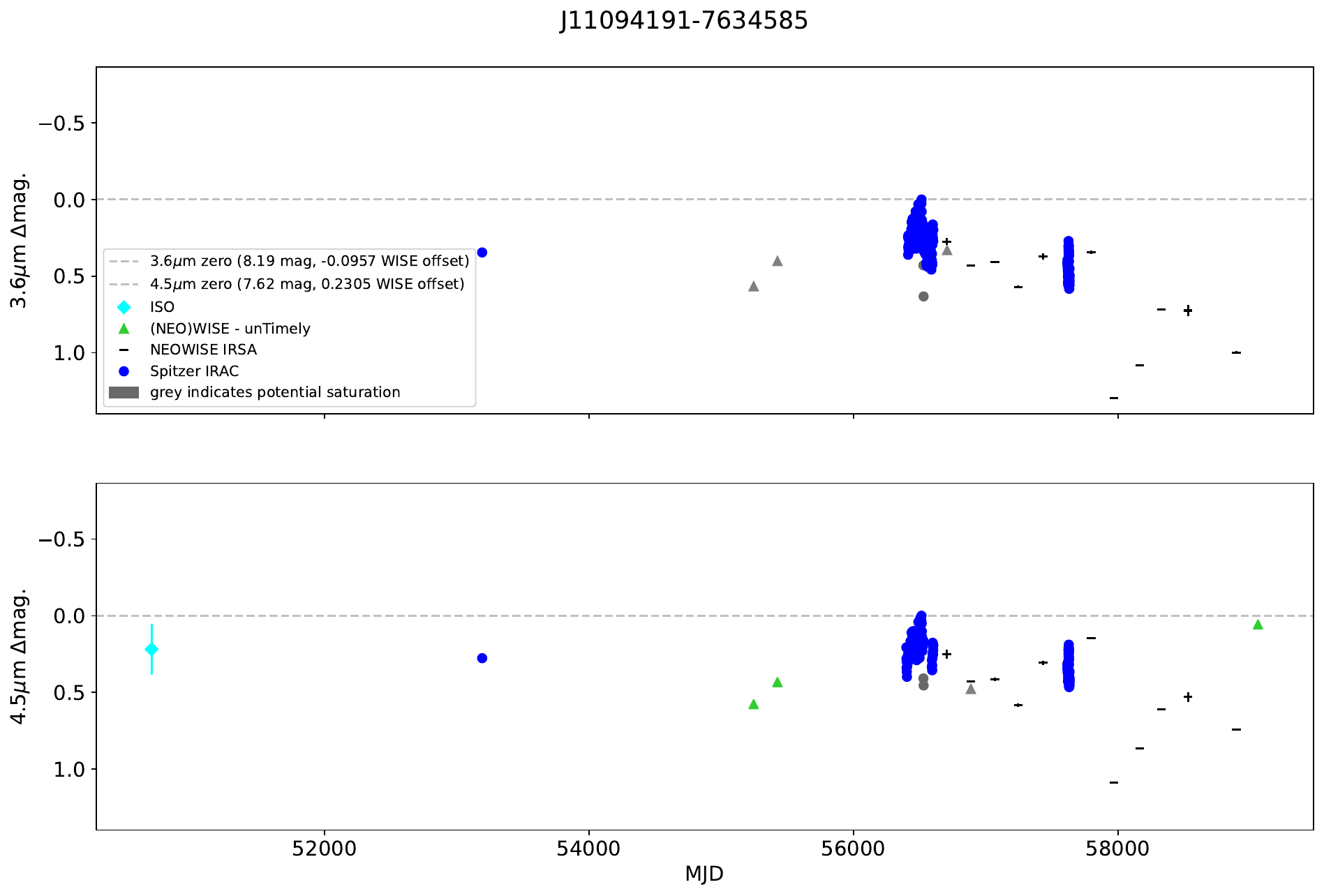}
    \includegraphics[width=15cm, height=10.5cm]{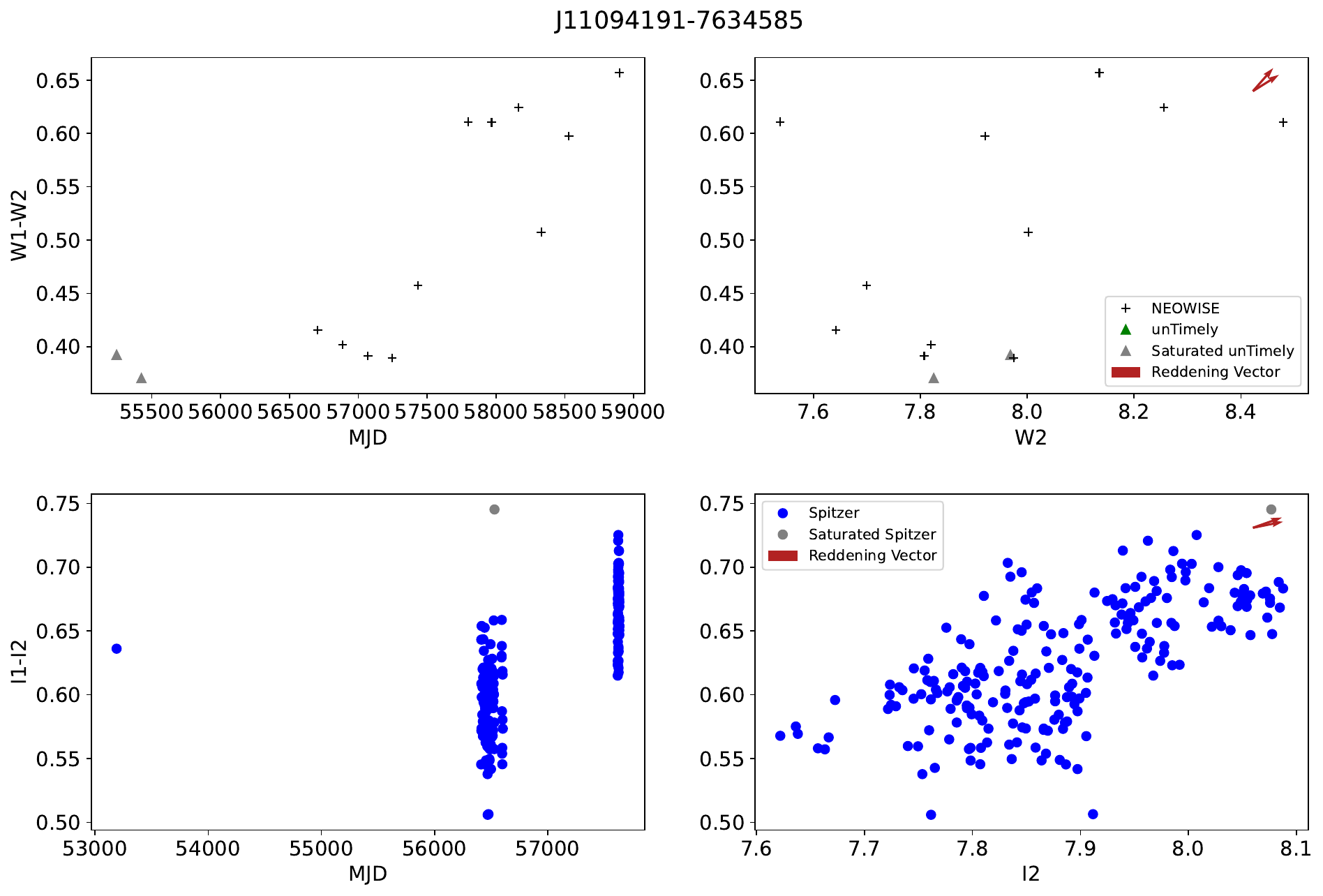}
    \caption{See above.}
\end{figure*}
% J11094191-7634585 X
\addtocounter{figure}{-1}

\begin{figure*}
    \centering
    \includegraphics[width=15cm, height=10.5cm]{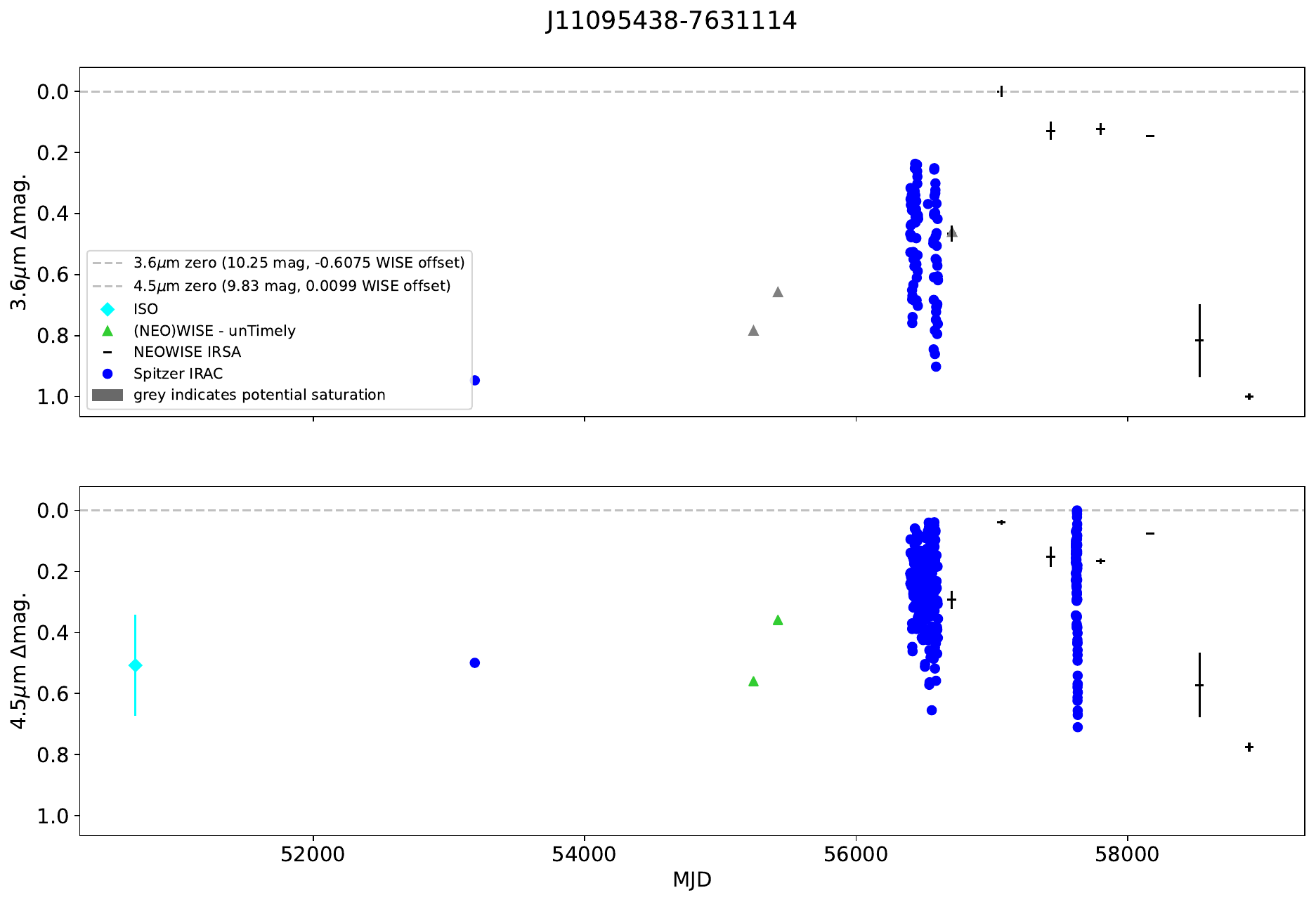}
    \includegraphics[width=15cm, height=10.5cm]{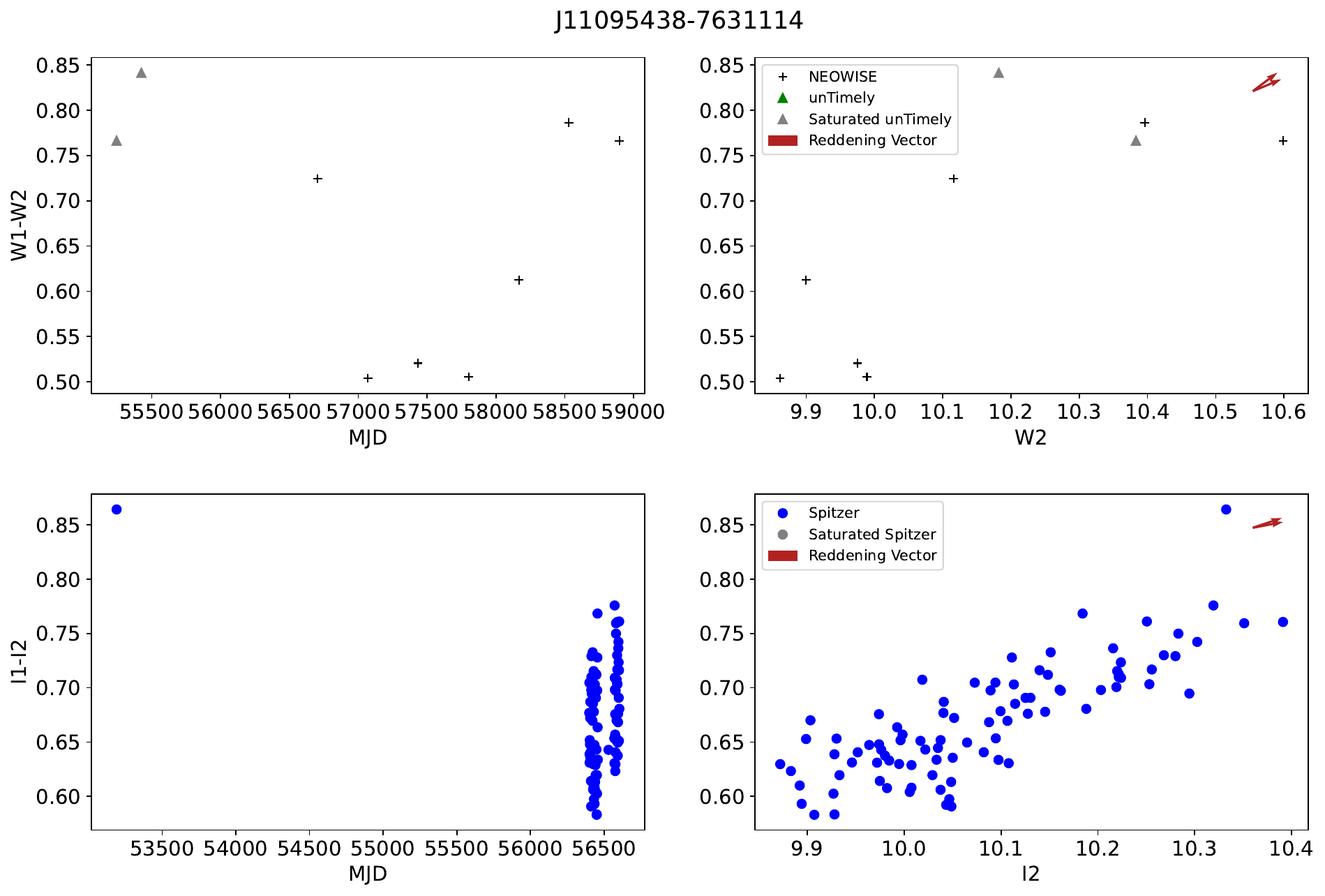}
    \caption{See above.}
\end{figure*}
% J11095438-7631114 X
\addtocounter{figure}{-1}

%\begin{figure*}
%    \centering
%    \includegraphics[width=15cm, height=10.5cm]{bursts/serpens/J18295225+0115476_DLC.pdf}
%    \caption{See above}
%\end{figure*}

%\begin{figure*}
%    \centering
%    \includegraphics[width=15cm, height=10.5cm]{bursts/NGC_1333/J03285841+3122176_DLC.pdf}
%    \caption{See above}
%\end{figure*}

%    \caption{See Above}
    %The light curves, color curves, and CMDs of the 26 fluctuating YSOs in our sample. These include 12 protostars: two in the NGC~1333 cluster (J03284529+3105419, J05460477-0014163), two in the Orion~B cloud (J05460363-0014492, J05471062+0021141), four in the Ophiuchus cloud (J16272180-2429536, J16264421-2434487, J16274162-2446450, J16264048-2427146), two in the Serpens Main cluster( J18295117+0116404, J18295954+0111583), and one in the Chameleon cloud (J11095438-7631114). These also include 11 pre-main sequence stars with disks: six in the NGC~1333 cluster (J03283452+3107055, J03285102+3118184, J03285120+3119549, J03285216+3122453, J03285955+3121467, J03292042+3118343), five in the Orion B cloud (J05405172-0226486, J05413972-0202241, J05415555-0223405, J05455630+0007085, J05460477-0014163, J05464741+0012594), two in the Ophiuchus cloud (J16262755-2441538, J16274709-2445353), one in the Corona Australis cluster (J19001555-3657578) and one in the Chamelon cloud (J11094191-7634585).}
%  XJ0328346+310704 in the NGC 1333 region.}

%\input{iso_cat_rotate_attempt} % rotated deluxetable

%%%%%%%%%%%%%%%%%%%%%%%%%%%%%%%

\end{document}